\documentclass[12pt]{article}
\usepackage{amsmath}
\usepackage{graphicx}
\usepackage{enumerate}
\usepackage{natbib}
\usepackage{url} 
\usepackage[flushleft]{threeparttable}

\DeclareMathOperator*{\argmin}{\arg\min}

\newtheorem{thm}{Theorem}

\newtheorem{lemma}{Lemma}

\newtheorem{assum}{Assumption}

\def\b{\boldsymbol}

\begin{document}

\def\spacingset#1{\renewcommand{\baselinestretch}%
{#1}\small\normalsize} \spacingset{1}

\title{Portfolio Analysis in High Dimensions with TE and Weight Constraints}
\author{\textsc{Mehmet Caner\thanks{
North Carolina State University, Nelson Hall, Department of Economics, NC 27695. Email: mcaner@ncsu.edu. }}
\and \textsc {Qingliang Fan%
\thanks{Department of Economics, The Chinese University of Hong Kong. Email: michaelqfan@gmail.com. 
The authors thank co-editor Jianqing Fan, an associate editor and two anonymous referees for comments that improved the paper.}}}
\maketitle


\begin{abstract}

This paper explores the statistical properties of forming constrained optimal portfolios within a high-dimensional set of assets. We examine portfolios with tracking error  constraints, those with simultaneous tracking error and weight restrictions, and portfolios constrained solely by weight. Tracking error measures portfolio performance against a benchmark (typically an index), while weight constraints determine asset allocation based on regulatory requirements or fund prospectuses. Our approach employs a novel statistical learning technique that integrates factor models with nodewise regression, named the {\it {\bf C}onstrained {\bf R}esidual Nodewise {\bf O}ptimal {\bf W}eight Regressio{\bf n}} (CROWN) method. We demonstrate its estimation consistency in large dimensions, even when assets outnumber the portfolio's time span. Convergence rate results for constrained portfolio weights, risk, and Sharpe Ratio are provided, and simulation and empirical evidence highlight the method's outstanding performance.

\end{abstract}


\newpage
\spacingset{1.9}
\section{Introduction}

A central concern in empirical portfolio management is the formation of optimal large portfolios, where the number of assets may exceed the time span of the portfolio. An equally significant challenge relates to the practical constraints imposed on large portfolios. 

One notable constraint pertains to tracking error (TE from now on), defined as the standard deviation between a portfolio's returns and those of its corresponding benchmark. Mutual fund reports are mandated to disclose performance against a benchmark, typically an index like the S\&P 500 in the U.S., or a sector/theme-specific index such as the healthcare sector index. The significance of  TE extends beyond index funds or Exchange Traded Funds (ETFs), as it serves as a crucial indicator for risk management and performance evaluation. The intricacies of TE restrictions have been investigated in previous studies within the finance literature, see, for example, \cite{roll1992} and \cite{jorion2003}.

Another pivotal aspect of optimal portfolio formation is the imposition of weight constraints. This restriction is typically outlined in a fund's prospectus, and fund companies routinely adhere to these constraints. Weight restrictions entail that fund managers can allocate only a specified percentage to a particular asset class. For instance, a fund might specify a maximum allocation of 20\% to non-U.S. stocks in the portfolio. Sector funds frequently incorporate weight restrictions on specific industries. The analysis of such constraints within the U.S. context is explored by \cite{abcc2004}. Additionally, regulatory bodies have the authority to impose constraints on portfolio formation. In France, for instance, bond and money market funds are generally restricted from holding more than 20\% of stocks. Notably, the consideration of weight constraints is sometimes intertwined with TE. These type of constraints are used since they  reduce risk by diversification and shrinking the universe of plausible portfolios.

In this paper, our primary focus is on addressing the question: ``{\it Can we construct a portfolio that is theoretically guaranteed to be or close to the optimal one given the aforementioned constraints?}'' The existence of optimal portfolios under constraints has been previously examined (see \cite{bbmp2011} and \cite{ces2022}). However, a notable gap persists in the literature regarding a high-dimensional statistical analysis of these constrained portfolios. Drawing on the factor model literature from \cite{fan2011} and integrating the residual nodewise regression technique as outlined in \cite{caner2022}, we conduct a comprehensive examination of constrained portfolios in a high-dimensional setting. We name our approach the  {{\bf C}onstrained {\bf R}esidual Nodewise {\bf O}ptimal {\bf W}eight Regressio{\bf n}}  (CROWN) method and analyze the estimated portfolio's out-of-sample risk and Sharpe Ratio under TE and weight constraints. We establish rates of convergence for our estimators of the optimal portfolio weights, risks, and Sharpe Ratio, demonstrating their consistency even when the number of assets in the portfolio exceeds the portfolio's time span in the constrained framework. Our framework accommodates the growth of factors, assets, and time span, with factors growing at a slower rate compared to time span and the number of assets.

There is  extensive literature on portfolio analysis that offers in-depth discussions on related measures such as risks and Sharpe Ratios. For instance,  \cite{fan2011}, \cite{fan2013} propose the Principal Orthogonal Complement Thresholding Estimator (POET), which examines unconstrained large portfolios using a principal components-based estimator. Additionally,  \cite{dlz2020} further  demonstrate the properties of minimum variance portfolio based on POET. \cite{lw2017} propose nonlinear shrinkage and single-factor shrinkage estimators to estimate the covariance matrix of returns for use in financial metrics within an unconstrained portfolio. These papers rely on the estimation of the covariance matrix of assets.  \cite{cai2020} explore the estimation of the precision matrix via CLIME (with $l_1$ constraint on the elements of the precision matrix) and investigate the potential gains in the performance of the minimum variance portfolio using high-frequency data. 
There is also the MaXimum Sharpe Ratio Estimated \& sparse Regression (MAXSER) proposed by \cite{ao2019}, which enables the consistent estimation of an unconstrained mean-variance efficient large portfolio.   \cite{kan2022} study distributional properties of the in-sample and out-sample Sharpe Ratios of estimated portfolios.  

\cite{caner2019} offer a nodewise regression estimate of the precision matrix of asset returns, demonstrating consistency results for portfolio weights and risk. More recently, \cite{caner2022} merged the factor model literature with residual-based nodewise regression to develop Sharpe Ratio estimators, showcasing their consistency in large portfolios.  
Note that our paper is not a simple extension of \cite{caner2022}. \cite{caner2022} is mainly about the precision matrix of returns estimation and analysis of Sharpe Ratios without weight or TE constraints. Also most of the analysis of Sharpe Ratio is in-sample as in Theorem 4 of  \cite{caner2022}. 
Our paper has three major differences  compared to \cite{caner2022}.
First, our paper analyzes $l_1$ estimation error of portfolio weights and variance of portfolios with constraints in a high dimensional context. Second, we are also interested in   out-sample Sharpe Ratio analysis with constraints, which is absent in \cite{caner2022}.
The third   difference between the two papers is the proof technique  difference. \cite{caner2022} analyzes the maximum out of sample Sharpe Ratio in their section 4.3 (without portfolio weights added to one) and can come up with a consistency result only the number of assets, $p$ is smaller than the time span of the portfolio $T$. Remark 2-Theorem 8 of \cite{caner2022} explains this issue, and the main reason for this limited result is the proof of their Theorem 8 crucially depends on the maximum eigenvalue of the covariance matrix of asset returns which diverge at $p$ rate. Here in this paper, our results in the unconstrained case (Lemma A.4), as well as constrained cases (Theorems 1-3, Lemma A.3) prove consistency when $p>T$, and our proof technique is different without using a maximal eigenvalue, but instead benefiting from a sup-norm of the covariance matrix of asset returns which diverge at a much slower rate than $p$.  Another related paper is \cite{bbmp2011}, where they find a formula for portfolio weights when there are TE and weight constraints.   Our paper focuses on high dimensions and estimation of metrics of large portfolios.
Compared to their paper, we also develop a data-based 
way of differentiating between binding and non-binding constraints and suggest how it can be done and estimated in our Theorem 3 here. That is a new result and a new idea.

Another strand of literature studies weight-constrained portfolios  as seen in works such as \cite{jag2003}, \cite{dgnu2009},   \cite{fan2012}. \cite{du22} considered a high-dimensional portfolio with cardinality constraints.  In contrast to other statistical-econometrics literature, our paper is the first one to  focus on TE and inequality constraints in the context of large portfolios. Our simulation and empirical results demonstrate the strong performance of the proposed method.

Section 2 introduces the model, while Section 3 considers the  TE constraints. Section 4 presents a joint analysis of TE and equality weights. In Section 5, a novel estimator for weights in a large portfolio, along with other metrics, is proposed for joint TE and inequality constraints.  Section 6 presents a subset of all simulations;  the rest is in  Appendix.
Section 7 offers empirical studies.    Appendix includes all proofs and addresses the scenario involving only weight constraints,  a comparison with an unconstrained portfolio, and  simulations-empirics comparing several methods.

\section{Model}\label{sec1}

In this section, we start with our factor model and then discuss our estimator for a precision matrix of asset returns. After that, we analyze assumptions. 
We start with the following model, for the $j$ th asset ($j=1,\cdots,p$) excess returns at time $t$ ($t=1,\cdots, T$)
\[ y_{j,t}= {\b b_j'} {\b f_t} + u_{j,t},\]
where ${\b b_j}$ is a $K \times 1$ vector of factor loadings, and ${\b f_t}$ is  $K \times 1$ vector of common factors to all asset returns, and $u_{j,t}$ is the idiosyncratic error term for the $j$ th asset at time $t$. All the factors are observed, and this model is used by \cite{fan2011}. All the vectors and matrices in the paper are in bold.
Define the covariance matrix of errors ${\b u_t}=(u_{1,t}, \cdots, u_{j,t}, \cdots, u_{p,t})': p \times 1$ vector, as ${\b \Sigma_u}:= E [{\b u_t} {\b  u_t}']$. 
We assume $\{( {\b f_t}, {\b u_t})\}_{t=1}^T$ to be a strictly stationary, ergodic, and strong mixing sequence of random vectors.
Also, let ${\cal F}_{-\infty}^0, {\cal F}_{T}^{\infty}$ be the $\b\Sigma$- algebras generated by $\{( {\b f_t}, {\b u_t})\}$, for $-\infty < t \le 0$, and $T \le t <\infty$, respectively. Denote the strong mixing coefficient as $\alpha(T):= \textnormal{sup}_{ {\cal A} \in {\cal F}_{-\infty}^0, {\cal B} \in {\cal F}_{T}^{\infty}}
| P ({\cal A}) P ({\cal B}) - P ( {\cal A} \cap {\cal B} )|.$  
Express the assets returns matrix as 
\[ {\b y}= {\b B} {\b  X} + {\b U},\]
with ${\b y}: p \times T$ matrix, and ${\b B} : p \times K$ factor loadings matrix, and  ${\b X}=(f_1, \cdots, f_T)$ as $K \times T$ matrix of factors, with ${\b U}$ as $p \times T$ error matrix.

We will assume the sparsity of the precision matrix of errors: ${\b \Omega}:= {\b \Sigma_u^{-1}}$, and let ${\b \Omega_j'}$ be the $ j$ th row vector of ${\b \Omega}$.
In ${\b \Omega_j}'$ the indices of non-zero elements of that $j$ th row is denoted as $S_j$, for $l=1,\cdots,p$,
$ S_j:= \{ l: \Omega_{j,l} \neq 0\},$
where $\Omega_{j,l}$ be the $j$ th row and $l$ th column element of ${\b \Omega}$. Let $s_j := | S_j|$ be the cardinality of nonzero elements in row ${\b \Omega_j'}$. $s_j$ will be nondecreasing in $T$. Denote the maximum number of nonzero elements across each row of ${\b \Omega}$ as $\bar{s}$, $\bar{s}:= \max_{1 \le j \le p} s_j$. We use the feasible residual-based nodewise regression to estimate ${\b \Omega}$ as shown in \cite{caner2022}. Denote this estimator as $\hat{\b \Omega}$. A detailed description is in Appendix B.2. $\hat{\b \Omega}$ is not symmetric. The covariance matrix of returns ${\b \Sigma_y}$ can be decomposed to
\[ {\b \Sigma_y}= {\b B} (\b \Sigma_f) {\b B'} + {\b \Sigma_u},\]
with $\b \Sigma_f$ as the covariance matrix of factors ($K \times K$). By Sherman-Morrison-Woodbury formula, with ${\bf \Theta}:={\bf \Sigma_y^{-1}}$
\[ \b \Theta:= \b \Omega - \b \Omega \b B [ \b \Sigma_f^{-1} + \b B' \b \Omega \b B]^{-1} \b B' \b \Omega,\]
and its estimate based on the feasible residual based nodewise estimate of errors is:
\begin{equation}
    \hat{\b \Theta}: = \hat{\b \Omega} - \hat{\b \Omega} \hat{\b B} [ \widehat{\b \Sigma_f}^{-1} + \hat{\b B}' \hat{\b \Omega}_{sym} \hat{\b B}]^{-1} \hat{\b B}' \hat{\b \Omega}, \label{thetahat}
\end{equation}
with 
$\hat{\b \Omega}_{sym}:= \frac{1}{2} [\hat{\b \Omega} + \hat{\b \Omega}']$, and the estimator of the covariance matrix of factors is defined by $\widehat{\b \Sigma_f}:= T^{-1} \b X \b X' - T^{-2} \b X \b 1_{T} \b 1_{T}' \b X'$, with $\b 1_{T}$ representing a $T \times 1$ vector of ones.  Since we need symmetricity of the square bracketed inverse term in (\ref{thetahat}) we use $\hat{\b \Omega}_{sym}$ there, and  $\hat{\b B}'= (\b X \b X')^{-1} \b X \b y: K \times p$ matrix. In the other parts of the formula, for (\ref{thetahat}) symmetricity of $\hat{\b \Theta}$ is not needed; see  \cite{caner2022} for this point.

Before assumptions, we allow both $p$ (number of assets) and $K$ (number of factors) to grow when $T$ increases. To save notation, we do not subscript $p, K$ with $T$. We also restrict $\min (p, T) > K$.

\subsection{Assumptions}

In this subsection, we provide the assumptions. But before that we need the following notation. We use the following notation in the paper regarding vector and matrix norms. Let $\b v$ be a generic $n \times 1$ vector and $\b H$ to be a generic $n \times n$ matrix. Then 
$\| \b v \|_1, \| \b v \|_2, \| \b v \|_{\infty}$ are the $l_1$, Euclidean, and sup norms of the vector $\b v$ respectively. Next, $\| \b H \|_{l_1}, \|\b H\|_{l_{\infty}}$ are the maximum column sum matrix norm and maximum row sum matrix norm, respectively. This means $\| \b H \|_{l_1}:= \max_{1 \le j \le n } \sum_{i=1}^n |H_{i,j}|$, and 
$\| \b H \|_{l_{\infty}}:= \max_{1 \le i \le n } \sum_{j=1}^n | H_{i,j}|$, where $H_{i,j}$ represents $(i,j)$ th element of matrix $H$. Let $\| \b H \|_{l_2}$ be the spectral norm of matrix H, and $\| \b H \|_{\infty}:= \max_{1 \le i \le n } \max_{1 \le j \le n} |H_{i,j}|$. First, $ \b u_{-j,t}$ is the $(p-1)\times 1$ vector of errors in $t$ th time period, except the $j$ th term in $ \b u_t$. Then define $\eta_{j,t}:= u_{j,t}-  \b u_{-j,t}' \b \gamma_j$. Also, define $Eigmin (\b A)$ to be the minimum eigenvalue of the matrix ${\b A}$.

\begin{assum}\label{as1}
(i). $\{ \b u_t \}_{t=1}^T, \{  \b f_t \}_{t=1}^T$ are sequences of (strictly) stationary and ergodic random vectors. Furthermore, $\{ \b u_t \}_{t=1}^T, \{ \b f_t \}_{t=1}^T$ are independent. $ \b u_t$ is a ($p \times 1$) zero mean random vector with covariance matrix $\b \Sigma_u$ ($p \times p$). $\textnormal{Eigmin} (\b \Sigma_u) \ge c > 0$, with $c$ a positive constant, and $\max_{1 \le j \le p} E\left[u_{j,t}^2\right] \le C < \infty$. 
 (ii). For the strong mixing vector of random variables $ \b f_t, \b u_t$:
$\alpha (t) \le \exp (-C t^{r_0})$, for a positive constant $r_0>0$.
\end{assum}

\begin{assum}\label{as2}
There exists positive constants $r_1$, $r_2$, $r_3 >0$ and another set of positive constants $c_1$, $c_2$, $c_3$, $s_1$, $s_2$, $s_3 >0$, and for $t=1,\cdots, T$, and
$j=1,\cdots, p$, with $k=1,\cdots, K$, (i).  
$
P\left[|u_{j,t} | > s_1\right] \le \exp[-(s_1/c_1)^{r_1}].
$ (ii). $
P\left[| \eta_{j,t} | > s_2 \right]\le\exp [-(s_2/c_2)^{r_2}].
$ (iii). $ 
P\left[|f_{k,t} | > s_3\right]\le \exp [-(s_3/c_3)^{r_3}].
$
(iv). There exists $0< \phi_1 < 1$ such that  $\phi_1^{-1} = 3 r_1^{-1} + r_0^{-1}$, and we also assume
$3 r_2^{-1} + r_0^{-1} > 1$, and $3 r_3^{-1} + r_0^{-1} >1$.
\end{assum}

Define $\phi_2^{-1}:= 1.5 r_1^{-1} + 1.5 r_2^{-1} + r_0^{-1}$, and $\phi_3^{-1}:= 1.5 r_1^{-1} + 1.5 r_3^{-1} + r_0^{-1}$, let
$\phi_{\min}:=\min (\phi_1, \phi_2, \phi_3)$.

\begin{assum}\label{as3}
(i). $[\ln (p)]^{(2/\phi_{min}) -1 } = o(T)$, and (ii). $K^2 = o(T)$, (iii). $K = o(p)$.
\end{assum}

\begin{assum}\label{as4}
(i). $ Eigmin[\b \Sigma_f] \ge c > 0$, with $\b \Sigma_f$ being the covariance matrix of the factors $ \b f_t$, $t=1,\cdots, T$.
(ii). $ \max_{1 \le k \le K } E \left[f_{kt}^2\right] \le C < \infty$, $\min_{1 \le k \le k} E \left[f_{kt}^2\right] \ge c > 0$.
(iii). $\max_{1 \le j \le p} E \left[\eta_{j,t}^2\right] \le C < \infty$.
\end{assum}

\begin{assum}\label{as5}

(i). $ \max_{1 \le j \le p} \max_{1 \le k \le K} | b_{jk} | \le C < \infty.$ (ii). $ \| p^{-1} \b B' \b B -\b  \Phi \|_{l_2} = o(1),$
for some $K \times K$ symmetric positive definite matrix $\b \Phi$ such that $Eigmin (\b \Phi)$ is bounded away from zero.
\end{assum}

\begin{assum}\label{as6}

 $Eigmax (\b \Sigma_u) \le C r_T$ with $C >0$ a positive constant, and $r_T \to \infty$ as $T\to \infty$ and $r_T /p \to 0$ and $r_T$ is a positive sequence.

\end{assum}

Define the rate via (16) and (22) of \cite{caner2022}
\begin{equation}
l_T:= r_T^2 K^{5/2} \max ( K^2 \bar{s}^{3/2} \frac{ln p}{T} , \bar{s} \sqrt{\frac{ln p}{T}}, \bar{s}^{1/2} K^{1/2} \sqrt{\frac{ \max [ lnp,ln T]}{T}}).\label{r-ln}
\end{equation}

We define the expected return,  the benchmark, and the portfolio of interest. This helps us understand the next assumption.
Define $\b \mu:= (\mu_1, \cdots, \mu_j, \cdots, \mu_p)'$ as the expected return vector and $\b m$ being the benchmark, and $\b w_d^*:= \b w^* - \b m $, with $\b w^*$ being the portfolio of interest.

\begin{assum}\label{as7}

(i). $\bar{s} l_T  K^{4} = o(1)$. (ii).  
$\frac{\b 1_p' \b \Theta \b  1_p }{p} \ge c >0$, with $c  $ being  a positive constant. (iii). $ \| \b m \|_1 = O ( \| \b w_d^* \|_1)$, $\b w^{*'} \b \Sigma_y \b w^* \ge c > 0$, $| \b w^{*'} \b \mu | \ge c >0$. 

\end{assum}

Assumptions \ref{as1}-\ref{as4} are used in \cite{caner2022} to provide the consistency of the precision matrix estimator for errors.
Also, we get $0 < \phi_2 < 1, 0 < \phi_3 < 1$ given Assumption \ref{as2}(iv). Furthermore, by Assumption 3, $\sqrt{\frac{\ln(p)}{T}} = o(1)$. Note that stationary GARCH models with finite second moments and continuous error distributions, as well as causal ARMA processes with continuous error distributions, and a certain class of stationary Markov chains satisfy our Assumptions \ref{as1}-\ref{as2} and are  discussed in p.61 of \cite{chang2019}. Assumption \ref{as5} is used in \cite{fan2011}. It puts some structure on the factor loadings, which is also used in \cite{caner2022}.

Next, we provide assumptions that are used to obtain consistency of the estimate of the precision matrix of returns in \cite{caner2022}. 
 Assumption \ref{as6} allows the maximal eigenvalue of $\b \Sigma_u$ to grow with T, and is also used in \cite{gagl2016}, and  \cite{caner2022}. 
Assumption \ref{as7}(i) is a tradeoff between sparsity, number of factors, and eigenvalue conditions. This is more restrictive than sparsity Assumption 8 in \cite{caner2022}. 

To get more intuition, we can do the following thought experiment: if we suspect that the financial markets are very noisy, $r_T$ being large, and if we suspect $\bar{s}$ to be large, it will not be a very good idea to have a growing number of factors in our model since the consistency of our estimators of variance and Sharpe Ratio cannot  be  attained.
 Large negative effects of a growing number of factors can be seen in estimation errors in Theorems 1-3 below. We should note that Assumption \ref{as7}(i) affects the consistency of variance and Sharpe Ratio of our estimators, but consistent weight estimation is slightly easier, as seen in Theorem 1(i) versus Theorem 1(ii)-(iii). In other words, if Assumption \ref{as7}(i) is violated, then the consistency of variance and Sharpe Ratio estimation is not plausible, but consistent weight estimation is plausible even when Assumption \ref{as7}(i) is violated.
 Theorem 1(i) estimation error rate on weights are $\bar{s} l_T  K^{3/2}$ versus Assumption \ref{as7}(i) rate of $\bar{s} l_T  K^4$. Another issue is the denseness of the precision matrix; what if $\bar{s}=p$? Then  we can have consistency only in case of $p<T$. To delve more into Assumption \ref{as7}(i), $\bar{s} l_T$ rate is coming from precision matrix estimation as shown in Theorem 2 of \cite{caner2022}, and $K^{3/2}$ is multiplied with precision matrix estimation rate when we have TE constraint in our portfolio weight estimation as in Theorem 1(i) below, to get the rate of convergence as $r_{w1}
 := \bar{s} l_T K^{3/2}$. Then, to estimate variance and Sharpe Ratio, there is additional multiplication of $K^{5/2}$ as seen in Theorem 1(ii)(iii) below, which provides the rate in Assumption \ref{as7}(i).

Then Assumption \ref{as7}(ii) imposes that Global Minimum Variance (unconstrained-scaled) portfolio variance is finite. Assumption \ref{as7}(ii) is used in \cite{caner2022}. Assumption \ref{as7}(iii) assigns the benchmark portfolio to be growing at most as the optimal  TE portfolio in $l_1$ norm in Section 3 below, and this is just needed to show that benchmark and optimal portfolios have similar behavior in $l_1$ norm. The rest of Assumption \ref{as7} constrains the optimal variance to be positive and absolute portfolio return to be bounded away from zero. We can have the following example to show Assumption \ref{as7}(i) is plausible.  Set $\bar{s}=\bar{s} (T), K= K(T)$ which are slowly varying functions of $T$. Set $r_T = O (T^{1/4- d_1})$, with $1/4> d_1>0$ a positive constant  and $p= aT, 0< a \le C < \infty$, 
 then, with first assuming that $\bar{s}(T)$, grows at the same rate as $K(T)$ or less, which implies the third term is the rate in (\ref{r-ln}), 
$
 l_T= O( T^{1/2 - 2 d_1}K(T)^3 \bar{s}(T)^{1/2} \sqrt{\frac{ln T}{T}}),
 $
then   Assumption 7(i) is satisfied
\[ \bar{s} (T)^{3/2} K(T)^{7} \frac{1}{T^{2 d_1}} \sqrt{ln T} \to 0.\]

If, in the previous example,  $\bar{s}(T)$, grows faster than $K(T)$, but still a slowly varying function in $T$, this implies the second term is the rate in (\ref{r-ln})
$
 l_T= O( T^{1/2- 2 d_1} K(T)^{5/2} \bar{s}(T) \sqrt{\frac{ln T}{T}})
 $,
then 
\[ \bar{s} (T)^{2} K(T)^{13/2} \frac{1}{T^{2 d_1}} \sqrt{ln T} \to 0.\]

In  the proof of Theorem 1(i), we use in  Appendix (A.20) an increasing  Sharpe Ratio with the number of assets in  maximum out-sample- Sharpe Ratio  portfolio, 
$\b \mu' \b \Theta \b \mu$. However, we also consider  two different bounded Sharpe Ratios in Appendix B.
 In all these two cases in Appendix B, we use a bounded Sharpe Ratio assumption  for the maximum Sharpe Ratio portfolio: $0 < c \le \b \mu' \b \Theta \b \mu \le C < \infty$. We discuss the differences after each theorem in Remarks.

\section{TE Constraint}\label{sec2}

The optimization problem is  to maximize the expected return with a TE constraint: 
\[ \max_{\b w \in R^p} \b \mu' \b w  \quad \mbox{subject to} \quad (\b w - \b m)' \b \Sigma_y (\b w - \b m) \le \text{TE}^2, \quad \b 1_p'  \b w =1,\]
with $\b m: p \times 1$ benchmark portfolio of weights, $\b w$ is the portfolio of weights that is tracking the benchmark portfolio $\b m$, $\b \mu: p \times 1$ vector of expected return on assets, and TE is the tracking error. Alternatively, we can write the optimization as 
\[ \b w_d^*:=\max_{\b w_d \in R^p} [ \b \mu' \b w_d - \frac{\Xi}{2} \b w_d' \b \Sigma_y \b w_d], \quad \mbox{subject to}\quad \b 1_p' \b w_d = 0,\]
with $\b w_d = \b w - \b m$, $\infty> \Xi>0$, where $\Xi$ can be roughly seen as a risk aversion parameter. The solution to above problem is, $\b \Theta:=\b \Sigma_y^{-1}$, as in (1) of \cite{bbmp2011}
\begin{equation}
\b w_d^*= \kappa \left[ \frac{\b \Theta \b \mu}{\b 1_p' \b \Theta \b \mu} - \frac{\b  \Theta \b 1_p}{\b 1_p' \b \Theta \b 1_p}
\right],\label{eqwd}
\end{equation}
where $\frac{\b \Theta \b \mu}{\b 1_p' \b \Theta \b \mu}$ is the portfolio that maximizes Sharpe Ratio, and $\frac{\b  \Theta \b 1_p}{\b 1_p' \b \Theta \b 1_p}$ is the standard global minimum variance portfolio. Let $c, C$ be positive constants. Then, $\kappa$ represents the risk tolerance parameter, which is described in equation (5)  of \cite{bbmp2011}, with a range of $0<  c \le |\kappa| \le C< \infty$. For risk-averse investors $0 <  c \le \kappa  \le C < \infty$.
 $\kappa, \b m$ are given by investors and not estimated. 
The relation between TE, $\kappa, \Xi$ is given in Appendix A.3.
The aim is to estimate optimal weights $\b w_d^*$, and $\b w^*:= \b w_d^* + \b m$. In that respect, we provide the following estimate
$ \hat{\b w}_d = \kappa \left[  \frac{\hat{\b \Theta}' \hat{\b \mu}}{\b 1_p' \hat{\b \Theta}' \hat{\b \mu}} - \frac{ \hat{\b \Theta}' \b 1_p}{\b 1_p' \hat{\b \Theta}' \b 1_p}
\right],$
with $\hat{\b \mu}= \frac{1}{T} \sum_{t=1}^T \b y_t$, with $\b y_t,$  $t=1,\cdots, T$,  being the columns of $\b y$ matrix. Note that  since 
\begin{equation}
\hat{\b w}= \hat{\b w}_d + \b m,\label{eqwh}
\end{equation}
 we have 
$ \| \hat{\b w}_d - \b w_d^* \|_1 = \| \hat{\b w} - \b w^* \|_1.$
 We provide the following Theorem \ref{t1}, which shows that portfolio weights can be consistently estimated in a large portfolio with TE constraint.

We also consider two other metrics: out-of-sample variance of a portfolio and out-of-sample Sharpe Ratio corresponding to  $\hat{\b w}$.
Specifically, the out-of-sample variance estimate of the constrained portfolio is
\begin{equation}
\hat{\b w}' \b \Sigma_y \hat{\b w},\label{var}
\end{equation}
and the estimate of the out-of-sample Sharpe Ratio of the constrained portfolio is
\begin{equation}
\widehat{SR}= \frac{\hat{\b w}' \b \mu}{\sqrt{\hat{\b w}' \b \Sigma_y \hat{\b w}}}.\label{SR}
\end{equation}
The Sharpe Ratio is $SR= \b w^{*'} \b \mu/\sqrt{\b w^{*'} \b \Sigma_y \b w^{*}}$. Define the rate $r_{w1}:= \bar{s}l_T K^{3/2}$. 
We provide consistent estimation of portfolio weights and out-of-sample variance of the constrained portfolio.

\begin{thm}\label{t1}
Under Assumptions \ref{as1}-\ref{as7}, with the following condition $|\b 1_p' \b \Theta \b \mu|/p \ge c > 0$, with $c$ being a positive constant, and $0 < c \le |\kappa | \le C < \infty$

(i). 
\[ \| \hat{\b w} - \b w^* \|_1= \| \hat{\b w}_d - \b w_d^* \|_1 = O_p (r_{w1})= O_p ( \bar{s} l_T  K^{3/2}) = o_p (1).\]

(ii). \[ \left| \frac{\hat{\b w}^{'} \b \Sigma_y \hat{\b w}}{\b w^{*'} \b \Sigma_y \b w^*} - 1 \right| = O_p (r_{w1} K^{5/2}) = O_p ( \bar{s} l_T K^{4}) = o_p (1).\]

(iii). \[ \left| \left[ \frac{\widehat{SR}}{SR}\right]^2 - 1 \right| = O_p ( r_{w1}  K^{5/2}) = O_p ( \bar{s} l_T  K^{4}) = o_p (1).\]

\end{thm}

\noindent {\bf Remarks}. 1. These are all  new results in the literature and allow $p>T$ due to $l_T$ definition in (\ref{r-ln}), when both $p,T$ diverge. An illustrative example of a representative rate is given at the end of Section 2.

2. $|\b 1_p' \b \Theta \b \mu/p| = p^{-1} |  \sum_{j=1}^p \sum_{k=1}^p \Theta_{j,k} \mu_k | \ge c >0$ is a very mild condition that prevents expected scaled-mean to be a very small number near zero. See also Remark 3 of Theorem 7 in \cite{caner2022} for this point.

3. If the precision matrix of errors is non-sparse, we can have $\bar{s}=p$. Then in this case $p<T$,  we need $p = o( T^{1/4})$, by using $l_T$ definition in (\ref{r-ln}) for Theorem \ref{t1}.

4. We can also have an estimate for TE as $\widehat{TE}_d:= \sqrt{\hat{\b w}_d' \b \Sigma_y \hat{\b w}_d}$, as an out-of-sample estimate. It can easily be shown that 
$|(\frac{\widehat{TE}_d}{TE})^2-1|$ will converge in probability to zero by following the proof of Theorem \ref{t1}(ii), and the rate will be the same as in Theorem \ref{t1}(ii).
This estimator is consistent.

5. Note that the variance and Sharpe Ratio  have a  slower rate of convergence compared to weight formation. This is clear by comparing Theorem \ref{t1}(i) with Theorem \ref{t1}(ii)-(iii).

6. In Appendix A.7, we analyze the differences analytically between the unconstrained portfolio weight and TE based portfolio weights in detail with proofs. In summary,  estimating a TE constrained portfolio may be easier than the unconstrained portfolio in the finite samples due to the benchmark portfolio not being estimated if the risk tolerance parameters of both portfolios are the same. However, asymptotically, there is no difference; the rate of convergence  of estimation errors to zero is the same and $r_{w1}$ in both cases, and the same analysis is true for variance and Sharpe Ratio estimation. 

7.   Our results use a diverging  Sharpe Ratio of  Maximum Sharpe Ratio (MSR) portfolio with a growing number of assets. This is a technical condition also used by \cite{caner2022}. If we  impose   a bounded Sharpe Ratio for MSR portfolio, $0< c \le \b \mu' \b \Theta \b \mu \le C < \infty$,  in Appendix B, Theorem B.1, Theorem B.3 clearly show we can have consistency but only with $p<T$ in  two  different cases with bounded Sharpe (MSR portfolio) ratio assumption.

8. Note that a possibility is to estimate the risk tolerance parameter $\kappa$ based on the lower bound for TE in Appendix A.3. We show that it is possible to estimate this parameter consistently in Appendix B.3. We provide a detailed proof. It can also be seen that since the rate matches Assumption 7(i), there will be no change in all the proofs  here, as well as Theorems 2-3 below,  if we use a constant $\kappa$.

\section{Joint TE and Weight Constraints}\label{s-jtew}

In this section we analyze weight constraints for a portfolio. Let $R$ represent the indices of restrictions and we define the cardinality of the set $R$ as  $r:=|R|$. 
We allow for at least one restricted asset, and also, not all the assets can be restricted. In other words, the number of restrictions is $r$, $1\le r<p$.
 To give an example, let $p=10$, and $R=\{ 1, 5,7\}$, this means assets 1, 5, 7 are restricted in a 10 asset portfolio, and $r=3$.  To make things a bit more clear in notation, define $\b 1_R$ as a $p \times 1$ vector where we have ones in the places of restricted assets and zeros in the unrestricted assets. To continue with the example above,
 $\b 1_R=(1,0,0,0,1,0,1,0,0,0)'$. The benchmark portfolio is given, usually with weight restrictions. E.g., $\b m=(0.1, 0, 0, 0.1, 0.2, 0, 0.3, 0.1, 0.1, 0.1)$. Define $ \omega:= \b 1_R' \b w - \b 1_R' \b m= \b 1_R' \b w_d$. Then $ \omega$ is the difference between the portfolio and the benchmark portfolio when both the benchmark and the portfolio are restricted. The constraints may take the form of certain stocks being excluded from the portfolio from the perspective of regulatory, ethical and religious considerations. We allow for $r \to \infty$ as $p \to \infty$ as long as $0<\frac{r}{p}<1$ where $\omega$ can be positive or negative. This representation is adopted by \cite{bbmp2011}. To give an example, $\omega=0$ may show the same restrictions applied to the benchmark, $\b m$, which should apply to a portfolio that we are forming, $\b w$. To continue with that simple example with 10 assets above, $\omega=0$, which implies $\b 1_R' \b w =0.6$, the weight of combined assets 1, 5, and 7 should be 0.6. If the weights are normalized to one and there is no short selling allowed, this means we form 60\% of our portfolio from these three assets. Inequality restrictions would be more common in portfolio formation either by funds themselves or  regulators. We will discuss this in the next section. The optimization problem with equality constraint is:

\[ \b w_{cp}^*:= \max_{\b w_d \in R^P} [\b \mu' \b w_d -  \frac{\b \Xi}{2} \b w_d' \b \Sigma_y \b w_d], \quad \mbox{subject to} \quad \b 1_p' \b w_d = 0, \quad \b 1_{R}' \b w_d = \omega.\]

The solution is given by Proposition 1 in \cite{bbmp2011}. This is basically the solution to TE portfolio added to a fraction of the portfolio $\b l$ where $\b l$ can be thought as the weighted difference between restricted and unrestricted minimum variance portfolios.

\begin{equation}
\b w_{cp}^*= (\omega - \kappa w_u) \b l + \b w_d^*,\label{ojtew}
\end{equation}
with $\b l:=\frac{\b k-\b a}{w_k - w_a},$ and 
$\b k := \frac{\b \Theta \b 1_R}{\b 1_p' \b \Theta \b 1_R}, \,  \b a:= \frac{\b \Theta \b 1_p}{\b 1_p' \b \Theta \b 1_p},$
which are fully invested portfolios, and $\b a$ is the global minimum variance portfolio, and $\b k$ can   roughly be considered as its restricted counterpart. 
See that 
 $ w_k := \frac{\b 1_R'  \b \Theta \b 1_R}{\b 1_R' \b \Theta \b 1_p}, \quad  w_a := \frac{\b 1_R' \b \Theta \b 1_p}{\b 1_p' \b \Theta \b 1_p}.$ Note that $w_k$ can be thought of  the addition of weights in the restricted minimum variance portfolio, and $w_a$  represents the addition of weights in the minimum variance portfolio in a less restricted way. 
 
Note $w_k \neq w_a$ since, by assumption, not all assets in the portfolio are restricted ($r<p$). We also define
\begin{equation}
 w_u = \b 1_R' \left [\frac{ \b \Theta \b \mu}{\b 1_p' \b \Theta \b \mu} - \frac{\b  \Theta \b 1_p}{\b 1_p' \b \Theta \b 1_p}\right ],\label{eqwu}
 \end{equation}
where $w_u$ can be thought as the subtraction of minimum variance portfolio weights from maximum Sharpe Ratio portfolio weights and that only restricted assets are taken into account.

The estimator can be obtained by using the estimate $\hat{\b \Theta}$, with 
\begin{equation}
 \widehat{\b w}_{cp}= (\omega - \kappa \hat{w}_u) \hat{\b l} + \hat{\b w}_d,\label{bind}
 \end{equation}
with 
$ \hat{\b l} = \frac{\hat{\b k} - \hat{\b a}}{\hat{w}_k - \hat{w}_a}$,
   \, and 
$ \hat{\b k}= \frac{\hat{\b \Theta}' \b 1_R}{\b 1_p' \hat{\b \Theta}' \b 1_R}, \quad \hat{\b a}= \frac{\hat{\b \Theta}' \b 1_p}{\b 1_p' \hat{\b \Theta}' \b 1_p}$, 
with 
$ \hat{w}_k - \hat{w}_a = \frac{\b 1_R' \hat{\b \Theta}' \b 1_R}{\b 1_R' \hat{\b \Theta}' \b 1_p} - \frac{\b 1_R' \hat{\b \Theta}' \b 1_p}{\b 1_p' \hat{\b \Theta}' \b 1_p},$
and $\hat{w_k} \neq \hat{w}_a$ since, by assumption, not all assets in the portfolio are restricted. We also have 
$ \hat{w}_u = \frac{\b 1_R' \hat{\b \Theta}' \b \mu}{\b 1_p' \hat{\b \Theta}' \b \mu} - \frac{\b 1_R' \hat{\b \Theta}' \b 1_p}{\b 1_p' \hat{\b \Theta}' \b 1_p}.$ To distinguish from the general benchmark $\b m$ we denote $\b m_R$ as the restricted benchmark portfolio.
Define $\b w_R^*:= \b w_{cp}^* + \b m_R$, and 
\begin{equation}
\hat{\b w}_R:= \hat{\b w}_{cp} + \b m_R.\label{eqwrh}
\end{equation}

Define the Sharpe Ratio associated with joint restrictions of TE and weight:
$ SR_R := \frac{\b w_{R}^{*'} \b \mu}{\sqrt{\b w_{R}^{*'} \b \Sigma_y \b w_{R}^{*} }},$
and estimate
$ \widehat{SR}_R  := \frac{\hat{\b w}_{R}' \b \mu}{\sqrt{ \hat{\b w}_R' \b \Sigma_y \hat{\b w}_{R}}}.$ We need the following  assumption which puts a lower bound on restricted portfolio terms.

\begin{assum}\label{as8}

(i). $\bar{s} l_T  K^{4} = o(1)$.

(ii). $|\frac{\b 1_p' \b \Theta \b \mu}{p} |\ge c > 0, \frac{\b 1_p' \b \Theta \b 1_p }{p} \ge c >0$,  $|\frac{\b 1_R' \b \Theta \b 1_p}{p}|\ge c >0$, with $c, C $ being  positive constants.

(iii). $0 < r/p < 1$, $| w_k - w_a | \ge c > 0$, with $c$ being a positive constant, and  $|\omega| \le C < \infty,$ $0< c \le  |\kappa| \le C < \infty.$

(iv). $|\b w_R^{*'} \b \mu | \ge c > 0, \b w_R^{*'} \b \Sigma_y \b w_R^{*} \ge c >0$, $\| \b m_R \|_1 = O (\b w_{cp}^*)$.

\end{assum}

Assumption 8(i) is the same as  Assumption 7(i).  Assumption  8(ii) provides a lower bound for returns scaled by variance, and they cannot be zero or converge to zero.  Global Minimum Variance  Portfolio variance has to be finite. The restricted version variance has to be finite. 
Assumption 8(iii) assumes the number of restrictions can be at least 1, and cannot be as large as $p$, so we can restrict $p-1$ assets at most. Technical term $w_k \neq w_a$ since we cannot have full restrictions on $p$ assets, but here we assume that  $| w_k - w_a|$ cannot converge to zero as well. Our equality restrictions have to be bounded by a positive finite constant in absolute value, and our risk tolerance parameter in absolute terms cannot be unbounded. It cannot converge to zero either, and has to be upper bounded by a positive constant. Assumption 8(iv) also does not allow restricted  portfolio return to be zero (nor converge to zero), the variance cannot converge to zero, and the constrained benchmark portfolio has to have the same order as the optimal joint TE and weight portfolio, $w_{cp}^*$.  We have the following theorem that shows it is possible to estimate a portfolio with weight and TE constraints in large dimensions. Also, consistent estimation of Sharpe Ratio is possible for $p>T$.

\begin{thm}\label{t2}
Under Assumptions \ref{as1}-\ref{as6}, \ref{as8}

(i).  
\[ \| \hat{\b w}_R - \b w_R^* \|_1 = 
\| \hat{\b w}_{cp} - \b w_{cp}^* \|_1=  O_p (r_{w1}) = O_p (\bar{s} l_T K^{3/2})= o_p (1).\]

(ii). \[ \left| \frac{\hat{\b w}_R'  \b \Sigma_y  \hat{\b w}_R}{\b w_R^{*'} \b \Sigma_y \b w_R^*} - 1 \right| = O_p ( r_{w1} K^{5/2}) = O_p (\bar{s} l_T K^4)=
o_p (1).\]

(iii).\[ \left| \left(\frac{\widehat{SR}_R}{SR_R}\right)^2 - 1
\right| = O_p (r_{w1}  K^{5/2}) = O_p ( \bar{s} l_T K^4)= o_p (1).\]
\end{thm}

\noindent {\bf Remarks}. 1. Comparing Theorem \ref{t1} with Theorem \ref{t2} shows that the rates with joint TE and weight constraints are the same.  So upper bounds on estimation errors asymptotically behave the same as in TE and joint TE plus weight constraints. But (\ref{bind}) shows that additional terms are estimated in joint TE plus weight-constrained portfolio compared to TE constrained portfolio; however, asymptotically, the differences disappear.
 Remark 4 below  compares  only TE and only weight-constrained portfolios to see the difference.

 2. We can also have an estimate for TE as $\widehat{TE}_{cp}:= \sqrt{\hat{\b w}_{cp}' \b \Sigma_y \hat{\b w}_{cp}}$, as an out-of-sample estimate. It can be shown that 
$|(\frac{\widehat{TE}_{cp}}{TE})^2-1|$ will converge in probability to zero by following the proof of Theorem \ref{t2}(ii), and the rate will be the same as in Theorem \ref{t2}(ii).
This estimator is consistent. 

3. There is also an issue of how TE constraints, jointly with weight constraints, affect the returns. Information Ratio (IR) is used to measure the effect of TE constraints on expected returns; however, \cite{bbmp2011} shows that IR in the case of weights joint with TE constraints is not an appropriate measure since it can decrease artificially with an increase in TE. They propose Adjusted Information Ratio (AIR). In (22) of \cite{bbmp2011}, it is clear that  AIR is measuring the returns of optimal portfolio minus constrained minimum TE portfolio satisfying the weight constraints with respect to risk of that portfolio. We will not be pursuing such an approach since our interest centers on optimal portfolios as in Theorem \ref{t2}.

4. The case of only weight constraints is analyzed after the proof of Theorem \ref{t2} in  Appendix A.6 in detail, and the rates are the same as in Theorem \ref{t2}.
Comparing TE only restrictions with only weight restriction portfolio is done in detail in Appendix A.6.1. To summarize, since TE has a benchmark portfolio and it does not have to be estimated, it will incur less estimation errors in $l_1$ norm for weights compared with only weight restricted portfolio in finite samples, but asymptotically upper bound estimation error rates are the same as $r_{w1}$.

5. Comparing unconstrained portfolio with joint TE and weight-constrained portfolio is shown via proofs in Appendix A.7 and A.7.3. In finite samples, it is not clear whether joint TE plus weight or unconstrained portfolio estimation is easier in $l_1$ norm estimation error-sense.
However, asymptotically, upper bounds in estimation errors in both portfolios are the same and $r_{w1}$.
Variance and Sharpe Ratio estimation errors behave like weight estimation errors. 

6.  Remark 7 in Theorem 1 applies here as well. We use diverging $\b \mu' \b \Theta \b \mu $ in our proofs, but if we impose bounded $\b \mu' \b \Theta \b \mu$, the consistency is only achieved when $p<T$. This can be seen in our proofs for the bounded Sharpe Ratio case in Appendix B, as in Theorem B.2, Theorem B.4 there.

7. Note that we use same $\kappa$, the risk tolerance parameter  both in this section with joint TE and weight constraints and only  TE constraint in the previous section. The reason is that $\kappa$ is tied to the TE constraint in both cases, and hence use the same Lagrange multiplier in the constrained problem in p.299, and p.304 of \cite{bbmp2011}.
However, the cases of unconstrained portfolio and only weight constrained portfolio is different. There our risk tolerance parameter is $\kappa_w$, defined in Appendix A.6 and after (A.116). 
In cases of unconstrained portfolio and only weight constrained portfolio, $\kappa_w$ is tied to the variance constraint in the portfolio, hence $\kappa \neq \kappa_w$. But we argue that before Lemma A.3, this does not change the proofs so we assume $\kappa = \kappa_w$ to simplify the notation in our proofs. Consistent estimation of $\kappa_w$ is possible and can be shown as in Section B.3 for $\kappa$.

\section{TE with Inequality Constraints  on the Weights of the Portfolio}\label{sec5}

In practice, inequality constraints on the weight of the portfolio are more common. Of course, one major issue is that how the inequality weight interacts with large number of assets in the portfolio? In this part, we assume $\omega \ge 0$; by that restriction, we exclude the case where the benchmark portfolio, $\b m$, does not satisfy the inequality constraint. So we exclude an unlikely event in practice.  The optimization problem is:
\[ \max_{\b w_d \in R^p } [ \b \mu' \b w_d - \frac{\Xi}{2} \b w_d' \b \Sigma_y \b w_d] \quad \mbox{subject to }  \quad \b 1_p'  \b w_d = 0, \quad \b 1_R'  \b w_d \le \omega.\]

Note that when the weight constraint is not binding, we should get TE  solution $\b w_d^*$ in Section \ref{sec2}. Otherwise, we should get the solution in Section \ref{s-jtew}, which is $\b w_{cp}^*$. This $\b w_{cp}^*$ is the joint TE and equality weight constraint based portfolio. We analyze the cases that may lead to binding or nonbinding weight constraint solutions now. We need $\b 1_R' \b w_d^* \ge \omega$ to have a binding solution. Equivalently, the weight constraint is binding when $\kappa w_u \ge \omega$ since $\b 1_R' \b w_d^* = \kappa w_u$ by (\ref{eqwd}) and (\ref{eqwu}).\footnote{As in \cite{bbmp2011} we use greater than equal constraint for binding rather than just equality in Section 4.} The nonbinding case will be $\kappa w_u < \omega$, as also seen in (17) of \cite{bbmp2011}.

We provide the following theoretical portfolio 
$\b w_{op}^*:= \b w_d^* 1_{ \{ \kappa w_u < \omega \}} + \b w_{cp}^* 1_{ \{ \kappa w_u > \omega \}}.$
$1_{\{\cdot\}}$ is an indicator function. This means, if the weight constraint is not binding, i.e. $\kappa w_u < \omega$, then the portfolio is $\b w_d^*$, which is TE (only) constrained portfolio. If the weight constraint is binding, i.e. $\kappa w_u > \omega$, then the portfolio is given by joint TE and weight constraint portfolio, which is $\b w_{cp}^*$. 
For technical reasons, we exclude $\kappa w_u = \omega$. The feasible estimator of this theoretical portfolio is given by
$ \hat{\b w}_{op}:=  \hat{\b w}_d 1_{ \{ \kappa \hat{w}_u < \omega \}} + \hat{\b w}_{cp} 1_{ \{ \kappa \hat{\b w}_u > \omega \}}.$

By (\ref{eqwh}) and (\ref{eqwrh}), the estimated portfolio with benchmarks taken into account is defined as
$ \hat{\b w}_{est}:=\hat{\b w} 1_{ \{\kappa \hat{\b w}_u < \omega \}} + \hat{\b w}_R 1_{ \{ \kappa \hat{\b w}_u > \omega \} }.$
The theoretical counterpart is given by \footnote{An equivalent way of writing the nonbinding constraint in the indicator function is by $\b 1_R' \b w < \omega + \b 1_R' \b m$, and the binding constraint in the indicator function can be written as $\b 1_R' \b w > \omega + \b 1_R' \b m$.}
$ \b w_{est}^*:=\b w^* 1_{\{  \kappa w_u < \omega \}} + \b w_R^* 1_{ \{ \kappa w_u > \omega \}}.$ We define the Sharpe Ratio  whether we are in the binding or nonbinding weights constraint case as follows
$SR_{est}^*:= \frac{\b w^{*'} \b \mu }{\sqrt{ \b w^{*'} \b \Sigma_y \b w^* }} 1_{ \{ \kappa w_u < \omega \} }
+ \frac{ \b w_R^{*'} \b \mu}{\sqrt{ \b w_R^{*'} \b \Sigma_y \b w_R^*}} 1_{ \{ \kappa w_u > \omega \}},$
and its estimator as 
$ \widehat{SR}_{est}:= \frac{\hat{\b w}' \b \mu }{\sqrt{ \hat{\b w}^{'} \b \Sigma_y \hat{\b w}}} 1_{ \{ \kappa \hat{w}_u < \omega \} } 
+ \frac{ \hat{\b w}_R^{'} \b \mu }{\sqrt{ \hat{\b w}_R^{'} \b \Sigma_y \hat{\b w}_R }} 1_{ \{ \kappa \hat{w}_u > \omega \}}.$

Next theorem shows that our weight, variance, and Sharpe Ratio estimates are consistent, regardless of whether inequality constraints are binding or not.  This is a new result in the literature. In reality, we do not know whether the theoretical constraints are binding or not. However, our data-based method selects whether the constraints are binding or not and our estimates converge in probability to the optimal weight, variance or Sharpe Ratio. Let $C, c$ be some  positive constants.

\begin{thm}\label{t3}
Under Assumptions \ref{as1}-\ref{as8} with $| \omega - \kappa w_u | \ge C > 2 \epsilon >0$, $\omega \ge 0$, $\kappa \ge c>  0$

(i). If $\kappa w_u < \omega$, (non-binding  weight restriction)  
we have the same rate of convergence
\[ \| \hat{\b w}_{est} - \b w_{est}^* \|_1=
\| \hat{\b w}_{op} - \b w_{op}^* \|_1 = \| \hat{\b w} - \b w^*\|_1= O_p (r_{w1}) = O_p( \bar{s} l_T K^{3/2})=o_p (1),\]
otherwise if $\kappa w_u > \omega$, (binding  weight restriction) then 
\[ \| \hat{\b w}_{est} - \b w_{est}^* \|_1=
\| \hat{\b w}_{op} - \b w_{op}^* \|_1 = \| \hat{\b w}_{R} - \b w_{R}^* \|_1=O_p (r_{w1}) = O_p ( \bar{s} l_T K^{3/2})= o_p (1).\]

(ii). If $\kappa w_u < \omega$, (non-binding  weight restriction) then 
\[ \left| \frac{\hat{\b w}_{est}' \b \Sigma_y \hat{\b w}_{est}}{\b w_{est}^{*'} \b \Sigma_y \b w_{est}^*} - 1 \right| =
\left| \frac{\hat{\b w}' \b \Sigma_y \hat{\b w}}{\b w^{*'} \b \Sigma_y \b w^*} - 1 \right|= O_p (r_{w1} K^{5/2}) = O_p ( \bar{s} l_T K^4)= o_p (1),\]
otherwise if $\kappa w_u > \omega$ (binding weight restriction), then 
\[ \left| \frac{\hat{\b w}_{est}' \b \Sigma_y \hat{\b w}_{est}}{\b w_{est}^{*'} \b \Sigma_y \b w_{est}^*} - 1 \right|  = 
\left| \frac{\hat{\b w}_{R}' \b \Sigma_y \hat{\b w}_{R}}{\b w_{R}^{*'} \b \Sigma_y \b w_{R}^*} - 1 \right|  =
O_p ( r_{w1}  K^{5/2}) = O_p ( \bar{s} l_T K^4)= o_p (1).\]

(iii). If $\kappa w_u < \omega$ (non-binding weight restriction) then 
\[ \left| \left[ \frac{\widehat{SR}_{est}}{SR_{est}^*}
\right]^2 - 1
\right| = 
\left| \left[ \frac{\widehat{SR}}{SR}
\right]^2 - 1
\right| =O_p ( r_{w1}  K^{5/2}) = O_p (\bar{s} l_T K^4)= o_p (1).\]
otherwise if $\kappa w_u  >\omega$ (binding weight restriction), then 
\[ \left| \left[ \frac{\widehat{SR}_{est}}{SR_{est}^*}
\right]^2 - 1
\right| = 
\left| \left[ \frac{\widehat{SR}_{R}}{SR_{R}^*}
\right]^2 - 1
\right| =O_p ( r_{w1}  K^{5/2}) = O_p (\bar{s} l_T K^4)= o_p (1).\]

\end{thm}

\noindent {\bf Remarks}. 1. $\epsilon$ is well defined as an upper bound that goes to zero. Specifically, 
$ | \kappa (\hat{w}_u - w_u ) | \le \epsilon$, with probability approaching 1 (wpa1),  and given that $\kappa$ is a positive constant, we have $\epsilon:= \bar{s}  l_T K^{3/2} \to 0$ by Assumption 
\ref{as8}(i). This can be seen in Lemma A.2(iv).

2. We exclude only a small ``local to zero'' neighborhood of zero in the indicator function $| \omega - \kappa w_u| > 2 \epsilon >0$, since $\epsilon \to 0$. 

3. Our theorem is new and encompasses portfolio decisions based on binding versus nonbinding constraints taken into account stochastically.

4. The main ingredient of the proof is when the constraint is binding, we have wpa1, $\kappa \hat{w}_u > \omega$; and the same is true for nonbinding constraint,  that $\kappa \hat{w}_u < \omega$, wpa1. We can select binding versus nonbinding constraints correctly, with probability approaching 1, and this decision is incorporated into our proof of  Theorem \ref{t3}.

5. One question is whether inequality constrained portfolios have less estimation errors compared with an unconstrained portfolio in $l_1$ norm based weight estimation.
The answer depends on  whether the constraints are binding or not. If the constraints are binding, Remark 5 of Section 4 will be in effect. On the other hand, if the constraints are not binding, Remark 6 of Section 3 will be the answer.

6. One key issue is that short-sale constraints. These take the form of $w_j \ge 0$, $j=1,\cdots,p$. Short sale constraint is a floor constraint on weight. 	In Appendix A.8, we show that floor  constraints on weights on assets in a portfolio can be used as cap constraints on weights on the remaining assets in the portfolio. In other words, if we want to put short-sales constraints on assets 1 and 5, these will be equivalent to cap constraints on all other assets (i.e., all assets except asset 1 and  asset 5) in our portfolio.
Our method can only work in two limited sub-cases but cannot cover a short sales constraint for each asset in a portfolio.  First, our short-sale constraint can be applied to only one asset in our portfolio, say $j$ th asset, $w_j \ge 0$. The second possibility is not a classic short sales constraint but a short sales constraint on a group of assets jointly in a portfolio. This group of assets can have their sum of total weights 
to be larger than equal to zero. Appendix  A.8  contains derivations of weights and more discussion about short sales constraints. Also, in empirics, we provide results on how much our portfolios have short positions.

7. One can note that all the rates in Theorems 1(i)-2(i)-3(i), and Theorems 1(ii)(iii), 2(ii)(iii), 3(ii)(iii) are the same. This is also true with Lemma A.3 and Lemma A.4. So even though there are different restrictions in the portfolios, such as TE and weight constraints, the rate of convergence is the same. The reason for this is estimation of unconstrained maximum Sharpe Ratio weights 
$\b \Theta \b \mu/\b1_p'  \b \Theta \b \mu$ derives the rates. 
The detailed discussion is in Appendix B.4. The practical implication of this result may be that, regardless of restrictions, if the portfolio manager wants to achieve a target return, there will be no difficulty 
stemming from restrictions to achieve that target return.

\section{Simulations}

 In this section, we provide a brief simulation study to illustrate the finite sample performance of CROWN. We only show simulations with TE constraint and certain $p,T$ combinations, and the rest of the exercise, with weight, TE+weight, and inequality constraints, with more $p,T$ combinations and DGPs are in  Appendix. In Tables \ref{tablete80}-\ref{tablete320} we report estimation errors of the weight, risk, and Sharpe Ratio. We also provide estimated Sharpe Ratio (SR), and TE.  All the estimation errors and estimated SR and TE are obtained by averaging across 200 iterations.

The excess return data $\{\b y_t\}_{t=1}^{T}$ is generated from a standard three-factor model whose parameters are calibrated from the actual data.  
 The excess asset return at time t, for j-th asset is:

\begin{equation}\label{simeq1}
y_{j,t} = b_{j,1}f_{1,t} + b_{j,2}f_{2,t} + b_{j,3}f_{3,t}+u_{j,t}, \quad j = 1,...,p; \ t=1,...T.   
\end{equation}
Specifically, we assume the factors $\b f_t=(f_{1,t}, f_{2,t}, f_{3,t})'$ follow a stationary VAR(1) model, $\b f_t=\b c_f+\b \Pi_f \b f_{t-1}+\b e_t,$ with errors following a normal distribution with zero mean and covariance matrix $\b cov (\b e_t) = \b \Sigma_f - \b \Pi_f \b \Sigma_f \b \Pi_f'$, which is proved in Appendix A.1.  
For factor loadings $\b b_j = \left( b_{j,1}, b_{j,2} , b_{j,3} \right)^\top$, $j = 1,2,\cdots,p$, and the idiosyncratic errors $\b u_{t} = (u_{1,t},\ldots,u_{p,t})^\top$, we draw their values from normal distributions. Specifically, $\b b_j \sim N_3(\b \mu_{b},\b \Sigma_{b})$; and $\b u_{t} \sim N(0,\b \Sigma_{u})$ with $\b \Sigma_u$ has the Toeplitz form such that the $(i,j)$ th element of $\b \Sigma_u$ is $0.25^{|i-j|}$.
Factor loadings have the following mean and variance-covariance matrix in Table \ref{bjsim}.
\begin{table}[htbp!]
	\scriptsize
	\centering
    	\caption{Parameters of $ \b b_j $}
    	\label{bjsim}
\begin{tabular}{c|ccc }
\hline $\b \mu_b$ & \multicolumn{3}{c}{$\b \Sigma_b$} \\
\hline 1.0166 & 0.0089 & 0.0013 & 0.0046 \\
0.5799 & 0.0013 & 0.2188 & -0.0134 \\
0.2937 & 0.0046 & -0.0134 & 0.1491 \\
\hline
\end{tabular} 
\end{table}

Then we set the parameters of the design in  factors  and errors as in Table \ref{ftsim} with $\b \Sigma_f$ as the covariance matrix of $\b f_t$.

\begin{table}[htbp!]
	\scriptsize
	\centering
    	\caption{Parameters of $\b f_t$}
    	\label{ftsim}
\begin{tabular}{c|ccc|ccc}
\hline $\b c_f$ & & $ \b \Sigma_f  $ & & \multicolumn{3}{c}{$\b \Pi_f$} \\
\hline 0.0445 & 1.5016 & 0.1338 & 0.1682 & -0.1204 & 0.1555 & -0.0324 \\
0.0060 & 0.1338 & 0.3667 & -0.0310 & -0.0074 & -0.0378 & 0.00318 \\
0.0021 & 0.1682 & -0.0310 & 0.6017 & -0.0027 & 0.0031 & 0.01669 \\
\hline
\end{tabular}
\end{table}

Tables \ref{bjsim}-\ref{ftsim} are obtained through the following  procedure. We utilize the daily data on returns of 100 industrial portfolios, which are intersections of 10 portfolios formed on size and 10 portfolios formed on book-to-market ratio, and the Fama/French 3 factors from the website of Kenneth French, at   https://mba.tuck.dartmouth.edu/pages/faculty/ken.french/. We use 20-year data from 2003-01-01 to 2022-12-31 for calibration.
 We use $p/T=2$ for $p=80, 320$ on the left panels of Tables \ref{tablete80}-\ref{tablete320} respectively with $TE=0.1, 0.2, 0.3$. Then we use $p/T=0.8$, with $p=80, 320$ on the right panels of Tables \ref{tablete80}-\ref{tablete320} respectively. The benchmark market index $\b m$ is an equal-weighted portfolio.We also report two more benchmark portfolios: 1), the ``Oracle'', which is the portfolio given the population expected return $\b \mu$ and covariance $\b \Sigma_y$ and plug in the analytical solution for CROWN, given the required constraints; 2), the portfolio with no constraint, ``NCON'', which is defined in  Appendix A.7. We also report other popular methods for covariance/precision matrix estimation, specifically, the nodewise estimator \cite{caner2019} will be called  NW, the  POET of  \cite{fan2013}, the  nonlinear shrinkage estimator and its single factor version  \cite{lw2017} will be abbreviated as  NLS, SFNL respectively. We construct these comparison portfolios using the covariance estimators from their respective methods.
To verify the main theoretical results and make comparisons, we report the following statistics
$TE = \sqrt{(\hat{\b w}-\b m)^\prime{\b \Sigma_y}(\hat{\b w}-\b m)}, $
is the TE estimate, where $\hat{\b w}$, $\b m$ are the estimated weight and weight of the benchmark portfolio, in Theorem 1,  respectively.  The other metrics and their estimates, risk, Sharpe Ratio, weights are defined as in Theorem 1.

Bold-faced numbers in Tables \ref{tablete80}-\ref{tablete320}  are the category winners among CROWN, NW, POET, NLS, and SFNL methods. First, in high dimensional Table \ref{tablete320}, CROWN is the winner in all metrics.  To give an example, at $p=320, T=160$ CROWN has SR estimation error of 0.0003 at TE=0.1, and the others range between 0.0059-0.0814. Then, as a second observation, in both Tables \ref{tablete80}-\ref{tablete320}, on the left and right panels, when we increase $p,T$ fourfold but keep $p/T$ same, CROWN gets lower estimation rates, in line with theory. To give an example, with $p=80, T=40$ and with $TE=0.2$, CROWN has an estimation error of risk at 0.1622 in Table \ref{tablete80}, and this error decreases in Table \ref{tablete320}, with $p=320, T=160$, to 0.0015.

\begin{table}
	\scriptsize
\centering
        \caption{Simulations with TE constraint, $p=80$ }
\begin{threeparttable}\label{tablete80}
        \begin{tabular}{lccccccccccc}		\noalign{\global\arrayrulewidth1pt}\hline \noalign{\global\arrayrulewidth0.4pt}
            & \multicolumn{5}{c}{p=80, T=40}            &  & \multicolumn{5}{c}{p=80, T=100}        \\
           \cline{2-6} \cline{8-12}
& \multicolumn{11}{c}{TE = 0.1} \\ \hline
               & TE	& Weight-ER	& Risk-ER & SR-ER &	SR & & TE	& Weight-ER	& Risk-ER & SR-ER &	SR\\
Oracle & 0.1000 & 0.0000 & 0.0000 & 0.0000 & 0.0350 &   & 0.1000 & 0.0000 & 0.0000 & 0.0000 & 0.0350\\
NCON & 0.4707 & 2.0415 & 0.1921 & 0.0248 & 0.0352 &   & 0.3698 & 1.1704 & 0.2394 & 0.0665 & 0.0361\\
Index & - & 0.4698 & 0.1665 & 0.0570 & 0.0339 &   & - & 0.4698 & 0.1697 & 0.0255 & 0.0345\\ \hline
NW & 0.4013 & 1.7347 & 11.5532 & 0.0150 & 0.0346 &   & 0.2844 & 0.8272 & 0.2427 & 0.0337 & 0.0344\\
CROWN & 0.1840 & 0.8981 & \textbf{0.0717} & \textbf{0.0055} & \textbf{0.0349} &   & \textbf{0.0984} & \textbf{0.1608} & \textbf{0.0027} & \textbf{0.0005} & \textbf{0.0349}\\
POET & 0.1192 & 0.8844 & 0.1316 & 0.0148 & 0.0347 &   & 0.1110 & 0.4889 & 0.0444 & 0.0046 & \textbf{0.0349}\\
NLS & \textbf{0.0961} & 0.6810 & 0.1056 & 0.0121 & 0.0347 &   & 0.1006 & 0.5363 & 0.0518 & 0.0049 & \textbf{0.0349}\\
SFNL & 0.1158 & \textbf{0.5463} & 0.0842 & 0.0113 & 0.0348 &   & 0.1077 & 0.4194 & 0.0404 & 0.0049 & \textbf{0.0349}\\ \hline
& \multicolumn{11}{c}{TE = 0.2} \\ \hline
               & TE	& Weight-ER	& Risk-ER & SR-ER &	SR & & TE	& Weight-ER	& Risk-ER & SR-ER &	SR\\
Oracle & 0.2000 & 0.0000 & 0.0000 & 0.0000 & 0.0350 &   & 0.2000 & 0.0000 & 0.0000 & 0.0000 & 0.0350\\
NCON & 0.4707 & 2.2349 & 0.2741 & 0.0255 & 0.0352 &   & 0.3698 & 1.4176 & 0.3165 & 0.0655 & 0.0361\\
Index & - & 0.9396 & 0.0483 & 0.0579 & 0.0339 &   & - & 0.9396 & 0.0511 & 0.0264 & 0.0345\\ \hline
NW & 0.4378 & 2.2713 & 12.4291 & \textbf{0.0166} & \textbf{0.0346} &   & 0.5688 & 1.6544 & 0.5052 & 0.0720 & 0.0337\\
CROWN & 0.3680 & 1.7961 & \textbf{0.1622} & 0.0293 & 0.0345 &   & \textbf{0.1967} & \textbf{0.3216} & \textbf{0.0048} & \textbf{0.0012} & \textbf{0.0349}\\
POET & 0.2383 & 1.7688 & 0.2771 & 0.0407 & 0.0342 &   & 0.2220 & 0.9778 & 0.0919 & 0.0119 & 0.0348\\
NLS & \textbf{0.1922} & 1.3621 & 0.2188 & 0.0319 & 0.0344 &   & 0.2012 & 1.0727 & 0.1069 & 0.0145 & 0.0347\\
SFNL & 0.2315 & \textbf{1.0927} & 0.1726 & 0.0282 &  0.0345  &   & 0.2155 & 0.8389 & 0.0824 & 0.0131 & 0.0347\\ \hline
& \multicolumn{11}{c}{TE = 0.3} \\ \hline
               & TE	& Weight-ER	& Risk-ER & SR-ER &	SR & & TE	& Weight-ER	& Risk-ER & SR-ER &	SR\\
Oracle & 0.3000 & 0.0000 & 0.0000 & 0.0000 & 0.0349 &   & 0.3000 & 0.0000 & 0.0000 & 0.0000 & 0.0349\\
NCON & 0.4707 & 2.5137 & 0.3459 & 0.0265 & 0.0352 &   & 0.3698 & 1.7773 & 0.3841 & 0.0675 & 0.0361\\
Index & - & 1.4094 & 0.0555 & 0.0037 & 0.0339 &   & - & 1.4094 & 0.0529 & 0.0246 & 0.0345\\ \hline
NW & 0.4816 & 2.4137 & 14.9933 & \textbf{0.0176} & \textbf{0.0346} &   & 0.8533 & 2.4817 & 0.7762 & 0.1068 & 0.0330\\
CROWN & 0.5520 & 2.6942 & 0.2641 & 0.0622 & 0.0338 &   & \textbf{0.2951} & \textbf{0.4824} & \textbf{0.0066} & \textbf{0.0021} & \textbf{0.0349}\\
POET & 0.3575 & 2.6533 & 0.4294 & 0.0701 & 0.0337 &   & 0.3330 & 1.4667 & 0.1407 & 0.0223 & 0.0346\\
NLS & \textbf{0.2883} & 2.0431 & 0.3349 & 0.0546 & 0.0340 &   & 0.3018 & 1.6090 & 0.1632 & 0.0273 & 0.0345\\
SFNL & 0.3473 & \textbf{1.6390} & \textbf{0.2622} & 0.0472 & 0.0341 &   & 0.3232 & 1.2583 & 0.1246 & 0.0233 & 0.0345\\ \noalign{\global\arrayrulewidth1pt}\hline \noalign{\global\arrayrulewidth0.4pt}    
        \end{tabular}
\begin{tablenotes}
\item[]  ``TE'' is the Tracking Error. ``Risk-ER'', ``SR-ER'', ``Weight-ER'' are the estimation errors (defined in  Appendix) of risk, SR and portfolio weights, respectively, as compared to the oracle. ``SR'' is the Sharpe Ratio. Due to space limitation we report the estimated return, risk in  Appendix. ``-'' means the statistic is not available.
\end{tablenotes}
\end{threeparttable}
\end{table}

\begin{table}
	\scriptsize
\centering
        \caption{Simulations with TE constraint, $p=320$ }
\begin{threeparttable}\label{tablete320}
        \begin{tabular}{lccccccccccc}		\noalign{\global\arrayrulewidth1pt}\hline \noalign{\global\arrayrulewidth0.4pt}
            & \multicolumn{5}{c}{p=320, T=160}            &  & \multicolumn{5}{c}{p=320, T=400}        \\
           \cline{2-6} \cline{8-12}
& \multicolumn{11}{c}{TE = 0.1} \\ \hline
               & TE	& Weight-ER	& Risk-ER & SR-ER &	SR & & TE	& Weight-ER	& Risk-ER & SR-ER &	SR\\
Oracle & 0.1000 & 0.0000 & 0.0000 & 0.0000 & 0.0353 &   & 0.1000 & 0.0000 & 0.0000 & 0.0000 & 0.0353\\
NCON & 0.3988 & 1.3717 & 0.3109 & 0.0705 & 0.0365 &   & 0.3975 & 1.3825 & 0.3109 & 0.0708 & 0.0365\\
Index & - & 0.6366 & 0.1593 & 0.0699 & 0.0340 &   & - & 0.6366 & 0.1624 & 0.0400 & 0.0346\\ \hline
NW & 0.6485 & 2.4803 & 0.8161 & 0.0814 & 0.0338 &   & 0.6661 & 2.6621 & 0.8475 & 0.0689 & 0.0340\\
CROWN & \textbf{0.0996} & \textbf{0.2142} & \textbf{0.0009} & \textbf{0.0003} & \textbf{0.0353} &   & \textbf{0.1001} & \textbf{0.1570} & \textbf{0.0008} & \textbf{0.0002} & \textbf{0.0353}\\
POET & 0.1752 & 5.1018 & 3.4305 & 0.0347 & 0.0345 &   & 0.1427 & 2.8335 & 1.8423 & 0.0199 & 0.0349\\
NLS & 0.1088 & 2.7118 & 0.2809 & 0.0092 & 0.0351 &   & 0.1209 & 2.1571 & 0.1368 & 0.0043 & 0.0352\\
SFNL & 0.1414 & 1.5372 & 0.1697 & 0.0059 & 0.0352 &   & 0.1236 & 1.0816 & 0.0797 & 0.0025 & 0.0352\\ \hline
& \multicolumn{11}{c}{TE = 0.2} \\ \hline
               & TE	& Weight-ER	& Risk-ER & SR-ER &	SR & & TE	& Weight-ER	& Risk-ER & SR-ER &	SR\\
Oracle & 0.2000 & 0.0000 & 0.0000 & 0.0000 & 0.0354 &   & 0.2000 & 0.0000 & 0.0000 & 0.0000 & 0.0354\\
NCON & 0.3988 & 1.8302 & 0.3878 & 0.0650 & 0.0365 &   & 0.3975 & 1.8374 & 0.3877 & 0.0653 & 0.0365\\
Index & - & 1.2733 & 0.0300 & 0.0747 & 0.0340 &   & - & 1.2733 & 0.0327 & 0.0450 & 0.0346\\ \hline
NW & 1.2969 & 4.9607 & 1.8131 & 0.1476 & 0.0326 &   & 1.3322 & 5.3243 & 1.8713 & 0.1241 & 0.0331\\
CROWN & \textbf{0.1992} & \textbf{0.4284} & \textbf{0.0015} & \textbf{0.0008} & \textbf{0.0353} &   & \textbf{0.1999} & \textbf{0.3140} & \textbf{0.0014} & \textbf{0.0005} & \textbf{0.0353}\\
POET & 0.2190 & 6.1584 & 5.2542 & 0.0386 & 0.0344 &   & 0.2039 & 3.5276 & 3.5383 & 0.0221 & 0.0349\\
NLS & 0.2176 & 5.4237 & 0.5927 & 0.0347 & 0.0347 &   & 0.2417 & 4.3142 & 0.2847 & 0.0215 & 0.0350\\
SFNL & 0.2829 & 3.0745 & 0.3447 & 0.0180 & 0.0350 &   & 0.2472 & 2.1633 & 0.1594 & 0.0086 & 0.0352\\ \hline
& \multicolumn{11}{c}{TE = 0.3} \\ \hline
               & TE	& Weight-ER	& Risk-ER & SR-ER &	SR & & TE	& Weight-ER	& Risk-ER & SR-ER &	SR\\
Oracle & 0.3000 & 0.0000 & 0.0000 & 0.0000 & 0.0354 &   & 0.3000 & 0.0000 & 0.0000 & 0.0000 & 0.0354\\
NCON & 0.3988 & 2.3658 & 0.4532 & 0.0616 & 0.0365 &   & 0.3975 & 2.3728 & 0.4531 & 0.0618 & 0.0365\\
Index & - & 1.9099 & 0.0800 & 0.0777 & 0.0340 &   & - & 1.9099 & 0.0766 & 0.0481 & 0.0346\\ \hline
NW & 1.9454 & 7.4410 & 2.9154 & 0.1940 & 0.0318 &   & 1.9983 & 7.9864 & 2.9966 & 0.1628 & 0.0324\\
CROWN & \textbf{0.2987} & \textbf{0.6426} & \textbf{0.0020} & \textbf{0.0012} & \textbf{0.0354} &   & \textbf{0.2999} & \textbf{0.4710} & \textbf{0.0019} & \textbf{0.0008} & \textbf{0.0354}\\
POET & 0.3128 & 8.4775 & 10.5361 & 0.0480 & 0.0342 &   & 0.3568 & 5.2820 & 10.4017 & 0.0275 & 0.0348\\
NLS & 0.3265 & 8.1355 & 0.9194 & 0.0644 & 0.0343 &   & 0.3626 & 6.4713 & 0.4374 & 0.0449 & 0.0346\\
SFNL & 0.4243 & 4.6117 & 0.5197 & 0.0343 & 0.0348 &   & 0.3707 & 3.2449 & 0.2373 & 0.0181 & 0.0351\\ \noalign{\global\arrayrulewidth1pt}\hline \noalign{\global\arrayrulewidth0.4pt}    
        \end{tabular}
\begin{tablenotes}
\item[]  See notes in Table \ref{tablete80}.
\end{tablenotes}
\end{threeparttable}
\end{table}

\section{Empirical Study}

In the empirical study, we show the performance of the proposed portfolio using monthly return data of S\&P 500 index component stocks. We consider the portfolio with 1) TE constraint only and 2) joint TE and weight inequality constraints, which are two leading examples in the  mutual fund industry. The following three tracked benchmark indices are considered: (1) S\&P 500 index. (2) S\&P 500 Health Care index and (3) S\&P 500 Energy index. Notice all three indices' component stocks change over time. To avoid the look-ahead survival bias, we follow the approach of \cite{bs2022} to obtain portfolio weight estimates. When constructing the portfolios, we consider all available S\&P 500 constituent stocks at each point of the out-of-sample period. In the Appendix, we also consider the S\&P 500 Industrials index as the benchmark, as well as the characteristics and industry-sorted portfolios. The portfolio is rebalanced monthly. 

Table \ref{table5} analyzes the S\&P 500 index with TE constraints, Tables \ref{tablehc}-\ref{tableeg} analyze joint weight and TE inequality constraints in the S\&P 500 Healthcare index and S\&P 500 Energy index, respectively.
Our forecasts are in a rolling-window setup, and the window size is set to 180.  The full sample period for S\&P 500 index study is 01/1981 to 12/2020, and the out-of-sample period is 01/1996 to 12/2020. The entire sample period for the two themed indices (healthcare and energy) is 01/1991 to 12/2020, and the out-of-sample period is 01/2006 to 12/2020, due to the availability of themed index under consideration. The in-sample period is from $1:T_I$, and the out-sample period is from $T_{I+1}:T$.
 Our first window is $1:T_{I}$, where $T_I=180$.   After getting estimates for the first 180 months, we use them to calculate metrics like Sharpe Ratio for the 181st month in the sample as out-of-sample forecast. Then we repeat the same process by ignoring the first period in-sample but with the same window length.  We repeat this process with a fixed window length until the end of out-sample, and we average all out-sample and report in the empirical results tables the average returns (AVR), TE (TE), standard deviation (Risk), and Sharpe Ratio (SR). We follow \cite{caner2022} to obtain out-of-sample portfolio return and variance of the portfolio.
After weights are estimated in the window, and defining the  weight estimate at the first window as $\hat{\b w}_{T_I}$, the out-of-sample portfolio return $\b w' \b \mu$ is estimated by 
$(\widehat{ \b w' \b \mu})_{oos}:= \frac{1}{T- T_I} \sum_{t= T_I}^{T-1} \hat{\b w}_t' \b y_{t+1}$,
and for example $\hat{\b w}_t$ at $t=T_I$ is the weight estimate in the window $1: T_I$. The out-of-sample variance of the portfolio, $\b w' \b \Sigma_y \b w$ 
is estimated by 
$\frac{1}{T- T_I -1} \sum_{t=T_I}^{T-1} [\hat{\b w}_t' \b y_{t+1} - (\widehat{ \b w' \b \mu})_{oos}]^2.$

Following \cite{ao2019} and \cite{caner2022}, we conduct hypothesis tests regarding the Sharpe Ratio to check the statistical difference between different models. For the Sharpe Ratio test, specifically, we test
\begin{equation}\label{SRtest}
H_0: S R_{CROWN} \le S R_0 \quad \text { vs } \quad H_a: S R_{CROWN}>S R_0
\end{equation}
where $S R_{CROWN}$ denotes the Sharpe Ratio of the CROWN portfolio, and $S R_0$ denotes the Sharpe Ratio of the portfolio under comparison such as POET, SFNL. We report the p-values of the Sharpe Ratio test  \citep{ledoit2008robust}, and the results were very similar to the test of \cite{memmel2003performance}.
We also consider  transaction costs. Following \cite{ao2019}, the excess portfolio return net of transaction cost is computed as  $y_{t}^{net} = \left(1 - \sum_{i=1}^pc\left|\widehat{w}_{t+1,i}-\widehat{w}_{t,i}^+\right| \right)(1+ y_{t}) -1,$ where $\widehat{w}_{t+1,i}$ is the $i$ th element of portfolio weight after rebalancing, and $\widehat{w}_{t,i}^+$ is the portfolio weight before rebalancing;	$\b y_{t}$ is the excess return of the portfolio without transaction cost and $c$ measures the level of the transaction cost, which is set to 50 basis points. Formally, the portfolio turnover is defined as  $TO = \frac{1}{RT}\sum_{l =1 }^{RT}\sum_{i=1}^p\left|\widehat{w}_{l+1,i}-\widehat{w}_{l,i}^+\right|,$
where $\widehat w_{l+1,i}$ is the desired portfolio weight at $(l+1)$ th rebalancing, and $\widehat w_{l,i}^+$ is portfolio weight before the $(l+1)$ th rebalancing, $RT$ is the number of rebalancing events.  Note that all our Tables \ref{table5}-\ref{tableeg} have taken into account transaction costs. At last, the targeted TE is 5\% annualized.

To evaluate the performance of those implementable portfolios, we compute the infeasible out-of-sample oracle portfolio.
For this oracle portfolio, instead of returns in the backward-looking sample period, the expected return $\b \mu$ and covariance $\b \Sigma_y$ are estimated with returns of the stocks in the out-of-sample periods (01/1996-12/2020 for S\&P 500 Index and 01/2006-12/2020 for S\&P 500 Sector indices). 
Our comparison group includes the benchmark,  INDEX, which is the market index portfolio, and the infeasible out-of-sample oracle portfolio.We also compare CROWN with other popular method for covariance/precision matrix estimation detailed in the simulation section. In the TE constraint case, CROWN achieves the best return/risk balance. As shown in Table \ref{table5}, the Sharpe Ratio of CROWN portfolio is 0.0938, whereas POET has 0.0828, and NLS has 0.0676. Notice that our proposed portfolio has the lowest turnover ratio, as it has the lowest TE to the benchmark index. The p-values of the SR comparison test suggest that the proposed CROWN portfolio  has statistically significantly better  performance at 1\% level among comparison portfolios, after considering the transaction cost. The result shows that as $p>T_I$,  in Table \ref{table5}, the proposed CROWN performs well as suggested by the theory. 

\begin{table}[h!]
	\scriptsize
	\centering
	\caption{Empirical results for TE constraint}
    	\label{table5}
 		\begin{threeparttable}
        \begin{tabular}{lccccccccc}		\noalign{\global\arrayrulewidth1pt}\hline \noalign{\global\arrayrulewidth0.4pt}
            & \multicolumn{8}{c}{S\&P 500 Index} \\
            \cline{2-10}
& AVR & TE & Risk & SR &  TO    &\textit{p-val} & Max & Min  & Total \\ 
                     &     &    &      &    &      &          &  Weight & Weight & Short \\ \hline
Oracle & 0.0180 & 0.0109 & 0.0432 & 0.4162 & 0.1072 & 0.0000 & 0.0512 & -0.0002 & 0.0012\\
\hline
INDEX & 0.0037 & - & 0.0418 & 0.0894 & - & 0.0000 & - & - & -\\
NW & 0.0034 & 0.0037 & 0.0417 & 0.0827 & 0.6960 & 0.0038 & 0.0944 & -0.0452 & 1.0634\\
CROWN & \textbf{0.0039} & \textbf{0.0021} & 0.0418 & \textbf{0.0938} & 0.6096 & - & 0.0417 & -0.0000 & 0.0000\\
POET & 0.0035 & 0.0032 & 0.0417 & 0.0828 & 0.6934 & 0.0000 & 4.6343 & -4.0492 & 160.0132\\
NLS & 0.0028 & 0.0046 & \textbf{0.0412} & 0.0676 & 0.7989 & 0.0000 & 0.1482 & -0.0385 & 1.6725\\
SFNL & 0.0033 & 0.0033 & 0.0414 & 0.0799 & 0.7129 & 0.0000 & 0.0369 & -0.0020 & 1.6129\\
\noalign{\global\arrayrulewidth1pt}\hline \noalign{\global\arrayrulewidth0.4pt}    
            \end{tabular}

   			\begin{tablenotes}
				\item[] Notations: average returns (AVR), tracking error (TE), standard deviation (Risk), Sharpe Ratio (SR), turnover ratio (TO). ``p-val'' is the p-value for the SR test. ``Max/Min weight'' is the maximum/minimum individual asset weight in the portfolio, ``Total short'' is the total shorting position of the portfolio. ``Oracle'' is the infeasible out-of-sample oracle portfolio, ``Index'' refers to S\&P 500 Index, S\&P 500 Healthcare Sector Index, and S\&P 500 Energy Sector Index in Table \ref{table5}, Table \ref{tablehc}, and Table \ref{tableeg}, respectively.  
			\end{tablenotes}
		\end{threeparttable}
\end{table}

\begin{table}[h!]
\scriptsize
	\centering
    \caption{Empirical results with TE and inequality weight constraint}
\label{tablehc}
\begin{threeparttable}
    \begin{tabular}{lcccccccccc}		\noalign{\global\arrayrulewidth1pt}\hline \noalign{\global\arrayrulewidth0.4pt}
        & \multicolumn{9}{c}{S\&P Healthcare Index} \\
        \cline{2-11}
& AVR & TE & Risk & SR &   TO    &\textit{p-val} & \%Themed & Max & Min  & Total \\ 
                     &     &    &      &    &        &        &  Assets  & Weight & Weight & Short \\ \hline
Oracle & 0.0120 & 0.0039 & 0.0385 & 0.3118 & 0.0408 & 0.0000 & 0.8000 & 0.1297 & -0.0002 & 0.0020\\
\hline
INDEX & \textbf{0.0081} & - & 0.0389 & 0.2088 & - & 0.4928 & 0.8000 & - & - & -\\
NW & 0.0080 & 0.0023 & 0.0384 & 0.2076 & 0.0673 & 0.4244 & 0.8031 & 0.1201 & -0.0019 & 0.0474\\
CROWN & 0.0078 & 0.0059 & \textbf{0.0373} & \textbf{0.2090} & 0.0853 & - & 0.8035 & 0.1201 & -0.0092 & 0.1097\\
POET & 0.0058 & 0.0057 & 0.0379 & 0.1541 & 0.5344 & 0.0000 & 0.8019 & 0.1212 & -0.0042 & 0.1105\\
NLS & 0.0077 & 0.0033 & 0.0382 & 0.2027 & 0.0879 & 0.1689 & 0.8036 & 0.1211 & -0.0052 & 0.1226\\
SFNL & 0.0079 & \textbf{0.0019} & 0.0384 & 0.2056 & 0.0703 & 0.3452 & 0.8017 & 0.1208 & -0.0030 & 0.0650\\
\noalign{\global\arrayrulewidth1pt}\hline \noalign{\global\arrayrulewidth0.4pt}    
            \end{tabular}
   			\begin{tablenotes}
				\item[] See notes in Table \ref{table5}.
			\end{tablenotes}
		\end{threeparttable}

\end{table}

\begin{table}[h!]
\scriptsize
	\centering
    \caption{Empirical results with TE and inequality weight constraint}
\label{tableeg}
\begin{threeparttable}
    \begin{tabular}{lcccccccccc}		\noalign{\global\arrayrulewidth1pt}\hline \noalign{\global\arrayrulewidth0.4pt}
        & \multicolumn{9}{c}{S\&P 500 Energy Index} \\
        \cline{2-11}
& AVR & TE & Risk & SR &   TO    &\textit{p-val} & \%Themed & Max & Min  & Total \\ 
                     &     &    &      &    &        &        &  Assets  & Weight & Weight & Short \\ \hline
Oracle & 0.0090 & 0.0167 & 0.0597 & 0.1504 & 0.0364 & 0.0000 & 0.8000 & 0.2945 & -0.0029 & 0.1261\\
\hline
INDEX & 0.0034 & - & 0.0723 & 0.0473 & - & 0.1586 & 0.8000 & - & - & -\\
NW & 0.0031 & 0.0157 & \textbf{0.0601} & 0.0513 & 0.1131 & 0.2337 & 0.8045 & 0.2868 & -0.0055 & 0.1903\\
CROWN & \textbf{0.0035} & 0.0135 & 0.0619 & \textbf{0.0569} & 0.0396 & - & 0.8001 & 0.2883 & -0.0000 & 0.0002\\
POET & 0.0031 & \textbf{0.0134} & 0.0624 & 0.0497 & 0.1327 & 0.0018 & 0.8001 & 0.2884 & -0.0006 & 0.0173\\
NLS & 0.0031 & 0.0147 & 0.0605 & 0.0508 & 0.0906 & 0.0711 & 0.8022 & 0.2877 & -0.0036 & 0.1477\\
SFNL & 0.0033 & 0.0140 & 0.0612 & 0.0534 & 0.0704 & 0.0723 & 0.8016 & 0.2880 & -0.0021 & 0.0793\\
\noalign{\global\arrayrulewidth1pt}\hline \noalign{\global\arrayrulewidth0.4pt}    
            \end{tabular}

   			\begin{tablenotes}
				\item[] See notes in Table \ref{table5}.
			\end{tablenotes}
		\end{threeparttable}

\end{table}

	Next, we conduct empirical exercises to study the portfolios with both TE and inequality weight constraints. Specifically, we select the healthcare/energy assets from the same S\&P 500 asset pool, and impose the weight constraint such that our portfolio invests at least 80\% of its assets in the themed assets. The number of energy sector stocks is 24, and the number of healthcare stocks is 49.
	The results are shown in Tables \ref{tablehc} and \ref{tableeg}. We can observe that the proposed portfolios (CROWN) show overall good performances in average return, risk, and Sharpe Ratio. First, the Sharpe Ratio of the proposed CROWN is the best among all portfolios in both cases. In the healthcare case, our method also has the lowest risk as well among all methods.
	For the healthcare index case, the SR tests generally do not reject the null of indifferent performance. For the energy index case, the p-values are mostly significant at the 10\% level, suggesting the statistical significance of the CROWN's better performance. The shrinkage-based portfolios also outperformed the energy index. Finally, the average returns of our proposed portfolio is also the top performer in the energy sector.

\section{Conclusion and Discussion}

In this paper, we pioneer the development of the  high-dimensional constrained portfolio weights estimator using a novel statistical  technique. Our analysis demonstrates the consistency of the estimator, even in scenarios where the number of assets surpasses the portfolio's time span. We provide rate of convergence results for scenarios involving TE constraints or joint TE and inequality-equality weight constraints.



{\bf Disclosure Statement}

The authors report there are no competing interests to declare.


{





\begin{center}
{\bf \Large Appendix}
\end{center}

We have two parts to Appendix. Specifically, Section A.1 covers simulations, 
Section A.2 covers the proof of Theorem 1 in the main text, Section A.3 covers the relation between TE and risk parameters, Section A.4 covers the proof of Theorem 2 in the main text, Section A.5 covers the proof of Theorem 3 in the main text. Section A.6 covers the case of only weight constraints. Section A.7 analyzes the unconstrained portfolio and compares with the constrained portfolios. Section A.8 considers the short-sales constraint. Section A.9 covers extra empirics. Appendix B covers two possibilities of bounded Sharpe Ratio, (maximum Sharpe Ratio portfolio), with bounded variance of Global Minimum Variance assumption, and with scaled Global Minimum variance, in Section B.1,  also we cover the feasible nodewise regression in Section B.2. These two sections involve extensive proofs. Section B.3 provides consistent estimation of the risk tolerance parameter. Section B.4 discusses the common rate across different restrictions. Section B.5 contains extra simulation results for dense precision matrix of the errors.





\newpage

\setcounter{section}{0}\renewcommand{\thesection}{A.\arabic{section}}
\setcounter{equation}{0}\setcounter{lemma}{0}\renewcommand{\theequation}{A.\arabic{equation}}\renewcommand{\thelemma}{A.\arabic{lemma}}
\setcounter{table}{0}\renewcommand{\thetable}{A.\arabic{table}}

\begin{center}
{\bf \Large Appendix A} 
\end{center}

\section{Extra Simulations}\label{simu}

In this section we try to analyze our estimator in relation to the Theorems, and also compare with existing methods. But we should emphasize that other methods do not have theories for constrained high dimensional portfolio formation.

 We provide simulations in this section  that is not covered in the main text. First, we  show simulations with TE constraint  at $p=160, T=80$, $p=160, T=200$. Then 
 we use  $p,T$ combinations such as $p=80, 160, 320$ with $p/T=2, p/T=0.8$ combinations  with  TE+weight, inequality constraints, and weight constraints. 
 

\subsection{Design}

The excess return data $\{\b y_t\}_{t=1}^{T}$ is generated from a standard three-factor model whose parameters are calibrated from the real data.  
 The excess asset return at time t, for j-th asset is:
\begin{equation}\label{simeq1}
y_{j,t} = b_{j,1}f_{1,t} + b_{j,2}f_{2,t} + b_{j,3}f_{3,t}+u_{j,t}, \quad j = 1,...,p; \ t=1,...T.   
\end{equation}
Specifically, we assume the factors $\b f_t=(f_{1,t}, f_{2,t}, f_{3,t})'$ follow a stationary VAR(1) model, $\b f_t=\b c_f+\b \Pi_f \b f_{t-1}+\b e_t,$ with errors following a normal distribution with zero mean with Covariance matrix of $\b e_t$ is $\b \Sigma_f - \b \Pi_f \b \Sigma_f \b \Pi_f'$. We show the proof below.
Note that we can also write the stationary VAR(1) model as in (10.1.4) and (10.1.8) of \cite{jh1994}, since $\b f_t$ is stationary so $ \b E \b f_t = \b E \b f_{t-1}:= \b \mu_1$
\[ \b f_t - \b \mu_1= \b \Pi_f (\b f_{t-1} - \b \mu_1) + \b e_t,\]

To derive covariance matrix of errors, 
\begin{eqnarray*}
E [ \b e_t \b e_t'] & = & E \left[ \left\{ (\b f_t -  \b \mu_1 ) - \b \Pi_f (\b f_{t-1} - \b \mu_1)]  \right\} 
\left\{ (\b f_t -  \b \mu_1 ) - \b \Pi_f (\b f_{t-1} - \b \mu_1)]  \right\} \right]' \\
& = & \b \Sigma_f - \b \Pi_f E \left[ (\b f_{t-1} - \b \mu_1) ( \b f_t - \b \mu_1)' \right] \\
& - & E [(\b f_t - \b \mu_1) (\b f_{t-1}- \b \mu_1)'] \b \Pi_f' \\
& + & \b \Pi_f \b \Sigma_f \b \Pi_f' \\
& = & \b \Sigma_f - 2 \b \Pi_f \b \Sigma_f \b \Pi_f' + \b \Pi_f \b \Sigma_f \b \Pi_f' = \b \Sigma_f - \b \Pi_f \b \Sigma_f \b \Pi_f' \end{eqnarray*}
where we use stationarity of $\b f_t$ with $\b f_t - \b \mu_1= \b \Pi_f (\b f_{t-1} - \b \mu_1) + \b e_t$ used to get third equality with $\b E \b f_{t-1} \b e_t =0$.

For factor loadings $\b b_j = \left( b_{j,1}, b_{j,2} , b_{j,3} \right)^\top$, $j = 1,2,\cdots,p$, and the idiosyncratic errors $\b u_{t} = (u_{1,t},\ldots,u_{p,t})^\top$, we draw their values from normal distributions. Specifically, $\b b_j \sim N_3(\b \mu_{b},\b \Sigma_{b})$; and $\b u_{t} \sim N(0,\b \Sigma_{u})$ with $\b \Sigma_u$ has the Toeplitz form such that the $(i,j)$-th element of $\b \Sigma_u$ is $0.25^{|i-j|}$.
Factor loadings have the following mean and variance-covariance matrix in Table \ref{bjsim}.
\begin{table}[htbp!]
	\scriptsize
	\centering
    	\caption{Parameters of $ \b b_j $}
    	\label{bjsim}
\begin{tabular}{c|ccc }
\hline $\b \mu_b$ & \multicolumn{3}{c}{$\b \Sigma_b$} \\
\hline 1.0166 & 0.0089 & 0.0013 & 0.0046 \\
0.5799 & 0.0013 & 0.2188 & -0.0134 \\
0.2937 & 0.0046 & -0.0134 & 0.1491 \\
\hline
\end{tabular} 
\end{table}

Then we set the parameters of the design in  factors  and errors as in Table \ref{ftsim} with $\b \Sigma_f$ as the covariance matrix of $\b f_t$.

\begin{table}[htbp!]
	\scriptsize
	\centering
    	\caption{Parameters of $\b f_t$}
    	\label{ftsim}
\begin{tabular}{c|ccc|ccc}
\hline $\b c_f$ & & $ \b \Sigma_f  $ & & \multicolumn{3}{c}{$\b \Pi_f$} \\
\hline 0.0445 & 1.5016 & 0.1338 & 0.1682 & -0.1204 & 0.1555 & -0.0324 \\
0.0060 & 0.1338 & 0.3667 & -0.0310 & -0.0074 & -0.0378 & 0.00318 \\
0.0021 & 0.1682 & -0.0310 & 0.6017 & -0.0027 & 0.0031 & 0.01669 \\
\hline
\end{tabular}
\end{table}

Tables \ref{bjsim}-\ref{ftsim} are obtained through the following  procedure. We utilize the daily data on returns of 100 industrial portfolios, which are intersections of 10 portfolios formed on size and 10 portfolios formed on book-to-market ratio, and the Fama/French 3 factors from Kenneth French's website. We use 20-year data from 2003-01-01 to 2022-12-31 for calibration.

\subsection{Results: Estimation Errors}

First, we start by introducing the various methods/indexes of comparison.
The benchmark market index $\b m$ is an equal-weighted portfolio. We also report two more benchmark portfolios, 1), the ``Oracle'', which is the portfolio given the population expected return $\b \mu$ and covariance $\b \Sigma_y$ and plug in the analytical solution for CROWN, given the required constraints; 2), the portfolio with no constraint, ``NCON'', which is defined in  Appendix A.7. We also report other popular methods for covariance/precision matrix estimation, specifically, they will be  denoted as the comparison portfolios, the nodewise estimator \cite{caner2019} will be called NW, the  POET of  \cite{fan2013}, the  nonlinear shrinkage estimator, and its single factor version  \cite{lw2017} will be abbreviated as  NLS, SFNL respectively. We construct these comparison portfolios using the covariance estimators from their respective methods.

It should be emphasized that the CROWN method uses different formulae under different constraints scenarios. 
To be more specific, for TE constraint only, we use $\hat{\b w}_1=\hat{\b w}_d+\b m$ and $\b w_1^*=\b w^*_d+\b m$; for TE and weight (equality) constraints, we use $\hat{\b w}_1=\hat{\b w}_{cp}+ \b m_R$ and $w_1^*=\b w^*_{cp}+\b m_R$.  For TE and weight inequality constraints, we use $\hat{\b w}_1 = \hat{\b w}_d + \b m$ if $\kappa \hat{w}_u < \omega$ and $\hat{\b w}_R = \hat{\b w}_{cp} + \b  m_R$ if $\kappa \hat{w}_u \ge \omega$. The theoretical $\b w_1^*=\b w^*_{d}+\b m$ if $\kappa w_u < \omega$, and $\b w_1^*= \b w_{cp}^* +  \b m_R$ if $\kappa w_u \ge \omega$. For the weight only constraint, $\hat{\b w}_1=\hat{\b w}_c$ defined in the Appendix A.6, and $\b w_1^*=\b w_c^*$, and notice in this case we do not track any index benchmark. For the TE constraints, we set the target TE to 0.1, 0.2, and 0.3, respectively; the weight restriction (if any) is imposed on the first $10$ stocks, that is, $1_R = (\underbrace{1,\dots,1}_{1\times 10},0,\ldots,0)'$, and the binding restricted weight is set such that $\b 1_R' \b w=w_x=0.2$. 

The other metrics and their estimates,  and estimation errors of risk, Sharpe Ratio, weights are defined as in Theorems 1-3. We show the best portfolio among the comparison portfolios (portfolios except Oracle, NCON, Index portfolios) with bold letter. Note that $TE$ represents the estimated TE, $SR$ represents the Sharpe Ratio estimate, and $Weight-ER, Risk-ER, SR-ER$ represents the weight estimation errors, risk estimation errors, and Sharpe Ratio estimation errors respectively. In the next subsection we also provide estimates of all metrics across 200 iterations. 

Table \ref{t160} shows that CROWN has the lowest estimation errors among the comparison portfolios. Also as discussed after Theorem 1, CROWN has lower errors than NCON, due to benchmark portfolio not being estimated. Tables \ref{t80w}-\ref{t320w} show the joint TE plus weight equality constraints effects on estimators of several metrics. Estimation errors of CROWN decrease when we move from $p=80, T=40$ in Table \ref{t80w} to Table \ref{t320w} with $p=320, T=160$. To give an example, at $TE=0.1$ risk estimation error is 0.0720 in Table \ref{t80w} to 0.0009 in Table \ref{t320w}. Also in Tables \ref{t160w}-\ref{t320w},
 CROWN has the lowest estimation errors in all metrics among comparison portfolios. Also by analyzing the average SR estimate, our method has the highest average SR among comparison portfolios in Tables \ref{t160w}-\ref{t320w}. Next, Tables \ref{t80iw}-\ref{t320iw} show the inequality constraints effect on estimators.
 Compared with other Tables, we have both binding (for weight constraint) and non-binding constraint cases. Our method performs very well at Tables \ref{t160iw}-\ref{t320iw} in terms of estimation errors. In Table \ref{t320iw}, we have both binding ($TE=0.2, 0.3$) cases, and non-binding case with $TE=0.1$. In the non-binding case with $p=320, T=160$ case, at $TE=0.1$, our Sharpe Ratio estimation error is 0.0003 and has the lowest one among comparison portfolios. Then in the binding case, with $TE=0.3$ at $p=320, T=160$, the estimation error for Sharpe Ratio is very low at 0.0012.  Also we observe that all our estimation errors decline when we move from  Table \ref{t80iw} to Table \ref{t320iw}. We see that similar patterns are present in only weight Tables \ref{80w}-\ref{320w}.

 \subsection{ Averages}
 
 In this part of simulations we report our average estimates for all metrics. These are in Tables \ref{et80}-\ref{ew320}.
 In Tables \ref{et160}-\ref{et320} CROWN has the highest SR on average, ranging between 0.0352-0.0354. In Table \ref{et80}, POET has the best average return at 0.0655 in TE=0.3, p=80, T=40. SFNL has the best average risk at 1.8282. We see very similar results in other tables.

\begin{table}
    	\scriptsize
        \centering
        \caption{Simulations with TE constraint, p=160}
        \label{t160}
\begin{tabular}{lccccccccccc}		\noalign{\global\arrayrulewidth1pt}\hline \noalign{\global\arrayrulewidth0.4pt}
            & \multicolumn{5}{c}{p=160, T=80}            &  & \multicolumn{5}{c}{p=160, T=200}        \\
           \cline{2-6} \cline{8-12}
& \multicolumn{11}{c}{TE = 0.1} \\ \hline
               & TE	& Weight-ER	& Risk-ER & SR-ER &	SR & & TE	& Weight-ER	& Risk-ER & SR-ER &	SR\\
Oracle & 0.1000 & 0.0000 & 0.0000 & 0.0000 & 0.0352 &   & 0.1000 & 0.0000 & 0.0000 & 0.0000 & 0.0352\\
NCON & 0.3844 & 1.2463 & 0.2998 & 0.0661 & 0.0363 &   & 0.3872 & 1.2403 & 0.2998 & 0.0665 & 0.0364\\
Index & - & 0.5633 & 0.1747 & 0.0910 & 0.0336 &   & - & 0.5633 & 0.1776 & 0.0619 & 0.0341\\ \hline
NW & 0.4134 & 1.4152 & 0.4388 & 0.0611 & 0.0341 &   & 0.4181 & 1.4383 & 0.4390 & 0.0501 & 0.0343\\
CROWN & \textbf{0.0994} & \textbf{0.1993} & \textbf{0.0012} & \textbf{0.0005} & \textbf{0.0352} &   & \textbf{0.0996} & \textbf{0.1728} & \textbf{0.0019} & \textbf{0.0005} & \textbf{0.0352}\\
\end{tabular}
        \end{table}

\clearpage

\begin{table}
    	\scriptsize
        \centering
        \caption{Simulations with TE and weight equality constraint, p=80}
        \label{t80w}
        \begin{tabular}{lccccccccccc}		\noalign{\global\arrayrulewidth1pt}\hline \noalign{\global\arrayrulewidth0.4pt}
            & \multicolumn{5}{c}{p=80, T=40}            &  & \multicolumn{5}{c}{p=80, T=100}        \\
           \cline{2-6} \cline{8-12}
& \multicolumn{11}{c}{TE = 0.1} \\ \hline
               & TE	& Weight-ER	& Risk-ER & SR-ER &	SR & & TE	& Weight-ER	& Risk-ER & SR-ER &	SR\\
Oracle & 0.1000 & 0.0000 & 0.0000 & 0.0000 & 0.0350 &   & 0.1000 & 0.0000 & 0.0000 & 0.0000 & 0.0350\\
NCON & 0.4769 & 2.0857 & 0.2446 & 0.0248 & 0.0351 &   & 0.3788 & 1.2718 & 0.2711 & 0.0651 & 0.0361\\
Index & - & 0.5124 & 0.1660 & 0.0570 & 0.0339 &   & - & 0.5124 & 0.1693 & 0.0255 & 0.0345\\ \hline
NW & 2.0613 & 6.8420 & 11.9091 & 0.3450 & 0.0275 &   & 0.2880 & 0.8725 & 0.2409 & 0.0345 & 0.0343\\
CROWN & 0.1879 & 0.9330 & \textbf{0.0720} & \textbf{0.0060} & \textbf{0.0349} &   & \textbf{0.1028} & \textbf{0.1725} & \textbf{0.0031} & \textbf{0.0006} & \textbf{0.0349}\\
\end{tabular}
        \end{table}

\clearpage

\begin{table}
    	\scriptsize
        \centering
        \caption{Simulations with TE and weight equality constraint, p=160}
        \label{t160w}
        \begin{tabular}{lccccccccccc}		\noalign{\global\arrayrulewidth1pt}\hline \noalign{\global\arrayrulewidth0.4pt}
            & \multicolumn{5}{c}{p=160, T=80}            &  & \multicolumn{5}{c}{p=160, T=200}        \\
           \cline{2-6} \cline{8-12}
& \multicolumn{11}{c}{TE = 0.1} \\ \hline
               & TE	& Weight-ER	& Risk-ER & SR-ER &	SR & & TE	& Weight-ER	& Risk-ER & SR-ER &	SR\\
Oracle & 0.1000 & 0.0000 & 0.0000 & 0.0000 & 0.0352 &   & 0.1000 & 0.0000 & 0.0000 & 0.0000 & 0.0352\\
NCON & 0.3844 & 1.3442 & 0.3007 & 0.0673 & 0.0363 &   & 0.3872 & 1.3386 & 0.3007 & 0.0676 & 0.0364\\
Index & - & 0.6914 & 0.1732 & 0.0900 & 0.0336 &   & - & 0.6914 & 0.1761 & 0.0609 & 0.0341\\ \hline
NW & 0.4170 & 1.5137 & 0.4341 & 0.0644 & 0.0340 &   & 0.4173 & 1.4530 & 0.4327 & 0.0491 & 0.0343\\
CROWN & 0.1125 & \textbf{0.2300} & \textbf{0.0017} & \textbf{0.0008} & \textbf{0.0352} &   & \textbf{0.1130} & \textbf{0.1999} & \textbf{0.0025} & \textbf{0.0008} & \textbf{0.0352}\\
\end{tabular}
        \end{table}

\clearpage

\begin{table}
    	\scriptsize
        \centering
        \caption{Simulations with TE and weight equality constraint, p=320}
        \label{t320w}
        \begin{tabular}{lccccccccccc}		\noalign{\global\arrayrulewidth1pt}\hline \noalign{\global\arrayrulewidth0.4pt}
            & \multicolumn{5}{c}{p=320, T=160}            &  & \multicolumn{5}{c}{p=320, T=400}        \\
           \cline{2-6} \cline{8-12}
& \multicolumn{11}{c}{TE = 0.1} \\ \hline
               & TE	& Weight-ER	& Risk-ER & SR-ER &	SR & & TE	& Weight-ER	& Risk-ER & SR-ER &	SR\\
Oracle & 0.1000 & 0.0000 & 0.0000 & 0.0000 & 0.0352 &   & 0.1000 & 0.0000 & 0.0000 & 0.0000 & 0.0352\\
NCON & 0.3988 & 1.4983 & 0.3129 & 0.0724 & 0.0365 &   & 0.3975 & 1.5081 & 0.3128 & 0.0727 & 0.0365\\
Index & - & 0.7911 & 0.1561 & 0.0682 & 0.0340 &   & - & 0.7911 & 0.1591 & 0.0383 & 0.0346\\ \hline
NW & 0.6557 & 2.5519 & 0.8209 & 0.0818 & 0.0338 &   & 0.6713 & 2.7238 & 0.8506 & 0.0685 & 0.0340\\
CROWN & 0.1199 & \textbf{0.2495} & \textbf{0.0009} & \textbf{0.0005} & \textbf{0.0352} &   & \textbf{0.1197} & \textbf{0.1872} & \textbf{0.0008} & \textbf{0.0004} & \textbf{0.0352}\\
\end{tabular}
        \end{table}

\clearpage

\begin{table}
    	\scriptsize
        \centering
        \caption{Simulations with TE constraint and inequality constraint, p=80}
        \label{t80iw}
        \begin{tabular}{lccccccccccc}		\noalign{\global\arrayrulewidth1pt}\hline \noalign{\global\arrayrulewidth0.4pt}
            & \multicolumn{5}{c}{p=80, T=40}            &  & \multicolumn{5}{c}{p=80, T=100}        \\
           \cline{2-6} \cline{8-12}
& \multicolumn{11}{c}{TE = 0.1 (Non-binding)} \\ \hline
               & TE	& Weight-ER	& Risk-ER & SR-ER &	SR & & TE	& Weight-ER	& Risk-ER & SR-ER &	SR\\
Oracle & 0.1000 & 0.0000 & 0.0000 & 0.0000 & 0.0349 &   & 0.1000 & 0.0000 & 0.0000 & 0.0000 & 0.0349\\
NCON & 0.4769 & 2.0968 & 0.2452 & 0.0257 & 0.0351 &   & 0.3788 & 1.2692 & 0.2716 & 0.0670 & 0.0361\\
Index & - & 0.4698 & 0.1665 & 0.0570 & 0.0339 &   & - & 0.4698 & 0.1697 & 0.0255 & 0.0345\\ \hline
NW & 0.4044 & 2.1621 & 10.4782 & 0.0166 & 0.0346 &   & 0.2971 & 0.8971 & 0.2522 & 0.0376 & 0.0343\\
CROWN & 0.1934 & 0.9920 & \textbf{0.0733} & \textbf{0.0064} & \textbf{0.0348} &   & 0.1105 & \textbf{0.1770} & \textbf{0.0019} & \textbf{0.0006} & \textbf{0.0349}\\
\end{tabular}
        \end{table}

\clearpage

\begin{table}
    	\scriptsize
        \centering
        \caption{Simulations with TE constraint and inequality constraint, p=160}
        \label{t160iw}
        \begin{tabular}{lccccccccccc}		\noalign{\global\arrayrulewidth1pt}\hline \noalign{\global\arrayrulewidth0.4pt}
            & \multicolumn{5}{c}{p=160, T=80}            &  & \multicolumn{5}{c}{p=160, T=200}        \\
           \cline{2-6} \cline{8-12}
& \multicolumn{11}{c}{TE = 0.1 (Non-binding)} \\ \hline
               & TE	& Weight-ER	& Risk-ER & SR-ER &	SR & & TE	& Weight-ER	& Risk-ER & SR-ER &	SR\\
Oracle & 0.1000 & 0.0000 & 0.0000 & 0.0000 & 0.0352 &   & 0.1000 & 0.0000 & 0.0000 & 0.0000 & 0.0352\\
NCON & 0.3844 & 1.2770 & 0.3001 & 0.0665 & 0.0363 &   & 0.3872 & 1.2710 & 0.3004 & 0.0668 & 0.0364\\
Index & - & 0.5633 & 0.1747 & 0.0910 & 0.0336 &   & - & 0.5633 & 0.1776 & 0.0619 & 0.0341\\ \hline
NW & 0.4168 & 1.439 & 0.4417 & 0.0616 & 0.0341 &   & 0.4193 & 1.4448 & 0.4400 & 0.0501 & 0.0343\\
CROWN & 0.1025 & \textbf{0.2041} & \textbf{0.0010} & \textbf{0.0004} & \textbf{0.0352} &   & \textbf{0.1025} & \textbf{0.1781} & \textbf{0.0017} & \textbf{0.0005} & \textbf{0.0352}\\
\end{tabular}
        \end{table}

\clearpage

\begin{table}
    	\scriptsize
        \centering
        \caption{Simulations with TE constraint and inequality constraint, p=320}
        \label{t320iw}
        \begin{tabular}{lccccccccccc}		\noalign{\global\arrayrulewidth1pt}\hline \noalign{\global\arrayrulewidth0.4pt}
            & \multicolumn{5}{c}{p=320, T=160}            &  & \multicolumn{5}{c}{p=320, T=400}        \\
           \cline{2-6} \cline{8-12}
& \multicolumn{11}{c}{TE = 0.1 (Non-binding)} \\ \hline
               & TE	& Weight-ER	& Risk-ER & SR-ER &	SR & & TE	& Weight-ER	& Risk-ER & SR-ER &	SR\\
Oracle & 0.1000 & 0.0000 & 0.0000 & 0.0000 & 0.0353 &   & 0.1000 & 0.0000 & 0.0000 & 0.0000 & 0.0353\\
NCON & 0.3988 & 1.3878 & 0.3109 & 0.0706 & 0.0365 &   & 0.3975 & 1.3995 & 0.3108 & 0.0708 & 0.0365\\
Index & - & 0.6366 & 0.1593 & 0.0699 & 0.0340 &   & - & 0.6366 & 0.1624 & 0.0400 & 0.0346\\ \hline
NW & 0.6475 & 2.4806 & 0.8141 & 0.0812 & 0.0338 &   & 0.6652 & 2.6548 & 0.8459 & 0.0687 & 0.0340\\
CROWN & \textbf{0.1009} & \textbf{0.2165} & \textbf{0.0009} & \textbf{0.0003} & \textbf{0.0353} &   & \textbf{0.1005} & \textbf{0.1593} & \textbf{0.0007} & \textbf{0.0002} & \textbf{0.0353}\\
\end{tabular}
        \end{table}

\clearpage

\begin{table}
    	\scriptsize
        \centering
        \caption{Simulation results for weight constraint, p=80}
        \label{80w}
        \begin{tabular}{lccccccccccc}		\noalign{\global\arrayrulewidth1pt}\hline \noalign{\global\arrayrulewidth0.4pt}
            & \multicolumn{5}{c}{p=80, T=40}            &  & \multicolumn{5}{c}{p=80, T=100}        \\
           \hline
& TE	& Weight-ER	& Risk-ER & SR-ER &	SR & & TE	& Weight-ER	& Risk-ER & SR-ER &	SR \\
Oracle & 0.5770 & 0.0000 & 0.0000 & 0.0000 & 0.0349 &   & 0.5018 & 0.0000 & 0.0000 & 0.0000 & 0.0355\\
NCON & 0.4769 & 2.2332 & 0.3878 & 0.0243 & 0.0351 &   & 0.3698 & 1.3761 & 0.3969 & 0.0333 & 0.0361\\
Index & - & 1.7159 & 1.4293 & 0.0528 & 0.0339 &   & - & 1.7159 & 1.1482 & 0.0599 & 0.0345\\ \hline
NW & 1.4912 & 5.7520 & 35.9307 & 0.2618 & 0.0295 &   & 1.8087 & 9.6189 & 7.5540 & 0.2245 & 0.0313\\
CROWN & 0.6963 & \textbf{2.9710} & \textbf{0.0905} & 0.2306 & 0.0311 &   & 0.4987 & \textbf{0.5734} & \textbf{0.0165} & \textbf{0.0030} & \textbf{0.0355}\\
POET & 1.2035 & 9.4276 & 3.9704 & 0.2855 & 0.0300 &   & 1.0459 & 5.9578 & 1.5047 & 0.2173 & 0.0314\\
NLS & 1.0737 & 8.1873 & 3.2651 & 0.2498 & 0.0307 &   & 0.9162 & 6.2965 & 1.5416 & 0.2236 & 0.0313\\
SFNL & 1.1701 & 7.2141 & 2.7521 & \textbf{0.2214} & \textbf{0.0313} &   & 0.9201 & 5.5130 & 1.3481 & 0.1857 & 0.0320\\ \noalign{\global\arrayrulewidth1pt}\hline \noalign{\global\arrayrulewidth0.4pt}    
        \end{tabular}
        \end{table}

\begin{table}
    	\scriptsize
        \centering
        \caption{Simulation results for weight constraint, p=160}
        \label{160w}
        \begin{tabular}{lccccccccccc}		\noalign{\global\arrayrulewidth1pt}\hline \noalign{\global\arrayrulewidth0.4pt}
            & \multicolumn{5}{c}{p=160, T=80}            &  & \multicolumn{5}{c}{p=160, T=200}        \\
           \hline
& TE	& Weight-ER	& Risk-ER & SR-ER &	SR & & TE	& Weight-ER	& Risk-ER & SR-ER &	SR \\
Oracle & 0.5293 & 0.0000 & 0.0000 & 0.0000 & 0.0359 &   & 0.5384 & 0.0000 & 0.0000 & 0.0000 & 0.0359\\
NCON & 0.3844 & 1.5666 & 0.4575 & 0.0270 & 0.0363 &   & 0.3872 & 1.5503 & 0.4576 & 0.0273 & 0.0364\\
Index & - & 1.9453 & 1.4454 & 0.1243 & 0.0336 &   & - & 1.9453 & 1.4514 & 0.0964 & 0.0341\\ \hline
NW & 3.0635 & 16.2695 & 19.1781 & 0.2888 & 0.0302 &   & 3.0506 & 16.4066 & 17.4211 & 0.2421 & 0.0312\\
CROWN & 0.5303 & \textbf{0.7040} & \textbf{0.0096} & \textbf{0.0040} & \textbf{0.0358} &   & 0.5374 & \textbf{0.6111} & \textbf{0.0114} & \textbf{0.0024} & \textbf{0.0358}\\
POET & 1.3021 & 15.4234 & 6.2715 & 0.3064 & 0.0299 &   & 1.1517 & 9.3312 & 2.2483 & 0.2542 & 0.0310\\
NLS & 1.0423 & 14.0260 & 5.6680 & 0.2724 & 0.0306 &   & 0.9432 & 11.063 & 2.6127 & 0.2790 & 0.0304\\
SFNL & 1.1963 & 11.0378 & 4.1314 & 0.2199 & 0.0317 &   & 0.9306 & 8.0252 & 1.8221 & 0.1912 & 0.0323\\ \noalign{\global\arrayrulewidth1pt}\hline \noalign{\global\arrayrulewidth0.4pt}    
        \end{tabular}
        \end{table}

\begin{table}
    	\scriptsize
        \centering
        \caption{Simulation results for weight constraint, p=320}
        \label{320w}
        \begin{tabular}{lccccccccccc}		\noalign{\global\arrayrulewidth1pt}\hline \noalign{\global\arrayrulewidth0.4pt}
            & \multicolumn{5}{c}{p=320, T=160}            &  & \multicolumn{5}{c}{P=320, T=400}        \\
           \hline
& TE	& Weight-ER	& Risk-ER & SR-ER &	SR & & TE	& Weight-ER	& Risk-ER & SR-ER &	SR \\
Oracle & 0.5539 & 0.0000 & 0.0000 & 0.0000 & 0.0360 &   & 0.5540 & 0.0000 & 0.0000 & 0.0000 & 0.0360\\
NCON & 0.3988 & 1.8975 & 0.5501 & 0.0244 & 0.0365 &   & 0.3975 & 1.8755 & 0.5503 & 0.0246 & 0.0365\\
Index & - & 2.3303 & 1.6079 & 0.1099 & 0.0340 &   & - & 2.3303 & 1.6148 & 0.0814 & 0.0346\\ \hline
NW & 5.1879 & 26.7324 & 44.3432 & 0.3128 & 0.0299 &   & 5.3522 & 28.4937 & 45.2845 & 0.2672 & 0.0309\\
CROWN & 0.5543 & \textbf{0.7506} & \textbf{0.0068} & \textbf{0.0025} & \textbf{0.0360} &   & 0.5536 & \textbf{0.5565} & \textbf{0.0050} & \textbf{0.0014} & \textbf{0.0360}\\
POET & 5.2928 & 127.7871 & 1.2E3 & 0.6890 & 0.0185 &   & 7.2152 & 95.7614 & 1.2E3 & 0.6268 & 0.0205\\
NLS & 1.0239 & 25.9676 & 10.255 & 0.2963 & 0.0302 &   & 0.9647 & 20.0888 & 4.3202 & 0.3397 & 0.0293\\
SFNL & 1.2177 & 16.5203 & 5.5261 & 0.2076 & 0.0321 &   & 0.9202 & 11.5878 & 2.2774 & 0.1910 & 0.0324\\ \noalign{\global\arrayrulewidth1pt}\hline \noalign{\global\arrayrulewidth0.4pt}    
        \end{tabular}
        \end{table}

\clearpage

\begin{table}
    \scriptsize
    \centering
    \caption{Simulations Average: TE constraint, p=80}
	\label{et80}
    \begin{tabular}{lccccccccc}		
    \noalign{\global\arrayrulewidth1pt}\hline \noalign{\global\arrayrulewidth0.4pt}
    
    & \multicolumn{4}{c}{p=80, T=40}            &  & \multicolumn{4}{c}{p=80, T=100}        \\
    \cline{2-5} \cline{7-10}
& \multicolumn{9}{c}{TE = 0.1} \\ \hline
           & Return      & Risk & SR   & SR-SD &  &  Return      & Risk & SR & SR-SD   \\
Oracle & 0.0512 & 1.4652 & 0.0350 & 0.0000 &  & 0.0512 & 1.4652 & 0.0350 & 0.0000\\
NCON & 0.0447 & 1.2736 & 0.0352 & 0.0005 &  & 0.0451 & 1.2512 & 0.0361 & 0.0000\\
Index & 0.0537 & 1.5825 & 0.0339 & 0.0000 &  & 0.0547 & 1.5847 & 0.0345 & 0.0000\\ \hline
NW & 0.0515 & 1.7622& 0.0346 & 0.0000 &   & \textbf{0.0561} & 1.6332 & 0.0344 & 0.0001\\
CROWN & 0.0529 & \textbf{1.5168} & \textbf{0.0349} & 0.0047 &   & 0.0511 & \textbf{1.4633} & \textbf{0.0349} & 0.0000\\
\end{tabular}
\end{table}

\clearpage

\begin{table}
\scriptsize
    \centering
    \caption{Simulations Average: TE constraint, p=160}
	\label{et160}
    \begin{tabular}{lccccccccc}		\noalign{\global\arrayrulewidth1pt}\hline \noalign{\global\arrayrulewidth0.4pt}
        & \multicolumn{4}{c}{p=160, T=80}            &  & \multicolumn{4}{c}{p=160, T=200}        \\
       \cline{2-5} \cline{7-10}
& \multicolumn{9}{c}{TE = 0.1} \\ \hline
           & Return      & Risk & SR & SR-SD  &  &  Return      & Risk & SR & SR-SD    \\
Oracle & 0.0519 & 1.4758 & 0.0352 & 0.0000 &  & 0.0519 & 1.4758 & 0.0352 & 0.0000\\
NCON & 0.0449 & 1.2349 & 0.0363 & 0.0000 &  & 0.0449 & 1.2349 & 0.0364 & 0.0000\\
Index & 0.0537 & 1.5995 & 0.0336 & 0.0000 &  & 0.0546 & 1.6015 & 0.0341 & 0.0000\\ \hline
NW & \textbf{0.0604} & 1.7697 & 0.0341 & 0.0002 &  & \textbf{0.0607} & 1.7702 & 0.0343 & 0.0001\\
CROWN & 0.0519 & \textbf{1.4749} & \textbf{0.0352} & 0.0000 &  & 0.0519 & \textbf{1.4743} & \textbf{0.0352} & 0.0000\\
\end{tabular}
        \end{table}

\clearpage

\begin{table}
    \scriptsize
    \centering
    \caption{Simulations Average: TE constraint, p=320}
    \label{et320}
    \begin{tabular}{lccccccccc}		
    \noalign{\global\arrayrulewidth1pt}\hline \noalign{\global\arrayrulewidth0.4pt}
    
    & \multicolumn{4}{c}{p=320, T=160}            &  & \multicolumn{4}{c}{p=320, T=400}        \\
    \cline{2-5} \cline{7-10}
& \multicolumn{9}{c}{TE = 0.1} \\ \hline
           & Return      & Risk & SR & SR-SD  &  &  Return      & Risk & SR & SR-SD    \\
Oracle & 0.0523 & 1.4832 & 0.0353 & 0.0000 &   & 0.0523 & 1.4832 & 0.0353 & 0.0000\\
NCON & 0.0449 & 1.2312 & 0.0365 & 0.0000 &  & 0.0449 & 1.2313 & 0.0365 & 0.0000\\
Index & 0.0543 & 1.5970 & 0.0340 & 0.0000 &  & 0.0552 & 1.5991 & 0.0346 & 0.0000\\ \hline
NW & \textbf{0.0675} & 1.9986 & 0.0338 & 0.0014 &  & \textbf{0.0686} & 2.0159 & 0.0340 & 0.0001\\
CROWN & 0.0523 & \textbf{1.4825} & \textbf{0.0353} & 0.0000 &  & 0.0523 & \textbf{1.4826} & \textbf{0.0353} & 0.0000\\
\end{tabular}
        \end{table}

\clearpage

\begin{table}
        \scriptsize
        \centering
        \caption{Simulations Average: TE and weight constraints, p=80}
	   \label{etw80}
        \begin{tabular}{lccccccccc}		
        \noalign{\global\arrayrulewidth1pt}\hline \noalign{\global\arrayrulewidth0.4pt}
        
        & \multicolumn{4}{c}{p=80, T=40}            &  & \multicolumn{4}{c}{p=80, T=100}        \\
        \cline{2-5} \cline{7-10}
& \multicolumn{9}{c}{TE = 0.1} \\ \hline
               & Return      & Risk & SR & SR-SD  &  &  Return      & Risk & SR & SR-SD    \\
Oracle & 0.0512 & 1.4655 & 0.0350 & 0.0000 &  & 0.0512 & 1.4655 & 0.0350 & 0.0000\\
NCON & 0.0447 & 1.2736 & 0.0351 & 0.0005 &  & 0.0451 & 1.2512 & 0.0361 & 0.0000\\
Index & 0.0537 & 1.5825 & 0.0339 & 0.0000 &  & 0.0547 & 1.5847 & 0.0345 & 0.0001\\ \hline
NW & \textbf{0.0656} & 3.1233 & 0.0275 & 0.0053 &  & \textbf{0.0561} & 1.6323 & 0.0343 & 0.0000\\
CROWN & 0.0529 & 1.5173 & \textbf{0.0349} & 0.0001 &  & 0.0511 & 1.4633 & \textbf{0.0349} & 0.0001\\
\end{tabular}
\end{table}

\clearpage

\begin{table}
        \scriptsize
        \centering
        \caption{Simulations Average: TE and weight constraints, p=160}
		\label{etw160}
        \begin{tabular}{lccccccccc}		
        \noalign{\global\arrayrulewidth1pt}\hline \noalign{\global\arrayrulewidth0.4pt}
        
        & \multicolumn{4}{c}{p=160, T=80}            &  & \multicolumn{4}{c}{p=160, T=200}        \\
        \cline{2-5} \cline{7-10}
& \multicolumn{9}{c}{TE = 0.1} \\ \hline
               & Return      & Risk & SR & SR-SD  &  &  Return      & Risk & SR & SR-SD    \\
Oracle & 0.0520 & 1.4767 & 0.0352 & 0.0000 &  & 0.052 & 1.4767 & 0.0352 & 0.0000 \\
NCON & 0.0449 & 1.2349 & 0.0363 & 0.0000 &  & 0.0449 & 1.2349 & 0.0364 & 0.0000\\
Index & 0.0537 & 1.5995 & 0.0336 & 0.0000 &  & 0.0546 & 1.6015 & 0.0341 & 0.0001\\ \hline
NW & \textbf{0.0601} & 1.7679 & 0.0340 & 0.0020 &   & \textbf{0.0606} & 1.7675 & 0.0343 & 0.0000\\
CROWN & 0.0519 & 1.4754 & \textbf{0.0352} & 0.0000 &  & 0.0519 & 1.4748 & \textbf{0.0352} & 0.0001\\
\end{tabular}
        \end{table}

\clearpage
        
\begin{table}
        \scriptsize
        \centering
        \caption{Simulations Average: TE and weight constraints, p=320}
		\label{etw320}
        \begin{tabular}{lccccccccc}		
        \noalign{\global\arrayrulewidth1pt}\hline \noalign{\global\arrayrulewidth0.4pt}
        
        & \multicolumn{4}{c}{p=320, T=160}            &  & \multicolumn{4}{c}{p=320, T=400}        \\
        \cline{2-5} \cline{7-10}
& \multicolumn{9}{c}{TE = 0.1} \\ \hline
               & Return      & Risk & SR & SR-SD  &  &  Return      & Risk & SR & SR-SD   \\
Oracle & 0.0523 & 1.4853 & 0.0352 & 0.0000 &  & 0.0523 & 1.4853 & 0.0352 & 0.0000\\
NCON & 0.0449 & 1.2312 & 0.0365 & 0.0000 &  & 0.0449 & 1.2313 & 0.0365 & 0.0000\\
Index & 0.0543 & 1.5970 & 0.0340 & 0.0000 &  & 0.0552 & 1.5991 & 0.0346 & 0.0000\\ \hline
NW & \textbf{0.0677} & 2.0040 & 0.0338 & 0.0000 &  & \textbf{0.0687} & 2.0205 & 0.0340 & 0.0001\\
CROWN & 0.0523 & \textbf{1.4846} & \textbf{0.0352} & 0.0001 &  & 0.0523 & \textbf{1.4847} & \textbf{0.0352} & 0.0000\\
\end{tabular}
\end{table}

\clearpage

\begin{table}
        \scriptsize
        \centering
        \caption{Simulations Average: TE and inequality weight constraint, p=80}
		\label{etiw80}
        \begin{tabular}{lccccccccc}		
        \noalign{\global\arrayrulewidth1pt}\hline \noalign{\global\arrayrulewidth0.4pt}
        
        & \multicolumn{4}{c}{p=80, T=40}            &  & \multicolumn{4}{c}{p=80, T=100}        \\
        \cline{2-5} \cline{7-10}
& \multicolumn{9}{c}{TE = 0.1 (Non-binding)} \\ \hline
               & Return      & Risk & SR & SR-SD  &  &  Return      & Risk & SR & SR-SD    \\
Oracle & 0.0512 & 1.4652 & 0.0350 & 0.0000 &   & 0.0512 & 1.4652 & 0.0350 & 0.0000\\
NCON & 0.0447 & 1.2736 & 0.0351 & 0.0005 &  & 0.0451 & 1.2512 & 0.0361 & 0.0000\\
Index & 0.0537 & 1.5825 & 0.0339 & 0.0000 &  & 0.0547 & 1.5847 & 0.0345 & 0.0000\\ \hline
NW & 0.0515 & 1.7627 & 0.0346 & 0.0020 &  & \textbf{0.0561} & 1.6332 & 0.0344 & 0.0001 \\
CROWN & 0.0529 & \textbf{1.5168} & \textbf{0.0349} & 0.0001 &  & 0.0511 & \textbf{1.4633} & \textbf{0.0349} & 0.0000\\
\end{tabular}
\end{table}

\clearpage

\begin{table}
    	\scriptsize
        \centering
        \caption{Simulations Average: TE and inequality weight constraint, p=160}
\label{etiw160}
        \begin{tabular}{lccccccccc}		\noalign{\global\arrayrulewidth1pt}\hline \noalign{\global\arrayrulewidth0.4pt}
            & \multicolumn{4}{c}{p=160, T=80}            &  & \multicolumn{4}{c}{p=160, T=200}        \\
           \cline{2-5} \cline{7-10}
& \multicolumn{9}{c}{TE = 0.1 (Non-binding)} \\ \hline
               & Return      & Risk & SR & SR-SD  &  &  Return      & Risk & SR & SR-SD    \\
Oracle & 0.0519 & 1.4758 & 0.0352 & 0.0000 &   & 0.0519 & 1.4758 & 0.0352 & 0.0000\\
NCON & 0.0449 & 1.2349 & 0.0363 & 0.0000 &  & 0.0449 & 1.2349 & 0.0364 & 0.0000\\
Index & 0.0543 & 1.5970 & 0.0340 & 0.0000 &  & 0.0552 & 1.5991 & 0.0346 & 0.0000\\ \hline
NW & \textbf{0.0604} & 1.7697 & 0.0341 & 0.0002 &  & \textbf{0.0607} & 1.7702 & 0.0343 & 0.0001\\
CROWN & 0.0519 & \textbf{1.4749} & \textbf{0.0352} & 0.0000 &  & 0.0519 & \textbf{1.4743} & \textbf{0.0352} & 0.0000\\
\end{tabular}
\end{table}

\clearpage
\newpage

\begin{table}
    	\scriptsize
        \centering
        \caption{Simulations Average: TE and inequality weight constraint, p=320}
\label{etiw320}
        \begin{tabular}{lccccccccc}		\noalign{\global\arrayrulewidth1pt}\hline \noalign{\global\arrayrulewidth0.4pt}
            & \multicolumn{4}{c}{p=320, T=160}            &  & \multicolumn{4}{c}{p=320, T=400}        \\
           \cline{2-5} \cline{7-10}
& \multicolumn{9}{c}{TE = 0.1 (Non-binding)} \\ \hline
               & Return      & Risk & SR & SR-SD  &  &  Return      & Risk & SR & SR-SD\\
Oracle & 0.0523 & 1.4832 & 0.0353 & 0.0000 &  & 0.0523 & 1.4832 & 0.0353 & 0.0000\\
NCON & 0.0449 & 1.2312 & 0.0365 & 0.0000 &  & 0.0449 & 1.2313 & 0.0365 & 0.0000\\
Index & 0.0543 & 1.5970 & 0.0340 & 0.0000 &  & 0.0552 & 1.5991 & 0.0346 & 0.0000\\ \hline
NW & \textbf{0.0675} & 1.9981 & 0.0338 & 0.0001 &  & \textbf{0.0686} & 2.0157 & 0.0340 & 0.0001\\
CROWN & 0.0523 & \textbf{1.4825} & \textbf{0.0353} & 0.0000 &  & 0.0523 & \textbf{1.4826} & \textbf{0.0353} & 0.0000\\
\end{tabular}
\end{table}

\clearpage

\begin{table}
        \scriptsize
        \centering
        \caption{Simulation Average: weight constraint, p=80}
\label{ew80}
        \begin{tabular}{lccccccccc}		
        \noalign{\global\arrayrulewidth1pt}\hline \noalign{\global\arrayrulewidth0.4pt}
        
        & \multicolumn{4}{c}{p=80, T=40}            &  & \multicolumn{4}{c}{p=80, T=100}        \\
        \hline
& Return      & Risk & SR & SR-SD  &   &  Return      & Risk & SR  & SR-SD   \\
Oracle & 0.0354 & 1.0153 & 0.0349 & 0.0000 &  & 0.0384 & 1.0812 & 0.0355 & 0.0000\\
NCON & 0.0447 & 1.2736 & 0.0351 & 0.0005 &  & 0.0451 & 1.2512 & 0.0361 & 0.0000\\
Index & 0.0537 & 1.5825 & 0.0339 & 0.0000 &  & 0.0547 & 1.5847 & 0.0345 & 0.0000\\ \hline
NW & 0.0440 & 2.2366 & 0.0295 & 0.0058 &  & \textbf{0.0987} & 3.1578 & 0.0313 & 0.0005 \\
CROWN & 0.0341 & \textbf{1.0979} & 0.0311 & 0.0013 &  & 0.0387 & \textbf{1.0901} & \textbf{0.0355} & 0.0000\\
POET & \textbf{0.0718} & 2.3952 & 0.0300 & 0.0010 &  & 0.0536 & 1.7052 & 0.0314 & 0.0007\\
NLS & 0.0682 & 2.2199 & 0.0307 & 0.0008 &  & 0.0537 & 1.7172 & 0.0313 & 0.0006 \\
SFNL & 0.0653 & 2.0837 & \textbf{0.0313} & 0.0009 &  & 0.0529 & 1.6521 & 0.0320 & 0.0005\\ 
\noalign{\global\arrayrulewidth1pt}\hline \noalign{\global\arrayrulewidth0.4pt}  
\end{tabular}
\end{table}

\begin{table}
        \scriptsize
        \centering
        \caption{Simulation Average: weight constraint, p=160}
 \label{ew160}
        \begin{tabular}{lccccccccc}		
        \noalign{\global\arrayrulewidth1pt}\hline \noalign{\global\arrayrulewidth0.4pt}
        
        & \multicolumn{4}{c}{p=160, T=80}            &  & \multicolumn{4}{c}{p=160, T=200}        \\
        \hline
& Return      & Risk & SR & SR-SD  &  &  Return      & Risk & SR & SR-SD    \\
Oracle & 0.0367 & 1.0229 & 0.0359 & 0.0000 &   & 0.0367 & 1.0229 & 0.0359 & 0.0000\\
NCON & 0.0449 & 1.2349 & 0.0363 & 0.0000 &  & 0.0449 & 1.2349 & 0.0364 & 0.0000\\
Index & 0.0537 & 1.5995 & 0.0336 & 0.0000 &  & 0.0546 & 1.6015 & 0.0341 & 0.0000\\ \hline
NW & \textbf{0.1340} & 4.5029 & 0.0302 & 0.0021 &  & \textbf{0.1370} & 4.3881 & 0.0312 & 0.0003\\
CROWN & 0.0368 & \textbf{1.0277} & \textbf{0.0358} & 0.0000 &  & 0.0368 & \textbf{1.0287} & \textbf{0.0358} & 0.0000\\
POET & 0.0817 & 2.7388 & 0.0299 & 0.0011 &  & 0.0569 & 1.8367 & 0.0310 & 0.0007\\
NLS & 0.0803 & 2.6253 & 0.0306 & 0.0007 &   & 0.0590 & 1.9366 & 0.0304 & 0.0005\\
SFNL & 0.0730 & 2.3061 & 0.0317 & 0.0006 &  & 0.0553 & 1.7137 & 0.0323 & 0.0003\\ 
\noalign{\global\arrayrulewidth1pt}\hline \noalign{\global\arrayrulewidth0.4pt}    
\end{tabular}
\end{table}

\begin{table}
        \scriptsize
        \centering
        \caption{Simulation Average: weight constraint, p=320}
\label{ew320}
        \begin{tabular}{lccccccccc}		
        \noalign{\global\arrayrulewidth1pt}\hline \noalign{\global\arrayrulewidth0.4pt}
        
        & \multicolumn{4}{c}{p=320, T=160}  &  & \multicolumn{4}{c}{P=320, T=400}        \\
        \hline
& Return      & Risk & SR & SR-SD  &  &  Return      & Risk & SR & SR-SD    \\
Oracle & 0.0356 & 0.9889 & 0.0360 & 0.0000 &  & 0.0356 & 0.9889 & 0.0360 & 0.0000\\
NCON & 0.0449 & 1.2312 & 0.0365 & 0.0000 &  & 0.0449 & 1.2313 & 0.0365 & 0.0000\\
Index & 0.0543 & 1.5970 & 0.0340 & 0.0000 &  & 0.0552 & 1.5991 & 0.0346 & 0.0000\\ \hline
NW & \textbf{0.1988} & 6.6544 & 0.0299 & 0.0004 &  & \textbf{0.2075} & 6.7263 & 0.0309 & 0.0003\\
CROWN & 0.0357 & \textbf{0.9923} & \textbf{0.0360} & 0.0000 &  & 0.0357 & \textbf{0.9914} & \textbf{0.0360} & 0.0000\\
POET & 0.1179 & 11.0076 & 0.0185 & 0.0078 &   & 0.0549 & 9.0600 & 0.0205 & 0.0082\\
NLS & 0.0999 & 3.3034 & 0.0302 & 0.0005 &   & 0.0667 & 2.2762 & 0.0293 & 0.0005\\
SFNL & 0.0808 & 2.5186 & 0.0321 & 0.0003 &  & 0.0580 & 1.7876 & 0.0324 & 0.0003\\ 
\noalign{\global\arrayrulewidth1pt}\hline \noalign{\global\arrayrulewidth0.4pt}    
        \end{tabular}
        \end{table}

\clearpage


\section{Proof of Theorem 1}

{\bf Proof of Theorem 1(i)}. See that, with $\b \Theta:= \b \Sigma_y^{-1}$, $\hat{\b \Theta}:= \hat{\b \Sigma}_y^{-1}$, $\b \Theta$ is symmetric,
\begin{equation}
\hat{\b w}_d - \b w_d^* = \kappa \left[ \left( \frac{\hat{\b \Theta}' \hat{\b \mu}}{\b 1_p' \hat{\b \Theta}' \hat{\b \mu}} - \frac{\b \Theta \b \mu}{\b 1_p' \b \Theta \b \mu}
\right) + \left( \frac{\b \Theta \b 1_p}{\b 1_p' \b \Theta \b 1_p } - \frac{\hat{\b \Theta}' \b 1_p}{\b 1_p'
 \hat{\b \Theta} 1_p}
\right)
\right].\label{pt1.1}
\end{equation}
Define the scalar $A:= \b 1_p' \b \Theta \b 1_p/p$, and $\hat{A}:= \b 1_p' \hat{\b \Theta}' \b 1_p/p$, and consider the following right side term in (\ref{pt1.1})

\begin{equation}
\left\| \frac{\b \Theta \b 1_p}{\b 1_p' \b \Theta \b 1_p } - \frac{\hat{\b \Theta}' \b 1_p}{\b 1_p'
 \hat{\b \Theta}' \b 1_p}
\right\|_1 =  \left\|  \frac{\hat{\b \Theta}' \b 1_p/p}{\hat{A}} - \frac{\b \Theta \b 1_p/p}{A}
\right\|_1 = \frac{ \| (\hat{\b \Theta}' \b 1_p/p) A - (\b \Theta \b 1_p/p) \hat{A}\|_1}{| \hat{A} A | }
\label{pt1.2}
\end{equation}

We consider the numerator and denominator of (\ref{pt1.2}) separately. First see that 
\begin{equation}
| \hat{A} A | = | (\hat{A}- A) A + A^2| \ge A^2 - A  | \hat{A} - A |.\label{pt1.3}
\end{equation}

By Assumption 7(ii), $A \ge c > 0$ for $c >0$ being a positive constant, and by Lemma B.5 of \cite{caner2022} $A \le C < \infty$, for a positive constant $C >0$. Then by Assumptions 1-7, 
\begin{eqnarray}
 | \hat{A} - A | & \le & \| \b 1_p \|_{\infty} \| (\hat{\b \Theta} - \b \Theta)' \b 1_p \|_1/p \nonumber \\
 & \le & \max_j \| \hat{\b \Theta}_j - \b \Theta_j \|_1 \nonumber \\
& = &  O_p ( \bar{s} l_T)=o_p (1),\label{pt1.3a}
 \end{eqnarray}
where $\hat{\b \Theta}_j: p \times 1$ is the $j$ th row of $\hat{\b \Theta}$ shown in column format, and for the rate we use Theorem 2(i) of \cite{caner2022}.
Combine all those to have 
\begin{equation}
| \hat{A} A | \ge c^2 - o_p (1).\label{pt1.4}
\end{equation}

For the numerator, add and subtract $(\b \Theta \b 1_p/p)A$ in (\ref{pt1.2})
\begin{equation}
\| (\hat{\b \Theta}' \b 1_p/p)A - (\b \Theta \b 1_p/p) \hat{A} \|_1 
\le \frac{1}{p} [\| (\hat{\b \Theta}- \b \Theta)' \b 1_p A \|_1 + \| (\b \Theta \b 1_p) ( A - \hat{A}) \|_1],\label{pt1.5}
\end{equation}
where $\b \Theta$ is symmetric. Consider each term in (\ref{pt1.5}), with $A \ge c >0$
\begin{eqnarray*}
\frac{1}{p} \| (\hat{\b \Theta} - \b \Theta )' \b 1_p  A \|_1 & = & A \| (\hat{\b \Theta} - \b \Theta)' \b 1_p/p \|_1 \\
& \le & A \| (\hat{\b \Theta} - \b \Theta)' \|_{l_1} \| \b 1_p /p\|_1 \\
& = & A \| \hat{\b \Theta} -\b \Theta \|_{l_{\infty}} \\
& = &  A \max_{1 \le j \le p} \| \hat{\b \Theta}_j - \b \Theta_j \|_1,
\end{eqnarray*}
where we use p.345 of \cite{hj2013} for the inequality. By Assumptions 1-7, by Theorem 2(i) of \cite{caner2022}
\begin{equation}
 \max_{1 \le j \le p} \| \hat{\b \Theta}_j - \b \Theta_j \|_1 = O_p ( \bar{s} l_T).\label{a6a}
 \end{equation}
So by Lemma B.5 of \cite{caner2022} since $ A  \le C < \infty$
\begin{equation}
\frac{1}{p}\| (\hat{\b \Theta} - \b \Theta)' \b 1_p A \|_1 = O_p ( \bar{s} l_T).\label{pt1.6}
\end{equation}
\noindent Next in (\ref{pt1.5}) consider 
\begin{equation}
\| (\b \Theta \b 1_p/p) (A - \hat{A}) \|_1 \le  | A - \hat{A}| \| \b \Theta \b 1_p/p\|_{1} \le  C | A - \hat{A}|,\label{pt1.7}
\end{equation}
by p.345 of \cite{hj2013} for the second inequality, and we use
 \begin{equation}
 p^{-1} \| \b \Theta \b 1_p\|_1 \le p^{-1/2} \| \b \Theta \b 1_p \|_2 \le p^{-1/2} \| \b 1_p \|_2 Eigmax (\b \Theta) \le C ,\label{pt1.7a}
 \end{equation}
where we use $\b \Theta$  symmetric and Exercise 7.53b of \cite{abamag2005} for the second inequality, and third inequality is by Lemma B.5 of \cite{caner2022}.

Combine (\ref{pt1.7}) (\ref{pt1.3a})
\begin{equation}
\| (\b \Theta \b 1_p/p) (A - \hat{A}) \|_1 = O (1) O_p (\bar{s} l_T) = O_p ( \bar{s} l_T ).\label{pt1.9}
\end{equation}
Combine rates (\ref{pt1.6})(\ref{pt1.9}), so the rate in (\ref{pt1.5}) is
\begin{equation}
\| (\hat{\b \Theta}' \b 1_p/p)A - (\b \Theta \b 1_p/p) \hat{A} \|_1 
= O_p ( \bar{s} l_T ).\label{pt1.10}
\end{equation}
Then by (\ref{pt1.4})(\ref{pt1.10}) in (\ref{pt1.2})
\begin{equation}
\frac{ \| (\hat{\b \Theta}' \b 1_p/p) A - (\b \Theta \b 1_p/p) \hat{A}\|_1}{| \hat{A} A | } = O_p ( \bar{s} l_T ) = o_p (1),
\label{pt1.11}
\end{equation}
by Assumption 7(i), $ \bar{s} l_T = o(1)$. Define scalar $F:= \b 1_p' \b \Theta \b \mu/p$, and $\hat{F}:= \b 1_p' \hat{\b \Theta}' \hat{\b \mu}/p$ on right side term in  (\ref{pt1.1})

\begin{equation}
\left\| \left( \frac{\hat{\b \Theta}' \hat{\b \mu}}{\b 1_p' \hat{\b \Theta}' \hat{\b \mu}} - \frac{\b \Theta \b \mu}{\b 1_p' \b \Theta \b \mu}
\right)  \right\|_1 = \frac{ \| (\hat{\b \Theta}' \hat{\b \mu}/p) F - (\b \Theta \b \mu /p) \hat{F}\|_1}{| F \hat{F}|}.\label{pt1.12}
\end{equation}
In (\ref{pt1.12}) consider the denominator first 
\[ | F \hat{F}| \ge F^2 - |F| | \hat{F} - F|,\]
with the same analysis in (\ref{pt1.3}). Then under Assumptions 1-7(i) 
\begin{equation}
 | \hat{F} - F | = O_p (K \bar{s} l_T) = o_p (1).\label{a13a}
 \end{equation}
We prove (\ref{a13a}) now.   First, consider the following term
\begin{eqnarray}
| \b 1_p' (\hat{\b \Theta} - \b \Theta)' (\hat{\b \mu } - \b \mu )|/p & \le & \| \b 1_p' (\hat{\b \Theta} - \b \Theta)' \|_1 \| (\hat{\b \mu } - \b \mu )\|_{\infty}/p \nonumber \\
& \le &( \| \b 1_p \|_1/p) \| (\hat{\b \Theta} - \b \Theta )'\|_{l_1} \| (\hat{\b \mu } - \b \mu )\|_{\infty},\label{pt1.12a}
\end{eqnarray}
where we use Holder's inequality for the first inequality, and p.345 of \cite{hj2013} for the second inequality. 
Before the next proof we have the following result.
 By symmetry $\| \b \Theta \|_{l_1} = \| \b \Theta \|_{l_{\infty}}$, with Assumptions 1-7, and defining 
$\b L:= \b B [ ({\bf \Sigma_f})^{-1} + \b B' \b \Omega \b B ]^{-1} \b B'$ and use $\b \Theta:= \b \Sigma_y^{-1}$ formula in Section 2
\begin{eqnarray}
\| \b \Theta \|_{l_{\infty}} &=& \| \b \Omega - \b \Omega \b L \b \Omega \|_{l_{\infty}}\nonumber \\
& \le & \|\b  \Omega \|_{l_{\infty}} + \| \b \Omega \|_{l_{\infty}} \| \b L \|_{l_{\infty}}  \| \b \Omega \|_{l_{\infty}} \nonumber \\
& = & O (\sqrt{\bar{s}}) + O (\bar{s} r_T K^{3/2}) = O (\bar{s} r_T K^{3/2}),\label{pt1.8}
\end{eqnarray}
where the rates are from (A.96)-(A.97) of \cite{caner2022}. 
Note that $\b \Theta$ is symmetric, and adding and subtracting
\begin{eqnarray}
\left| \frac{\b 1_p' \hat{\b \Theta}' \hat{\b \mu}}{p} - \frac{\b 1_p' \b \Theta \b \mu}{p}
\right| & \le & \frac{1}{p} \left| \b 1_p' (\hat{\b \Theta} - \b \Theta)' (\hat{\b \mu} - \b \mu)
\right| +\frac{1}{p} \left| \b 1_p (\hat{\b \Theta}- \b \Theta)' \b \mu 
\right| +\frac{1}{p} \left| \b 1_p' \b \Theta (\hat{\b \mu} - \b \mu)
\right| \nonumber \\
& = & O_p ( \bar{s} l_T) O_p ( max ( K \frac{\sqrt{ln T}}{\sqrt{T}}, \frac{\sqrt{ln p}}{\sqrt{T}})) 
+ O_p ( \bar{s} l_T) O (K)  \nonumber \\
&+& O_p (\bar{s} r_T K^{3/2}) O_p ( max ( K \frac{\sqrt{ln T}}{\sqrt{T}}, \frac{\sqrt{ln p}}{\sqrt{T}})) 
\nonumber \\
& = & O_p ( \bar{s} l_T K),\label{A17A}
\end{eqnarray} 
where we use  (\ref{pt1.12a}) and the same technical analysis in (\ref{pt1.3a}), and the rates  are by using Holder's inequality and from (\ref{a6a}) here and then Theorem 2(ii) of \cite{caner2022} for the first term on the right side, and for the second term rates are by (\ref{a6a}) here and then (B.8) of \cite{caner2022} which is $\| \b \mu \|_{\infty} = O (K)$ , and for the third term, rates are by (\ref{pt1.8}) here and $\| \b \Theta' \|_{l_1}= \|\b  \Theta \|_{l_{\infty}}$, and Theorem 2(ii) of \cite{caner2022}. Second rate is the slowest by $l_T$ definition in (2) in main text.
 
By Cauchy Schwartz inequality
\begin{equation}
 | F | \le \left| \frac{\b 1_p' \b \Theta \b 1_p}{p}
\right|^{1/2} \left| \frac{\b \mu' \b \Theta \b \mu }{p}
\right|^{1/2}
= O (K^{1/2}),\label{a.18a}
\end{equation}
where the last inequality is by Lemma B.5 of \cite{caner2022}.
Then  by the condition $| F | \ge c >0$, we have by Assumption 7 and by (\ref{a13a})
\begin{equation}
| F \hat{F} | \ge c^2 - o_p (1).\label{pt1.13}
\end{equation}
Now consider the numerator in (\ref{pt1.12}), by adding and subtracting $(\b \Theta \b \mu/p)F$ and triangle inequality
\begin{equation}
\| \frac{ \hat{\b \Theta}' \hat{\b \mu}}{p} F - \frac{\b \Theta \b \mu}{p} \hat{F} \|_1 \le 
\| \frac{ \hat{\b \Theta}' \hat{\b \mu}}{p} F - \frac{\b \Theta \b \mu}{p} F \|_1 + \| \frac{\b \Theta \b \mu}{p}  (\hat{F} - F) \|_1.\label{pt1.14}
\end{equation}
In (\ref{pt1.14}) by adding and subtracting, $\b \Theta$ being symmetric
\begin{equation}
 \hat{\b \Theta}' \hat{\b \mu} - \b \Theta \b \mu = (\hat{\b \Theta} - \b \Theta)' \hat{\b \mu} + \b \Theta (\hat{\b \mu} - \b \mu).\label{pt1.14a}
 \end{equation}
Use the first term on the right side above
\begin{eqnarray}
\frac{1}{p} \| (\hat{\b \Theta} - \b \Theta)' \hat{\b \mu} \|_1 & \le & \frac{1}{p} \| (\hat{\b \Theta} - \b \Theta)' \|_{l_1} \| \hat{\b \mu} \|_1 \nonumber \\
& \le & \| \hat{\b \Theta} - \b \Theta \|_{l_{\infty}} \| \hat{\b \mu} \|_{\infty} = O_p ( \bar{s} l_T) O_p (K),\label{pt1.15}
\end{eqnarray}
where we use p.345 of \cite{hj2013} for the first inequality and for the rates, we use Theorem 2 of \cite{caner2022} and (B.8) of \cite{caner2022} by Assumptions 1-7. Now consider the second term on the right side of (\ref{pt1.14a})
\begin{eqnarray}
\frac{1}{p} \| \b \Theta (\hat{\b \mu} - \b \mu ) \|_1 &\le & \| \b \Theta \|_{l_1} \| \hat{\b \mu} - \b \mu \|_{\infty} \nonumber \\
& = & \| \b \Theta \|_{l_{\infty}} \| \hat{\b \mu} - \b \mu \|_{\infty}  \nonumber \\
&=& O (\bar{s} r_T K^{3/2}) O_p (max (K \sqrt{\frac{ln T}{T}}, \sqrt{\frac{lnp}{T}}))\label{pt1.16})
\end{eqnarray}
where we use the same analysis as in (\ref{pt1.15}) for the inequality, and the rates are by (\ref{pt1.8}) here and Theorem 2(ii) of \cite{caner2022}. Then (\ref{pt1.15}) is slower as a rate than in (\ref{pt1.16}) due to $l_T$ definition in (2). So with $| F| = O (\sqrt{K})$, in (\ref{pt1.14}) the first right side term is 
\begin{equation}
\| \frac{\hat{\b \Theta}' \hat{\b \mu}}{p} - \frac{\b \Theta \b \mu}{p} \|_1 |F| = O_p ( \bar{s} l_T K^{3/2}).\label{pt1.17}
\end{equation}
Next in (\ref{pt1.14}), the second right side term is
\begin{eqnarray}
\| \frac{\b \Theta \b \mu}{p}  (\hat{F} - F) \|_1 & = & \| \frac{\b \Theta \b \mu}{p} \|_1 | \hat{F} - F | \nonumber \\
& \le &  \frac{1}{p^{1/2}}  \| \b \Theta \b \mu \|_2 | \hat{F} - F | \le [Eigmax (\b \Theta )]^{1/2} \left|  \frac{\b \mu' \b \Theta \b \mu}{p}
\right|^{1/2} | \hat{F} - F|
 \nonumber \\
 & = & O (K^{1/2}) O_p (K \bar{s} l_T) = O_p ( \bar{s} l_T K^{3/2}),\label{pt1.18}
\end{eqnarray}
where we use Exercise 7.53b of \cite{abamag2005} and and Lemma B.5 of \cite{caner2022}.
 Clearly the rate in  (\ref{pt1.14})
\begin{equation}
\| \frac{ \hat{\b \Theta}' \hat{\b \mu}}{p} F - \frac{\b \Theta \b  \mu}{p} \hat{F} \|_1 = O_p (\bar{s} l_T  K^{3/2}).\label{pt1.19}
\end{equation}
Use (\ref{pt1.19})(\ref{pt1.13}) in (\ref{pt1.12}) to have 
\begin{equation}
\left\| \left( \frac{\hat{\b \Theta}' \hat{\b \mu}}{1_p' \hat{\b \Theta}' \hat{\b \mu}} - \frac{\b \Theta \b \mu}{\b 1_p' \b \Theta \b \mu}
\right)  \right\|_1 =O_p (\bar{s} l_T  K^{3/2}) = o_p (1),\label{pt1.20}
\end{equation}
by Assumption 7(i) to have the last equality. Use (\ref{pt1.11})(\ref{pt1.20}) in the weights definition in (\ref{pt1.1}), with $r_{w1}:= \bar{s} l_T K^{3/2}$,
\[ \| \hat{\b w}_d - \b w_d^* \|_1 = O_p (\bar{s} l_T  K^{3/2}) =  O_p (r_{w1})= o_p (1)\].{\bf Q.E.D.}

{\bf Proof of Theorem 1(ii)}. The difference between the out-of sample large portfolio variance estimate and its theoretical counterpart is:
\[ \hat{\b w}' \b \Sigma_y  \hat{\b w} - \b w^{*'} \b \Sigma_y \b w^*.\]
 Note that $\b w^*= \b w_d^* + \b m$, and $\hat{\b w}= \hat{\b w}_d + \b m$. We can simplify
 \begin{equation}
 \hat{\b w}' \b \Sigma_y \hat{\b w} - \b w^{*'} \b \Sigma_y \b w^* = (\hat{\b w}_d' \b \Sigma_y \hat{\b w}_d - \b w_d^{*'} \b \Sigma_y \b w_d^*) + 2 \b m' \b \Sigma_y (\hat{\b w}_d - \b w_d^*).\label{pt1a-1}
  \end{equation}
We consider each term in (\ref{pt1a-1}). First,
\begin{equation}
(\hat{\b w}_d' \b \Sigma_y \hat{\b w}_d - \b w_d^{*'} \b \Sigma_y \b w_d^*) = (\hat{\b w}_d - \b w_d^*)' \b \Sigma_y (\hat{\b w}_d - \b w_d^*) + 2 \b w_d^{*'} \b \Sigma_y (\hat{\b w}_d - \b w_d^*).\label{pt1a-2}
\end{equation}

\noindent Before the analysis of two terms on the right side of (\ref{pt1a-2}), we see
\begin{eqnarray}
\| \b \Sigma_y \|_{\infty} &\le& \| \b B {\bf \Sigma}_f \b B' \|_{\infty} + \| \b \Sigma_u \|_{\infty} \nonumber \\
& \le & K^2 \| \b B \|_{\infty}^2 \| {\bf \Sigma}_f \|_{\infty} + \| \b \Sigma_u \|_{\infty} = O (K^2),\label{pt1a-3}
\end{eqnarray}
where we obtain the second inequality using Lemma A.1(ii) of \cite{caner2022} which is $\| \b A_1 \b B_1 \b A_1' \|_{\infty} \le K^2 \| \b A_1 \|_{\infty}^2 \| \b B_1 \|_{\infty}$, where $\b A_1: p \times K$ matrix, $\b B_1: K \times K $ matrix, and the rates are by Assumption 1, 4, 5(i).

Consider each term in (\ref{pt1a-2}), and see that by Holder's inequality, and $\| \b A \b x \|_{\infty} \le \| \b A \|_{\infty} \| \b x \|_1$ for generic $\b A, \b x$ matrix-vector pair,
\begin{eqnarray}
(\hat{\b w}_d - \b w_d^*)' \b \Sigma_y (\hat{\b w}_d - \b w_d^*) & \le & \| \hat{\b w}_d - \b w_d^* \|_1 \| \b \Sigma_y ( \hat{\b w}_d - \b w_d^* )\|_{\infty} \nonumber \\
& \le & \| \hat{\b w}_d - \b w_d^* \|_1^2 \| \b \Sigma_y \|_{\infty} \nonumber \\
& =& O_p (r_{w1}^2) O (K^2),\label{pt1a-4}
\end{eqnarray}
where we use Theorem 1(i), since $r_{w1}$ is the rate of convergence in Theorem 1(i), and (\ref{pt1a-3}).

Before considering second right side  term  in (\ref{pt1a-2}), we analyze $l_1$ norm of $w_d^*$.

\begin{equation}
\| \b w_d^* \|_1 = | \kappa | \| \frac{\b \Theta \b \mu}{\b 1_p' \b \Theta \b \mu} - \frac{\b \Theta \b 1_p}{\b 1_p' \b \Theta \b 1_p} \|_1.\label{pt1a-4a}
\end{equation}

In (\ref{pt1a-4a})

\begin{equation}
\| \frac{\b \Theta \b \mu}{\b 1_p' \b \Theta \b \mu} - \frac{\b \Theta \b 1_p}{\b 1_p' \b \Theta 1_p} \|_1
\le \| \frac{\b \Theta \b \mu/p}{\b 1_p' \b \Theta \b \mu/p} \|_1 + \| \frac{\b \Theta \b 1_p/p}{\b 1_p' \b \Theta \b 1_p/p} \|_1.\label{pt1a-5}
\end{equation}

First we consider the denominators in (\ref{pt1a-5}).
By Assumption 7(ii) and $|\b 1_p' \b \Theta \b \mu |/p \ge c >0$ in the statement of Theorem 1, denominators are bounded away from zero. For the numerators in (\ref{pt1a-5}), as in (\ref{pt1.18})
\begin{eqnarray}
\| \b \Theta \b \mu/p \|_1 = O ( K^{1/2}).\label{pt1a-6}
\end{eqnarray}
In the same way as (A.10) 
\begin{equation}
\| \b \Theta \b 1_p/p\|_1 = O (1).\label{pt1a-7}
\end{equation}
Clearly by (\ref{pt1a-6})(\ref{pt1a-7}) into (\ref{pt1a-4a}), and since $ \kappa$ is a constant,
\begin{equation}
\| \b w_d^* \|_1 = O (K^{1/2}).\label{pt1a-8}
\end{equation}
Clearly we have growing exposure case in the weights differenced from the benchmark. See that the second right side term in (\ref{pt1a-2}) can be upper bounded as , by Holder's inequality,
\begin{eqnarray}
| \b w_d^{*'} \b \Sigma_y (\hat{\b w}_d - \b w_d^*) | & \le & \| \b w_d^* \|_1 \| \b \Sigma_y ( \hat{\b w}_d - \b w_d^* ) \|_{\infty} \nonumber \\
& \le & \| \b w_d^* \|_1  \| \b \Sigma_y \|_{\infty} \| \hat{\b w}_d - \b w_d^* \|_1,\label{pt1a-9}
\end{eqnarray}
where we use $\| \b A \b x \|_{\infty} \le \| \b A \|_{\infty} \| \b x \|_1$ for generic matrix $\b A$ and generic vector $\b x$.

Combine Theorem 1(i), (\ref{pt1a-3})(\ref{pt1a-8}) in (\ref{pt1a-9}) to have 
\begin{equation}
| \b w_d^{*'} \b \Sigma_y (\hat{\b w}_d - \b w_d^*) |  = O_p (r_{w1}   K^{5/2})=o_p (1),\label{pt1a-10}
\end{equation}
by Assumption 7. Combine (\ref{pt1a-4})(\ref{pt1a-10}) in (\ref{pt1a-2}) to have 
\begin{equation}
| \hat{\b w}_d' \b \Sigma_y \hat{\b w}_d - \b w_d^{*'} \b \Sigma_y \b w_d^* | = O_p ( r_{w1}  K^{5/2}) =o_p (1),\label{pt1a-11}
\end{equation}
by $r_{w1}$ rate definition and Assumption 7(i). Since we assume $\| \b m \|_1 = O ( \| \b w_d^* \|_1)$, we have 
\begin{equation}
| \b m' \b \Sigma_y (\hat{\b w}_d - \b w_d^*) | = O_p (r_{w1}  K^{5/2}) = o_p (1),\label{pt1a-12}
\end{equation}
where we use exactly the same analysis in (\ref{pt1a-10}). Next combine (\ref{pt1a-11})(\ref{pt1a-12})(\ref{pt1a-4}) in (\ref{pt1a-1}), and Assumption 7(iii) to have the desired result.{\bf Q.E.D}

{\bf Proof of Theorem 1(iii)}. Note that 

\[ \left[ \frac{\widehat{SR}}{SR}\right]^2 -1= \frac{[(\frac{\hat{\b w}' \b \mu)^2}{\hat{\b w}' \b \Sigma_y \hat{\b w}}]}{[ \frac{(\b w^{*'} \b \mu)^2 }{\b w^{*'} \b \Sigma_y \b w^*}]} -1 
= \left( \frac{(\hat{\b w}' \b \mu)^2}{(\b w^{*'} \b \mu)^2 }
\right) \left( \frac{\b w^{*'} \b \Sigma_y \b w^*}{\hat{\b w}' \b \Sigma_y \hat{\b w}} \right)
-1.\]
Next by adding and subtracting one from each term on the right side of the expression above
\[ \left[\frac{\widehat{SR}}{SR}\right]^2 -1 = \left[ \frac{(\hat{\b w}' \b \mu)^2 - (\b w^{*'} \b \mu)^2}{(\b w^{*'} \b \mu)^2 } +1 
\right] \left[ \frac{\b w^{*'} \b \Sigma_y \b w^* - \hat{\b w}' \b \Sigma_y \hat{\b w}}{\hat{\b w}' \b \Sigma_y \hat{\b w}} +1
\right] -1.\]

Clearly we can simplify the right side above
\begin{eqnarray}
\left| \left[\frac{\widehat{SR}}{SR}\right]^2 -1 \right| & \le & \left| \frac{(\hat{\b w}' \b \mu)^2 - (\b w^{*'} \b \mu)^2}{(\b w^{*'} \b \mu)^2 }
\right|
\left| \frac{\b w^{*'} \b \Sigma_y \b w^* - \hat{\b w}' \b \Sigma_y \hat{\b w}}{\hat{\b w}' \b \Sigma_y \hat{\b w}}\right| \nonumber \\
& + & \left| \frac{(\hat{\b w}' \b \mu)^2 - (\b w^{*'} \b \mu)^2}{(\b w^{*'} \b \mu)^2 }
\right| + \left| \frac{\b w^{*'} \b \Sigma_y \b w^* - \hat{\b w}' \b \Sigma_y \hat{\b w}}{\hat{\b w}' \b \Sigma_y \hat{\b w}}\right|.\label{pt1a-13}
\end{eqnarray}
Note that 
\begin{equation}
\frac{ | (\hat{\b w}' \b \mu )^2 - (\b w^{*'} \b \mu)^2|}{(\b w^{*'}\b \mu)^2} = \frac{| [ (\hat{\b w} + \b w^*)' \b \mu][(\hat{\b w}- \b w^*)' \b \mu]|}{(\b w^{*'} \b \mu)^2}.\label{pt1a-14}
\end{equation}
Next, 
\begin{eqnarray}
| (\hat{\b w}+ \b w^*)' \b \mu| & \le & \| \hat{\b w} + \b w^* \|_1 \| \b \mu \|_{\infty} \nonumber \\
& = & \| \hat{\b w} - \b w^* + \b w^* + \b w^* \|_1 \| \b \mu \|_{\infty} \nonumber \\
& \le & \| \hat{\b w} - \b w^* \|_1 \| \b \mu \|_{\infty} + 2 \| \b w^* \|_1 \| \b \mu \|_{\infty} \nonumber \\
& = & O_p (r_{w1} K ) + O (  K^{3/2}) = o_p (1) + O( K^{3/2}) \nonumber \\
&=& O_p ( K^{3/2}),\label{pt1a-15}
\end{eqnarray}
where for the first inequality we use Holder's, and then the second inequality is by triangle inequality,
for the rates we use Theorem 1(i) here and (B.8) of \cite{caner2022} in which $\| \b \mu \|_{\infty} = O(K)$, and $\b w^*:= \b w_d^* + \b m$, (\ref{pt1a-8}), Assumption 7(iv), and the last rate is by Assumption 7(i) with $r_{w1}=o(1)$ definition, and $\bar{s}, K$ as nondecreasing functions of $T$.

First consider, by Theorem 1(ii), and Holder's inequality, and by assuming $|\b w^{*'} \b \mu| \ge c >0$, and (B.8) of \cite{caner2022} 
\begin{equation}
\frac{|\hat{\b w}' \b \mu - \b w^{*'} \b \mu|}{ ( \b w^{*'} \b \mu)^2} \le \frac{\| \hat{\b w} - \b w^* \|_1 \| \b \mu \|_{\infty}}{ (\b w^{*'} \b \mu)^2}= O_p (r_{w1}K ).\label{pt1a-16}
\end{equation}

So combine (\ref{pt1a-15})(\ref{pt1a-16}) with (\ref{pt1a-14})
\begin{equation}
\frac{ | (\hat{\b w}' \b \mu )^2 - (\b w^{*'} \b \mu)^2|}{(\b w^{*'} \b \mu)^2} = O_p (r_{w1} K ) O_p ( K^{3/2}) = O_p (\bar{s}  l_T K^4) = o_p (1),\label{pt1a-17}
\end{equation}
by Assumption 7(i) and $r_{w1}$ definition in Theorem 1(i).

Then
\begin{eqnarray}
\left| \frac{\b w^{*'} \b \Sigma_y \b w^* - \hat{\b w}' \b \Sigma_y \hat{\b w}}{\hat{\b w}' \b \Sigma_y \hat{\b w}}\right|
& \le & \left| \frac{\b w^{*'} \b \Sigma_y \b w^* - \hat{\b w}' \b \Sigma_y \hat{\b w}}{\b w^{*'} \b \Sigma_y \b w^*- | \hat{\b w}' \b \Sigma_y \hat{\b w} - \b w^{*'} \b \Sigma_y \b w^*|}\right| \nonumber \\
& = & O_p (r_{w1}  K^{5/2})= O_p (\bar{s}  l_T K^4),\label{pt1a-17a}
\end{eqnarray}
by Theorem 1(ii) proof. Use (\ref{pt1a-17})(\ref{pt1a-17a}) in (\ref{pt1a-13}) to have the desired result.
{\bf Q.E.D.}

\section{ Relation Between Tracking Error and Risk Aversion}

Here we provide the relation between TE and risk aversion parameter $\Xi$.

$T E \ge \sqrt{(\b w^*-\b m)^{\prime} \b \Sigma_y (\b w^*-\b m)}=\sqrt{\b w_d^{*\prime} \b \Sigma_y  \b w_d^*}$. From equations (1)-(5) of \cite{bbmp2011} we have, $\b \Theta:= \b \Sigma_y^{-1}$:

$$
\b w_d^*=\frac{{ \b 1_p}^{\prime} \b \Theta \b \mu}{\Xi}\left(\frac{\b \Theta \b \mu}{{\b 1_p}'\b  \Theta \b \mu}-\frac{\b \Theta { \b 1_p}}{{ \b 1_p}' \b \Theta \b 1_p}\right)=\frac{1}{\Xi}\left(\b \Theta \b \mu-\frac{{\b 1_p}' \b \Theta \b \mu}{{\b 1_p}' \b \Theta { \b 1_p}} \cdot \b \Theta { \b 1_p}\right)
$$
Define $\zeta:=\frac{{ \b 1_p}' \b \Theta \b \mu}{{ \b 1_p}' \b \Theta {\b 1_p}}$, then $\b w_d^*=\frac{\b \Theta}{\Xi}(\b \mu- \zeta { \b 1_p})$.
Now we plug in TE and by symmetry of $\b \Theta$
$$
\begin{aligned}
TE & \ge \sqrt{\b w_d^{*\prime} \b \Sigma_y  \b w_d^*}=\sqrt{\frac{1}{\Xi} \cdot(\b \mu-\zeta { \b 1_p})^{\prime}\b  \Theta \cdot \b \Sigma_y  \cdot \frac{1}{\Xi} \b \Theta \cdot(\b \mu-\zeta { \b 1_p})} \\
& =\frac{1}{\Xi} \cdot \sqrt{(\b \mu-\zeta {\b  1_p})^{\prime} \cdot \b \Theta \cdot(\b \mu-\zeta { \b 1_p})}
\end{aligned}
$$

See also that by (5) of \cite{bbmp2011}  risk aversion parameter, $\Xi$,  and risk tolerance parameter $\kappa$ are related 
\[ \kappa = \frac{\b 1_p' \b \Theta \b \mu}{\Xi}.\]

\section{ Proofs for Joint Tracking Error and Binding Weight Constraints}

This part of Appendix contains proofs related to joint tracking error and weight constraints (binding, equality constraints).

\begin{lemma}\label{la1}
Under Assumptions 1-6, 8,

(i).\[ \| \hat{\b k} - \b k \|_1 =  O_p (\bar{s} l_T ) = o_p (1),\]

(ii).\[ \| \hat{\b a} - \b a \|_1 = O_p (\bar{s} l_T ) = o_p (1).\]

(iii).\[ |(\hat{w}_k - \hat{w}_a) - (w_k - w_a) | = O_p (\bar{s} l_T ) = o_p (1).\]

(iv).\[ | w_k - w_a | = O (1).\]

(v). \[ \| \b k - \b a \|_1 = O(1).\]

\end{lemma}

{\bf Proof of Lemma \ref{la1}}.

(i).  Add and subtract 
\[ \hat{\b k}- \b k = \frac{\hat{\b \Theta}' \b 1_R }{\b 1_p' \hat{\b \Theta}' \b 1_R } - \frac{\b \Theta \b 1_R}{\b 1_p' \hat{\b \Theta}' \b 1_R} + \frac{\b \Theta \b 1_R}{\b 1_p' \hat{\b \Theta}' \b 1_R}
- \frac{\b \Theta \b 1_R}{\b 1_p' \b \Theta \b 1_R}.\]
Next by triangle inequality
\begin{equation}
\| \hat{\b k} - \b k \|_1 \le \left\| \frac{(\hat{\b \Theta}-  \b \Theta)' \b 1_R/p}{\b 1_p' \hat{\b \Theta}' \b 1_R/p }\right\|_1 + 
\left\| \b \Theta \b 1_{R}/p \right\|_1 \left| \frac{1}{1_p' \hat{\b \Theta}' \b 1_R/p} - \frac{1}{\b 1_p' \b \Theta \b 1_R/p}
\right|.\label{pl1-1}
\end{equation} 
Consider the first term on the right side of (\ref{pl1-1})
\begin{eqnarray}
\| (\hat{\b \Theta}-  \b \Theta)' \b 1_R/p\|_1 & \le & \| (\hat{\b \Theta}- \b \Theta)' \|_{l_1} \| \b 1_R/p\|_1 \nonumber \\
& = & \| \hat{\b \Theta } - \b \Theta \|_{l_{\infty}} r/p \nonumber \\
& \le & \max_{1 \le j \le p} \| \hat{\b \Theta}_j - \b \Theta_j\|_1  \nonumber \\
& = & O_p (\bar{s} l_T),\label{pl1-2} 
\end{eqnarray}
by p.345 of \cite{hj2013} for the first inequality, and we use $l_1$ norm of transpose of  a matrix is equal to $l_{\infty}$ norm of a matrix for the first equality, and for the second inequality, we use $l_{\infty}$ norm matrix definition, and for the rate we use Theorem 2 of \cite{caner2022}, and $0< r/p \le C_1 <1$ with $C_1$ a positive constant below 1. Then by reverse triangle inequality
\begin{equation}
 | \frac{\b 1_p' \hat{\b \Theta}' \b 1_R}{p}| \ge | \frac{\b 1_p' \b \Theta \b 1_R}{p}| - | \frac{\b 1_p' \hat{\b \Theta} ' \b 1_R}{p} - \frac{\b 1_p' \b \Theta \b 1_R}{p}|.\label{pl1-2aa}
 \end{equation}
By Assumption  8 we know that $|\b 1_p' \b \Theta \b 1_R/p| \ge c > 0$, and 
\begin{eqnarray}
| \frac{\b 1_p' (\hat{\b \Theta} - \b \Theta)' \b 1_R}{p}| &\le& \| \b 1_p \|_{\infty} \| (\hat{\b \Theta} - \b \Theta)' \b 1_R \|_1 \left(\frac{1}{p}\right) \nonumber \\
& = & O_p (\bar{s} l_T)=o_p(1),\label{pl1-2a}
\end{eqnarray}
where we use Holder's inequality and then the same analysis in (\ref{pl1-2}) with Assumption 8. Next use (\ref{pl1-2a}) in (\ref{pl1-2aa}) to have 
\begin{equation}
| \b 1_p' \hat{\b \Theta}' \b 1_R/p| \ge c - o_p(1).\label{pl1-3}
\end{equation}
Use (\ref{pl1-2})(\ref{pl1-3}) in (\ref{pl1-1}) first right-side term
\begin{equation}
\left\| \frac{(\hat{\b \Theta}-  \b \Theta)' \b 1_R/p}{\b 1_p' \hat{\b \Theta} \b 1_R/p }\right\|_1 =  O_p (\bar{s} l_T).\label{pl1-4}
\end{equation}

Next, consider the second term on the right-side of (\ref{pl1-1})
\[
\left| \frac{1}{\b 1_p' \hat{\b \Theta}' \b 1_R/p} - \frac{1}{\b 1_p' \b \Theta \b 1_R/p}
\right| = \left| \frac{\b 1_p' \b \Theta \b 1_R/p - \b 1_p' \hat{\b \Theta}' \b 1_R/p}{(\b 1_p' \hat{\b \Theta}' \b 1_R/p)(\b 1_p' \b \Theta \b 1_R/p)}
\right|.
\]

Then clearly by (\ref{pl1-2aa})-(\ref{pl1-3}), and by assuming $|\b 1_p' \b \Theta \b 1_R|/p \ge c >0$, Assumption 8
\begin{equation}
\left| \frac{1}{\b 1_p' \hat{\b \Theta}' \b 1_R/p} - \frac{1}{\b 1_p' \b \Theta \b 1_R/p}
\right| = O_p (\bar{s} l_T).\label{pl1-6}
\end{equation}

Next, in the second term in (\ref{pl1-1}) by the analysis in (\ref{pt1.7a}), and $0<r/p <1$
\begin{equation}
\| \b \Theta \b 1_R/p \|_1 = O (1).\label{pl1-7}
\end{equation}
Use (\ref{pl1-6})(\ref{pl1-7})  for the second right side term in (\ref{pl1-1})
\begin{equation}
\left\| \b \Theta \b 1_{R}/p \right\|_1 \left| \frac{1}{\b 1_p' \hat{\b \Theta}' \b 1_R/p} - \frac{1}{\b 1_p' \b \Theta \b 1_R/p}
\right| = O_p (\bar{s} l_T) O (1)
.\label{pl1-8}
\end{equation}

By (\ref{pl1-4})(\ref{pl1-8}), (\ref{pl1-1}) has the rate in (\ref{pl1-8})
\[ \| \hat{\b k} - \b k \|_1 = O_p (\bar{s} l_T ) = o_p (1),\]
via Assumption 8.{\bf Q.E.D.}

(ii). The analysis in (\ref{pt1.2}) that leads to (\ref{pt1.11}) provides
\[ \| \hat{\b a} - \b a \|_1 = O_p (\bar{s} l_T ) = o_p (1).\]
{\bf Q.E.D.}

(iii). We can rewrite

\begin{eqnarray}
(\hat{w}_k - \hat{w}_a) - (w_k - w_a) & = & \left[ \frac{\b 1_R' \hat{\b \Theta}' \b 1_R}{\b 1_R' \hat{\b \Theta}' \b 1_p} - \frac{\b 1_R' \b \Theta \b 1_R}{\b 1_R' \b \Theta \b 1_p}
\right] \nonumber \\
& + &  \left[ \frac{\b 1_R' \b \Theta \b 1_p}{\b 1_p' \b \Theta \b 1_p} - \frac{\b 1_R' \hat{\b \Theta}' \b 1_p}{\b 1_p' \hat{\b \Theta}' \b 1_p}
\right].\label{3.1}
\end{eqnarray}

Consider the following first right side term in (\ref{3.1}), add and subtract
\begin{eqnarray}
\left[ \frac{\b 1_R' \hat{\b \Theta}' \b 1_R}{\b 1_R' \hat{\b \Theta}' \b 1_p} - \frac{\b 1_R' \b \Theta \b 1_R}{\b 1_R' \b  \Theta \b 1_p}
\right]  & = & \left[  \frac{\b 1_R' \hat{\b \Theta}' \b 1_R}{\b 1_R' \hat{\b \Theta}' \b 1_p} - \frac{\b 1_R' \b \Theta \b 1_R}{\b 1_R' \hat{\b \Theta}' \b 1_p}
\right] \nonumber \\
& + & \left[ \frac{\b 1_R' \b \Theta \b 1_R}{\b 1_R' \hat{\b \Theta}' \b 1_p} - \frac{\b 1_R' \b \Theta \b 1_R}{\b 1_R' \b \Theta \b 1_p}
\right].\label{3.2}
\end{eqnarray}

Use triangle inequality
\begin{eqnarray}
\left| \left[ \frac{\b 1_R' \hat{\b \Theta}' \b 1_R}{\b 1_R' \hat{\b \Theta}' \b 1_p} - \frac{\b 1_R' \b \Theta \b 1_R}{\b 1_R' \b \Theta \b 1_p}
\right] \right| & \le  & \left| \frac{(\b 1_R' (\hat{\b \Theta} - \b \Theta)' \b 1_R)/p}{\b 1_R' \hat{\b \Theta}' \b 1_p/p}
\right| \nonumber \\
& + & \left| \frac{\b 1_R' \b \Theta \b 1_R}{p} \right| \left| \frac{1}{\b 1_R' \hat{\b \Theta}' \b 1_p/p} - \frac{1}{\b 1_R' \b \Theta \b 1_p/p}
\right| .\label{3.3}
\end{eqnarray}

First  by the inequality 
\[ \| \b M \b x \|_1 \le p \max_{1 \le j \le p} \| \b M_j \|_1 \| \b x \|_{\infty},\]
where $\b M: p \times p$ generic matrix and $\b x: p \times 1$ generic vector, $\b M_j$ as the jth row of $M$, (written in column form: $\b M_j : p \times 1$) we have, by Holder's inequality  and the inequality immediately above
\begin{equation}
| \b 1_R' (\hat{\b \Theta} - \b \Theta)' \b 1_R/p | \le \| \b 1_R \|_{\infty} \| (\hat{\b \Theta} - \b \Theta )' \b 1_R/p \|_1 \le \max_{1 \le j \le p } \| \hat{\b \Theta}_j - \b \Theta_j \|_1 = O_p (\bar{s} l_T),\label{3.4}
\end{equation}
by Theorem 2 of \cite{caner2022}. In the same way as in (\ref{pl1-3})
\begin{equation}
| \frac{\b 1_R' \hat{\b \Theta}' \b 1_p }{p} | \ge c - o_p (1).\label{3.5}
\end{equation}

Now consider the second term in (\ref{3.3}). First
\begin{equation}
\left| \frac{\b 1_R' \b \Theta \b 1_R}{p} 
\right| \le C < \infty,\label{3.6}
\end{equation}
where we use Cauchy-Schwartz inequality with (B.22) of \cite{caner2022}.
 Then combine (\ref{3.4})(\ref{3.5})(\ref{3.6}) (\ref{pl1-6}) in first right side term in (\ref{3.1}),
\begin{equation}
\left|  \frac{\b 1_R' \hat{\b \Theta}' \b 1_R}{\b 1_R' \hat{\b \Theta}' \b 1_p} - \frac{\b 1_p' \b \Theta \b 1_R}{\b 1_R' \b \Theta \b 1_p}
 \right| = O_p ( \bar{s} l_T ).\label{3.7}
\end{equation}
The same analysis in (\ref{3.7}) follows for the second term on the right side of (\ref{3.1})
\begin{equation}
\left|  \frac{\b 1_R' \b \Theta \b 1_p}{\b 1_p' \b \Theta \b 1_p} - \frac{\b 1_R' \hat{\b \Theta}' \b 1_p}{\b 1_p' \hat{\b \Theta}' \b 1_p}
 \right| = O_p ( \bar{s} l_T ).\label{3.8}
\end{equation}
Then by (\ref{3.7})(\ref{3.8}) in (\ref{3.1})
\[ |(\hat{w}_k - \hat{w}_a) - (w_k - w_a) | = O_p (\bar{s} l_T ).\]
{\bf Q.E.D.}

(iv). By definitions
\[ | w_k - w_a | = \left| \frac{\b 1_R' \b \Theta \b 1_R/p}{\b 1_R' \b \Theta \b 1_p/p} - \frac{\b 1_R' \b \Theta \b 1_p/p}{\b 1_p' \b \Theta \b 1_p/p}
\right|
.\]
By triangle inequality
\begin{equation}
| w_k - w_a | \le  \left| \frac{\b 1_R' \b \Theta \b 1_R/p}{\b 1_R' \b \Theta \b 1_p/p}\right|  + \left|  \frac{\b 1_R' \b \Theta \b 1_p/p}{\b 1_p' \b \Theta \b 1_p/p}
\right|.\label{3.9}
\end{equation}
Consider the first right side term in (\ref{3.9}) by assumption we have $|\b 1_R' \b \Theta \b 1_p /p| \ge c > 0$ and (\ref{3.6})
\begin{equation}
\left| \frac{\b 1_R' \b \Theta \b 1_R/p}{\b 1_R' \b \Theta \b 1_p/p}\right| = O ( 1).\label{3.10}
\end{equation}
In the same way as in (\ref{3.10}) above
\begin{equation}
\left|  \frac{\b 1_R' \b \Theta \b 1_p/p}{\b 1_p' \b \Theta \b 1_p/p}
\right| = O ( 1).\label{3.11}
\end{equation}
So by (\ref{3.10})(\ref{3.11}) we have 
\[ | w_k - w_a | = O (1).\]
{\bf Q.E.D.}

(v).  By triangle inequality
\[ \| \b k - \b a \|_1 \le \frac{ \|\b  \Theta \b 1_R \|_1/p}{| \b 1_p' \b \Theta \b 1_R|/p} + \frac{\|\b  \Theta \b 1_p\|_1/p}{|\b 1_p' \b \Theta \b 1_p|/p}
= O (1) + O (1),\]
by the analysis in (\ref{pt1.7a}) and (\ref{pl1-7}) and $|\b 1_p' \b \Theta \b 1_R|/p \ge c >0$ Assumption 8.{\bf Q.E.D.}

\begin{lemma}\label{la2}
Under Assumptions 1-6, 8

(i). \[ \| \hat{\b l} - \b l \|_1 = O_p ( \bar{s} l_T) = o_p (1).\]

(ii). \[ | w_u | = O ( K^{1/2}).\]

(iii). \[ \|\b  l \|_1 = O (1).\]

(iv). \[ | \hat{w}_u - w_u | = O_p (\bar{s} l_T K^{3/2}).\]

\end{lemma}

{\bf Proof of Lemma \ref{la2}}. 

(i).
\[ \| \hat{\b l} - \b l \|_1 = \| \frac{ \hat{\b k} - \hat{\b a}}{\hat{w}_k - \hat{w}_a} - \frac{\b k-\b a}{w_k - w_a} \|_1.\]

Then set for easing the derivations and proofs $\b y:=(\b k-\b a): p \times 1, \hat{\b y}:=(\hat{\b k}-\hat{\b a}): p \times 1, x:=w_k - w_a, \hat{x}:=\hat{w}_k - \hat{w}_a$, where $\hat{x},x$ are both scalars.
So we have 
\[ \| \hat{\b l} - \b l \|_1 = \| \frac{\hat{\b y}}{\hat{ x}} - \frac{\b y }{x} \|_1 = \| \frac{\hat{\b y}x}{\hat{x}x}- \frac{\hat{x}\b y}{\hat{x}x}\|_1.\]
By adding and subtracting and then triangle inequality
\begin{equation}
\| \frac{\hat{\b y}x}{\hat{x}x}- \frac{\hat{x}\b y}{\hat{x}x}\|_1 \le \| \frac{\hat{\b y}x}{\hat{x}x} - \frac{\hat{\b y}\hat{x}}{\hat{x}x} \|_1 + 
\| \frac{\hat{\b y}\hat{x}}{\hat{x}x} - \frac{\hat{x} \b y }{\hat{x} x} \|_1.\label{pl2.0}
\end{equation}
For the first right side term in (\ref{pl2.0}), adding and subtracting and triangle inequality
\begin{eqnarray}
\| \frac{\hat{\b y}x}{\hat{x}x} - \frac{\hat{\b y}\hat{x}}{\hat{x}x} \|_1& =& \| \frac{\hat{\b y} (\hat{x}-x)}{\hat{x}x}\|_1 \nonumber \\
& \le & \| \frac{(\hat{\b y} - \b y)(\hat{x}-x)}{\hat{x} x } \|_1 + \| \frac{\b y (\hat{x}- x)}{\hat{x}x}\|_1.\label{pl2.1}
\end{eqnarray}

For the denominator 
\begin{equation}
| \hat{x} x | = | x^2 + x (\hat{x} - x)| \ge x^2 - x | \hat{x} - x|.\label{pl2.2}
\end{equation}
Now use Lemma \ref{la1}(iii)-(iv) and by definition of $\hat{x}:= \hat{w}_k - \hat{w}_a, x:=w_k - w_a$, and by Assumption 8
$ | w_k - w_a | \ge c >0$, for a positive constant $c$,
\begin{equation}
| \hat{x} x | \ge c^2 - O (1)  O_p (\bar{s} l_T ) \ge c^2 - o_p (1),\label{pl2.2a}
\end{equation}
by Assumption 8, $\bar{s}  l_T =o(1)$. Then consider the numerator of the first right side in (\ref{pl2.1}), by $\hat{\b y},\b y,\hat{x},x$ definitions at the beginning of the proof
\begin{eqnarray}
\| (\hat{\b y} - \b y) (\hat{x} - x ) \|_1 & \le & \| \hat{\b y} - \b y \|_1 | \hat{x} - x | \nonumber \\
& = & \| (\hat{\b k} - \hat{\b a}) - (\b k - \b a)\|_1 | (\hat{w}_k - \hat{w}_a) - (w_k - w_a)| \nonumber \\
& \le & [ \| \hat{\b k } - \b k \|_1 + \| \hat{\b a} - \b a \|_1] | (\hat{w}_k - \hat{w}_a) - (w_k - w_a)| \nonumber \\
& = & O_p ( \bar{s}l_T ). O_p (\bar{s} l_T ) = o_p (1),\label{pl2.3}
\end{eqnarray}
by Lemma \ref{la1}(i)-(iii) and triangle inequality, and Assumption 8, $\bar{s} l_T  = o(1)$. Next, we consider the numerator of the second right side term in (\ref{pl2.1}) by $\b y, \hat{x}-x$ definitions
\begin{eqnarray}
\| \b y (\hat{x} - x) \|_1 & = & \| (\b k- \b a) [ (\hat{w}_k -\hat{w}_a) - (w_k - w_a)] \|_1 \nonumber \\
& \le & \| \b k - \b a \|_1 | (\hat{w}_k -\hat{w}_a) - (w_k - w_a)| \nonumber \\
& = & O (1) O_p (\bar{s} l_T ) \nonumber \\
& = & O_p ( \bar{s} l_T) = o_p (1),\label{pl2.4}
\end{eqnarray}
where we use Lemma \ref{la1}(iii),(v) and Assumption 8. Combine (\ref{pl2.2})-(\ref{pl2.4}) in (\ref{pl2.1})  and use definitions of $\hat{\b y},\b y, \hat{x},x$ at the beginning of the proof here
\begin{eqnarray}
\| \frac{\hat{\b y}x}{\hat{x}x} - \frac{\hat{\b y}\hat{x}}{\hat{x}x} \|_1& =&  \| \frac{(\hat{\b k}- \hat{\b a})(w_k - w_a)}{(\hat{w}_k - \hat{w}_a) (w_k - w_a)} - 
\frac{(\hat{\b k}- \hat{\b a})(\hat{w}_k - \hat{w}_a)}{(\hat{w}_k - \hat{w}_a)(w_k - w_a)} \|_1 \nonumber \\
& = & O_p (\bar{s}l_T ) = o_p (1),\label{pl2.5}
\end{eqnarray}
where the rate is determined by (\ref{pl2.4}) by Assumption 8. Consider the second term on the right side of (\ref{pl2.0}), with $\hat{\b y}- \b y := (\hat{\b k}- \hat{\b a}) - (\b k -\b a)$
$\hat{x} - x := (\hat{w}_k - \hat{w}_a) - (w_k - w_a)$, and $x:=w_k - w_a$
\begin{eqnarray}
\| \frac{\hat{\b y}\hat{x}}{\hat{x}x} - \frac{\hat{x} \b y }{\hat{x} x} \|_1 & = & \| \frac{(\hat{\b y}- \b y) \hat{x}}{\hat{x} x } \|_1 \nonumber \\
& \le  & \| \frac{(\hat{\b y}- \b y ) (\hat{x} - x)}{\hat{x} x } \|_1 + \| \frac{(\hat{\b y}- \b y )x}{\hat{x} x }\|_1 \nonumber \\
& = & [O_p ( \bar{s} l_T )]^2 + O_p ( \bar{s}l_T ) O (1) \nonumber \\
& = & O_p ( \bar{s} l_T ) = o_p (1),\label{pl2.6}
\end{eqnarray}
by (\ref{pl2.3}), and (\ref{pl2.2}), and Lemma \ref{la1}(i)(ii)(iv) and Assumption 8. Combine (\ref{pl2.5})(\ref{pl2.6}) in (\ref{pl2.0}) to have 
\[ \| \frac{\hat{\b k} - \hat{\b a}}{\hat{w}_k - \hat{w}_a} - \frac{\b k-\b a}{w_k - w_a} \|_1 = O_p (\bar{s} l_T )=o_p (1).\]

(ii). We have 
\[ w_u := \frac{\b 1_R' \b \Theta \b \mu/p}{\b 1_p' \b \Theta \b \mu/p} - \frac{\b 1_R' \b \Theta \b 1_p/p}{\b 1_p' \b \Theta \b 1_p/p}.\]
Then clearly
\begin{equation}
| w_u | \le \left| \frac{\b 1_R' \b \Theta \b \mu/p}{\b 1_p' \b \Theta \b \mu/p}\right| + \left|\frac{\b 1_R' \b \Theta \b 1_p/p}{\b 1_p' \b \Theta \b 1_p/p}\right|.\label{w.0}
\end{equation}
By the same analysis in (\ref{a.18a}), $r<p$
\begin{equation}
| \b 1_R' \b \Theta \b \mu |/p = O ( K^{1/2}),\label{w.1}
\end{equation}
By Assumption 8,  $|\b 1_p' \b \Theta \b \mu /p| \ge c >0$
\begin{equation}
\left| \frac{\b 1_R' \b \Theta \b \mu/p}{\b 1_p' \b \Theta \b \mu/p}\right| = O ( K^{1/2}).\label{w.2}
\end{equation}
Then in the same way by using (\ref{3.6}) and assuming $| \b 1_p' \b \Theta \b 1_p/p| \ge c > 0$ as in Assumption 8
\begin{equation}
\left|\frac{\b 1_R' \b \Theta \b 1_p/p}{\b 1_p' \b \Theta \b 1_p/p}\right| = O (1).\label{w.3}
\end{equation}
Use (\ref{w.1})-(\ref{w.3}) in (\ref{w.0}) to have the desired result.{\bf Q.E.D}

(iii). 
\[ \| \b l \|_1 = \| \frac{\b k-\b a}{w_k - w_a} \|_1 = O (1),\]
by Lemma \ref{la1}(v) and Assumption 8(iii) $|w_k - w_a| \ge  c >0$.

(iv). By using $\hat{w}_u, w_u$ definitions

\[ | \hat{w}_u - w_u | = \left| \left( \frac{\b 1_R' \hat{\b \Theta}' \hat{\b \mu}/p}{\b 1_p' \hat{\b \Theta}' \hat{\b \mu}/p} - \frac{\b 1_R' \hat{\b \Theta}' \b 1_p/p}{\b 1_p' \hat{\b \Theta}' \b 1_p/p}
\right) - \left( \frac{\b 1_R' \b \Theta \b \mu/p}{\b 1_p' \b \Theta \b \mu /p} - \frac{\b 1_R' \b \Theta \b 1_p/p}{\b 1_p' \b \Theta \b 1_p/p}
\right)
\right|.\]
Rewriting above and by triangle inequality
\begin{equation}
| \hat{w}_u - w_u | \le \left|  \frac{\b 1_R' \hat{\b \Theta}' \hat{\b \mu}/p}{\b 1_p' \hat{\b \Theta}' \hat{\b \mu}/p} - \frac{\b 1_R' \b \Theta \b \mu/p}{\b 1_p' \b \Theta \b \mu /p}
\right| + \left| \frac{\b 1_R' \hat{\b \Theta}' \b 1_p/p}{\b 1_p' \hat{\b \Theta}' \b 1_p/p} - \frac{\b 1_R' \b \Theta \b 1_p/p}{\b 1_p' \b \Theta \b 1_p/p}\right|.\label{pl3.1}
\end{equation}
Then consider the first right side term in (\ref{pl3.1}), by adding and subtracting and triangle inequality
\begin{eqnarray}
\left|  \frac{\b 1_R' \hat{\b \Theta}' \hat{\b \mu}/p}{\b 1_p' \hat{\b \Theta}' \hat{\b \mu}/p} - \frac{\b 1_R' \b \Theta \b \mu/p}{\b 1_p' \b \Theta \b \mu /p}
\right| 
& \le & \left|  \frac{\b 1_R' \hat{\b \Theta}' \hat{\b \mu}/p}{\b 1_p' \hat{\b \Theta}' \hat{\b \mu}/p} -  \frac{\b 1_R' \b \Theta \b \mu/p}{\b 1_p' \hat{\b \Theta}' \hat{\b \mu}/p} \right|
\nonumber \\
& + & \left| \frac{\b 1_R' \b \Theta \b \mu/p}{\b 1_p' \hat{\b \Theta}' \hat{\b \mu} /p}-\frac{\b 1_R' \b \Theta \b \mu/p}{\b 1_p' \b \Theta \b \mu /p} \right|.\label{pl3.0}
\end{eqnarray}

Consider the denominator on the first term on the right side of (\ref{pl3.0})
\begin{equation}
|\b 1_p' \hat{\b \Theta}' \hat{\b \mu}/p| \ge | \b 1_p' \b \Theta \b \mu/p| - | \b 1_p' \hat{\b \Theta}' \hat{\b \mu}/p - \b 1_p' \b \Theta \b \mu/p|.\label{pl3.2}
\end{equation}

Since $\b \Theta$ is symmetric, and by Holder's inequality
\begin{eqnarray}
\left| \frac{\b 1_R' \hat{\b \Theta}' \b \mu}{p} - \frac{\b 1_R' \b \Theta \b \mu}{p}
\right| &\le& \frac{1}{p} \| \b 1_R \|_{\infty} \| \hat{\b \Theta}' \hat{\b \mu} - \b \Theta \b \mu \|_1  \nonumber \\
& \le &\frac{1}{p} \| \b 1_R \|_{\infty} \| (\hat{\b \Theta} - \b \Theta)' \hat{\b \mu}  \|_1 
 +  \frac{1}{p}  \| \b 1_R \|_{\infty} \| \b \Theta (\hat{\b \mu} - \b \mu ) \|_1\nonumber \\
& = & O_p ( \bar{s} l_T) O_p (K) + O_p (\bar{s} r_T K^{3/2}) O_p ( \max (K \sqrt{ln T/T}, \sqrt{ln p/T})) \nonumber \\
& = & O_p (\bar{s} l_T K) = o_p (1),\label{pl3.3}
\end{eqnarray}
where we use add and subtract $\b \Theta \hat{\b \mu}$ and triangle inequality to get the second inequality,  and the rates are by (\ref{pt1.15})(\ref{pt1.16}) and the (\ref{pt1.15}) rate is slower due to $l_T$ definition in (2).

Use (\ref{pl3.3}) with Assumption 8  $|\b 1_p' \b \Theta \b \mu |/p\ge c > 0$ in (\ref{pl3.2})
to have 
\begin{equation}
|\b 1_p' \hat{\b \Theta} \hat{\b \mu}/p| \ge c - o_p (1).\label{pl3.3a}
\end{equation}
Using (\ref{pl3.3}) for the numerator of the first right side term in (\ref{pl3.0}) in combination with (\ref{pl3.3a})
 \begin{equation}
  \left|  \frac{\b 1_R' \hat{\b \Theta} \hat{\b \mu}/p}{\b 1_p' \hat{\b \Theta} \hat{\b \mu}/p} -  \frac{\b 1_R' \b \Theta \b \mu/p}{\b 1_p' \hat{\b \Theta} \hat{\b \mu}/p} \right|
= O_p ( \bar{s} l_T K).\label{pl3.4}
\end{equation}
Consider the second term on the right side of (\ref{pl3.0})
\begin{equation}
 \left| \frac{\b 1_R' \b \Theta \b \mu/p}{\b 1_p' \hat{\b \Theta}' \hat{\b \mu} /p}-\frac{\b 1_R' \b \Theta \b \mu/p}{\b 1_p' \b \Theta \b \mu /p} \right|
\le | \b 1_R' \b \Theta \b \mu /p| \left| \frac{\b 1_p' \hat{\b \Theta}' \hat{\b \mu}/p - \b 1_p' \b \Theta \b \mu/p}{(\b 1_p' \hat{\b \Theta}' \hat{\b \mu}/p)(\b 1_p' \b \Theta \b \mu/p)}
\right|.\label{pl3.5}
\end{equation}
In (\ref{pl3.5}), use (\ref{w.1}) (\ref{pl3.3a}) with same analysis in (\ref{pl3.3}) with $\| \b 1_p \|_{\infty} = \| \b 1_R \|_{\infty} =1$ with Assumption 8
\begin{equation}
| \b 1_R' \b \Theta \b \mu /p| \left| \frac{\b 1_p' \hat{\b \Theta}' \hat{\b \mu}/p - \b 1_p' \b \Theta \b \mu/p}{(\b 1_p' \hat{\b \Theta}' \hat{\b \mu}/p)(\b 1_p' \b \Theta \b \mu/p)}
\right| = O ( K^{1/2}) O_p ( \bar{s} l_T K) = O_p ( \bar{s}  l_T K^{3/2}).\label{pl3.7}
\end{equation}

Consider (\ref{pl3.4})(\ref{pl3.7}) in (\ref{pl3.0}) to have 
the first right side term in (\ref{pl3.1}) as 
\begin{equation}
\left|  \frac{\b 1_R' \hat{\b \Theta}' \hat{\b \mu}/p}{\b 1_p' \hat{\b \Theta}' \hat{\b \mu}/p} - \frac{\b 1_R' \b \Theta \b \mu/p}{\b 1_p' \b \Theta \b \mu /p}
\right| =O_p ( \bar{s}  l_T K^{3/2}).\label{pl3.7a}
\end{equation}

We consider the second right side term in (\ref{pl3.1}) by adding and subtracting, $\frac{\b 1_R' \b \Theta \b 1_p/p}{\b 1_p' \hat{\b \Theta}' \b 1_p/p}$, and via a triangle inequality

\begin{eqnarray}
&&\left| \frac{\b 1_R' \hat{\b \Theta}' \b 1_p/p}{\b 1_p' \hat{\b \Theta}' \b 1_p/p} - \frac{\b 1_R' 
\b \Theta \b 1_p/p}{\b 1_p' \b \Theta \b 1_p/p}\right|
\le \left| \frac{\b 1_R' (\hat{\b \Theta} - \b \Theta)' \b 1_p /p}{\b 1_p' \hat{\b \Theta}' \b 1_p/p}
\right| \nonumber \\
& + & \left| \frac{\b 1_R' \b \Theta \b 1_p}{p}
\right| \left| \frac{1}{\b 1_p' \hat{\b \Theta}' \b 1_p/p} - \frac{1}{\b 1_p' \b \Theta \b 1_p/p}
\right|.\label{pl3.7b}
\end{eqnarray}
Consider the first term on the right side of (\ref{pl3.7b}), and its denominator specifically
\begin{equation}
| \b 1_p' \hat{\b \Theta}' \b 1_p/p| \ge | \b 1_p' \b \Theta \b 1_p/p| - | \b 1_p' (\hat{\b \Theta}- \b \Theta)' \b 1_p/p| \ge c- o_p (1),\label{pl3.7c}
\end{equation}
where we use Double Holder's inequality, Theorem 2 of \cite{caner2022} to have the asymptotically negligible result, and the lower bound constant is by Assumption 8. Next, by Holder's inequality and then using matrix norm inequality in p.345 of \cite{hj2013}

\begin{eqnarray}
\left| \b 1_R' (\hat{\b \Theta}- \b \Theta)' \b 1_p/p
\right| & \le & \frac{1}{p} \| \b 1_R \|_{\infty} \| (\hat{\b \Theta} - \b \Theta)' \b 1_p \|_1 \nonumber \\
& \le & \| (\hat{\b \Theta}- \b \Theta)' \|_{l_1} \nonumber \\
& = & \| \hat{\b \Theta} - \b \Theta \|_{l_{\infty}} = O_p ( \bar{s} l_T),\label{pl3.7d}
\end{eqnarray}
and the rate is by Theorem 2 of \cite{caner2022}. Use (\ref{pl3.7c})(\ref{pl3.7d}) for the first right side term in (\ref{pl3.7b})
\[ \frac{ | \b 1_R' (\hat{\b \Theta} - \b \Theta )' \b 1_p/p|}{|\b 1_p' \hat{\b \Theta}' \b 1_p /p|} = O_p (\bar{s} l_T).\]
Then 
\begin{equation}
\left| \frac{\b 1_R' \b  \Theta  \b 1_p}{p}
\right| \left| \frac{1}{\b 1_p' \hat{\b \Theta}' \b 1_p/p} - \frac{1}{\b 1_p' \b \Theta \b 1_p/p}
\right| = O (1) O_p (\bar{s} l_T)= O_p ( \bar{s} l_T ),\label{pl3.7e}
\end{equation}
where we use the analysis in (\ref{3.6}), Theorem 2 of \cite{caner2022} with the same analysis in (\ref{pl3.7c})(\ref{pl3.7d}). 
This rate is also the rate for second right side term in (\ref{pl3.1}).
Clearly the rate in (\ref{pl3.7a}) is slower than the one in (\ref{pl3.7e}). So the rate for (\ref{pl3.1}) is (\ref{pl3.7a}),

\[ | \hat{w}_u - w_u | = O_p ( \bar{s} l_T K^{3/2}).\]
{\bf Q.E.D.}

{\bf Proof of Theorem 2(i)}. Optimal portfolio (with TE and constrained weight) by Proposition 1 of \cite{bbmp2011}
\[ \b w_{cp}^*= (\omega - \kappa w_u) \b l + \b w_d^*,\]
and its estimate is 
\[ \hat{\b w}_{cp}= (\omega - \kappa \hat{w}_u) \hat{\b l } + \hat{\b w}_d.\]
We have the estimation errors that are bounded 
\begin{equation}
\| \hat{\b w}_{cp} - \b w_{cp}^* \|_1 \le \| [ (\omega - \kappa \hat{w}_u) \hat{\b l}] - [(\omega - \kappa w_u) \b l ] \|_1 
+ \| \hat{\b w}_d - \b w_d^* \|_1.\label{pt2.1}
\end{equation}
Consider the first term in (\ref{pt2.1}) by adding and subtracting
\[ (\omega - \kappa \hat{w}_u) \hat{\b l} - (\omega - \kappa w_u ) \b l 
= (\omega - \kappa \hat{w}_u) \hat{\b l} - (\omega - \kappa w_u) \hat{\b l} + 
(\omega - \kappa w_u) \hat{\b l} - (\omega - \kappa w_u) \b l.\]

Then apply triangle inequality
\begin{equation}
 \|(\omega - \kappa \hat{w}_u) \hat{\b l} - (\omega - \kappa w_u )  \b l \|_1 \le \| \kappa (\hat{w}_u - w_u) \hat{\b l} \|_1 + 
 \| (\omega - \kappa w_u) (\hat{\b l} - \b l ) \|_1.\label{pt2.2}
 \end{equation}
 Next, in the first right side term in (\ref{pt2.2})
 \begin{equation}
 \| \kappa (\hat{w}_u - w_u) \hat{\b l}\| \le \| \kappa (\hat{w}_u - w_u) (\hat{\b l} - \b l )\|_1+
 \| \kappa (\hat{w}_u - w_u ) \b l \|_1.\label{pt2.3}
  \end{equation} 
  Combine (\ref{pt2.2})(\ref{pt2.3})
  \begin{equation}
 \|(\omega - \kappa \hat{w}_u) \hat{\b l} - (\omega - \kappa w_u )\b l \|_1
 \le \| \kappa (\hat{w}_u - w_u) (\hat{\b l} - \b l )\|_1+
 \| \kappa (\hat{w}_u - w_u ) \b l \|_1 + \| (\omega - \kappa w_u ) (\hat{\b l} -\b l )\|_1.\label{pt2.4}
 \end{equation}
 
 Consider each term in (\ref{pt2.4}) by Lemma \ref{la2}
\begin{equation}
 \| \kappa (\hat{w}_u - w_u) (\hat{\b l} -\b  l )\|_1  
\le | \kappa | |\hat{w}_u - w_u | \| \hat{\b l} - \b l \|_1 = O_p ( \bar{s} l_T K^{3/2}) O_p (\bar{s} l_T).\label{pt2.5}
\end{equation}
Next, analyze the second right side term in (\ref{pt2.4}) by Lemma \ref{la2}
\begin{equation}
\| \kappa (\hat{w}_u - w_u) \b l \|_1 \le | \kappa | |\hat{w}_u - w_u | \| \b l \|_1 
= O_p ( \bar{s} l_T K^{3/2}) O (1) = O_p (\bar{s}  l_T K^{3/2}).\label{pt2.6} 
\end{equation}
Then analyze the third term on the right side of (\ref{pt2.4}) by Lemma \ref{la2}
\begin{equation}
\| (\omega - \kappa w_u ) (\hat{\b l} - \b l )\|_1 \le [ | \omega| + | \kappa | | w_u |] \| \hat{\b l} - \b l \|_1
=O ( K^{1/2}) O_p (\bar{s} l_T )
= O_p (\bar{s} l_T K^{1/2}).\label{pt2.7}
\end{equation}
Given Assumption 8, the slowest term is (\ref{pt2.6})
hence

\begin{equation}
 \|(\omega - \kappa \hat{w}_u) \hat{\b l} - (\omega - \kappa w_u )  \b l \|_1
 = O_p (\bar{s}  l_T K^{3/2}).\label{pt2.8}
\end{equation}
Then use (\ref{pt2.8}) and Theorem 1(i)
\begin{equation}
\| \hat{\b w}_{cp} - \b w_{cp}^* \|_1 = O_p ( \bar{s}  l_T K^{3/2}) + O_p ( \bar{s} l_T K^{3/2}) = O_p (\bar{s}  l_T K^{3/2})= O_p (r_{w1}).\label{pt2.90}
\end{equation}
{\bf Q.E.D.}

{\bf Proof of Theorem 2(ii)}. Define $\hat{\b w}_R:= \hat{\b w}_{cp} + \b m_R$ and $\b w_R^*:= \b w_{cp}^* + \b m_R$. Then 
\begin{eqnarray*}
\hat{\b w}_R' \b \Sigma_y \hat{\b w}_R &-& \b w_R^{*'} \b \Sigma_y \b w_R^*  \nonumber \\
& = & (\hat{\b w}_{cp}' \b \Sigma_y \hat{\b w}_{cp}) - (\b w_{cp}^{*'} \b \Sigma_y \b w_{cp}^*) + 2 \b m_R' \b \Sigma_y (\hat{\b w}_{cp} - \b w_{cp}^*).
\end{eqnarray*}

Next consider the first term on the right side above by adding and subtracting
 \begin{eqnarray}
 (\hat{\b w}_{cp}' \b \Sigma_y \hat{\b w}_{cp})& -& (\b w_{cp}^{*'} \b \Sigma_y \b w_{cp}^*) = (\hat{\b w}_{cp} - \b w_{cp}^* )' \b \Sigma_y (\hat{\b w}_{cp} - \b w_{cp}^*) \nonumber \\
 & + & 2 \b w_{cp}^{*'} \b \Sigma_y (\hat{\b w}_{cp} - \b w_{cp}^*).\label{pt2.9}
 \end{eqnarray}

 Consider the first term on the right side of (\ref{pt2.9}) and using Theorem 2(i) and (\ref{pt1a-3}) by  Holder's inequality  and $\| \b A \b x \|_{\infty} \le \| \b A \|_{\infty} \| \b x \|_1$ inequality for generic matrix $\b A$, and generic vector $\b x$
 \begin{equation}
 (\hat{\b w}_{cp} - \b w_{cp}^* )' \b \Sigma_y (\hat{\b w}_{cp} - \b w_{cp}^*) \le \| \hat{\b w}_{cp} - \b w_{cp}^* \|_1^2 \| \b \Sigma_y \|_{\infty}
 = O_p (r_{w1}^2) O (K^2).\label{pt2.10}
 \end{equation}

 We consider $l_1$ norm of $w_{cp}^*$ 
 \begin{eqnarray}
 \| \b w_{cp}^* \|_1 & = & \| (\omega - \kappa w_u ) \b l + \b w_d^* \|_1 \le   \| (\omega - \kappa w_u ) \b l \|_1 + \| \b w_d^* \|_1 \nonumber \\
 & \le & \left[ | \omega | + | \kappa | | w_u| 
 \right]  \| \b l \|_1 + \| \b w_d^* \|_1 \nonumber \\
 & = & O (K^{1/2}) O (1) + O (K^{1/2}) = O ( K^{1/2}),\label{pt2.11}
  \end{eqnarray}
 where $\omega, \kappa$ are bounded by definition, and we use (\ref{pt1a-8}), Lemma \ref{la2}(ii)(iii) for the rates. The second term on the right side of (\ref{pt2.9}) becomes
 
 \begin{eqnarray}
| \b w_{cp}^{*'} \b \Sigma_y (\hat{\b w}_{cp} - \b w_{cp}^*)|  & \le & \| \b w_{cp}^* \|_1 \| \b \Sigma_y \|_{\infty} \| \hat{\b w}_{cp} - \b w_{cp}^* \|_1 \nonumber \\
& = & O (K^{1/2}) O (K^2) O_p (r_{w1}) = O_p ( r_{w1} K^{5/2}),\label{pt2.12}
\end{eqnarray}
where we use the  Holder inequality and $\| \b A \b x \|_{\infty} \le \| \b A \|_{\infty} \| \b x \|_1$ inequality for generic matrix $\b A$, and generic vector $\b x$ for the inequality and the rates are from (\ref{pt1a-3}), and (\ref{pt2.11}) with Theorem 2(i). Combine (\ref{pt2.10})(\ref{pt2.12})  in (\ref{pt2.9}) to have, since $r_{w1}=o(1)$ by Assumption 8  
 
 \begin{equation}
 (\hat{\b w}_{cp}'\b  \Sigma_y \hat{\b w}_{cp}) - (\b w_{cp}^{*'} \b \Sigma_y \b w_{cp}^*) = O_p ( r_{w1}  K^{5/2}).\label{pt2.13}
 \end{equation}
 Then given that $\|  \b m_R \|_1 = O (\| \b w_{cp}^* \|_1)$, and by the  Holder inequality and $\| \b A \b x \|_{\infty} \le \| \b A \|_{\infty} \| \b x \|_1$ inequality for generic matrix $\b A$, and generic vector $\b x$ for the inequality, 
  \begin{eqnarray}
 | \b m_R'  \b \Sigma_y (\hat{\b w}_{cp} - \b w_{cp}^* )| &\le& \| \b m_R \|_1 \| \b \Sigma_y \|_{\infty} \| \hat{\b w}_{cp} - \b w_{cp}^* \|_1 \nonumber \\
 & = & O ( K^{1/2}) O (K^2) O_p (r_{w1}) = O_p ( r_{w1} K^{5/2}),\label{pt2.14}
  \end{eqnarray}
 and the rates are by (\ref{pt1a-3}) and Theorem 2(i), Assumption 8(iv) and (\ref{pt2.11}). Combine (\ref{pt2.13})(\ref{pt2.11})  in the first equation of this proof, and by $\b w_{R}^{*'} \b \Sigma_y \b w_{R}^* \ge c > 0$ assumption (i.e this imposes that there will be no local to zero variance) and we have the result
 \[ \| \frac{\hat{\b w}_R^{'} \b \Sigma_y \hat{\b w}_R}{\b w_R^{*'} \b \Sigma_y \b w_R^*} - 1 \|_1 = O_p ( r_{w1} K^{5/2}) = o_p (1).\]
 {\bf Q.E.D.}

 {\bf Proof of Theorem 2(iii)}. 
 
 We follow the proof of Theorem 1(iii) given Theorem 2(i)-(ii).  Hence we have  as in (\ref{pt1a-13})
 \begin{eqnarray}
\left| \left[\frac{\widehat{SR_R}}{SR_R}\right]^2 -1 \right| & \le & \left| \frac{(\hat{\b w}_R' \b \mu)^2 - (\b w_R^{*'} \b \mu)^2}{(\b w_R^{*'} \b \mu)^2 }
\right|
\left| \frac{\b w_R^{*'} \b \Sigma_y \b w_R^* - \hat{\b w}_R' \b \Sigma_y \hat{\b w}_R}{\hat{\b w}_R' \b \Sigma_y \hat{\b w}_R}\right| \nonumber \\
& + & \left| \frac{(\hat{\b w}'_R \b \mu)^2 - (\b w_R^{*'} \b \mu)^2}{(\b w_R^{*'} \b \mu)^2 }
\right| + \left| \frac{\b w_R^{'*} \b  \Sigma_y \b w_R^* - \hat{\b w}_R' \b \Sigma_y \hat{\b w}_R}{\hat{\b w}_R' \b \Sigma_y \hat{\b w}_R}\right|.\label{pt2.15}
\end{eqnarray}

To move forward we need the following results from Theorem 2(i) above and $\| \b \mu \|_{\infty} = O (K)$ from (B.8) of \cite{caner2022} 
\[ \| \hat{\b w}_R - \b w_R^* \|_1 \| \b \mu \|_{\infty} = O_p ( r_{w1} K).\]
Also by $\b w_R^*:= \b w_{cp}^* + \b m_R$ and (\ref{pt2.11}) and Assumption 8
\[ \| \b w_R^* \|_1 \| \b \mu \|_{\infty} = O (  K^{1/2}) O (K) = O (  K^{3/2}).\]
 
 Follow (\ref{pt1a-14})-(\ref{pt1a-17}) analysis, (just replacing $\hat{\b w}$ with $\hat{\b w}_R$ and $\b w^*$ with $\b w_R^*$ with rates in Theorems 2(i)(ii) and (\ref{pt2.11})
 
 \begin{equation}
 \left| \frac{(\hat{\b w}'_R \b \mu)^2 - (\b w_R^{*'} \b \mu)^2}{(\b w_R^{*'} \b \mu)^2 }
\right|  = O_p (r_{w1}  K^{5/2}).\label{pt2.16}
  \end{equation}
  
  Next by using proof of Theorem 2(ii), and $\b w_R' \b \Sigma_y \b w_R \ge c > 0$ assumption
 
 \begin{eqnarray}
 \left| \frac{\b w_R^{'*} \b \Sigma_y \b w_R^* - \hat{\b w}_R' \b \Sigma_y \hat{\b w}_R}{\hat{\b w}_R' \b \Sigma_y \hat{\b w}_R}\right| & \le & 
 \left| \frac{\b w_R^{*'} \b \Sigma_y \b w_R^* - \hat{\b w}_R' \b \Sigma_y \hat{\b w}_R}{\b w_R^{*'} \b \Sigma_y \b w_R^* - | \hat{\b w}_R' \b \Sigma_y \hat{\b w}_R - \b w_R^{*'} \b \Sigma_y \b w_R^*|} \right|\nonumber \\
   & = & O_p (r_{w1}  K^{5/2}).\label{pt2.17}
\end{eqnarray}
Combine (\ref{pt2.16})(\ref{pt2.17}) in (\ref{pt2.15}) to have the desired result.
\[ \left| \left(\frac{\widehat{SR}_R}{SR_R}\right)^2 - 1
\right| = O_p (r_{w1}  K^{5/2}) = o_p (1).\]
{\bf Q.E.D}

\section{Proofs for Joint Tracking Error and Inequality Constraints}
 
 {\bf Proof of Theorem 3}.

 First, related to indicator functions and constraints
 \[ | (\omega - \kappa w_u) - (\omega - \kappa \hat{w}_u)| = | \kappa (\hat{w}_u - w_u)|.\]
 Then define $\epsilon:= \bar{s} l_T K^{3/2}$, with $\epsilon >0$, and by Assumption 8(iii), $0 < c \le |\kappa | \le C < \infty$. So by Lemma A.2(iv)
 \begin{equation}
 | (\omega - \kappa w_u) - (\omega - \kappa \hat{w}_u)| \le C \epsilon,\label{ipt3}
 \end{equation}
 wpa1.  By Assumption 8, $\epsilon \to 0$. Now we define the event $E_1:= \{ | \kappa (\hat{w}_u - w_u)| \le \epsilon\}$ and we let $\epsilon \to 0$, and base our proofs on the event $E_1$. Clearly  by (\ref{ipt3}) and the sentence immediately above, $P (E_1) \to 1$.  
 
 Now we proceed as follows. First, we show that based on $E_1$, when $\omega > \kappa w_u$, this implies that $\omega > \kappa \hat{w}_u$. Similarly, then we also show, based on $E_1$, when $\omega < \kappa w_u$, this implies $\omega < \kappa \hat{w}_u$.  We will link these inequalities to optimal theoretical, $\b w_{op}^*$ and estimated, $\hat{\b w}_{op}$ weights, and 
 since $P (E_1) \to 1$, the proof will conclude.

 Start with by assuming $\omega - \kappa w_u \ge C > 2 \epsilon > 0$, for a positive constant $C >0$, then 
 \begin{eqnarray*}
 (\omega - \kappa \hat{w}_u) & = & (\omega - \kappa \hat{w}_u) - (\omega - \kappa w_u) + (\omega - \kappa w_u) \\
 & \ge & (\omega - \kappa w_u) - | (\omega - \kappa \hat{w}_u) - (\omega - \kappa w_u)| \\
 & = & (\omega - \kappa w_u) - | \kappa (\hat{w}_u - w_u)| \ge C - \epsilon > 2 \epsilon - \epsilon = \epsilon >0,
  \end{eqnarray*}
 where the first equality is by adding and subtracting and the first inequality is by triangle inequality, and the second inequality is by using $E_1$ definition, and the last inequality is by assumption. So we have that when $E_1$ is true with $\omega - \kappa w_u \ge C > 2 \epsilon >0$ we have $\omega - \kappa \hat{w}_u > \epsilon > 0$. Now let us go in the reverse direction. We are given $\omega - \kappa w_u \le - C < - 2\epsilon < 0$.  Start with 
 \begin{eqnarray*}
 \omega - \kappa w_u & = & \omega - \kappa w_u -( \omega - \kappa \hat{w}_u) + (\omega - \kappa \hat{w}_u) \\
 & \ge & (\omega - \kappa \hat{w}_u) - | (\omega - \kappa \hat{w}_u ) - ( \omega - \kappa \hat{w}_u)| \\
 & = & (\omega - \kappa \hat{w}_u ) - | \kappa (\hat{w}_u - w_u)|\\
 & \ge & (\omega - \kappa \hat{w}_u) - \epsilon,
  \end{eqnarray*}
 where the first equality is by adding and subtracting, and the first inequality is by triangle inequality and the last inequality is by $E_1$. Then combine the last inequality above by assumption 
 \[ 0 > - 2 \epsilon > - C \ge (\omega - \kappa w_u) \ge (\omega - \kappa \hat{w}_u) - \epsilon,\]
 and add $\epsilon$ to all sides 
 \[ - \epsilon > - ( C - \epsilon) \ge ( \omega - \kappa \hat{w}_u).\]
 This last point  clearly shows that when $(\omega - \kappa w_u) \le - C < - 2 \epsilon < 0$ we have $(\omega - \kappa \hat{w}_u) < 0$ since $-\epsilon <0$.

(i).  Now see that 
 \begin{eqnarray*}
 \| \hat{\b w}_{op} - \b w_{op}^* \|_1 & = & \| (\hat{\b w}_d 1_{ \{ \kappa \hat{w}_u < \omega \} }- \b w_d^* 1_{ \{ \kappa w_u < \omega \}} )
 + ( \hat{\b w}_{cp} 1_{ \{ \kappa \hat{w}_u > \omega \}} - \b w_{cp}^* 1_{ \{ \kappa w _u > \omega \} } \|_1 \\
 & \le & \| \hat{\b w}_d 1_{ \{ \kappa \hat{w}_u < \omega \} } - \b w_d^* 1_{ \{ \kappa w_u < \omega \}} \|_1 \\
 & + & \| \hat{\b w}_{cp} 1_{ \{ \kappa \hat{w}_u > \omega \} } - \b w_{cp}^* 1_{ \{ \kappa w_u > \omega \}} \|_1. 
  \end{eqnarray*}
  Based on $E_1$ and case 1 (nonbinding case above) with $\kappa w_u < \omega$, which implies $\kappa \hat{w}_u < \omega$
  \[ \hat{\b w}_d 1_{ \{ \kappa \hat{w}_u < \omega \} } - \b w_d^* 1_{ \{ \kappa w_u < \omega \}} = \hat{\b w}_d -  \b w_d^*
  = \hat{\b w} - \b w^*,
  \]
  since $\hat{\b w}:= \hat{\b w}_d + \b m$, and $\b w^*:= \b w_d^* + \b m$.
  Based on $E_1$, the second case (binding case) with $\kappa w_u > \omega$ implying $\kappa \hat{w}_u > \omega $
  \[ \hat{\b w}_{cp} 1_{ \{ \kappa \hat{w}_u > \omega \}} - \b w_{cp}^* 1_{ \{ \kappa w_u > \omega \}} = \hat{\b w}_{cp} - \b w_{cp}^*
  = \hat{\b w}_R - \b w_R^*,
  \]
  since $\hat{\b w}_R:= \hat{\b w}_{cp}+  \b m_R$ and $ \b w_R^*:= \b w_{cp}^* + \b m_R$.
  Then Theorems 1(i) and 2(ii) provide the desired result given $P (E_1) \to 1$.{\bf Q.E.D}
 
 (ii).  We base our proofs on $E_1$, and at the end of the proof we relax this condition.  First see that by definition of $\b w_{est}^*, \hat{\b w}_{est}$
 \begin{eqnarray*}
 \hat{\b w}_{est}' \b \Sigma_y \hat{\b w}_{est} - \b w_{est}^{*'} \b \Sigma_y \b w_{est}^* & = & 
 (\hat{\b w}' \b \Sigma_y \hat{\b w} 1_{ \{ \kappa \hat{w}_u < \omega \}} - \b w^{*'} \b \Sigma_y \b w^* 1_{ \{ \kappa w_u < \omega \}}) \\
 & + & ( \hat{\b w}_R' \b \Sigma_y \hat{\b w}_R 1_{ \{ \kappa \hat{w}_u > \omega \} } - \b w_{R}^{*'} \b \Sigma_y  \b w_{R}^* 1_{ \{ \kappa w_u > \omega \}}).
  \end{eqnarray*}

  Given $E_1$, as in the proof of Theorem 3(i) above with $\{ \kappa w_u < \omega \}$ we have $\{ \kappa \hat{w}_u < \omega \}$ 
  \[ \hat{\b w}^{'} \b \Sigma_y \hat{\b w} 1_{ \{ \kappa \hat{w}_u < \omega \}} - \b w^{*'} \b \Sigma_y \b w^* 1_{ \{ \kappa w_u < \omega \} }
  =  \hat{\b w}^{'}\b  \Sigma_y  \hat{\b w}  - \b w^{*'} \b \Sigma_y \b w^*.\]
  In the same way, give $E_1$, if $\{ \kappa w_u > \omega \}$ we have $\{ \kappa \hat{w}_u > \omega \}$, 
  so 
  \[ \hat{\b w}_R' \b \Sigma_y \hat{\b w}_R 1_{ \{ \kappa \hat{w}_u > \omega \}} - \b w_R^{*'} \b \Sigma_y \b w_R^* 1_{ \{ \kappa w_u > \omega \}}
  = \hat{\b w}_R' \b \Sigma_y \hat{\b w}_R - \b w_R^{*'} \b \Sigma_y \b w_R^*.\]
  Apply Theorem 1(ii), Theorem 2(ii) and $P (E_1 ) \to 1$ to have the desired result with Assumptions 7(iv), 8(iv).{\bf Q.E.D.}
  
  (iii). Given $E_1$ if $ \kappa w_u < \omega $ then as in the proof here we get $\kappa \hat{w}_u < \omega $
  so 
  \[ \frac{\widehat{SR}_{est}}{SR_{est}^*} = \frac{\left(\frac{\hat{\b w}' \b \mu }{\sqrt{ \hat{\b w}' \b \Sigma_y \hat{\b w}}}\right)}{\left( \frac{\b w^{*'} \b \mu}{\sqrt{\b w^{*'} \b \Sigma_y \b w^*}}
  \right)},\]
  which is covered by the proof of Theorem 1(iii).
  
  Given $E_1$ if $\kappa w_u > \omega$ we have $\kappa \hat{w}_u > \omega $ by the proof. So 
  \[ \frac{\widehat{SR}_{est}}{SR_{est}^*} =\frac{ \left( \frac{\hat{\b w}_R' \b \mu}{\sqrt{ \hat{\b w}_R' \b \Sigma_y \hat{\b w}_R}} \right)}{\left( 
  \frac{\b w_R'\b  \mu}{\sqrt{ \b w_R^{*'} \b \Sigma_y \b w_R^*}}
  \right) },\]
  and the proof is by Theorem 2(iii). Then since $P (E_1) \to 1$ we have the desired result.{\bf Q.E.D.}

 \section{ Only Weight Constraints}

In this subsection we only analyze weight constraints. The portfolio will be formed in a way to maximize returns subject to a risk constraint as well as weight constraint on the portfolio. There will be no tracking error constraint. So there is no benchmark portfolio involved in this part.  So, with the weight constraint defined as $w_x>0$  
\[ \max_{\b w \in R^p} \b \mu' \b w \quad  \, \mbox{subject to} \quad \b w' \b \Sigma_y \b w \le \sigma^2, \quad \b 1_p'  \b w=1, \quad \b 1_R' \b w= w_x,\]
where $\sigma^2$ is the risk constraint put by the investor. We can write the same optimization  problem
with $\Delta>0$ 
\begin{equation}
\b w_c^*:=\max_{\b w \in R^p} [ \b \mu' \b w - \frac{\Delta}{2} \b w' \b \Sigma_y \b w] \quad \mbox{subject to} \quad \b 1_p'  \b w=1, \quad \b 1_R' \b w= w_x.\label{opw}
\end{equation}

The following proposition provides the optimal weight in this subsection, as far as we know this is new and can be very helpful also in practice to setup constrained portfolios. Denote $B_1:= \b 1_p' \b \Theta \b \mu$, $B_2:= \b 1_p' \b \Theta \b 1_p$, $B_3:= \b 1_R' \b \Theta \b 1_p$, $B_4:= \b 1_R' \b \Theta \b 1_R$, $B_5:= \b 1_R' \b \Theta \b \mu$.  Assume $B_1 \neq 0, B_2 \neq 0, B_3 \neq 0$. Define also $\kappa_w$ as a positive risk tolerance parameter, and its relation to $\Delta$ will be given below in the proofs. Also define $\hat{B}_1:= 1_p' \hat{\b \Theta} \hat{\b \mu}$, $\hat{B}_2:= \b 1_p' \hat{\b \Theta}' \b 1_p$, $\hat{B}_3:= \b 1_R' \hat{\b \Theta}' \b 1_p$.

{\bf Proposition A.1}. {\it The optimal weight for (\ref{opw}) is given by the following, with $B_4 B_2 \neq B_3^2$, $B_1 \neq 0$ 
\[ \b w_c^*= \left (\kappa_w ( \frac{\b \Theta \b \mu}{B_1} - \b a) + (w_x - \kappa_w w_u) \b l \right )
 + \left( \b a - \b l \left(\frac{B_3}{B_2}\right)\right).\]}

Remark. The first term, $\kappa_w ( \frac{\b \Theta \b \mu}{B_1} - \b a) + (w_x - \kappa_w w_u) \b l$, in the optimal portfolio is  same as  $\b w_{cp}^*$ in (7) in main text (apart from constants $\kappa_w \neq \kappa, w_x \neq \omega$), the joint tracking error and weight portfolio. The additional term above is  $\left( \b a - \b l \left(\frac{B_3}{B_2}\right)\right)$ which is the close to global minimum variance portfolio weights when we have few restrictions.

Note that an estimator is, with $\hat{B}_2, \hat{B}_3$ definitions used in the last term below 
\[ \hat{\b w}_c = \kappa_w ( \frac{\hat{\b \Theta} \hat{\b \mu}}{\hat{B}_1} -  \hat{\b a}) + (w_x - \kappa_w \hat{w}_u) \hat{\b l}+ 
\left( \hat{\b a} - \hat{\b l}\left( \frac{\b 1_R' \hat{\b \Theta}' \b 1_p}{\b 1_p' \hat{\b \Theta}' \b 1_p}
\right)\right).\]
We show that this estimator is consistent in Lemma A.3 in Appendix and has the rate of convergence, $r_{w1}$ in Theorem 2, so all the analysis and rate results in Theorem 2 carry over to this case.

 We provide the proofs on only weights scenario, which analyzes the only weight constraint optimization without benchmarking and tracking errors but with a risk constraint on the portfolio.

 {\bf Proof of Proposition A.1}.
 
 First order conditions are, with $\delta >0, \nu >0$ are the lagrange multipliers 
 \begin{equation}
 \b \mu - \Delta \b \Sigma_y \b w_c^* - \delta \b 1_p - \nu \b 1_R =0,\label{pp31-1}
 \end{equation}
 \begin{equation}
 \b 1_p' \b w_c^* =1,\label{pp31-2}
  \end{equation}
 \begin{equation}
 \b 1_R' \b w_c^* = w_x.\label{pp31-3}
 \end{equation}
 
 Multiply (\ref{pp31-1})  by $\b \Theta:= \b \Sigma_y^{-1}$, and $\kappa_w:= \b 1_p' \b \Theta \b \mu/\Delta$

 \begin{eqnarray}
 \b w_c^* & = & \frac{1}{\Delta} \left( \b \Theta \b  \mu - \delta \b  \Theta \b 1_p - \nu \b \Theta \b 1_R
 \right) \nonumber \\
 & = & \frac{\kappa_w}{\b 1_p' \b \Theta \b \mu} \left( \b \Theta \b \mu - \delta\b \Theta \b 1_p - \nu \b \Theta \b 1_R \right)\nonumber \\
 & = & \kappa_w \left( \frac{\b \Theta \b \mu}{\b 1_p' \b \Theta \b  \mu} - \frac{\delta \b 1_p' \b \Theta \b 1_p}{\b 1_p' \b \Theta \b \mu} \b a - \frac{\nu \b 1_R' \b \Theta \b 1_p}{\b 1_p' \b  \Theta \b  \mu} \b k 
 \right),\label{pp31-4}
  \end{eqnarray}
 where in the last equality we use 
 \[ \b \Theta \b 1_p = \b 1_p'  \b \Theta \b 1_p (\frac{\b \Theta \b 1_p}{\b 1_p' \b \Theta \b 1_p}) = (\b 1_p' \b \Theta \b 1_p) \b a,\]
 by $\b a:=\b \Theta \b 1_p/\b 1_p'  \b \Theta \b 1_p$ definition in Section 3, and
 \[ \b \Theta \b 1_R = \b 1_R' \b \Theta \b 1_p (\frac{\b \Theta \b 1_R}{\b 1_R' \b \Theta \b 1_p})= (\b 1_R' \b \Theta \b 1_p)\b  k,\]
 with $\b k:= \b \Theta \b 1_R/\b 1_R' \b \Theta \b 1_p$ definition in Section 4.
 
 Now use (\ref{pp31-4}) in (\ref{pp31-2})
 \[ 1 - \delta \left(\frac{\b 1_p' \b \Theta \b 1_p}{\b 1_p' \b \Theta \b \mu}\right) (\b 1_p' \b a) - \nu \left(\frac{\b 1_R' \b \Theta \b 1_p}{\b 1_p' \b \Theta \b \mu}\right) (\b 1_p' \b k) = \frac{1}{\kappa_w}.\] 
 
 See that $\b 1_p' \b a =1$ by $\b a$ definition as well as $\b 1_p'\b k=1$ by $\b k$ definition. Then above equation simplifies  by multiplying each side with $\b 1_p' \b \Theta \b \mu$
 \begin{equation}
 (\b 1_p' \b \Theta \b \mu) (1 - \frac{1}{\kappa_w}) = \delta(\b 1_p' \b \Theta \b 1_p) + \nu (\b 1_R' \b \Theta \b 1_p).\label{pp31-5}
 \end{equation}
  Now use (\ref{pp31-4}) in (\ref{pp31-3}) 
 \[ \b 1_R' \b w_c^*= \kappa_w [ \frac{\b 1_R' \b \Theta \b \mu}{\b 1_p' \b  \Theta \b \mu} - \delta \frac{\b 1_p' \b \Theta \b 1_p}{\b 1_p' \b \Theta \b \mu} (\b 1_R' \b a) - 
 \nu \frac{\b 1_R' \b \Theta \b 1_p}{\b 1_p' \b \Theta \b \mu}  (\b 1_R' \b k)]=w_x.\] 
 This last equation implies, by $\b 1_R'\b a=\b 1_R' \b \Theta \b 1_p/\b 1_p' \b \Theta \b 1_p, \b 1_R'\b k=\b 1_R' \b \Theta \b 1_R/\b 1_p' \b \Theta \b 1_R$ definitions of $\b a,\b k$ and 
 multiply $\b 1_p' \b \Theta \b \mu $ on each side
 \begin{equation}
 (\b 1_R' \b \Theta \b \mu)- \frac{w_x}{\kappa_w} (\b 1_p' \b \Theta \b \mu) = \delta (\b 1_R'\b  \Theta \b 1_p) + \nu (\b 1_R' \b \Theta \b 1_R).\label{pp31-6}
  \end{equation}

 Rewrite (\ref{pp31-5})(\ref{pp31-6})
 \begin{equation}
 B_1 (1 - 1/\kappa_w) = \delta B_2 + \nu B_3.\label{pl31-6a}
 \end{equation}
 \begin{equation}
 B_5 - \frac{w_x}{\kappa_w}  B_1 = \delta B_3 + \nu B_4.\label{pl31-6b}
  \end{equation}

 In that respect multiply (\ref{pl31-6a}) by $B_3$ and multiply (\ref{pl31-6b}) by $B_2$, and subtract (\ref{pl31-6a}) from (\ref{pl31-6b}) after that multiplication to have 
 \begin{equation}
 \nu = \frac{B_2 B_5}{B_4 B_2 - B_3^2} - (w_x \frac{B_1}{\kappa_w}) (\frac{B_2}{B_4 B_2 - B_3^2})
 - \frac{B_1 B_3 (1-1/\kappa_w)}{B_4 B_2 - B_3^2}.\label{pp31-7}
  \end{equation}
  Then by (\ref{pl31-6a})
  \begin{equation}
  \delta = \frac{B_1}{B_2} (1- 1/\kappa_w) - \nu \frac{B_3}{B_2}.\label{pp31-8}
  \end{equation}
  
 Rewrite (\ref{pp31-4})
 \[ \b w_c^*= \kappa_w \left[ \frac{\b \Theta \b \mu}{B_1} - \left( \frac{\delta B_2}{B_1}
 \right) \b a - \left( \frac{\nu B_3}{B_1}
 \right) \b k 
 \right].\]

  Substitute (\ref{pp31-8}) into above equality and simplify
  \begin{equation}
  \b w_c^*= \kappa_w \left(\frac{\b \Theta \b \mu}{B_1}-\b a \right) + \b a + \kappa_w \left[ \frac{\nu B_3}{B_1} (\b a-\b k)\right].\label{pp31-9}
  \end{equation}
  We simplify on the right side of  (\ref{pp31-9}), the scalar term $\nu B_3/B_1$ by using (\ref{pp31-7})
  \begin{equation}
  \frac{\nu B_3}{B_1}= \left[ \frac{B_2 B_5}{B_4 B_2 - B_3^2}
  \right] \left( \frac{B_3}{B_1}
  \right) - \frac{w_x}{\kappa_w} \left( \frac{B_2 B_3}{B_4 B_2 - B_3^2}
  \right) - \left[ \frac{B_3^2 (1- 1/\kappa_w)}{B_4 B_2 - B_3^2}
  \right].\label{pp31-10}
  \end{equation}
  
  Take the second right side term in (\ref{pp31-10}) and multiply by $(\b a-\b k) \kappa_w$
  \begin{equation}
  \kappa_w \left[ - \frac{w_x}{\kappa_w} \left[\frac{B_2 B_3}{B_4 B_2- B_3^2}\right] (\b a - \b k)
  \right] = w_x \left[\frac{B_2 B_3}{B_4 B_2- B_3^2} \right]( \b k-\b a)  = w_x \left[\frac{\b k-\b a}{w_k -w_a}\right] = w_x\b  l,\label{pp31-10a}
   \end{equation} 
  and to get last equality we use $\b l:= \frac{\b k-\b a}{w_k - w_a}$ and $w_k, w_a$ definitions from Section 4
  \[ w_k - w_a:= \frac{B_4}{B_3} - \frac{B_3}{B_2} = \frac{B_4 B_2 - B_3^2}{B_3 B_2}.\]
  
   Then consider the first and third terms on the right side of (\ref{pp31-10}) 
  multiplying by $(\b a-\b k) \kappa_w$
  \begin{eqnarray}
  -\kappa_w \left( \frac{B_2 B_5}{B_4 B_2 - B_3^2}
  \right) \left( \frac{B_3}{B_1}
  \right) (\b k-\b a) &+& (\b k-\b a) \left(\frac{B_3^2 (1 - 1 /\kappa_w) \kappa_w}{B_4 B_2 - B_3^2}\right)\nonumber \\
&  = & -\kappa_w \left[ \frac{B_2 B_5}{B_4 B_2 - B_3^2} 
\right] \left[ \frac{B_3}{B_1}
\right] (\b k-\b a) \nonumber \\
&+& \frac{(\b k-\b a) B_3^2 \kappa_w}{B_4 B_2 - B_3^2} - \frac{(\b k-\b a) B_3^2}{B_4 B_2 - B_3^2}
\nonumber \\
&=&-\kappa_w w_u \b l - (\b k-\b a) \frac{B_3^2}{B_4 B_2 - B_3^2},\label{pp31-11}
  \end{eqnarray}
 and to get last equality above by $w_u$ and $\b l$ definitions in Section 4
 \begin{eqnarray}
 w_u \b l & = & \left[ \frac{B_5 B_2 - B_3 B_1}{B_1 B_2} 
 \right] \left[\frac{\b k-\b a}{B_4 B_2 - B_3^2} (B_3 B_2)\right]\nonumber  \\
 & = & \frac{B_5 B_2}{B_4 B_2 - B_3^2} \left( \frac{B_3}{B_1}
 \right) (\b k-\b a) \nonumber \\
 & - & \frac{(\b k-\b a) B_3^2}{B_4 B_2 - B_3^2}.\label{pp31-11a}
 \end{eqnarray}

  Next combine (\ref{pp31-10})-(\ref{pp31-11}) in (\ref{pp31-9}) to have 
 \[ \b w_c^* = \kappa_w ( \frac{\b \Theta \b \mu}{B_1} - \b a ) + \b a + (w_x - \kappa w_u) \b l - (\b k-\b a) \frac{B_3^2}{B_4 B_2 - B_3^2},\]
 The last term on right side can be written as, by $\b l:= \frac{(\b k-\b a) B_3 B_2}{B_4 B_2 - B_3^2}$ definition 
 \[ (\b k-\b a) \frac{B_3^2}{B_4 B_2 - B_3^2} = \b l \frac{B_3}{B_2}.\]
 Substituting this in optimal portfolio for the third term in $\b w_c^*$ provides the desired result.
  {\bf Q.E.D} 
 
 Note that by using (3)(7) definition of $\b w_{cp}^*$ we can write the optimal portfolio as 
 \[\b  w_c^*= \b w_{cp}^* - (\kappa - \kappa_w) (\frac{\b \Theta \b \mu}{B_1} - \b a) -
 [ (\omega - w_x) - (\kappa - \kappa_w) w_u] \b l + (\b a- \b l \frac{B_3}{B_2}).\] 
 
 For the rate analysis, because of $\b w_{cp}^*$ definition with $\kappa, \kappa_w, \omega, w_x$ being constants, the second and third terms on the right hand side will not change
 the  rate of convergence analysis (this is due to rate of second and third terms rate will be equal to rate of $\b w_{cp}^*$), so to simplify the proof without losing any generality, we set $\kappa=\kappa_w, \omega=w_x$ to make the second and third terms on the right side above to be zero. Let $C,c$ be  positive constants below.
  
 {\bf Lemma A.3}. {\it Under Assumptions 1-6, 8, $0 <  c \le   \kappa_w \le C <  \infty, | w_x | \le C < \infty$, $B_4 B_3 \neq B_2^2$, 
 we have
 \[ \| \hat{\b w}_c - \b w_c^* \|_1 = O_p (r_{w1})=o_p(1).\]
 }
 
 Remark. Note that this is also the rate in Theorem 2, which is joint tracking error and weight constraint.

 {\bf Proof of Lemma A.3}
 
 We take $\kappa=\kappa_w, \omega = w_x$ in $\b w_c^*$ above to get a simplified analysis without losing any generality.
 
 To show the consistency we need the rate and consistency of 
 $ \hat{\b a} - \hat{\b l} \frac{\hat{B_3}}{\hat{B_2}}$ with $\hat{B}_2:= \b 1_p' \hat{\b \Theta}' \b 1_p, \hat{B}_3:=\b 1_R' \hat{\b \Theta}'\b 1_p$. In that respect we set an upper bound and simplify that 
 \begin{equation}
 \| (\hat{\b a} - \hat{\b l} \frac{\hat{B_3}}{\hat{B_2}}) - ( \b a - \b l \frac{B_3}{B_2}) \|_1 
 \le \| \hat{\b a} - \b a \|_1 + \| \hat{\b l} \frac{\hat{B_3}}{\hat{B}_2} - \b l \frac{B_3}{B_2}\|_1.\label{pla3-1}
  \end{equation}
  We need to simplify the second term on the right side of (\ref{pla3-1}) above, by adding and subtracting $\b l B_3/\hat{B}_2$ and triangle inequality
  \begin{equation}
  \| \hat{\b l} \frac{\hat{B_3}}{\hat{B}_2} - \b l \frac{B_3}{B_2}\|_1 \le \| \hat{\b l} \frac{\hat{B_3}}{\hat{B}_2} - \b l \frac{B_3}{\hat{B}_2}\|_1
  + \| \b l \frac{ B_3}{\hat{B}_2} - \b l \frac{l B_3}{B_2}\|_1.\label{pla3-2}  
    \end{equation}

    Then in (\ref{pla3-2}), the first term on the right side, first by adding and subtracting $\b l \hat{B}_3/\hat{B}_2$, and then add and subtract $(\hat{\b l}-\b l ) B_3/\hat{B}_2$
    \begin{eqnarray*}
   \hat{\b l}  \frac{\hat{B}_3}{\hat{B}_2} - \b l \frac{B_3}{\hat{B}_2} & = & 
   \frac{\hat{\b l} \hat{B}_3}{\hat{B}_2} - \frac{\b l \hat{B}_3}{\hat{B}_2} + \frac{\b l \hat{B}_3}{\hat{B}_2} - \frac{\b l B_3}{B_2} \nonumber \\
   & = & \frac{(\hat{\b l} - \b l) \hat{B}_3}{\hat{B}_2} + \frac{\b l (\hat{B}_3 - B_3)}{\hat{B}_2} \nonumber \\
   & = &  \frac{(\hat{\b l} - \b l) (\hat{B}_3 - B_3)}{\hat{B}_2} + \frac{(\hat{\b l} - \b l) B_3}{\hat{B}_2} + 
   \frac{\b l (\hat{B}_3 - B_3)}{\hat{B}_2}
       \end{eqnarray*}

  By triangle inequality  
 \begin{equation} 
 \| \hat{\b l} \frac{\hat{B_3}}{\hat{B}_2} - \b l \frac{B_3}{\hat{B}_2}\|_1 \le \| \hat{\b l} - \b l \|_1 \left| \frac{\hat{B}_3 - B_3}{\hat{B}_2}
 \right| + \| \b l \|_1 \left| \frac{\hat{B}_3 - B_3}{\hat{B}_2}
 \right| + \left | \frac{B_3}{\hat{B}_2}\right| \| \hat{\b l} - \b l \|_1
 .\label{pla3-3}
\end{equation}
See that 
\[ \hat{B_2} = \hat{B}_2 - B_2 + B_2 \ge B_2 - | \hat{B}_2 - B_2|.\]

Then consider the second term on the right side of (\ref{pla3-2}), by above inequality

\begin{equation}
\| \frac{\b l B_3}{\hat{B}_2} -  \frac{\b l B_3}{B_2}\|_1 \le \frac{ \|\b  l \|_1 | B_3| | \hat{B_2} - B_2|}{B_2^2 - | \hat{B}_2 - B_2 | |B_2|}.\label{pla3-5}
\end{equation}

Use (\ref{pla3-3})(\ref{pla3-5}) on the right side of (\ref{pla3-2})
\begin{eqnarray}
 \| \hat{\b l} \frac{\hat{B_3}}{\hat{B}_2} - \b l \frac{B_3}{B_2}\|_1 & \le & \| \hat{\b l} - \b l \|_1 \left| \frac{\hat{B}_3 - B_3}{\hat{B}_2}
 \right| + \| \b l \|_1 \left| \frac{\hat{B}_3 - B_3}{\hat{B}_2}
 \right| + \left | \frac{B_3}{\hat{B}_2}\right| \| \hat{\b l} -\b  l \|_1 \nonumber \\
 & + & \frac{ \| \b l \|_1 | B_3| | \hat{B_2} - B_2|}{B_2^2 - | \hat{B}_2 - B_2 | |B_2|}.\label{pla3-6} 
\end{eqnarray}

We consider each one of the terms on the right side of (\ref{pla3-6}). By (\ref{pt1.3})(\ref{pt1.3a})(\ref{pl1-2a})
\begin{equation}
p^{-1} | \hat{B_3} - B_3 | = O_p ( \bar{s} l_T) = o_p (1) , \quad p^{-1} | \hat{B}_2 - B_2 | = O_p ( \bar{s} l_T) = o_p (1).\label{pla3-8}
\end{equation}
By Assumption 8(ii) $B_2:=\b 1_p' \b \Theta \b 1_p \ge c p > 0$.Then use (\ref{pla3-8}) and Lemma \ref{la2}(i), the first term on the right side of (\ref{pla3-6})
\begin{equation}
\| \hat{\b l} - \b l \|_1 \left| \frac{\hat{B}_3 - B_3}{\hat{B}_2}
 \right|  \le \frac{\| \hat{\b l}- \b l \|_1 | \hat{B}_3 - B_3 |/p}{(B_2 - | \hat{B}_2 - B_2|)/p} =O_p (\bar{s}^2 l_T^2 ).\label{pla3-10}
\end{equation}
Consider the second term on the right side of (\ref{pla3-6}) by Assumption 8(ii), Lemma \ref{la2}(iii),(\ref{pla3-8})

\begin{equation}
\| \b l \|_1 \left| \frac{\hat{B}_3 - B_3}{\hat{B}_2}
 \right| \le \frac{(|\hat{B}_3 - B_3|/p) \|  \b l \|_1}{B_2/p - | \hat{B}_2 - B_2|/p} = O_p ( \bar{s} l_T ).\label{pla3-11}
 \end{equation}

 Consider the third term on the right side of (\ref{pla3-6}), by Lemma \ref{la2}, (\ref{pla3-8})(\ref{3.6})
 \begin{equation}
\left | \frac{B_3}{\hat{B}_2}\right| \| \hat{\b l} - \b l \|_1 \le \frac{\| \hat{\b l} - \b l \|_1 |  B_3 |/p}{B_2/p - | \hat{B}_2- B_2|/p} = O_p (\bar{s} l_T ).\label{pla3-12}
\end{equation}
Then we consider the fourth term on the right side of (\ref{pla3-6}) by Lemma \ref{la2}(iii), (\ref{pla3-8})(\ref{3.6}), Assumption 8(ii)
\begin{eqnarray}
\frac{ \| \b l \|_1 | B_3| | \hat{B_2} - B_2|}{B_2^2 - | \hat{B}_2 - B_2 | |B_2|} &\le &
\frac{ \| \b l \|_1 |B_3/p| | \hat{B}_2 - B_2|/p}{B_2^2/p - (|\hat{B}_2 - B_2|/p)(B_2/p)  }\nonumber \\
& = & O (1) O_p ( 1) O_p ( \bar{s} l_T) \nonumber \\
& = & O_p ( \bar{s} l_T).\label{pla3-13}
\end{eqnarray}
 So combine the rate with Lemma \ref{la1}(ii) in (\ref{pla3-1})
\begin{equation}
\| (\hat{\b a} - \frac{\hat{\b l} \hat{B}_3}{\hat{B}_2}) - ( \b a -\b  l \frac{B_3}{B_2}) \|_1 = O_p ( \bar{s} l_T ).\label{pla3-14}
\end{equation}
Now remember that by Theorem 2(i), 
\[ \| \hat{\b w}_{cp} - \b w_{cp} \|_1 = O_p (r_{w1}),\]
where $r_{w1}= O (\bar{s} l_T K^{3/2})$, then clearly $r_{w1}$ is slower than   the rate in (\ref{pla3-14}), so the rate for 
\[ \| \hat{\b w}_c - \b w_c^*\|_1 = O_p (r_{w1})= o_p (1).\]
{\bf Q.E.D.}

Now we show that optimal weights in joint tracking error and weight case, $\b w_{cp}^*$ with only weight case, $\b w_c^*$, here have the same order in $l_1$ norm hence variance of the portfolio estimation as well as Sharpe Ratio estimation will be consistent and will have the same exact rates in Theorems 2(ii)-(iii). We prove that the orders are the same.
 We take $\kappa=\kappa_w, \omega = w_x$ in $\b w_c^*$ above to get a simplified analysis without losing any generality in rate of convergence. First note that 
\[ \b w_{c}^* = \b w_{cp}^* + (\b a - \b l \frac{B_3}{B_2}).\]
See that by (\ref{pt2.11})
\[ \| \b w_{cp}^* \|_1 = O (K^{1/2}).\]
Then we can use the triangle inequality to have 
\[\| \b a - \b l \frac{B_3}{B_2} \|_1 \le \| \b a \|_1 + \| \b l \|_1 \left| \frac{B_3}{B_2}
\right|.\]
By (\ref{pt1a-7}), $\b a:=\b \Theta \b 1_p/\b 1_p' \b \Theta \b 1_p$,  and Assumption 8, $\b 1_p' \b \Theta \b 1_p/p \ge c >0$
\[ \| \b a \|_1 = O (1).\]
By Lemma A.2(iii)
\[ \|\b  l \|_1 = O (1).\]
Also by definitions of $B_3, B_2$, and using the same analysis in (\ref{3.6}) for $B_3$ with Assumption 8 and (\ref{3.6}),  $\b 1_p' \b \Theta \b 1_p/p \ge c > 0$ so using the triangle inequality
\[ \|\b  a - \b l \frac{B_3/p}{B_2/p}\|_1 = O (1).\]
This last result implies
\begin{eqnarray*}
  \| \b w_c^* \|_1 &=& \| \b w_{cp}^* - (\b a - \b l \frac{B_3}{B_2}) \|_1 \\
  & \le & \| \b w_{cp}^* \|_1 + \| \b a \|_1 + \| \b l \|_1 \frac{|B_3|}{|B_2|} \\
  & = & O ( K^{1/2})= O (\b w_{cp}^*).
   \end{eqnarray*}
Hence all Theorem 2(ii)(iii) proofs will follow. We get the same rate of convergence as in Theorem 2(ii)(iii) for the only weight constraint case.

\subsection{Differences between TE constraint and only weight constraint portfolio}

TE constrained optimal portfolio is given in the main text as, with $\b a$ as the Global Minimum Variance Portfolio 
\begin{equation}
\b w^* = \b w_d^* + \b m = \kappa \left[ \frac{\b \Theta \b \mu}{\b 1_p' \b \Theta \b \mu} - \b a
\right] + \b m,\label{ow1}
\end{equation}
with $\b m $ as the benchmark portfolio. Then only weight constrained optimal portfolio is (given $\kappa = \kappa_w$, the risk tolerance parameter of investors are the same in both cases) by Proposition A.1

\begin{equation}
\b w_c^* = \left( \kappa \left( \frac{\b \Theta \b \mu }{\b 1_p' \b \Theta \b \mu} - \b a
\right) + \left( w_x - \kappa w_u \right) \b l 
\right) + \left( \b a - \b l \frac{\b 1_R' \b \Theta \b 1_p}{\b 1_p' \b \Theta \b 1_p} 
\right).\label{ow2}
\end{equation}
Using the estimators, by replacing $\b \mu$ with $\hat{\b \mu}$, and $\b \Theta $ with $\hat{\b \Theta}$, using (\ref{ow1})(\ref{ow2})
\begin{eqnarray}
(\hat{\b w}_c - \hat{\b w} ) - (\b w_c^*- \b w^*) & = & 
[ ( w_x - \kappa \hat{w}_u) \hat{\b l} - ( w_x - \kappa w_u) \b l ] \nonumber \\
& + & [ \hat{\b a} - \hat{\b l} ( \frac{\b 1_R' \hat{\b \Theta} \b 1_p}{\b 1_p' \hat{\b \Theta} \b 1_p}) - \b a - \b l ( \frac{\b 1_p' \b \Theta \b 1_p}{\b 1_p' \b \Theta \b 1_p})].\label{ow3}
\end{eqnarray}
Note that on the right side of (\ref{ow3}) comparing with (\ref{ow1}), there are no terms coming from TE portfolio in the difference above. Since
$ \| (\hat{\b w}_c - \hat{\b w} ) - (\b w_c^*- \b w^*) \|_1 \ge 0$, TE weight estimation errors are less compared with only weight restricted portfolio in $l_1$ norm.
Same issues will be present in variance and Sharpe ratio estimation too. But we see that upper bound on the estimation errors on the difference between only weight and TE portfolio converges to zero in probability by (\ref{pla3-14}) via Assumption 8(i). So asymptotically there is no difference between estimation errors, however in finite samples we expect TE based portfolio to be estimated better.

\section{Unconstrained Case and Differences with Constrained Portfolios}

We consider an unconstrained (without upper bound weight constraints and no tracking error constraints) portfolio here. Here we maximize returns subject to risk and weights added to one. In this way, it will be directly comparable with respect to our TE constraint only, weight constraint only, and TE and weight constraints.
Note that this is different than \cite{caner2022}.  \cite{caner2022} considers an unconstrained portfolio without weights added to one. We consider 
$ \max_{\b w \in R^p} \b \mu' \b w \quad  \, \mbox{subject to} \quad \b w' \b \Sigma_y \b w \le \sigma^2, \quad \b 1_p'  \b w=1,$
where $\sigma^2$ is the risk constraint put by the investor. We can write the same optimization  problem
with $\Delta>0$ 
\begin{equation}
\b w_{n}^*:=\max_{\b w \in R^p} [ \b \mu' \b w - \frac{\Delta}{2} \b w' \b \Sigma_y \b w] \quad \mbox{subject to} \quad \b 1_p'  \b w=1.\label{unw}
\end{equation}
Clearly (\ref{unw}) is a subcase of optimization in (\ref{opw}) with $\nu=0$ in (\ref{pp31-1}).

{\bf Proposition A.2}. {\it The optimal weight for (\ref{unw}) is given by the following, with $\b 1_p' \b \Theta \b \mu \neq 0$,
\[ \b w_{n}^*= \kappa_w \left( \frac{\b \Theta \b \mu}{\b 1_p' \b  \Theta \b \mu} - \b a \right) + \b a .\]}

Remark. The difference between the weight constrained portfolio in  Proposition A.1 and Proposition A.2 is:
$(w_x - \kappa_w w_u) \b l 
  - \b l \left(\frac{\b 1_R' \b \Theta \b 1_p}{\b 1_p' \b \Theta \b 1_p}\right)$.  The difference is all due to optimal weight constrained portfolio terms in Proposition A.1.   Note that the portfolio weights $\b l$ is the difference between unconstrained GMV and weight constrained only GMV portfolio. In our case since $\b l =0$ with no weight constraints, so these extra terms do not appear in unconstrained portfolio weight $\b w_{n}^*$. We show a brief proof of this below.

  {\bf Proof of Proposition A.2}. Start with First Order Conditions in (\ref{pp31-1}). See that this is a special case of Proposition A.1 where we set $\nu=0$ (this is the lagrange multiplier attached to weight constraint). The solution is immediate when we set $\nu=0$ in (\ref{pp31-9}).
  {\bf Q.E.D.}

  The population variance be in this unconstrained case is $\b w_n^{*'} \b \Sigma_y \b w_n^*$ and its estimated by $\hat{\b w}_n' \b \Sigma_y \hat{\b w}_n$, and note that 
  \[
  \hat{\b w}_n:= \kappa_w ( \frac{\hat{\b \Theta} \hat{\b \mu}}{\b 1_p' \hat{\b \Theta}' \hat{\b \mu}} - \hat{\b a}) + \hat{\b a}
  .\] The out-of-sample-unconstrained-Sharpe  Ratio is
  $ SR_n = \b w_n^{*'} \b \mu / \sqrt{\b w_n^{*'} \b \Sigma_y \b w_n^*}.$
  The estimator is:
  $ \widehat{SR}_u = \hat{\b w}_n' \b \mu / \sqrt{ \hat{\b w}_n' \b \Sigma_y \hat{\b w}_n}.$
  Here we provide consistency results.

 {\bf Lemma A.4}. {\it Under Assumptions 1-6, 8, with $\b w_n^{*'} \b \mu \ge c > 0$, and $\b w_n^{*'} \b \Sigma_y \b w_n^* \ge c > 0$  
 
 (i). \[ \| \hat{\b w}_n - \b w_n^* \|_1 = O_p (r_{w1}) =o_p (1).\]
 
 (ii). \[ | \frac{\hat{\b w}_n' \b \Sigma_y \hat{\b w}_n}{  \b w_n^{*'} \b \Sigma_y \b w_n^*} -1  | = O_p ( r_{w1}  K^{5/2}) = O_p ( \bar{s} l_T  K^4) = o_p (1).\]

 (iii). \[ \left| \left[ \frac{\widehat{SR}}{SR}\right]^2 -1 
 \right|  = O_p (r_{w1} K^{5/2} ) = O_p ( \bar{s} l_T  K^4)=o_p(1). \]
 
 }
  
 Remark. These are the same rates as TE portfolio, unlike the specific case of our unconstrained portfolio in Section 4.3 of \cite{caner2022} we allow $p>T$ and still get consistency. The proof technique here is different from \cite{caner2022}. \cite{caner2022} uses $Eigmax (\b \Sigma_y)$ in the proof which is diverging at a rate $p$ in that paper, as in Remark 2-Theorem 8 of that paper. Our alternative proof technique does not need this and hence we obtain $p>T$ and consistency together.
  
  {\bf Proof of Lemma A.4}
  
  (i). By triangle inequality and for the rates use (\ref{pt1.20}) and Lemma \ref{la1}(ii)
  \begin{eqnarray}
  \|  \hat{\b w}_n - \b w_n^* \|_1 &\le &| \kappa_w | \| \frac{ \hat{\b \Theta} \hat{\b \mu} }{\b 1_p' \hat{\b \Theta}' \hat{\b \mu} } - \frac{ \b \Theta \b \mu}{\b 1_p' \b \Theta \b \mu} \|_1 
  + | 1 - \kappa_w | \| \hat{\b a} - \b a \|_1\nonumber \\
  & = & O_p ( \bar{s} l_T K^{3/2}) + O_p ( \bar{s} l_T ) = O_p (r_{w1}).
   \label{pla4-1}
  \end{eqnarray}
  
  (ii).  Note that by adding and subtracting and triangle inequality
  \begin{equation}
  | \hat{\b w}_n'  \b \Sigma_y \hat{\b w}_n - \b w_n^{*'} \Sigma_y \b w_n^* | 
  \le | (\hat{\b w}_n - \b w_n^* )' \Sigma_y (\hat{\b w}_n - \b w_n^*)| + 2 | \b w_n^* \Sigma_y (\hat{\b w}_n - \b w_n^* )|.\label{pla4-2}
  \end{equation}
  Clearly by using Holder's inequality twice for the first right side term for (\ref{pla4-2})
  \begin{eqnarray}
  | (\hat{\b w}_n - \b w_n^* )' \Sigma_y (\hat{\b w}_n - \b w_n^*)|  
  & \le & \| \hat{\b w}_n - \b w_n^* \|_1^2 \| \b \Sigma_y \|_{\infty} \nonumber \\
  & = & O_p (r_{w1}^2) O (K^2),\label{pla4-3}
  \end{eqnarray}
  by Lemma A.4(i) and (\ref{pt1a-3}). Next, for the second right side term on (\ref{pla4-2}), by again using Holder's inequality twice
 \begin{equation}
  | \b w_n^* \b \Sigma_y (\hat{\b w}_n - \b w_n^* )| \le \| \b w_n^* \|_1 \| \hat{\b w}_n - \b w_n^* \|_1  \| \b \Sigma_y \|_{\infty}.\label{pla4-4} 
\end{equation}
To evaluate (\ref{pla4-4}) we need Proposition A.2
\[ \| \b w_n^* \|_1 \le | \kappa_w | \| \frac{\b \Theta \b \mu }{\b 1_p' \b \Theta \b \mu} \|_1 + \| \b a \|_1 | 1 - \kappa_w|.\]
By (\ref{pt1a-6}) and $|\b 1_p' \b \Theta \b \mu/p | \ge c >0$ 
\begin{equation}
\| \frac{ \b \Theta \b \mu }{\b 1_p' \b \Theta \b \mu} \|_1 = O (K^{1/2}).\label{pla4-5}
\end{equation}
 and then by assuming $\b 1_p' \b \Theta \b 1_p/p \ge c > 0$ by (A.11)(\ref{pt1.9})
 \begin{equation}
\| \b a \|_1 = \| \frac{ \b \Theta \b 1_p}{\b 1_p' \b \Theta \b 1_p} \|_1 = O (1).\label{pla4-6}
\end{equation}

Clearly (\ref{pla4-5}) is larger as a rate than the one in (\ref{pla4-6}) so given that $| 1 - \kappa_w | \le C < \infty$
\begin{equation}
\| \b w_n^* \|_1 = O (K^{1/2}).\label{pla4-7}
\end{equation}
Use (\ref{pla4-5}) in (\ref{pla4-4})
\begin{eqnarray}
  | \b w_n^* \b \Sigma_y (\hat{\b w}_n - \b w_n^* )| & \le &  \| \b w_n^* \|_1 \| \hat{\b w}_n - \b w_n^* \|_1  \| \b \Sigma_y \|_{\infty} \nonumber \\
  & = & O (K^{1/2}) O_p (r_{w1}) O (K^2) = O_p (r_{w1}  K^{5/2}) \nonumber \\
  & = & O_p (\bar{s} l_T  K^4) = o_p (1).\label{pla4-8}
  \end{eqnarray}
  Use (\ref{pla4-3})(\ref{pla4-8}) and $\b w_n^{*'} \b \Sigma_y \b w_n^* \ge c > 0$ to have the desired result.
{\bf Q.E.D.}

(iii). The analysis of Sharpe Ratio  is the same as in Theorem 1(iii) given Lemma A.4(i)(ii) here.
{\bf Q.E.D}

\subsection{Difference between Unconstrained Case and TE restriction}

In TE only weight restriction in Theorem 1, the optimal weight is:
\[ \b w^*= \kappa (\frac{\b \Theta \b \mu}{\b 1_p' \b \Theta \b \mu} - \b a ) + \b m.\]
Comparison of unconstrained optimal weight $\b w_n^*$ above in Proposition A.2 and $\b w^*$ immediately above shows that if $\kappa=\kappa_w$ (the risk tolerance parameter being equal)
then the difference is that the last term in $\b w_n^*$ in Proposition A.2 (unconstrained portfolio weights) is $\b a$ which itself is Global Minimum Variance (GMV) portfolio weights and in $\b w^*$ in Theorem 1 (TE restricted portfolio weights) is $\b m$ which is the benchmark portfolio. Since $\b m$ is not estimated, estimating $\b w_n^*$ involves of estimation of $\b a$, which is $\hat{\b a}$ in Lemma A.1. But $\b  a$ will be estimated with sampling errors , as shown in Lemma A.1. So clearly unconstrained portfolio weight estimation in $l_1$ norm, estimation error will have an extra term to estimate. 
To see this, note $\b w_n^* - \b w^* = \b a - \b m$, and $\hat{\b w}_n - \hat{\b w} = \hat{\b a} - \b m$. So 
\[ \| (\hat{\b w}_n - \b w_n^*) - ( \hat{\b w} - \b w^*) \|_1 = \| \hat{\b a} - \b a \|_1 \ge 0 ,\]
where the difference on the right side is all due to estimation errors from unconstrained portfolio. One question is that what is the rate of this extra estimation error in unconstrained portfolio weights, by Lemma A.1(ii), the rate is
\[ \| \hat{\b a} - \b a \|_1 = O_p ( \bar{s} l_T ) = o_p (1).\]
But this rate is converging  to zero in probability by Assumption 8, so asymptotically  there is  no difference between TE and unconstrained weight estimation. Variance and Sharpe ratio analysis differences are the same as portfolio weight estimation that is analyzed here.

\subsection{ The Difference between Unconstrained and Only Weight Constraint Case}

We compare the results of Proposition A.1 and A.2. We should also note that unconstrained case is a subcase of only weight constraint case in terms of optimization, and this point is seen in the proofs of Propositions A.1-A.2, and discussed there. In that respect
\begin{equation}
w_c^* - w_n^* = (w_x - \kappa_w w_u ) \b l - \b l \frac{B_3}{B_2}.\label{w.000}
\end{equation}
So the weight constrained case has two additional terms that we show immediately above on the right side of  (\ref{w.000}). The only weight constraint case has to estimate these extra two terms. In this respect, compared to unconstrained case, we expect $l_1$ estimation errors of the only weight constraint- portfolio weights to be larger in $l_1$ norm.
Remembering that the two right hand side terms, which is the difference between weight constrained and unconstrained portfolios, in (\ref{w.000}) only belong to  weight constrained portfolio so it is clear that 
\[ \| (\hat{\b w}_c - \b w_c^*) - ( \hat{\b w}_n  - \b w_n^*) \|_1 \ge 0,\]
and hence only weight-constrained portfolio estimation error in $l_1$ norm is always larger than equal to unconstrained portfolio estimation errors. 

To make an asymptotic case, consider the upper bound of this error
\begin{eqnarray}
\| (\hat{\b w}_c - \hat{\b w}_n ) - (\b w_c^* - \b w_n^*) \|_1 & \le & \| (w_x - \kappa_w \hat{w}_u ) \hat{\b l} - (w_x - \kappa_w w_u ) \b l \|_1 \nonumber \\
& + & \| \hat{\b l } \frac{\hat{B}_3}{\hat{B}_2} - \b l \frac{B_3}{B_2} \|_1.\label{w.100}
\end{eqnarray}
 Clearly, exactly as in (\ref{pt2.8})
\begin{equation}
\| (w_x - \kappa_w \hat{w}_u ) \hat{\b l} - (w_x - \kappa_w w_u ) \b l \|_1 = O_p ( \bar{s} l_T K^{3/2}).\label{w.300}
\end{equation}

Now we consider the second right side term in (\ref{w.100})

\begin{equation}
\| \hat{\b l } \frac{\hat{B}_3}{\hat{B}_2} - \b l \frac{B_3}{B_2} \|_1 = O_p ( \bar{s}  l_T),\label{w.400}
\end{equation}
by (\ref{pla3-8})-(\ref{pla3-13}) via Lemma A.2-A.3, Assumption 8. Then clearly by (\ref{w.100})(\ref{w.300})(\ref{w.400})

\begin{equation}
\| (\hat{w}_c - \hat{w}_n ) - (w_c^* - w_n^*) \|_1 = O_p (\bar{s} l_T K^{3/2}) = O_p (r_{w1}).\label{w.500}
\end{equation}
So these extra two terms in (\ref{w.000}) for only weight constrained portfolio compared with unconstrained one introduces an estimation error  with rate $r_{w1}$. 
Asymptotically, via Assumption 8, this rate will go to zero, so asymptotically there may be no difference in estimating two portfolios, however in finite samples, in $l_1$ norm sense, estimating unconstrained portfolio is easier.

\subsection{ Difference Between Unconstrained Portfolio and Joint Tracking Error and Weight Constrained Portfolio}

 First, we analyze the $l_1$ estimation error difference between these two portfolios. Then we want to understand whether there are additional terms to be estimated in one portfolio versus the other one. Next, we want to understand whether there are any asymptotic differences  between two portfolio weights estimation errors in $l_1$ norm. Also, we discuss briefly variance and Sharpe Ratio differences between two portfolios in the light of whether estimating these two metrics will add to errors in comparison  to unconstrained one.

First equation (7)  and discussion just before (10) in the main text provides
\[ \b w_R^*= (\omega - \kappa w_u) \b l + \b w_d^* + \b m_R,\]
with $\b w_R^*$ representing the portfolio weights corresponding to TE and weight constraints. $\b m_R$ is the weight restricted benchmark portfolio, hence benchmark also follows the weight restrictions. $\b w_d^*$ is the tracking error portfolio weights. One question that is asked is what is the difference between these weights and the unconstrained one. To answer that 
see that by (3) in the main text 
\[ \b w_d^*= \kappa \left[ \frac{\b \Theta \b \mu}{\b 1_p' \b \Theta \b \mu }- \b a 
\right],\]
since $\b a:= \b \Theta \b 1_p / \b 1_p' \b \Theta \b 1_p$ which are the weights of Global Minimum Variance (GMV) Portfolio. Since $\b  w_n^*$ in Proposition A.2
\[ \b w_n^* = \kappa_w ( \frac{ \b \Theta \b \mu}{\b 1_p' \b \Theta \b \mu} - \b a ) + \b a,\]
The difference in the portfolio weights between $\b w_R^* - \b w_n^* $ given $\kappa = \kappa_w$, and $\b w_d^*$ formula used in $\b w_R^*$ 
\[ \b w_R^* - \b w_n^* = [( \omega - \kappa w_u) \b l + \b m_R] - [\b a],\]
and the estimated difference is 
\[ \hat{\b w}_R - \hat{\b w}_n = [(\omega - \kappa \hat{w}_u) \hat{\b l} + \b m_R] -[ \hat{\b a}].\]
Clearly
\[ \| (\hat{\b w}_R - \b w_R^*) - ( \hat{\b w}_n - \b w_n^*) \|_1 
= \| [(\omega - \kappa \hat{w}_u )  \hat{\b l} - ( \omega - \kappa w_u) \b l] -[ \hat{\b a} - \b a] \|_1.\]
The first square bracketed estimation error term on the right side is due to joint TE and weight constraint, and the second square bracketed one $(\hat{\b a} - \b a)$ is due to unconstrained portfolio. This implies that it is not clear whether joint TE and weight estimation errors is large compared with unconstrained portfolio in $l_1$ sense. Asymptotically the difference will converge in probability to zero by (\ref{w.300}), Lemma A.1(ii) and Assumption 8(i).  Note that also the same effects will occur in variance and Sharpe Ratio estimation as well.

  \section{Short-sale constraints}

  First, we show that cap constraints as we use in this paper, are equivalent to floor constraints.  The restricted index indicator is defined in the main text as $\b 1_R$, where $R$ represent the indices with restrictions (i.e. assets with restrictions). For example if $R={2, 5,7}$ out of $p=100$ assets, $1_R= (0,1,0,0,1,0,1,0,\cdots,0)':100\times 1, $where $1$'s are in positions 2, 5, 7 out of 100 asset indicator function. We define $R^c$ as the complement of $R$ as the indices of the unrestricted assets.
  In this example, $\b 1_{R^c}=(1,0,1,1,0,1,0,1,1,1,\cdots1)':100 \times 1$. Note that 
  \[ \b 1_R' \b w + \b 1_{R^c}' \b w =1.\]
  So $\b 1_{R^c}' \b w \le w_x $ is equivalent to $\b 1_{R}' \b w \ge 1 - w_x$ as $w_x$ is defined in Section A.6. So a floor constraint like short-sale, where $w_x=1$, is equivalent to a cap on the remaining assets. So all our optimization will be the same but only where we use $\b 1_{R}$ will be replaced by $\b 1_{R^c}$. 
  Specifically, portfolios $\b l , \b k$ and scalars $w_k, w_a, w_{u}$ will be replaced with the ones that use $\b 1_{ R^c}$ instead of $\b 1_{R}$ in their formulas, and they are $\b l_s, \b k_s$ and $w_{k_s}, w_{a_s}, w_{u_s}$ respectively, and are shown below.  For the inequality constraint case we replace
  $\b w_{op}^*$ in Section 5 of main text with the following
  \[ \b w_{nsi}:= \b w_d^*  1_{ \{ \kappa w_{u_s} < 1\}} + \b w_{on} 1_{ \{ \kappa w_{u_s} > 1\}},\]
 where $\b w_d^*$ is the same as in (3) in the main text
 \[ \b w_{on}:= (\omega - \kappa w_{u_s}) \b l_s + \b w_d^*,\]
 \[ \b l_s:= \frac{\b k_s - \b a }{w_{k_s} - w_{a_s}},\]
  and 
  \[ \b k_s:= \frac{\b \Theta \b 1_{R^c}}{\b 1_p' \b \Theta \b 1_{R^c}},\]
  with 
  \[ w_{k_s}:= \frac{\b 1_{R^c}' \b \Theta \b 1_{R^c}}{\b 1_{R^c}' \b \Theta \b 1_p}, \quad
  w_{a_s}:= \frac{\b 1_{R^c}' \b \Theta \b 1_p}{\b 1_p' \b \Theta \b 1_p},\]
  \[  w_{u_s}:= \b 1_{R^c}' \left[ \frac{\b \Theta \b \mu}{\b 1_p' \b \Theta \b \mu } - 
  \frac{\b \Theta \b 1_p}{\b 1_p' \b \Theta \b 1_p} \right].\]
  The estimators of $w_{k_s}, w_{a_s}, w_{u_s}$,  are obtained by using $\hat{\b \Theta}, \hat{\b \mu}$ in the terms above in this subsection. To save from notation, we have not provided them explicitly. 
But the estimator of the portfolio weights in the inequality case is, which replaces $\hat{\b w}_{op}$ in Section 5-main text:     
  \begin{equation}
   \hat{\b w}_{nsi}:= \hat{\b w}_d 1_{ \{ \kappa \hat{w}_{u_s} <1\}} + \hat{\b w}_{on} 1_{ \{ \kappa \hat{\b w}_{u_s} > 1 \}}.\label{nos}
   \end{equation}

  We subdivide the issue of short-sales constraint into two cases. First, we consider what will happen if there is only asset that we want to be constrained, and the constraint type will be only short-sale constraint. Assume without losing any generality, we have a short-sale constraint in third asset. There we have $\b 1_{R^c}
  := (1,1,0,1,\cdots, 1)'$, where $0$ is placed in third cell in a $p \times 1$ vector of ones. Then we  estimate weights according to (\ref{nos}). 
  As a second case, we can handle a joint short-sales constraint for a group of assets. This may be forming say a health related portfolio out of many assets in the portfolio, and imposing a short sale constraint for the health related portfolio. This may amount to 
  total of weights in health portfolio to be larger than equal to zero. This means, if health related portfolio, without losing any generality, are the first $4$ assets, then 
  $\b 1_{R^c}:= (0,0,0,0,1,1,1,\cdots, 1)'$ which are zeroes in the first 4 cells, and the rest is one. The estimation is done via (\ref{nos}). \\

  \section{Additional Empirics}

 In this section, we have additional empirics. Namely, we consider empirics without transaction costs for $S\&P$ 500 with TE restriction, whereas in the main text we use Table 5 for TE restriction with transaction costs. Also in the main text we have energy and healthcare sectors with transaction costs in Tables 6-7. We provide tables for the healthcare sector in weight restrictions only, TE restriction only and the joint TE plus weight inequality restrictions as well as the unconstrained one without transaction costs. 
 We also have healthcare sector tables with transaction costs with TE,  equality weight, and unconstrained portfolio.  All the remaining tables in energy-industrials sectors, and industry sorted portfolios as well as characteristic sorted portfolios will be with transaction costs. The versions without transaction costs are ready and be requested from authors on demand. Same as in the main text, we report the performance of the following portfolios: ``Oracle'' is the infeasible out-of-sample oracle portfolio that contains look-ahead information, ``Index'' refers to S\&P 500 Index, S\&P 500 Healthcare Sector Index, and S\&P 500 Energy Sector Index in corresponding context when the benchmark is such index as indicated by the table caption. 
 
 AVR represents average returns, TE represents Tracking Error, Risk represents standard deviation of the portfolio, SR represents the Sharpe Ratio, TO represents Turnover ratio. "p-val" is the p-value of the SR test. Max/Min weight are the maximum/minimum individual asset weights in the portfolio. Total Short is the total shorting position of the portfolio. Index and methods are described in the main text in simulation-empirics sections. For equality weight constraints 80\% of the themed assets are used, and for the inequality weight constraints at least 80\% of the themed assets are used. Note that the unconstrained oracle portfolios based on S\&P 500 stocks have zero TO  because the unconstrained solution does not change across the out-of-sample period. Also note that the oracle portfolios based on the industry and characteristic sorted portfolios have zero TO regardless of the type of constraints imposed because, different from the adaptive themed assets in S\&P 500 Sector indices, the themed assets here do not change across the out-of-sample period, so the solution remains the same.
 
   For $S\&P \, 500$, the in-sample period is from 01/1981-12/1995, and the out-sample is 01/1996-12/2020. For the health-energy-industrials sectors in-sample period is: 01/1991-12/2005, and the out-sample is 01/2006-12/2020. Window length is 180 months.
 
 Table \ref{te500} clearly shows that CROWN has the highest SR at 0.1682 without transaction costs.   Tables \ref{hte}-\ref{hu} show tables without transaction costs, with TE constraint, with only weight constraint, joint TE plus weight inequality and the unconstrained one (where weights are added to one). Clearly, CROWN has the highest Sharpe Ratio in all Tables \ref{hte}-\ref{hu}, the results are mixed in the other metrics without transaction costs. Tables \ref{tehc}-\ref{nchc} provide results for the health sector with transaction costs. CROWN still has the best Sharpe ratio among all methods even with transaction costs.  Tables \ref{ete}-\ref{enc} cover energy sector with transaction costs. Note that  for both energy and health sectors we  have the inequality tables in the main text. Again our method has the best Sharpe Ratio here among all methods. But to compare health care versus energy sector we see that in the equality weighted restrictions, Table \ref{whc} and Table \ref{eew}, CROWN has Sharpe ratio of 0.2090 versus 0.0632 respectively. It seems like health care sector delivered more than triple the Sharpe Ratio compared with energy sector. The main reason is the constraints.  To see this more clearly, at Table \ref{enc}, the Sharpe Ratio for the unconstrained portfolio is 0.2093.
 The other methods such as NLS has the best   risk in Table \ref{eew}, and POET has the best return. Tables \ref{ite}-\ref{inc} consider industrials sector. The results show that CROWN has the best Sharpe Ratio in the out-sample that we considered. 
 
 The industry sorted portfolios Tables \ref{48te}-\ref{48nc} report datasets of industry-sorted portfolios, from Fama-French data library, and specifically, the 48 industry portfolios. The data is monthly. The sample period is 1991/01 to 2005/12, and the out of sample period is 2006/01 to 2020/12.  For 48 industry portfolios under an equality (inequality) weight constraint, we allocate exactly (no less than) 0.8 of the weight to the top 24 portfolios and the remaining weight to the other 24 sorted portfolios.   The results show that in Tables \ref{48te}-\ref{48nc} CROWN has the best Sharpe Ratio. To give an example, with inequality constraints, CROWN has 0.1520 Sharpe Ratio, and the second best method is NLS with 0.1510 in Table \ref{48ic}. In terms of risk, CROWN has the best risk with 0.0605 in Table \ref{48ic}, but POET has the best return with 0.0094. 
 
 The characteristic sorted portfolios tables report datasets of characteristic-sorted portfolios, from Fama-French data library, and specifically, the 100 ($10\times10$) Size and Book-to-Market sorted portfolios. The data is monthly. The sample period is 1991/01 to 2005/12, and the out of sample period is 2006/01 to 2020/12.  We only have TE restricted case and unconstrained case in Tables \ref{10te}-\ref{10nc}. The results show our method has the largest SR among all methods in TE constrained case. Also our method is not affected by much from the constraint in terms of SR.
 
 In summary, our CROWN method delivers very good Sharpe Ratio results, but in terms of average return we see POET is doing very well, in terms of risk both shrinkage based methods NLS, SFNL and our CROWN does well.


\begin{table}
        \scriptsize
        \centering
        \caption{TE constraint, S\&P 500 Index: Without Transaction Cost}
	\label{te500}
\begin{tabular}{lcccccccc}	
	
        \noalign{\global\arrayrulewidth1pt}\hline \noalign{\global\arrayrulewidth0.4pt}
        
        & \multicolumn{8}{c}{S\&P 500 Index} \\
        \cline{2-9}
& AVR & TE & Risk & SR & \textit{p-val} & Max & Min  & Total \\ 
                     &     &    &      &    &                &  Weight & Weight & Short \\ \hline
Oracle & 0.0185 & 0.0110 & 0.0432 & 0.4286 & 0.0000 & 0.0512 & -0.0002 & 0.0012\\
\hline
INDEX & 0.0037 & - & 0.0418 & 0.0894 & 0.0000 & - & - & -\\
NW & \textbf{0.0069} & 0.0030 & 0.0415 & 0.1669 & 0.2967 & 0.0424 & -0.0021 & 0.0403\\
CROWN & \textbf{0.0069} & 0.0017 & \textbf{0.0411} & \textbf{0.1682} & - & 0.0420 & -0.0006 & 0.0082\\
POET & \textbf{0.0069} & \textbf{0.0013} & 0.0416 & 0.1664 & 0.2782 & 0.0418 & -0.0008 & 0.0122\\
NLS & 0.0068 & 0.0025 & 0.0412 & 0.1662 & 0.1600 & 0.0425 & -0.0022 & 0.0555\\
SFNL &\textbf{0.0069} & 0.0023 & 0.0413 & 0.1667 & 0.2130 & 0.0424 & -0.0020 & 0.0487\\
\noalign{\global\arrayrulewidth1pt}\hline \noalign{\global\arrayrulewidth0.4pt}    
            \end{tabular}
            \end{table}

\begin{table}
        \scriptsize
        \centering
        \caption{TE constraint, S\&P 500 Healthcare Sector: Without Transaction Cost} 
	\label{hte}
        \begin{tabular}{lcccccccc}		
	\noalign{\global\arrayrulewidth1pt}\hline \noalign{\global\arrayrulewidth0.4pt}
            & \multicolumn{8}{c}{S\&P 500 Healthcare Sector} \\
            \cline{2-9}
& AVR & TE & Risk & SR & \textit{p-val} & Max & Min  & Total \\ 
                     &     &    &      &    &                &  Weight & Weight & Short \\ \hline
Oracle & 0.0122 & 0.0039 & 0.0385 & 0.3173 & 0.0000 & 0.1296 & -0.0002 & 0.0020\\
\hline
INDEX & 0.0081 & - & 0.0389 & 0.2088 & 0.0646 & - & - & -\\
NW & 0.0083 & \textbf{0.0013} & 0.0386 & 0.2156 & 0.2447 & 0.1203 & -0.0011 & 0.0220\\
CROWN & 0.0083 & 0.0034 & \textbf{0.0378} & \textbf{0.2185} & - & 0.1201 & -0.0069 & 0.0536\\ %
POET & \textbf{0.0084} & 0.0072 & 0.0389 & 0.2150 & 0.3018 & 0.4816 & -0.4592 & 0.5604\\ %
NLS & 0.0082 & 0.0025 & 0.0384 & 0.2145 & 0.1359 & 0.1208 & -0.0048 & 0.0811\\
SFNL & 0.0083 & 0.0014 & 0.0386 & 0.2146 & 0.2104 & 0.1206 & -0.0022 & 0.0410\\
\noalign{\global\arrayrulewidth1pt}\hline \noalign{\global\arrayrulewidth0.4pt}    
            \end{tabular}
            \end{table}

\begin{table}
    \scriptsize
    \centering
    \caption{Equality weight constraint, S\&P 500 Healthcare Sector: Without Transaction Cost}
\label{hw}
    \begin{tabular}{lccccccccc}		\noalign{\global\arrayrulewidth1pt}\hline \noalign{\global\arrayrulewidth0.4pt}
        & \multicolumn{9}{c}{S\&P 500 Healthcare Sector} \\
        \cline{2-10}
& AVR & TE & Risk & SR & \textit{p-val} & \%Themed & Max & Min  & Total \\ 
                     &     &    &      &    &                &  Assets  & Weight & Weight & Short \\ \hline
Oracle & 0.0075 & 0.0452 & 0.0363 & 0.2072 & 0.6505 & 0.8000 & 0.0715 & -0.0198 & 1.1041\\
\hline
INDEX & 0.0081 & - & 0.0389 & 0.2088 & 0.4991 & 0.8000 & - & - & -\\
NW & 0.0093 & 0.0323 & 0.0452 & 0.2066 & 0.4287 & 0.8872 & 0.0664 & -0.0108 & 0.1493\\
CROWN & 0.0076 & 0.0386 & 0.0364 & \textbf{0.2090} & - & 0.8000 & 0.1318 & -0.0104 & 0.5970\\
POET & \textbf{0.1376} & 2.0684 & 2.0633 & 0.1909 & 0.3947 & 0.8000 & 0.5951 & -0.5687 & 2.7100\\
NLS & 0.0068 & 0.0329 & 0.0347 & 0.1955 & 0.2991 & 0.8000 & 0.1330 & -0.0251 & 1.4439\\
SFNL & 0.0071 & 0.0285 & \textbf{0.0341} & 0.2070 & 0.4115 & 0.8000 & 0.1407 & -0.0268 & 1.4261\\
\noalign{\global\arrayrulewidth1pt}\hline \noalign{\global\arrayrulewidth0.4pt}    
            \end{tabular}
            \end{table}

 \begin{table}
    \scriptsize
    \centering
    \caption{TE and inequality weight constraints, S\&P 500 Healthcare Sector: Without Transaction Cost}
	\label{htwi}
    \begin{tabular}{lccccccccc}		\noalign{\global\arrayrulewidth1pt}\hline \noalign{\global\arrayrulewidth0.4pt}
        & \multicolumn{9}{c}{S\&P 500 Healthcare Sector} \\
        \cline{2-10}
& AVR & TE & Risk & SR & \textit{p-val} & \%Themed & Max & Min  & Total \\ 
                     &     &    &      &    &                &  Assets  & Weight & Weight & Short \\ \hline
Oracle & 0.0122 & 0.0039 & 0.0385 & 0.3171 & 0.0000 & 0.8000 & 0.1297 & -0.0002 & 0.0020\\
\hline
INDEX & 0.0081 & - & 0.0389 & 0.2088 & 0.0568 & 0.8000 & - & - & -\\
NW & 0.0083 & \textbf{0.0013} & 0.0386 & 0.2156 & 0.2519 & 0.8018 & 0.1203 & -0.0011 & 0.0220\\
CROWN & 0.0083 & 0.0034 & \textbf{0.0379} & \textbf{0.2185} & - & 0.8020 & 0.1203 & -0.0052 & 0.0531\\ %
POET & \textbf{0.0084} & 0.0072 & 0.0389 & 0.2150 & 0.2731 & 0.9891 & 0.4816 & -0.4592 & 0.5604\\ %
NLS & 0.0082 & 0.0025 & 0.0384 &  0.2144 & 0.1431 & 0.8027 & 0.1210 & -0.0040 & 0.0880\\
SFNL & 0.0083 & 0.0015 & 0.0386 & 0.2147 & 0.2176 & 0.8013 & 0.1208 & -0.0023 & 0.0457\\
\noalign{\global\arrayrulewidth1pt}\hline \noalign{\global\arrayrulewidth0.4pt}    
            \end{tabular}
            \end{table}

\begin{table}
        \scriptsize
        \centering
        \caption{No constraint, S\&P 500 Healthcare Sector: Without Transaction Cost}
	\label{hu}
        \begin{tabular}{lcccccccc}		\noalign{\global\arrayrulewidth1pt}\hline \noalign{\global\arrayrulewidth0.4pt}
            & \multicolumn{8}{c}{S\&P 500 Healthcare Sector} \\
            \cline{2-9}
& AVR & TE & Risk & SR & \textit{p-val} & Max & Min  & Total \\ 
                     &     &    &      &    &                &  Weight & Weight & Short \\ \hline
Oracle & 0.0116 & 0.0494 & 0.0352 & 0.3308 & 0.0000 & 0.0505 & -0.0231 & 0.8701\\
\hline
INDEX & 0.0081 & - & 0.0389 & 0.2088 & 0.4237 & - & - & -\\
NW & 0.0094 & 0.0232 & 0.0434 & 0.2164 & 0.4865 & 0.0118 & -0.0005 & 0.0028\\
CROWN & 0.0071 & 0.0348 & 0.0326 & \textbf{0.2192} & - & 0.0298 & -0.0087 & 0.4312\\
POET & \textbf{0.5874} & 2.7888 & 2.6832 & 0.2189 & 0.4359 & 0.1576 & -0.1449 & 5.2152\\
NLS & 0.0069 & 0.0340 & \textbf{0.0318} & 0.2166 & 0.4453 & 0.0390 & -0.0217 & 1.0434\\ %
SFNL & 0.0070 & 0.0318 & 0.0325 & 0.2165 & 0.4479 & 0.0438 & -0.0241 & 1.0993\\ %
\noalign{\global\arrayrulewidth1pt}\hline \noalign{\global\arrayrulewidth0.4pt}    
            \end{tabular}
            \end{table}

\begin{table}
        \scriptsize
        \centering
        \caption{TE constraint, S\&P 500 Healthcare Sector: Transaction Cost}
	\label{tehc}
        \begin{tabular}{lccccccccc}		\noalign{\global\arrayrulewidth1pt}\hline \noalign{\global\arrayrulewidth0.4pt}
            & \multicolumn{8}{c}{S\&P 500 Healthcare Sector} \\
            \cline{2-10}
& AVR & TE & Risk & SR &  TO    &\textit{p-val} & Max & Min  & Total \\ 
                     &     &    &      &    &      &          &  Weight & Weight & Short \\ \hline
Oracle & 0.0120 & 0.0039 & 0.0385 & 0.3119 & 0.0409 & 0.0007 & 0.1296 & -0.0002 & 0.0020\\
\hline
INDEX & \textbf{0.0081} & - & 0.0389 & 0.2088 & - & 0.4889 & - & - & -\\
NW & 0.0080 & 0.0023 & 0.0384 & 0.2076 & 0.0673 & 0.4184 & 0.1201 & -0.0019 & 0.0474\\
CROWN & 0.0078 & 0.0059 & \textbf{0.0373} & \textbf{0.2091} & 0.0854 & - & 0.1198 & -0.0122 & 0.1102\\
POET & 0.0054 & 0.0072 & 0.0387 & 0.1389 & 0.6432 & 0.0000 & 0.1218 & -0.0051 & 0.1354\\
NLS & 0.0077 & 0.0033 & 0.0381 & 0.2027 & 0.0878 & 0.1619 & 0.1209 & -0.0068 & 0.1230\\
SFNL & 0.0079 & \textbf{0.0019} & 0.0384 & 0.2056 & 0.0702 & 0.3379 & 0.1207 & -0.0040 & 0.0653\\
\noalign{\global\arrayrulewidth1pt}\hline \noalign{\global\arrayrulewidth0.4pt}    
            \end{tabular}
            \end{table}

\begin{table}
    \scriptsize
    \centering
    \caption{Equality weight constraint, S\&P 500 Healthcare Sector: Transaction Cost}
\label{whc}
    \begin{tabular}{lcccccccccc}		\noalign{\global\arrayrulewidth1pt}\hline \noalign{\global\arrayrulewidth0.4pt}
        & \multicolumn{9}{c}{S\&P 500 Healthcare Sector} \\
        \cline{2-11}
& AVR & TE & Risk & SR &   TO    &\textit{p-val} & \%Themed & Max & Min  & Total \\ 
                     &     &    &      &    &        &        &  Assets  & Weight & Weight & Short \\ \hline
Oracle & 0.0075 & 0.0452 & 0.0363 & 0.2062 & 0.0077 & 0.4790 & 0.8000 & 0.0715 & -0.0198 & 1.1041\\
\hline
INDEX & 0.0081 & - & 0.0389 & 0.2088 & - & 0.4989 & - & - & - & -\\
NW & 0.0070 & 0.0298 & 0.0417 & 0.1683 & 0.3954 & 0.3076 & 0.8872 & 0.0664 & -0.0108 & 0.1493\\
CROWN & 0.0067 & 0.0353 & 0.0322 & \textbf{0.2090} & 0.0960 & - & 0.8000 & 0.0296 & -0.0085 & 0.4363\\
POET & \textbf{5.2115} & 28.5690 & 28.6000 & 0.1822 & 2.9888 & 0.3811 & 0.8000 & 0.5951 & -0.5687 & 2.7100\\
NLS & 0.0053 & 0.0334 & \textbf{0.0321} & 0.1659 & 0.3120 & 0.1747 & 0.8000 & 0.1330 & -0.0251 & 1.4439\\
SFNL & 0.0054 & 0.0310 & 0.0326 & 0.1660 & 0.3247 & 0.2059 & 0.8000 & 0.1407 & -0.0268 & 1.4261\\
\noalign{\global\arrayrulewidth1pt}\hline \noalign{\global\arrayrulewidth0.4pt}    
            \end{tabular}
            \end{table}

\begin{table}
        \scriptsize
        \centering
        \caption{No constraint, S\&P 500 Healthcare Sector: Transaction Cost}
	\label{nchc}
        \begin{tabular}{lccccccccc}		\noalign{\global\arrayrulewidth1pt}\hline \noalign{\global\arrayrulewidth0.4pt}
            & \multicolumn{8}{c}{S\&P 500 Healthcare Sector} \\
            \cline{2-10}
& AVR & TE & Risk & SR &  TO    &\textit{p-val} & Max & Min  & Total \\ 
                     &     &    &      &    &      &          &  Weight & Weight & Short \\ \hline
Oracle & 0.0116 & 0.0494 & 0.0352 & 0.3308 & - & 0.0000 & 0.0505 & -0.0231 & 0.8701\\
\hline
INDEX & 0.0081 & - & 0.0389 & 0.2088 & - & 0.4543 & - & - & -\\
NW & 0.0089 & 0.0232 & 0.0439 & 0.2024 & 0.0921 & 0.4630 & 0.0115 & -0.0005 & 0.0027\\
CROWN & 0.0068 & 0.0342 & 0.0324 & \textbf{0.2093} & 0.0913 & - & 0.0341 & -0.0083 & 0.4172\\
POET & \textbf{0.5603} & 2.7888 & 2.7861 & 0.2011 & 11.4042 & 0.3820 & 0.1384 & -0.1434 & 5.9364\\
NLS & 0.0053 & 0.0341 & \textbf{0.0320} & 0.1668 & 0.3105 & 0.1842 & 0.0389 & -0.0216 & 1.0435\\
SFNL & 0.0055 & 0.0317 & 0.0327 & 0.1669 & 0.3227 & 0.1971 & 0.0436 & -0.0240 & 1.0983\\
\noalign{\global\arrayrulewidth1pt}\hline \noalign{\global\arrayrulewidth0.4pt}    
            \end{tabular}
            \end{table}

\begin{table}
        \scriptsize
        \centering
        \caption{TE constraint, S\&P 500 Energy Sector: Transaction Cost}
	\label{ete}
        \begin{tabular}{lccccccccc}		\noalign{\global\arrayrulewidth1pt}\hline \noalign{\global\arrayrulewidth0.4pt}
            & \multicolumn{8}{c}{S\&P 500 Energy Sector} \\
            \cline{2-10}
& AVR & TE & Risk & SR &  TO    &\textit{p-val} & Max & Min  & Total \\ 
                     &     &    &      &    &      &          &  Weight & Weight & Short \\ \hline
Oracle & 0.0090 & 0.0172 & 0.0593 & 0.1526 & 0.0363 & 0.0000 & 0.2917 & -0.0029 & 0.1201\\
\hline
INDEX & 0.0034 & - & 0.0723 & 0.0473 & - & 0.1584 & - & - & -\\
NW & 0.0031 & 0.0157 & \textbf{0.0601} & 0.0514 & 0.1131 & 0.2375 & 0.2868 & -0.0055 & 0.1902\\
CROWN & \textbf{0.0035} & \textbf{0.0135} & 0.0619 & \textbf{0.0569} & 0.0396 & - & 0.2883 & -0.0000 & 0.0002\\
POET & 0.0033 & \textbf{0.0135} & 0.0619 & 0.0539 & 0.0795 & 0.0000 & 0.2884 & -0.0003 & 0.0041\\
NLS & 0.0031 & 0.0147 & 0.0605 & 0.0508 & 0.0906 & 0.0711 & 0.2877 & -0.0036 & 0.1476\\
SFNL & 0.0033 & 0.0140 & 0.0612 & 0.0534 & 0.0704 & 0.0718 & 0.2880 & -0.0021 & 0.0793\\
\noalign{\global\arrayrulewidth1pt}\hline \noalign{\global\arrayrulewidth0.4pt}    
            \end{tabular}
            \end{table}

\begin{table}
    \scriptsize
    \centering
    \caption{Equality weight constraint, S\&P 500 Energy Sector: Transaction Cost} 
	\label{eew}
    \begin{tabular}{lcccccccccc}		\noalign{\global\arrayrulewidth1pt}\hline \noalign{\global\arrayrulewidth0.4pt}
        & \multicolumn{9}{c}{S\&P 500 Energy Sector} \\
        \cline{2-11}
& AVR & TE & Risk & SR & TO & \textit{p-val} &\%Themed & Max & Min  & Total \\ 
                     &     &    &      &   &          &       &  Assets  & Weight & Weight & Short \\ \hline
Oracle & 0.0048 & 0.0671 & 0.0500 & 0.0968 & 0.0094 & 0.8133 & 0.8000 & 0.1571 & -0.0247 & 1.2508\\                     
\hline
INDEX & 0.0034 & - & 0.0723 & 0.0473 & -  & 0.4429 & 0.8000 & - & - & -\\
NW & 0.0051 & 0.0888 & 0.0859 & 0.0599 & 0.1916 &0.4843 & 0.8358 & 0.0938 & -0.0732 & 3.3531\\
CROWN & 0.0032 & 0.0346 & 0.0512 & \textbf{0.0632} & 0.1468 & - & 0.8000 & 0.1102 & -0.0171 & 0.7221\\
POET & \textbf{0.1328} & 2.2457 & 2.2503 & 0.0590 & 0.6238 &0.4832 & 0.8000 & 0.1768 & -0.0376 & 1.2378\\
NLS & 0.0027 & 0.0588 & \textbf{0.0435} & 0.0617 & 0.4758 &0.4901 & 0.8000 & 0.1441 & -0.0380 & 1.9440\\
SFNL & 0.0027 & 0.0587 & 0.0454 & 0.0590 & 0.5207 &0.4723 & 0.8000 & 0.1471 & -0.0365 & 2.0385\\
\noalign{\global\arrayrulewidth1pt}\hline \noalign{\global\arrayrulewidth0.4pt}    
\end{tabular}
\end{table}

\begin{table}
        \scriptsize
        \centering
        \caption{No constraint, S\&P 500 Energy Sector: Transaction Cost} 
	\label{enc}
        \begin{tabular}{lccccccccc}		\noalign{\global\arrayrulewidth1pt}\hline \noalign{\global\arrayrulewidth0.4pt}
            & \multicolumn{8}{c}{S\&P 500 Energy Sector} \\
            \cline{2-10}
& AVR & TE & Risk & SR &  TO    &\textit{p-val} & Max & Min  & Total \\ 
                     &     &    &      &    &      &          &  Weight & Weight & Short \\ \hline
Oracle & 0.0116 & 0.0859 & 0.0352 & 0.3308 & - & 0.0000 & 0.0505 & -0.0231 & 0.8701\\
\hline
INDEX & 0.0034 & - & 0.0723 & 0.0473 & - & 0.0494 & - & - & -\\
NW & 0.0089 & 0.0517 & 0.0439 & 0.2024 & 0.0921 & 0.4630 & 0.0115 & -0.0005 & 0.0027\\
CROWN & 0.0068 & 0.0739 & 0.0324 & \textbf{0.2093} & 0.0913 & - & 0.0341 & -0.0083 & 0.4172\\
POET & \textbf{0.5603} & 2.3908 & 2.7861 & 0.2011 & 11.4042 & 0.3820 & 0.1384 & -0.1434 & 5.9364\\
NLS & 0.0053 & 0.0684 & \textbf{0.0320} & 0.1668 & 0.3105 & 0.1842 & 0.0389 & -0.0216 & 1.0435\\
SFNL & 0.0055 & 0.0629 & 0.0327 & 0.1669 & 0.3227 & 0.1971 & 0.0436 & -0.0240 & 1.0983\\
\noalign{\global\arrayrulewidth1pt}\hline \noalign{\global\arrayrulewidth0.4pt}    
            \end{tabular}
            \end{table}

\begin{table}
        \scriptsize
        \centering
        \caption{TE constraint, S\&P 500 Industrials Sector: Transaction Cost} 
	\label{ite}
        \begin{tabular}{lccccccccc}		\noalign{\global\arrayrulewidth1pt}\hline \noalign{\global\arrayrulewidth0.4pt}
            & \multicolumn{8}{c}{S\&P 500 Industrials Sector} \\
            \cline{2-10}
& AVR & TE & Risk & SR & \textit{p-val}  &  TO  & Max & Min  & Total \\ 
                     &     &    &      &    &      &          &  Weight & Weight & Short \\ \hline
Oracle & 0.0118 & 0.0031 & 0.0525 & 0.2243 & 0.0000 & 0.0408 & 0.1545 & -0.002 & 0.0021\\
\hline
INDEX & \textbf{0.0078} & - & 0.0529 & 0.1484 & 0.3965 & - & - & - & -\\
NW & 0.0076 & 0.0036 & 0.0514 & 0.1482 & 0.3263 & 0.0802 & 0.1443 & -0.0029 & 0.0815\\
CROWN & 0.0073 & 0.0089 & \textbf{0.0483} & \textbf{0.1514} & - & 0.1065 & 0.1434 & -0.0039 & 0.1716\\
POET & 0.0074 & 0.0031 & 0.0525 & 0.1407 & 0.1908 & 0.1458 & 0.1452 & -0.0009 & 0.0199\\
NLS & 0.0072 & 0.0050 & 0.0505 & 0.1433 & 0.1197 & 0.1109 & 0.1449 & -0.0048 & 0.1954\\
SFNL & 0.0075 & \textbf{0.0030} & 0.0515 & 0.1457 & 0.2582 & 0.0836 & 0.1450 & -0.0028 & 0.1065\\
\noalign{\global\arrayrulewidth1pt}\hline \noalign{\global\arrayrulewidth0.4pt}    
            \end{tabular}
            \end{table}

\begin{table}
    \scriptsize
    \centering
    \caption{Equality weight constraint, S\&P 500 Industrials Sector: Transaction Cost}
	\label{iew}
    \begin{tabular}{lcccccccccc}		\noalign{\global\arrayrulewidth1pt}\hline \noalign{\global\arrayrulewidth0.4pt}
        & \multicolumn{9}{c}{S\&P 500 Industrials Sector} \\
        \cline{2-11}
& AVR & TE & Risk & SR &  \textit{p-val}    &  TO & \%Themed & Max & Min  & Total \\ 
                     &     &    &      &    &        &        &  Assets  & Weight & Weight & Short \\ \hline
Oracle & 0.0093 & 0.0637 & 0.0362 & 0.2580 & 0.8550 & 0.0098 & 0.8000 & 0.0566 & -0.0249 & 1.1934\\
\hline
INDEX & 0.0078 & - & 0.0529 & 0.1484 & 0.3396 & - & 0.8000 & - & - & -\\
NW & \textbf{0.0085} & 0.0135 & 0.0482 & 0.1772 & 0.4574 & 0.1734 & 0.8508 & 0.0459 & -0.0113 & 0.2163\\
CROWN & 0.0070 & 0.0563 & 0.0376 & \textbf{0.1863} & - & 0.1639 & 0.8000 & 0.0394 & -0.0138 & 0.8264\\
POET & 0.0082 & 0.0679 & 0.0522 & 0.1565 & 0.1052 & 0.4807 & 0.8000 & 0.0860 & -0.0299 & 1.8490\\
NLS & 0.0041 & 0.0548 & 0.0368 & 0.1105 & 0.0583 & 0.3641 & 0.8000 & 0.0512 & -0.0291 & 1.4759\\
SFNL & 0.0046 & 0.0470 & \textbf{0.0352} & 0.1296 & 0.1627 & 0.3403 & 0.8000 & 0.0540 & -0.0295 & 1.4104\\
\noalign{\global\arrayrulewidth1pt}\hline \noalign{\global\arrayrulewidth0.4pt}    
            \end{tabular}
            \end{table}

 \begin{table}
    \scriptsize
    \centering
    \caption{TE and inequality weight constraint, S\&P 500 Industrials Sector: Transaction Cost} 
	\label{iic}
    \begin{tabular}{lcccccccccc}		\noalign{\global\arrayrulewidth1pt}\hline \noalign{\global\arrayrulewidth0.4pt}
        & \multicolumn{9}{c}{S\&P 500 Industrials Sector} \\
        \cline{2-11}
& AVR & TE & Risk & SR &  \textit{p-val}    &  TO  & \%Themed & Max & Min  & Total \\ 
                     &     &    &      &    &        &        &  Assets  & Weight & Weight & Short \\ \hline
Oracle & 0.0118 & 0.0031 & 0.0525 & 0.2243 &  0.0000 & 0.0408  & 0.8002 & 0.1545 & -0.0002 & 0.0021\\
\hline
INDEX & \textbf{0.0078} & - & 0.0529 & 0.1484 & 0.3983  & - & 0.8000 & - & - & -\\
NW & 0.0076 & 0.0036 & 0.0514 & 0.1481 & 0.3280 & 0.0803 & 0.8019 & 0.1444 & -0.0029 & 0.0821\\
CROWN & 0.0073 & 0.0089 & \textbf{0.0483} & \textbf{0.1514} & - & 0.1065 & 0.8073 & 0.1435 & -0.0039 & 0.1728\\
POET & 0.0070 & 0.0037 & 0.0525 & 0.1342 & 0.0897 & 0.2204 & 0.8008 & 0.1452 & -0.0014 & 0.0362\\
NLS & 0.0072 & 0.0050 & 0.0505 & 0.1432 & 0.1209 & 0.1109 & 0.8056 & 0.1451 & -0.0048 & 0.1956\\
SFNL & 0.0075 & \textbf{0.0030} & 0.0515 & 0.1457 & 0.2601 & 0.0837 & 0.8038 & 0.1450 & -0.0028 & 0.1071\\
\noalign{\global\arrayrulewidth1pt}\hline \noalign{\global\arrayrulewidth0.4pt}    
            \end{tabular}
            \end{table}

 \begin{table}
        \scriptsize
        \centering
        \caption{No constraint, S\&P 500 Industrials Sector: Transaction Cost} 
	\label{inc}
        \begin{tabular}{lccccccccc}		\noalign{\global\arrayrulewidth1pt}\hline \noalign{\global\arrayrulewidth0.4pt}
            & \multicolumn{8}{c}{S\&P 500 Industrials Sector} \\
            \cline{2-10}
& AVR & TE & Risk & SR &   \textit{p-val}   & TO & Max & Min  & Total \\ 
                     &     &    &      &    &      &          &  Weight & Weight & Short \\ \hline
Oracle & 0.0116 & 0.0654 & 0.0352 & 0.3308 & 0.0000 & -  & 0.0505 & -0.0231 & 0.8701\\
\hline
INDEX & 0.0078 & - & 0.0529 & 0.1484 & 0.3123 & -  & - & - & -\\
NW & 0.0089 & 0.0178 & 0.0439 & 0.2024 & 0.4630 & 0.0921 & 0.0115 & -0.0005 & 0.0027\\
CROWN & 0.0068 & 0.0523 & 0.0324 & \textbf{0.2093} & - & 0.0913 & 0.0341 & -0.0083 & 0.4172\\
POET & \textbf{0.5603} & 2.5911 & 2.7861 & 0.2011 & 0.3820 & 11.4042 & 0.1384 & -0.1434 & 5.9364\\
NLS & 0.0053 & 0.0507 & \textbf{0.0320} & 0.1668 & 0.1842 & 0.3105 & 0.0389 & -0.0216 & 1.0435\\
SFNL & 0.0055 & 0.0454 & 0.0327 & 0.1669 & 0.1971 & 0.3227 & 0.0436 & -0.0240 & 1.0983\\
\noalign{\global\arrayrulewidth1pt}\hline \noalign{\global\arrayrulewidth0.4pt}    
            \end{tabular}
            \end{table}

\begin{table}
        \scriptsize
        \centering
        \caption{TE constraint, 48 Industry Portfolios: Transaction Cost} 
	\label{48te}
        \begin{tabular}{lccccccccc}		\noalign{\global\arrayrulewidth1pt}\hline \noalign{\global\arrayrulewidth0.4pt}
            & \multicolumn{8}{c}{48 Industry Portfolios} \\
            \cline{2-10}
& AVR & TE & Risk & SR &  TO    &\textit{p-val} & Max & Min  & Total \\ 
                     &     &    &      &    &      &          &  Weight & Weight & Short \\ \hline
Oracle & 0.0110 & 0.0059 & 0.0629 & 0.1743 & - & 0.9709 & 0.0924 & -0.0235 & 0.0820\\
\hline
INDEX & \textbf{0.0088} & - & 0.0632 & 0.1389 & - & 0.3032 & - & - & -\\
NW & 0.0086 & \textbf{0.0040} & 0.1107 & 0.1396 & 0.0697 & 0.2607 & 0.0429 & -0.0040 & 0.0051\\
CROWN & 0.0084 & 0.0122 & \textbf{0.0573} & \textbf{0.1458} & 0.1284 & - & 0.0898 & -0.0454 & 0.1535\\
POET & \textbf{0.0088} & 0.0043 & 0.0612 & 0.1440 & 0.0366 & 0.4202 & 0.0444 & -0.0070 & 0.0109\\
NLS & 0.0084 & 0.0110 & 0.0580 & 0.1442 & 0.1180 & 0.3867 & 0.0937 & -0.0633 & 0.2132\\
SFNL & 0.0085 & 0.0083 & 0.0592 & 0.1430 & 0.0916 & 0.3357 & 0.0742 & -0.0436 & 0.1185\\
\noalign{\global\arrayrulewidth1pt}\hline \noalign{\global\arrayrulewidth0.4pt}    
            \end{tabular}
            \end{table}

\begin{table}
    \scriptsize
    \centering
    \caption{Equality weight constraint, 48 Industry Portfolios: Transaction Cost} 
\label{48ew}
    \begin{tabular}{lcccccccccc}		\noalign{\global\arrayrulewidth1pt}\hline \noalign{\global\arrayrulewidth0.4pt}
        & \multicolumn{9}{c}{48 Industry Portfolios} \\
        \cline{2-11}
& AVR & TE & Risk & SR &   TO    &\textit{p-val} & \%Themed & Max & Min  & Total \\ 
                     &     &    &      &    &        &        &  Assets  & Weight & Weight & Short \\ \hline
Oracle & 0.0074 & 0.0736 & 0.0320 & 0.2302 & - & 0.0034 & 0.8000 & 0.3695 & -0.0004 & 1.6672\\
\hline
INDEX & \textbf{0.0094} & - & 0.0636 & 0.1481 & - & 0.4998 & 0.8000 & - & - & -\\
NW & 0.0065 & 0.0247 & 0.0465 & 0.1400 & 0.5918 & 0.4530 & 0.4021 & 0.2055 & -0.0624 & 0.2911\\
CROWN & 0.0054 & 0.0593 & \textbf{0.0364} & \textbf{0.1482} & 0.2908 & - & 0.8000 & 0.4079 & -0.1793 & 1.2464\\
POET & 0.0078 & 0.0909 & 0.0556 & 0.1406 & 0.5029 & 0.4673 & 0.8000 & 0.4237 & -0.3525 & 2.6325\\
NLS & 0.0055 & 0.0720 & 0.0382 & 0.1428 & 0.3310 & 0.4668 & 0.8000 & 0.4188 & -0.2416 & 1.8226\\
SFNL & 0.0055 & 0.0724 & 0.0389 & 0.1416 & 0.3691 & 0.4613 & 0.8000 & 0.4002 & -0.2319 & 1.9119\\
\noalign{\global\arrayrulewidth1pt}\hline \noalign{\global\arrayrulewidth0.4pt}    
            \end{tabular}
            \end{table}

\begin{table}
    \scriptsize
    \centering
    \caption{TE and inequality weight constraint, 48 Industry Portfolios}
\label{48ic} 
    \begin{tabular}{lcccccccccc}		\noalign{\global\arrayrulewidth1pt}\hline \noalign{\global\arrayrulewidth0.4pt}
        & \multicolumn{9}{c}{48 Industry Portfolios} \\
        \cline{2-11}
& AVR & TE & Risk & SR &   TO    &\%Themed  & \textit{p-val} & Max & Min  & Total \\ 
                     &     &    &      &    &        &   Assets     &    & Weight & Weight & Short \\ \hline
Oracle & 0.0112 & 0.0047 & 0.0634 & 0.1760 & - & 0.8634 & 0.9983 & 0.0656 & -0.0271 & 0.0783\\
\hline
INDEX & \textbf{0.0094} & - & 0.0636 & 0.1481 & - & 0.8000 & 0.2687 & - & - & -\\
NW & 0.0089 & 0.0032 & 0.0629 & 0.1409 & 0.1905 & 0.8403 & 0.0255 & 0.0471 & -0.0086 & 0.0319\\
CROWN & 0.0092 & 0.0060 & \textbf{0.0605} & \textbf{0.1520} & 0.0640 & 0.8021 & - & 0.0623 & -0.0248 & 0.0514\\
POET & \textbf{0.0094} & \textbf{0.0021} & 0.0626 & 0.1507 & 0.0181 & 0.8012 & 0.3837 & 0.0442 & -0.0044 & 0.0052\\
NLS & 0.0092 & 0.0054 & 0.0608 & 0.1510 & 0.0582 & 0.8012 & 0.3616 & 0.0612 & -0.0270 & 0.0639\\
SFNL & 0.0093 & 0.0041 & 0.0615 & 0.1503 & 0.0454 & 0.8009 & 0.2975 & 0.0557 & -0.0193 & 0.0353\\
\noalign{\global\arrayrulewidth1pt}\hline \noalign{\global\arrayrulewidth0.4pt}    
            \end{tabular}
            \end{table}  
\newpage
            
\begin{table}
       \scriptsize
        \centering
        \caption{No constraint, 48 Industry Portfolios: Transaction Cost} 
	\label{48nc}
        \begin{tabular}{lccccccccc}		\noalign{\global\arrayrulewidth1pt}\hline \noalign{\global\arrayrulewidth0.4pt}
            & \multicolumn{8}{c}{48 Industry Portfolios} \\
            \cline{2-10}
& AVR & TE & Risk & SR &  TO    &\textit{p-val} & Max & Min  & Total \\ 
                     &     &    &      &    &      &          &  Weight & Weight & Short \\ \hline
Oracle & 0.0086 & 0.0745 & 0.0323 & 0.2654 & - & 0.9641 & 0.3956 & -0.2542 & 1.7330\\
\hline
INDEX & \textbf{0.0088} & - & 0.0632 & 0.1389 & - & 0.4242 & - & - & -\\
NW & 0.0071 & 0.0228 & 0.0460 & 0.1539 & 0.1920 & 0.4900 & 0.2077 & -0.0437 & 0.1770\\
CROWN & 0.0055 & 0.0621 & 0.0350 & \textbf{0.1558} & 0.2759 & - & 0.3935 & -0.1874 & 1.1605\\
POET & 0.0075 & 0.0847 & 0.0490 & 0.1536 & 0.4376 & 0.4892 & 0.3861 & -0.2952 & 2.2535\\
NLS & 0.0051 & 0.0571 & \textbf{0.0339} & 0.1514 & 0.1985 & 0.4562 & 0.4553 & -0.1777 & 1.2014\\
SFNL & 0.0053 & 0.0538 & 0.0345 & 0.1538 & 0.2156 & 0.4805 & 0.4215 & -0.1576 & 1.0877\\
\noalign{\global\arrayrulewidth1pt}\hline \noalign{\global\arrayrulewidth0.4pt}    
            \end{tabular}
            \end{table}

\begin{table}
        \scriptsize
        \centering
        \caption{TE constraint, 10-10 BM \& Size Sorted Portfolios: Transaction Cost}
	\label{10te}
        \begin{tabular}{lccccccccc}		\noalign{\global\arrayrulewidth1pt}\hline \noalign{\global\arrayrulewidth0.4pt}
            & \multicolumn{8}{c}{10-10 BM \& Size Sorted Portfolios} \\
            \cline{2-10}
& AVR & TE & Risk & SR &  TO    &\textit{p-val} & Max & Min  & Total \\ 
                     &     &    &      &    &      &          &  Weight & Weight & Short \\ \hline
Oracle & 0.0167 & 0.0127 & 0.0610 & 0.2732 & - & 0.0000 & 0.0971 & -0.0997 & 1.1557\\
\hline
INDEX & 0.0071 & - & 0.0611 & 0.1169 & - & 0.3023 & - & - & -\\
NW & 0.0076 & \textbf{0.0026} & 0.0606 & 0.1260 & 0.0797 & 0.4644 & 0.0235 & -0.0022 & 0.0036\\
CROWN & \textbf{0.0081} & 0.0175 & 0.0635 & \textbf{0.1278} & 0.3634 & - & 0.1313 & -0.0970 & 0.8860\\
POET & 0.0064 & 0.0054 & 0.0612 & 0.1039 & 0.3270 & 0.1246 & 0.0284 & -0.0092 & 0.0677\\
NLS & 0.0075 & 0.0043 & 0.0603 & 0.1251 & 0.1375 & 0.4426 & 0.0571 & -0.0294 & 0.1972\\
SFNL & 0.0076 & 0.0034 & \textbf{0.0602} & 0.1257 & 0.0975 & 0.4581 & 0.0477 & -0.0196 & 0.1042\\
\noalign{\global\arrayrulewidth1pt}\hline \noalign{\global\arrayrulewidth0.4pt}    
            \end{tabular}
            \end{table}   

\begin{table}
        \scriptsize
        \centering
        \caption{No constraint, 10-10 BM \& Size Sorted Portfolios: Transaction Cost}
        \label{10nc}
        \begin{tabular}{lccccccccc}		\noalign{\global\arrayrulewidth1pt}\hline \noalign{\global\arrayrulewidth0.4pt}
            & \multicolumn{8}{c}{10-10 BM \& Size Sorted Portfolios} \\
            \cline{2-10}
& AVR & TE & Risk & SR &  TO    &\textit{p-val} & Max & Min  & Total \\ 
                     &     &    &      &    &      &          &  Weight & Weight & Short \\ \hline
Oracle & 0.0158 & 0.0779 & 0.0408 & 0.3874 & - & 0.0000 & 0.5679 & -0.2735 & 4.0179\\
\hline
INDEX & 0.0071 & - & 0.0611 & 0.1169 & 0.4146 & - & - & - & -\\
NW & 0.0101 & 0.0223 & 0.0511 & 0.1981 & 0.7771 & 0.5059 & 0.0880 & -0.0659 & 1.0508\\
CROWN & 0.0090 & 0.0777 & 0.0657 & 0.1367 & - & 0.9168 & 0.4195 & -0.2167 & 1.1771\\
POET & \textbf{0.0114} & 0.0956 & 0.0955 & 0.1195 & 0.4203 & 5.8565 & 1.5188 & -1.4386 & 21.711\\
NLS & 0.0105 & 0.0636 & 0.0426 & \textbf{0.2624} & 0.9451 & 0.9434 & 0.3894 & -0.2956 & 1.6583\\
SFNL & 0.0081 & 0.0531 & \textbf{0.0362} & 0.2237 & 0.8898 & 0.5402 & 0.3069 & -0.2039 & 1.6183\\
\noalign{\global\arrayrulewidth1pt}\hline \noalign{\global\arrayrulewidth0.4pt}    
            \end{tabular}
            \end{table}
 
\clearpage

  \vspace{1in}
  
\setcounter{section}{0}\renewcommand{\thesection}{B.\arabic{section}}
\setcounter{equation}{0}\setcounter{lemma}{0}\renewcommand{\theequation}{B.\arabic{equation}}\renewcommand{\thelemma}{B.\arabic{lemma}}
\setcounter{table}{0}\renewcommand{\thetable}{B.\arabic{table}}
  \begin{center}
  
  {\bf \Large APPENDIX B}
  \end{center}
  
  Section \ref{b1} imposes a different assumption instead of Assumption 7(ii) in the main text.  The analysis is extensive with proofs. Section \ref{b2}, we cover feasible residual nodewise regression. Section B.3 provides consistent estimation of the risk tolerance parameter. Section B.4 discusses the common rate across different restrictions. Section B.5 contains extra simulation results for dense precision matrix of the errors.

  \section{ Bounded Sharpe Ratio}\label{b1}

  This section considers the case of bounded Sharpe Ratio.    Namely, we want to impose, with $c, C$ are positive constants,
  $0 < c \le \b \mu' \b \Theta \b \mu \le C < \infty$ in Assumption 7(ii).  First, let us denote $j,k$ th element of matrix $\b \Theta$ as $\Theta_{j,k}$.  This boundedness can happen since 
  \begin{equation}
   0 < c \le \sum_{j=1}^p \sum_{k=1}^p \mu_j \mu_k \Theta_{j,k} \le C < \infty,\label{bsrr}
   \end{equation}
  there may be  different signs in $\mu_j, \mu_k$ for $j \neq k$, and hence can balance the sum to be bounded. Note that $\mu_j = \b b_j' E \b f_t, \mu_k = \b b_k' E\b  f_t$, hence a linear combination of factor loadings may give different signs, so different asset returns may be linked to factor loadings in an opposite way.  There are two possibilities with bounded Sharpe Ratio. Each will be discussed in subsections \ref{b1-1}, and \ref{b1-2} respectively.

  \subsection{Bounded Sharpe Ratio with Bounded Variance of GMV}\label{b1-1}
    
  First, it is possible that also the variance of the global minimum variance  portfolio  (we are not considering such a portfolio in this paper) is bounded too, hence in Assumption 7(ii), instead of 
  $\b 1_p' \b \Theta \b 1_p \ge c p >0$, we can have $ 0 < c \le \b 1_p' \b \Theta \b 1_p \le C < \infty$, which is equal to 
 \begin{equation}
  0 < c \le \sum_{j=1}^p \sum_{k=1} \Theta_{j,k} \le C < \infty,\label{bgmv}
  \end{equation}
  and this  may happen due to different signs in $\Theta_{j,k}$ with $j \neq k$.  The second possibility will be discussed in the sub-section \ref{b1-2}.
  

  We started analyzing Theorems 1-2 of \cite{caner2022} where we base our precision matrix estimator. There is no change in those results, so we start with Theorem 1 proof here. Start with the new Assumption 7. 
  
  {\bf Assumption 7*}. {\it (i). $p^2 \bar{s} l_T  K^3 = o(1)$.
  
  (ii). $ \infty> C \ge \b \mu' \b \Theta \b \mu \ge c > 0$, $\infty> C \ge \b 1_p' \b \Theta \b 1_p \ge c > 0$, with $C,c>0$ are  positive constants.
  
  (iii). $\| \b m \|_1 = O (\| \b w_d^* \|_1)$, $\b w^{*'} \b \Sigma_y \b w^* \ge c > 0$, and $ | \b w^{*'} \b \mu | \ge c > 0$.}
  
  Our sparsity assumption is stronger with extra $p^2$ compared to Assumption 7 but we have $K$ factors less in our new assumption.
  Also now we have bounded Sharpe Ratio assumption for maximum Sharpe Ratio portfolio, in part (ii) of new assumption.  Also we do not scale 
  $|\b 1_p' \b \Theta \b \mu |$ with $p$ as the statement of Theorem 1.  Define the rate  $r_{w1b}:= p^{3/2} \bar{s} l_T K$ in Appendix B.
  
  {\bf Theorem B.1}.{\it Under Assumptions 1-6, $7^*$, with the following condition $| \b 1_p' \b \Theta \b  \mu | \ge c >0$ and $c$ being a positive constant and $0 < c \le | \kappa | \le 
  C < \infty$
  
  (i). \[ \| \hat{\b w} - \b w^* \|_1 = O_p ( p^{3/2} \bar{s} l_T  K) = o_p (1).\]
  
  (ii). \[ \left| \frac{ \hat{\b w}' \b \Sigma_y \hat{\b w}}{\b w^{*'} \b \Sigma_y \b w^*} - 1 
  \right| = O_p ( p^2 \bar{s} l_T  K^3) = o_p (1).\]
  
  (iii). \[ \left|  \frac{\widehat{SR}^2}{SR^2} - 1
  \right| = O_p ( p^2 \bar{s} l_T K^3) = o_p (1).\]  
    }
  
  Note that compared to Theorem 1(i), we have extra $p^{3/2}$ rate, and compared to Theorem 1(ii)-(iii) we have extra $p^2$ rate in the rate of convergence. So we cannot claim when $p>T$, we have consistency, but still with $p<T$, and both $p,T$ growing we can have consistency. In terms of factors compared to Theorem 1, for example in Theorem B.1(iii), our rate here involves $K^3$ instead of Theorem 1(iii) in $K^4$. 
  
  {\bf Proof of Theorem B.1}.
  
  (i). 
  First, we highlight the differences compared to proof of Theorem 1(i) here. Now we have to define terms $A, \hat{A}$ differently without $p$ scale, hence
  $A:= \b 1_p' \b \Theta \b 1_p, \hat{A}:= \b 1_p' \hat{\b \Theta}' \b 1_p$. Since left side term in (\ref{pt1.2}) and (\ref{pt1.3}) does not change, we start with (\ref{pt1.3a}), and now assuming $p \bar{s} l_T \to 0$
  
  \begin{equation}
 | \hat{A} - A | = O_p ( p \bar{s} l_T)=o_p (1).\label{A5}
 \end{equation}
 
 Then given (\ref{pt1.4})-(\ref{a6a}) we have the following change in (\ref{pt1.6})
 \begin{equation}
 \| (\hat{\b \Theta} - \b \Theta)' 1_p A \|_1 = O_p ( p \bar{s} l_T).\label{A9}
 \end{equation}
 By Exercise 7.53b of \cite{abamag2005} and (B.22) of \cite{caner2022} which is $Eigmax (\b \Theta) \le C < \infty$ and by Assumption $7^{*}(i)$
 \begin{equation}
 \| \b \Theta \b 1_p \|_1 \le p^{1/2} \| \b \Theta \b 1_p \|_2 \le p^{1/2} [ Eigmax (\b \Theta)]^{1/2} (\b 1_p' \b \Theta \b 1_p)^{1/2} \le C p^{1/2}.
 \label{*1}
 \end{equation}

 \noindent Then see that modifying (\ref{pt1.7})(A.11) by (\ref{A5}), (\ref{pt1.9}) changes to 
 \begin{equation}
 \| (\b \Theta \b 1_p) ( A - \hat{A} ) \|_1 \le C  p^{1/2} | \hat{A} - A |  = O_p ( p^{3/2} \bar{s} l_T) 
 .\label{A12} 
  \end{equation}
Since (\ref{A9}) is faster than (\ref{A12})
\begin{equation}
\| (\hat{\b \Theta}' \b 1_p) A - (\b \Theta \b 1_p) \hat{A} \|_1 = O_p ( p^{3/2} \bar{s} l_T ), \label{A13} 
\end{equation}
  replacing (\ref{pt1.10}). Combine (\ref{A13}) with $A, \hat{A}$ definitions here, replacing (\ref{pt1.11})
 \begin{equation}
\frac{ \| (\hat{\b \Theta}' \b 1_p) A - (\b \Theta \b 1_p) \hat{A}\|_1}{| \hat{A} A | } = O_p ( p^{3/2} \bar{s} l_T ) = o_p (1),
\label{A14}
\end{equation}
by new Assumption $7^*$, $ p^{3/2} \bar{s} l_T  = o(1)$.  Now consider $F = \b 1_p' \b \Theta \b \mu$ and $\hat{F}= \b 1_p'  \hat{\b \Theta}'  \hat{\b \mu}$, and note that since $\b 1_p' \b \Theta \b 1_p$ and $\b \mu' \b \Theta \b \mu$ are upper bounded by a constant at this part of the Appendix, by Cauchy-Schwartz inequality $|F|$ is upper bounded by a constant as well. Then by the assumption on the statement of Theorem B.1 we can put a lower bound on $|F|$, where $|F | \ge c >0$.   Take (\ref{pt1.12}) and  
  
\begin{equation}
\left\| \left( \frac{\hat{\b \Theta}' \hat{\b \mu}}{\b 1_p' \hat{\b \Theta}' \hat{\b \mu}} - \frac{\b \Theta \b \mu}{\b 1_p' \b \Theta \b \mu}
\right)  \right\|_1 = \frac{ \| (\hat{\b \Theta}' \hat{\b \mu}) F - (\b \Theta \b \mu ) \hat{F}\|_1}{| F \hat{F}|}.\label{A15}
\end{equation}

Analyze the denominator and (\ref{a13a}) changes to, via new Assumption 7*(i) 
\begin{equation}
| \hat{F} - F | = O_p ( p \bar{s} l_T K) = o_p (1).\label{A16}
\end{equation}

Next clearly
\[ | F \hat{F} | \ge F^2 - F | \hat{F} - F | \ge c^2 - o_p (1).\]

Now consider the numerator in (\ref{A15}), by adding and subtracting $(\b \Theta \b \mu/p)F$ and triangle inequality
\begin{equation}
\|  \hat{\b \Theta}' \hat{\b \mu} F - \b \Theta \b \mu \hat{F} \|_1 \le 
\|  \hat{\b \Theta}' \hat{\b \mu} F - \b \Theta \b \mu F \|_1 + \| \b \Theta \b \mu (\hat{F} - F) \|_1.\label{A19}
\end{equation}
We use the same analysis in (A.22)-(A.26) without scaling by $p$, and since now $| F | = O(1)$
\begin{equation}
\| \hat{\b \Theta}' \hat{\b \mu} - \b \Theta \b \mu \|_1 |F| = O_p ( p \bar{s} l_T K).\label{A20}
\end{equation}

Next in (\ref{A19}), by Lemma B.5  or (B.22) of of \cite{caner2022}  we have $Eigmax (\b \Theta ) \le C $
\begin{eqnarray}
\| \b \Theta \b \mu  (\hat{F} - F) \|_1 & = & \| \b \Theta \b \mu \|_1 | \hat{F} - F | \nonumber \\
& \le & p^{1/2} \| \b \Theta \b \mu \|_2 | \hat{F} - F | \nonumber \\
& \le &p^{1/2} [ Eigmax (\b \Theta )]^{1/2}    (\b \mu' \b \Theta \b \mu)^{1/2}   | \hat{F} - F | \nonumber \\
& = &  O ( p^{1/2} ) O_p (p K \bar{s} l_T) = O_p ( p^{3/2} \bar{s} l_T K),\label{A24}
\end{eqnarray}
where we use Exercise 7.53b of \cite{abamag2005}, and $\b \Theta$ symmetric matrix for the second inequality.
The last rate is slower compared to (\ref{A20}) hence
by (\ref{A19})
\begin{equation}
\|  \hat{\b \Theta}' \hat{\b \mu} F - \b \Theta \b  \mu \hat{F} \|_1 = O_p (p^{3/2} \bar{s} l_T K).\label{A25}
\end{equation}
Use (\ref{A25}) in (\ref{A15})  to have 
\begin{equation}
\left\| \left( \frac{\hat{\b \Theta}' \hat{\b \mu}}{1_p' \hat{\b \Theta}' \hat{\b \mu}} - \frac{\b \Theta \b \mu}{\b 1_p' \b \Theta \b \mu}
\right)  \right\|_1 =O_p (p^{3/2} \bar{s} l_T  K) = o_p (1),\label{A26}
\end{equation}
by new Assumption 7*(i) to have the last equality. Use (\ref{A14})(\ref{A26}) in the weights definition in (\ref{pt1.1})
\[ \| \hat{\b w}_d - \b w_d^* \|_1 = O_p (p^{3/2} \bar{s} l_T  K) = O_p (r_{w1b})= o_p (1).\]
{\bf Q.E.D.}

(ii).
We start the proof of Theorem 1(ii) with our new Assumptions, the bounded Sharpe Ratio with Bounded GMV variance. Let us set 
\begin{equation}
r_{w1b}:= p^{3/2} \bar{s} l_T  K,\label{nrw1}
\end{equation}
 and this has extra $p^{3/2}$ rate compared with Theorem 1 in the main text due to not scaling by $p$ and but with bounded ratios we have a rate with $K$ here instead of $K^{3/2}$  in Theorem 1(i).
The proof of Theorem 1(ii) stays the same until (A.35) by (\ref{nrw1}) $r_{w1b}$ here in this section replacing $r_{w1}$ in Theorem 1.
Then the left side of (A.35) can be upper bounded
\begin{eqnarray}
\| \frac{\b \Theta \b \mu}{\b 1_p' \b \Theta \b \mu} - \frac{\b \Theta \b 1_p}{\b 1_p' \b \Theta \b 1_p} \|_1
&\le& \frac{\| \b \Theta \b \mu\|_1}{| \b 1_p' \b \Theta \b \mu|} + \frac{ \| \b \Theta \b 1_p\|_1}{| \b 1_p' \b \Theta \b 1_p}
\nonumber \\
& \le & C p^{1/2} + C p^{1/2} = O (p^{1/2}),\label{b.15a}
\end{eqnarray}
by (\ref{*1}), and by the proof of (\ref{A24}). Then use the  definition in (A.34) with (\ref{b.15a})
\begin{equation}
\| \b w_d^* \|_1 = O ( p^{1/2}),\label{A35}
\end{equation}
which has extra $p^{1/2}$ rate compared to $K^{1/2}$ in (\ref{pt1a-8}).

Then use (\ref{A35}) and Theorem B.1(i),  and (A.40) becomes with (A.32)
 \begin{equation}
| \b w_d^{*'} \b \Sigma_y (\hat{\b w}_d - \b w_d^*) |  = O_p (p^{1/2} r_{w1b}  K^{2})= O_p ( p^2 \bar{s} l_T K^3)=
o_p (1),\label{A37}
\end{equation}
by new Assumption 7*(i) and (\ref{nrw1}).
Since (\ref{A37}) rate is the slowest in Theorem 1(ii) and by the same analysis in (\ref{pt1a-10})
we have the desired result.
\[ \left| \frac{\hat{\b w}' \b \Sigma_y \hat{\b w}}{\b w^{*'} \b \Sigma_y \b w^*} - 1 
\right| = O_p ( p^{1/2} r_{w1b} K^{2})= O_p ( p^2 \bar{s} l_T  K^3) 
=o_p (1).\]
Note that this rate has extra $p^2$ compared to Theorem 1(ii) but instead of $K^4$ in Theorem 1(ii) we have $K^3$ in our rate.{\bf Q.E.D}

(iii).
Here in the proof of Theorem B.1(iii), with new Assumption 7*, we only show the changes. So (\ref{pt1a-13})(\ref{pt1a-14}) stays the same. But the following (\ref{pt1a-15}) has the new rate, by new Assumption 7* with $ r_{w1b} K \to 0$ and since $r_{w1b} K $ then go to zero with  (\ref{nrw1})(\ref{A35})
\begin{eqnarray}
| (\hat{\b w}+ \b w^*)' \b \mu| & \le & \| \hat{\b w} + \b w^* \|_1 \| \b \mu \|_{\infty} \nonumber \\
& \le & \| \hat{\b w} - \b w^* \|_1 \| \b \mu \|_{\infty} + 2 \| \b w^* \|_1 \| \b \mu \|_{\infty} \nonumber \\
& = & O_p (r_{w1b} K ) + O ( p^{1/2} K) = o_p (1) + O(p^{1/2}  K) \nonumber \\
&=& O_p (p^{1/2} K),\label{A42}
\end{eqnarray}  
We have the same analysis for (\ref{pt1a-16}) but with our new $r_{w1b}$ rate in (\ref{nrw1}). Next, use (\ref{A42})(\ref{pt1a-16}) in (\ref{pt1a-17})
  \begin{equation}
\frac{ | (\hat{\b w}' \b \mu )^2 - (\b w^{*'} \b \mu)^2|}{(\b w^{*'} \b \mu)^2} = O_p (r_{w1b} K ) O_p ( p^{1/2} K) = O_p (p^2 \bar{s} l_T K^3) = o_p (1),\label{A44}
\end{equation}
by Assumption 7*(i) and $r_{w1b}$ definition in (\ref{nrw1}).  
  
Then change (\ref{pt1a-17a}) to the following  
\begin{eqnarray}
\left| \frac{\b w^{*'} \b \Sigma_y \b w^* - \hat{\b w}' \b \Sigma_y \hat{\b w}}{\hat{\b w}' \b \Sigma_y \hat{\b w}}\right|
& \le & \left| \frac{\b w^{*'} \b \Sigma_y \b w^* - \hat{\b w}' \b \Sigma_y \hat{\b w}}{\b w^{*'} \b \Sigma_y \b w^*- | \hat{\b w}' \b \Sigma_y \hat{\b w} - \b w^{*'} \b \Sigma_y \b w^*|}\right| \nonumber \\
& = & O_p (r_{w1b} p^{1/2}   K^{2})= O_p (p^2 \bar{s}  l_T K^3),\label{A45}
\end{eqnarray}
by Theorem 1(ii) proof. Use (\ref{A44})(\ref{A45}) in (A.48) to have the desired result. Note that we extra $p^2$ in the rate of convergence due to not scaling and with bounded ratio assumptions we have $K^3$ rate here in (\ref{A45}) instead of $K^4$ in Theorem 1(ii).
{\bf Q.E.D.}

We replace Assumption 8 in the main text with Assumption $8^*$.

{\bf Assumption $8^*$}.{\it 

(i). $p^2 \bar{s} l_T  K^{3} =o(1)$.

(ii). $ \infty > C \ge  \b \mu' \b \Theta \b \mu  \ge c >0$,  $ \infty > C \ge  \b 1_p' \b \Theta \b 1_p  \ge c >0$,
 $ \infty > C \ge | \b 1_R' \b \Theta \b 1_p  | \ge c >0$, $ \infty > C \ge  \b 1_R' \b \Theta \b 1_R   \ge c >0$, $|\b 1_p' \b \Theta \mu | \ge c >0$.

 (iii). $0 < r/p <1$, $| w_k - w_a | \ge c > 0$, and $| \omega | \le C < \infty$, $ 0 < c \le | \kappa | \le C < \infty$.
 
 (iv). $| \b w_R^{*'} \b \mu | \ge c > 0$, $ \b w_R^{*'} \b \Sigma_y \b w_R^* \ge c > 0$, $ \| \b m_R \|_1 = O (\b w_{cp}^*)$.

}

Note that part (i) has a different  rate compared with Assumption 8(i), and has $p^2$ term. But we have $K$ factor less compared with Assumption 8(i).
Also (ii) has bounded terms without scaling by $p$, compared with scaled (divided by $p$) and bounded terms in Assumption 8(ii).

The following replaces Lemma \ref{la1}.

\begin{lemma}\label{lb1}
Under Assumptions 1-6, $8^*$

(i). \[ \| \hat{\b k } - \b k \|_1 = O_p (p^{3/2} \bar{s} l_T  ).\]

(ii). \[ \| \hat{\b a} - \b a \|_1 = O_p ( p^{3/2} \bar{s} l_T ).\]

(iii). \[ | (\hat{w}_k - \hat{w}_a) - (w_k - w_a) | = O_p ( p \bar{s} l_T).\]

(iv). \[ | w_k - w_a | = O(1).\]

(v). \[ \| \b k - \b a \|_1 = O (p^{1/2}).\]

\end{lemma}

Note that compared with its counterparts in Lemma \ref{la1}, Lemma \ref{lb1}(i) has extra $p^{3/2}$, (ii) has extra $p^{3/}2$ terms, hence the rates are slower here definitely. Lemma \ref{lb1}(iii) has an extra $p$  rate compared to Lemma \ref{la1}(iii).  The rate in Lemma \ref{lb1}(iv) is   the same as in  Lemma \ref{la1}(iv). Lemma \ref{lb1}(v) rate here has extra $p^{1/2}$ term compared with Lemma \ref{la1}(v).

{\bf Proof of Lemma \ref{lb1}}. The left side of (\ref{pl1-1}) is the same, and the terms on the right side is not scaled by $p$.  
Consider the first term on the right side of (\ref{pl1-1})
\begin{eqnarray}
\| (\hat{\b \Theta}-  \b \Theta)' \b 1_R\|_1 
&\le & \| \hat{\b \Theta } - \b \Theta \|_{l_{\infty}} p\nonumber \\
& = & O_p (p \bar{s} l_T),\label{A47} 
\end{eqnarray}
where we use Theorem 2 of \cite{caner2022}, and $r<p$. Then consider (\ref{pl1-2aa})
\begin{equation}
 | \b 1_p' \hat{\b \Theta}' \b 1_R| \ge | \b 1_p' \b \Theta \b 1_R| - | \b 1_p' \hat{\b \Theta} ' \b 1_R - \b 1_p' \b \Theta \b 1_R|.\label{A48}
 \end{equation}
By new Assumption  $8^{*}$ we know that $|\b 1_p' \b \Theta \b 1_R| \ge c > 0$, and 
\begin{eqnarray}
| \b 1_p' (\hat{\b \Theta} - \b \Theta)' \b 1_R| &\le& \| \b 1_p \|_{\infty} \| (\hat{\b \Theta} - \b \Theta)' \b 1_R \|_1  \nonumber \\
& = & O_p (p \bar{s} l_T)=o_p(1),\label{A49}
\end{eqnarray}
where we use Theorem 2 of \cite{caner2022}. Next in (\ref{A48}) use (\ref{A49}) and $p \bar{s} l_T \to 0$
\begin{equation}
 | \b 1_p' \hat{\b \Theta}' \b 1_R| \ge  c - o_p (1),\label{A50}
 \end{equation}
by Assumption $8^*$. 
Use (\ref{A47})(\ref{A50}) in (A.49) first right-side term, without scaled by $p$ as in the proof of Lemma \ref{la1},
\begin{equation}
\left\| \frac{(\hat{\b \Theta}-  \b \Theta)' \b 1_R}{\b 1_p' \hat{\b \Theta}' \b 1_R } \right\|_1 =  O_p (p \bar{s} l_T).\label{A51}
\end{equation}  
We consider the following second term on the right side of (A.49), without scaling by $p$, using (\ref{A48})-(\ref{A50})
 by assuming $|\b 1_p' \b \Theta \b 1_R| \ge c >0$, Assumption $8^*$
\begin{equation}
\left| \frac{1}{\b 1_p' \hat{\b \Theta}' \b 1_R} - \frac{1}{\b 1_p' \b \Theta \b 1_R}
\right| = O_p (p \bar{s} l_T).\label{A52}
\end{equation}  

Next, in the second term in (\ref{pl1-1}), without scaled by $p$,  by the analysis in (\ref{*1}), and $0<r/p <1$,
\begin{equation}
\| \b \Theta \b 1_R \|_1 = O_p (p^{1/2}).\label{A53}
\end{equation}
 Use (\ref{A52})(\ref{A53})  for the second right side term in (\ref{pl1-1}), without $p$ scaling,
\begin{equation}
\left\| \b \Theta \b 1_{R} \right\|_1 \left| \frac{1}{\b 1_p' \hat{\b \Theta}' \b 1_R} - \frac{1}{\b 1_p' \b \Theta \b 1_R}
\right| = O_p (p \bar{s} l_T) O (p^{1/2})
.\label{A54}
\end{equation}   
Between (\ref{A51}) and (\ref{A54}), clearly (\ref{A54}) has the slowest rate. So 
\[ \| \hat{\b k} - \b k \|_1 = O_p ( p^{3/2} \bar{s} l_T ) = o_p (1),\]
by Assumption $8^*$. Here we have extra $p^{3/2}$ term compared with our previous Lemma \ref{la1}(i).

(ii). We use (\ref{A14}), $\hat{\b a}, \b a$ definitions,  and get 
\[ \| \hat{\b a} - \b a \|_1 = O_p ( p^{3/2} \bar{s} l_T ) = o_p (1).\]

(iii). Note that (A.58)-(A.59) are the same and we do not scale (A.60) by $p$ anymore. Then (A.61) changes to 

\begin{equation}
| \b 1_R' (\hat{\b \Theta} - \b \Theta)' \b 1_R | \le \| \b 1_R \|_{\infty} \| (\hat{\b \Theta} - \b \Theta )' \b 1_R \|_1 \le  r \max_{1 \le j \le p } \| \hat{\b \Theta}_j - \b \Theta_j \|_1 = O_p (p \bar{s} l_T),\label{A58}
\end{equation}
Then (\ref{A50}) is used to get 
\begin{equation}
| \b 1_R' \hat{\b \Theta}' \b 1_p  | \ge c - o_p (1).\label{A59}
\end{equation}

Now consider the second term in (A.60). By new Assumption $8^*$ we have 
\begin{equation}
| \b 1_R' \b \Theta \b 1_R| \le C < \infty,\label{A60}
\end{equation}
so we have 
a change in (A.63), and now the upper bound is a constant. Then consider (\ref{A58})(\ref{A59})(\ref{A60}) and (\ref{A52}),  the first right side term in (A.58), via (A.60) right side (without scaling by $p$),
\begin{equation}
\left| \frac{\b 1_R' \hat{\b \Theta}' \b 1_R}{\b 1_R' \hat{\b \Theta}' \b 1_p} - \frac{\b 1_R' \b \Theta \b 1_R}{\b 1_R' \b \Theta \b 1_p}
\right| = O_p ( p \bar{s} l_T).\label{A61}
\end{equation}
Analysis of the second right side term in (A.58) follows the same analysis as in (\ref{A61}) but instead of (\ref{A58}) we use (\ref{A5})
so we have 
\begin{equation}
\left| \frac{\b 1_R' \hat{\b \Theta}' \b 1_p}{\b 1_p' \hat{\b \Theta}' \b 1_p} - \frac{\b 1_R' \b \Theta \b 1_p}{\b 1_p' \b \Theta \b 1_p}
\right| = O_p ( p \bar{s} l_T).\label{A62}
\end{equation}
Then by (\ref{A61})(\ref{A62}) in (A.58) 
\[ | (\hat{w}_k - \hat{w}_a) - (w_k - w_a ) | = O_p ( p \bar{s} l_T).\]
{\bf Q.E.D.}

(iv). With our new Assumptions $| \b 1_R' \b \Theta \b 1_p | \le C < \infty$, and $| \b 1_R' \b \Theta \b 1_R | \le C < \infty$, we have 
\[ | w_k - w_a | = O(1).\]
{\bf Q.E.D.}

(v). By triangle inequality
\[ \| \b k - \b a \|_1 \le \frac{ \|\b  \Theta \b 1_R \|_1}{| \b 1_p' \b \Theta \b 1_R|} + \frac{\|\b  \Theta \b 1_p\|_1}{|\b 1_p' \b \Theta \b 1_p|}
= O (p^{1/2}) + O (p^{1/2}),\]
by the analysis in (\ref{A53}) and $|\b 1_p' \b \Theta \b 1_R| \ge c >0$, and  new Assumption $8^*$.{\bf Q.E.D.}

\begin{lemma}\label{lb2}
{\it Under Assumptions 1-6, and $8^*$

(i). \[ \| \hat{\b l} - \b l \|_1 = O_p ( p^{3/2} \bar{s} l_T ).\]

(ii). \[ | w_u | = O(1).\]

(iii). \[ \| \b l \|_1 = O (p^{1/2}).\]

(iv). \[ | \hat{w}_u - w_u | = O_p (p \bar{s} l_T K).\]}

\end{lemma}

Note that the rates are different than Lemma A.2. Part (i), we have extra $p^{3/2}$ rate compared to Lemma A.2(i).
 For part (ii) Lemma \ref{lb2} rate is better, (not diverging) compared with Lemma A.2(ii) rate which is diverging with $n$. In Lemma \ref{lb2}(iii), our rate here is slower by $p^{1/2}$ compared with Lemma A.2(iii) rate. Lemma \ref{lb2}(iv), we have extra $p$ rate compared with Lemma A.2(iv) but we have $K^{1/2}$ factor less here.

{\bf Proof of Lemma \ref{lb2}}.

(i). All the equations until (\ref{pl2.2a}) are the same. Then with $\hat{\b y}:= \hat{\b k} - \hat{\b a}, \b y:= \b k - \b a$, $\hat{x}:= \hat{w}_k - \hat{w}_a, x:= w_k - w_a$
\begin{equation}
| \hat{x} x | \ge c^2 - O (1)  O_p (p \bar{s} l_T) \ge c^2 - o_p (1),\label{A69}
\end{equation}
by Lemma \ref{lb1}(iii)-(iv), and Assumption $8^*$.
Then 
\begin{eqnarray}
\| (\hat{\b y} - \b y) (\hat{x} - x ) \|_1 
& \le & [ \| \hat{\b k } - \b k \|_1 + \| \hat{\b a} - \b a \|_1] | (\hat{w}_k - \hat{w}_a) - (w_k - w_a)| \nonumber \\
& = & O_p ( p^{3/2} \bar{s} l_T ) O_p (p \bar{s} l_T) \nonumber \\
&=& o_p (1),\label{A70}
\end{eqnarray}
by Lemma \ref{lb1}(i)-(iii) and triangle inequality, and Assumption $8^*$.
Next,
\begin{eqnarray}
\| \b y (\hat{x} - x) \|_1 & = & \| (\b k- \b a) [ (\hat{w}_k -\hat{w}_a) - (w_k - w_a)] \|_1 \nonumber \\
& \le & \| \b k - \b a \|_1 | (\hat{w}_k -\hat{w}_a) - (w_k - w_a)| \nonumber \\
& = & O (p^{1/2}) O_p (p \bar{s} l_T) \nonumber \\
& = & O_p ( p^{3/2} \bar{s} l_T ) = o_p (1),\label{A71}
\end{eqnarray}
where we use Lemma \ref{lb1}(iii),(v) and Assumption $8^*$. Combine (\ref{A69})-(\ref{A71}) in (A.70), and the slower rate is by (\ref{A71})
\begin{equation}
\| \frac{\hat{\b y}x}{\hat{x}x} - \frac{\hat{\b y}\hat{x}}{\hat{x}x} \|_1
 =  O_p (p^{3/2} \bar{s}  l_T ) = o_p (1),\label{A72}
\end{equation}
by Assumption $8^*$. Consider the second term on the right side of (\ref{pl2.0})
\begin{eqnarray}
\| \frac{\hat{\b y}\hat{x}}{\hat{x}x} - \frac{\hat{x} \b y }{\hat{x} x} \|_1 
& \le  & \| \frac{(\hat{\b y}- \b y ) (\hat{x} - x)}{\hat{x} x } \|_1 + \| \frac{(\hat{\b y}- \b y )x}{\hat{x} x }\|_1 \nonumber \\
& = & O_p (p^{5/2} \bar{s}^2 l_T^2 ) + O_p ( p^{3/2} \bar{s} l_T )  \nonumber \\
& = & O_p ( p^{3/2} \bar{s} l_T ) = o_p (1),\label{A73}
\end{eqnarray}
by (\ref{A69})(\ref{A70}) and Lemma \ref{lb1}(i)(ii)(iv) and Assumption $8^*$. Combine (\ref{A72})(\ref{A73}) in (\ref{pl2.0}) to have 
\[ \|  \hat{\b l} - \b l \|_1 = O_p ( p^{3/2} \bar{s} l_T).\]
{\bf Q.E.D.}

(ii). Through our new Assumption $8^*$, via Cauchy-Schwartz inequality $| w_u | = O(1)$.
{\bf Q.E.D}

(iii). By Lemma \ref{lb1}(v) and Assumption $| w_k - w_a | \ge c >0$
\[ \| \b \l \|_1 = O (p^{1/2}).\]
{\bf Q.E.D.}

(iv). 
By triangle inequality
\begin{equation}
| \hat{w}_u - w_u | \le \left|  \frac{\b 1_R' \hat{\b \Theta}' \hat{\b \mu}}{\b 1_p' \hat{\b \Theta}' \hat{\b \mu}} - \frac{\b 1_R' \b \Theta \b \mu}{\b 1_p' \b \Theta \b \mu }
\right| + \left| \frac{\b 1_R' \hat{\b \Theta}' \b 1_p}{\b 1_p' \hat{\b \Theta}' \b 1_p} - \frac{\b 1_R' \b \Theta \b 1_p}{\b 1_p' \b \Theta \b 1_p}\right|.\label{A78}
\end{equation}
Then consider the first right side term in (\ref{A78}), by adding and subtracting and triangle inequality
\begin{eqnarray}
\left|  \frac{\b 1_R' \hat{\b \Theta}' \hat{\b \mu}}{\b 1_p' \hat{\b \Theta}' \hat{\b \mu}} - \frac{\b 1_R' \b \Theta \b \mu}{\b 1_p' \b \Theta \b \mu }
\right| 
& \le & \left|  \frac{\b 1_R' \hat{\b \Theta}' \hat{\b \mu}}{\b 1_p' \hat{\b \Theta}' \hat{\b \mu}} -  \frac{\b 1_R' \b \Theta \b \mu}{\b 1_p' \hat{\b \Theta}' \hat{\b \mu}} \right|
\nonumber \\
& + & \left| \frac{\b 1_R' \b \Theta \b \mu}{\b 1_p' \hat{\b \Theta}' \hat{\b \mu} }-\frac{\b 1_R' \b \Theta \b \mu}{\b 1_p' \b \Theta \b \mu } \right|.\label{A79}
\end{eqnarray}

Consider the denominator on the first term on the right side of (\ref{A79})
\begin{equation}
|\b 1_p' \hat{\b \Theta}' \hat{\b \mu}| \ge | \b 1_p' \b \Theta \b \mu| - | \b 1_p' \hat{\b \Theta}' \hat{\b \mu} - \b 1_p' \b \Theta \b \mu|.\label{A80}
\end{equation}

Since $\b \Theta$ is symmetric, by $r<p$
\begin{equation}
\left| \b 1_R' \hat{\b \Theta}' \hat{\b \mu} - \b 1_R' \b \Theta \b \mu
\right| 
 =  O_p (p \bar{s} l_T K) = o_p (1),\label{A81}
\end{equation}
by (A.19) without scaling by $p$ there on the left side.
Use the analysis in (\ref{A81}) with Assumption $8^*$,  $|\b 1_p' \b \Theta \b \mu |\ge c > 0$ in (\ref{A80})
to have 
\begin{equation}
|\b 1_p' \hat{\b \Theta}' \hat{\b \mu}| \ge c - o_p (1).\label{A82}
\end{equation}

Using (\ref{A81}) for the numerator of the first right side term in (\ref{A79}) in combination with (\ref{A82})
 \begin{equation}
  \left|  \frac{\b 1_R' \hat{\b \Theta} \hat{\b \mu}}{\b 1_p' \hat{\b \Theta}' \hat{\b \mu}} -  \frac{\b 1_R' \b \Theta \b \mu}{\b 1_p' \hat{\b \Theta}' \hat{\b \mu}} \right|
= O_p ( p \bar{s} l_T K).\label{A83}
\end{equation}

Consider the second term on the right side of (\ref{A79}), which is upper bounded by 
\begin{equation}
| \b 1_R' \b \Theta \b \mu | \left| \frac{\b 1_p' \hat{\b \Theta}' \hat{\b \mu}/p - \b 1_p' \b \Theta \b \mu}{(\b 1_p' \hat{\b \Theta}' \hat{\b \mu})(\b 1_p' \b \Theta \b \mu)}
\right| = O (1) O_p ( p \bar{s} l_T K) = O_p ( p \bar{s} l_T K),\label{A85}
\end{equation}
where we use  $| \b 1_R' \b \Theta \b \mu | \le C < \infty$ by Assumption $8^*$ and Cauchy Schwartz inequality, and same analysis in (\ref{A81}).

Consider (\ref{A83})(\ref{A85}) in  first right side term in (\ref{A78}) 
\begin{equation}
\left|  \frac{\b 1_R' \hat{\b \Theta}' \hat{\b \mu}}{\b 1_p' \hat{\b \Theta}' \hat{\b \mu}} - \frac{\b 1_R' \b \Theta \b \mu}{\b 1_p' \b \Theta \b \mu }
\right| =O_p (p \bar{s} l_T K. ).\label{A86}
\end{equation}

We consider the second right side term in (\ref{A78}) by adding and subtracting, $\frac{\b 1_R' \b \Theta \b 1_p}{\b 1_p' \hat{\b \Theta}' \b 1_p}$, and via a triangle inequality

\begin{eqnarray}
&&\left| \frac{\b 1_R' \hat{\b \Theta}' \b 1_p}{\b 1_p' \hat{\b \Theta}' \b 1_p} - \frac{\b 1_R' 
\b \Theta \b 1_p}{\b 1_p' \b \Theta \b 1_p}\right|
\le \left| \frac{\b 1_R' (\hat{\b \Theta} - \b \Theta)' \b 1_p }{\b 1_p' \hat{\b \Theta}' \b 1_p}
\right| \nonumber \\
& + & \left| \b 1_R' \b \Theta \b 1_p
\right| \left| \frac{1}{\b 1_p' \hat{\b \Theta}' \b 1_p} - \frac{1}{\b 1_p' \b \Theta \b 1_p}
\right|.\label{A87}
\end{eqnarray}
Consider the first term on the right side of (\ref{A87}), and its denominator specifically
\begin{equation}
| \b 1_p' \hat{\b \Theta}' \b 1_p| \ge | \b 1_p' \b \Theta \b 1_p| - | \b 1_p' (\hat{\b \Theta}- \b \Theta)' \b 1_p| \ge c- O_p ( p \bar{s} l_T) = c - o_p (1),\label{A88}
\end{equation}
by (\ref{A5}), and Assumption $8^*$. Then for the numerator of the first term in (\ref{A87})
\begin{eqnarray}
\left| \b 1_R' (\hat{\b \Theta}- \b \Theta)' \b 1_p
\right| & \le &  \| \b 1_R'  (\hat{\b \Theta} - \b \Theta)' \b \|_1 \| 1_p \|_{\infty}  \nonumber \\
& =  & \| (\hat{\b \Theta}- \b \Theta) \|_{l_1} \|  \b 1_R \|_{1} \nonumber \\
& = & \| (\hat{\b \Theta} - \b \Theta)' \|_{l_{\infty}} p  = O_p ( p \bar{s} l_T),\label{A89}
\end{eqnarray}
and the rate is by Theorem 2 of \cite{caner2022}. Use (\ref{A88})(\ref{A89}) for the first right side term in (\ref{A87})
\begin{equation}
 \frac{ | \b 1_R' (\hat{\b \Theta} - \b \Theta )' \b 1_p|}{|\b 1_p' \hat{\b \Theta}' \b 1_p |} = O_p (p \bar{s} l_T).\label{A89A}
 \end{equation}
 Then evaluate the second term on the right side of (\ref{A87}) 
\begin{equation}
\left| \b 1_R' \b  \Theta  \b 1_p
\right| \left| \frac{1}{\b 1_p' \hat{\b \Theta}' \b 1_p} - \frac{1}{\b 1_p' \b \Theta \b 1_p}
\right| =    O_p (p \bar{s} l_T),\label{A90}
\end{equation}
where we use (\ref{A5}) and Assumption $8^*$.  So (\ref{A78}) second right side term is 
\begin{equation}
\left|  \frac{\b 1_R' \hat{\b \Theta}' \b 1_p}{\b 1_p' \hat{\b \Theta}' \b 1_p} - \frac{ \b 1_R' \b \Theta \b 1_p}{\b 1_p' \b \Theta \b 1_p}
\right| = O_p ( p \bar{s} l_T).\label{A91}
\end{equation}
Clearly rate in (\ref{A86}) is slower than (\ref{A91}) rate. 
So 
\[ | \hat{w}_u - w_u | = O_p ( p \bar{s} l_T K).\]
{\bf Q.E.D}

{\bf Theorem B.2}. {\it Under Assumptions 1-6, $8^*$, then 

(i). \[ \| \hat{\b w}_R - \b w_R^* \|_1 = \| \hat{\b w}_{cp} - \b w_{cp}^* \|_1 = O_p ( p^{3/2} \bar{s} l_T K)=o_p (1).\]

(ii). \[ \| \frac{ \hat{\b w}_R' \b \Sigma_y \hat{\b w}_R }{\b w_R^{*'} \b \Sigma_y \b w_R^* }-1 \|_1 = O_p ( p^2 \bar{s} l_T  K^3)=o_p (1).\]

(iii). \[ \left| \left( \frac{\widehat{SR}_R}{SR_R}
\right)^2 
\right| = O_p ( p^2 \bar{s} l_T  K^3)=o_p (1).\]

}

Remarks. We compare the rates in Theorem B.2(i) with Theorem 2(i) above. The ratio  is: $p^{3/2}/ K^{1/2}$, so even though we have $p^{3/2} $ here with bounded Sharpe Ratio and Bounded GMV variance assumption, we have  better number of factor conditions here.  For Theorems B.2(ii)(iii), the difference with Theorems 2(ii)(iii) does depend on the ratio $p^2/ K$. So again we have extra $p^2$ in Theorems B.2(ii)(iii) compared with Theorems 2(ii)(iii), but we have better number of factor conditions. However, all our rates are slower, since $K<p$, and we can only have consistency with $p<T$.

{\bf Proof of Theorem B.2}

(i). See that (A.94)-(A.97) are the same as in Theorem 2(i) proof. 
 Consider each term in (A.97) by Lemma \ref{lb2}
\begin{equation}
 \| \kappa (\hat{w}_u - w_u) (\hat{\b l} -\b  l )\|_1  
\le | \kappa | |\hat{w}_u - w_u | \| \hat{\b l} - \b l \|_1 = O(1) O_p ( p \bar{s} l_T K) O_p (p^{3/2} \bar{s} l_T).\label{A95}
\end{equation}

Next, analyze the second right side term in (\ref{pt2.4}) by Lemma \ref{lb2}
\begin{equation}
\| \kappa (\hat{w}_u - w_u) \b l \|_1 \le | \kappa | |\hat{w}_u - w_u | \| \b l \|_1 
=  O(1) O_p ( p \bar{s}  l_T K) O (p^{1/2}) = O_p ( p^{3/2} \bar{s} l_T  K)  .\label{A96} 
\end{equation}

Then analyze the third term on the right side of (\ref{pt2.4}) by Lemma \ref{lb2}
\begin{equation}
\| (\omega - \kappa w_u ) (\hat{\b l} - \b l )\|_1 \le [ | \omega| + | \kappa | | w_u |] \| \hat{\b l} - \b l \|_1
= O(1) O (p^{3/2} \bar{s} l_T  ).\label{A97}
\end{equation}

Since $p^{3/2} \bar{s} l_T K \to 0$, the slowest rate is (\ref{A96}) among (\ref{A95})-(\ref{A97}).
So 
\begin{equation}
 \|(\omega - \kappa \hat{w}_u) \hat{\b l} - (\omega - \kappa w_u )  \b l \|_1
 = O_p (p^{3/2} \bar{s} l_T K).\label{A98}
\end{equation}
Then use (\ref{A98}) and Theorem B.1(i)
\begin{equation}
\| \hat{\b w}_{cp} - \b w_{cp}^* \|_1 = O_p ( p^{3/2} \bar{s} l_T K) + O_p ( p^{3/2} \bar{s} l_T K) = O_p (p^{3/2} \bar{s} l_T K).\label{A99}
\end{equation}
{\bf Q.E.D.}

(ii). Note that (A.103) is the same. Consider the first term on the right side of (A.103) and using Theorem B.2(i) and (A.32) by  Holder's inequality  and $\| \b A \b x \|_{\infty} \le \| \b A \|_{\infty} \| \b x \|_1$ inequality for generic matrix $\b A$, and generic vector $\b x$
 \begin{equation}
 (\hat{\b w}_{cp} - \b w_{cp}^* )' \b \Sigma_y (\hat{\b w}_{cp} - \b w_{cp}^*) \le \| \hat{\b w}_{cp} - \b w_{cp}^* \|_1^2 \| \b \Sigma_y \|_{\infty}
 = [O_p (p^{3/2} \bar{s} l_T  K)]^2 O (K^2).\label{A101}
 \end{equation}
 
 We consider $l_1$ norm of $w_{cp}^*$ 
 \begin{eqnarray}
 \| \b w_{cp}^* \|_1 & = & \| (\omega - \kappa w_u ) \b l + \b w_d^* \|_1 \le   \| (\omega - \kappa w_u ) \b l \|_1 + \| \b w_d^* \|_1 \nonumber \\
 & \le & \left[ | \omega | + | \kappa | | w_u| 
 \right]  \| \b l \|_1 + \| \b w_d^* \|_1 \nonumber \\
 & = & O (1) O (p^{1/2}) + O (p^{1/2} ) = O (p^{1/2}),\label{A102}
  \end{eqnarray}
 where $\omega, \kappa$ are bounded by definition, and we use (\ref{A35}), Lemma \ref{lb2}(ii)(iii) for the rates. The second term on the right side of (\ref{pt2.9}) becomes
 
 \begin{eqnarray}
| \b w_{cp}^{*'} \b \Sigma_y (\hat{\b w}_{cp} - \b w_{cp}^*)|  & \le & \| \b w_{cp}^* \|_1 \| \b \Sigma_y \|_{\infty} \| \hat{\b w}_{cp} - \b w_{cp}^* \|_1 \nonumber \\
& = & O (p^{1/2}) O (K^2) O_p (p^{3/2} \bar{s} l_T  K) \nonumber \\
&=& O_p ( p^2 \bar{s} l_T  K^3),\label{A103}
\end{eqnarray}
where we use the  Holder inequality and $\| \b A \b x \|_{\infty} \le \| \b A \|_{\infty} \| \b x \|_1$ inequality for generic matrix $\b A$, and generic vector $\b x$ for the inequality and the rates are from (A.32), and (\ref{A102}) with Theorem B.2(i). Clearly the slowest rate is (\ref{A103}) compared to (\ref{A101}) given Assumption $8^*$. So first term on the right side of (\ref{pt2.9})
 
 \begin{equation}
 (\hat{\b w}_{cp}'\b  \Sigma_y \hat{\b w}_{cp}) - (\b w_{cp}^{*'} \b \Sigma_y \b w_{cp}^*) = O_p ( p^2 \bar{s} l_T  K^3).\label{A104}
 \end{equation}

 The second term on the right side of (\ref{pt2.9})
 
 \begin{eqnarray}
 | \b m_R'  \b \Sigma_y (\hat{\b w}_{cp} - \b w_{cp} )| &\le& \| \b m_R \|_1 \| \b \Sigma_y \|_{\infty} \| \hat{\b w}_{cp} - \b w_{cp}^* \|_1 \nonumber \\
 & = & O (p^{1/2}) O (K^2) O_p (p^{3/2} \bar{s} l_T  K)\nonumber \\
 & =& O_p ( p^2 \bar{s} l_T  K^{3}),\label{A105}
  \end{eqnarray}
 and the rates are by (A.32) and Theorem B.2(i), Assumption $8^*$ and (\ref{A102}).  Then combine (\ref{A104})(\ref{A105})
 and by $\b w_{R}^{*'} \b \Sigma_y \b w_{R}^* \ge c > 0$ assumption (i.e this imposes that there will be no local to zero variance) and we have the result
 \[ \| \frac{\hat{\b w}_R^{'} \b \Sigma_y \hat{\b w}_R}{\b w_R^{*'} \b \Sigma_y \b w_R^*} - 1 \|_1 = O_p ( p^2 \bar{s} l_T  K^3) = o_p (1).\]
 {\bf Q.E.D.}

 (iii). We try to obtain rates for the right side of (A.109).
To move forward we need the following results from Theorem B.2(i) above and $\| \b \mu \|_{\infty} = O (K)$ from (B.8) of \cite{caner2022} 
\[ \| \hat{\b w}_R - \b w_R^* \|_1 \| \b \mu \|_{\infty} = O_p ( p^{3/2} \bar{s} l_T   K^{2}).\]
Also by $\b w_R^*:= \b w_{cp}^* + \b m_R$ and (\ref{A102}) and Assumption $8^*$
\[ \| \b w_R^* \|_1 \| \b \mu \|_{\infty} =  O (p^{1/2} K).\]
We need to consider third term on the right side of (\ref{pt2.15}), so first we analyze
\begin{eqnarray}
| (\hat{\b w}_R + \b w_R^*)' \b \mu | & \le & \| \hat{\b w}_R - \b w_R^* \|_1 \| \b \mu \|_{\infty} 
+ 2 \| \b w_R^* \|_1 \| \b \mu \|_{\infty} \nonumber \\
& = & O_p ( p^{3/2} \bar{s} l_T  K^{2}) + O ( p^{1/2}  K) \nonumber \\
& = & O_p ( p^{1/2} K),\label{A106}
\end{eqnarray}
where we use Theorem B.2(i), Assumption $8^*$ and (\ref{A102}). Then 
\[ | \hat{\b w}_R' \b \mu - \b w_R^{*'} \b \mu | \le \| \hat{\b w}_R - \b w_R^* \|_1 \| \b \mu \|_{\infty} 
= O_p ( p^{3/2} \bar{s} l_T  K^{2}).\]
Using the last result immediately above and (\ref{A106}) 
\begin{eqnarray}
 \left| \frac{(\hat{\b w}'_R \b \mu)^2 - (\b w_R^{*'} \b \mu)^2}{(\b w_R^{*'} \b \mu)^2 }
\right|  &=& O_p (p^{3/2}  \bar{s} l_T  K^{2}) O_p ( p^{1/2}  K) \nonumber \\
& = & O_p ( p^2 \bar{s}  l_T K^3)
.\label{A107}
  \end{eqnarray}
  Use (\ref{A107}) and use Theorem B.2(ii) on the right side of (\ref{pt2.15}) to have 
  \[ \left| \left(\frac{\widehat{SR}_R}{SR_R}\right)^2 - 1 
  \right| = O_p ( p^2 \bar{s}  l_T K^3).\]
  {\bf Q.E.D.}

  Proposition A.1 follows through since there are no assumptions related to maximum sharpe ratio portfolio, Sharpe Ratio being bounded or not, and no assumptions related to GMV variance is bounded away from zero or infinity. We obtain the counterpart of Lemma A.3 above by changing Assumption 8 to Assumption $8^*$.

   {\bf Lemma B.3}. {\it  Under Assumptions 1-6, $8^*$ with $0< c \le \kappa_w \le C < \infty$, $|w_x | \le C < \infty$, $B_4 B_3 \neq B_2^2$, we have 
   \[ \| \hat{\b w}_c - \b w_c^* \|_1= O_p (p^{3/2} \bar{s} l_T  K)=o_p(1).\] 
   
   }
     
  {\bf Proof of Lemma B.3}. The proof follows through Lemma A.3 proof. All equations are unchanged until and including (A.133).
 We consider each one of the terms on the right side of (A.133). By (\ref{A5})
\begin{equation}
 | \hat{B_3} - B_3 | = O_p ( p \bar{s} l_T) = o_p (1) , \quad  | \hat{B}_2 - B_2 | = O_p ( p \bar{s} l_T) = o_p (1).\label{A130}
\end{equation}
Consider the first right side term in (A.133), by (\ref{A130}) Lemma B.2(i), with $r<p$
 \begin{eqnarray}
\| \hat{\b l} - \b l \|_1 \left| \frac{\hat{B}_3 - B_3}{\hat{B}_2}
 \right|  &\le& \frac{\| \hat{\b l}- \b l \|_1 | \hat{B}_3 - B_3 |}{(B_2 - | \hat{B}_2 - B_2|)} \nonumber \\
 &
 =& O_p (p^{3/2} \bar{s} l_T )
 O_p ( p \bar{s} l_T ).\label{A131}
\end{eqnarray} 
 
 Consider the second term on the right side of (A.133) by Assumption $8^*$, Lemma \ref{lb2}(iii),(\ref{A130}), with $r<p$

\begin{eqnarray}
\| \b l \|_1 \left| \frac{\hat{B}_3 - B_3}{\hat{B}_2}
 \right| &\le &\frac{(|\hat{B}_3 - B_3|) \|  \b l \|_1}{B_2 - | \hat{B}_2 - B_2|} \nonumber \\
& = & O (p^{1/2})  O_p ( p \bar{s} l_T)\nonumber \\
& = & O_p ( p^{3/2} \bar{s}  l_T ).\label{A132}
 \end{eqnarray}

 Consider the third term on the right side of (A.133), by Lemma \ref{lb2}, (\ref{A130}), $| B_3 | \le C < \infty$ by Assumption
 \begin{equation}
\left | \frac{B_3}{\hat{B}_2}\right| \| \hat{\b l} - \b l \|_1 \le \frac{\| \hat{\b l} - \b l \|_1 |  B_3 |}{B_2 - | \hat{B}_2- B_2|} = O_p (
p^{3/2} \bar{s} l_T).\label{A133}
\end{equation} 
 
 Then we consider the fourth term on the right side of (A.133) by Lemma \ref{lb2}(iii), (\ref{pla3-8})(\ref{A130}), Assumption $8^*$
\begin{eqnarray}
\frac{ \| \b l \|_1 | B_3| | \hat{B_2} - B_2|}{B_2^2 - | \hat{B}_2 - B_2 | |B_2|} &\le &
\frac{ \| \b l \|_1 |B_3| | \hat{B}_2 - B_2|}{B_2^2 - (|\hat{B}_2 - B_2|)(B_2)  }\nonumber \\
& = & O_p (p^{1/2}) O_p ( p \bar{s} l_T)  \nonumber \\
& = & O_p ( p^{3/2} \bar{s} l_T  ).\label{A134}
\end{eqnarray} 
By (\ref{A131})-(\ref{A134}), since $p^{3/2} \bar{s} l_T \to 0$ by Assumption $8^*$ via Lemma B.1
\begin{equation}
\| (\hat{\b a} - \frac{\hat{\b l} \hat{B}_3}{\hat{B}_2}) - ( \b a -\b  l \frac{B_3}{B_2}) \|_1 = O_p ( p^{3/2} \bar{s} l_T).\label{A135}
\end{equation}

Now remember that by Theorem B.2(i), 
\[ \| \hat{\b w}_{cp} - \b w_{cp} \|_1 = O_p (p^{3/2} \bar{s} l_T  K),\]
 which is slower than (\ref{A135}) so the rate for 
\[ \| \hat{\b w}_c - \b w_c^*\|_1 = O_p (p^{3/2} \bar{s} l_T  K)= o_p (1).\]
{\bf Q.E.D.}

We follow the analysis after Lemma A.3, and show that the orders of $\b w_c^*$ and $\b w_{cp}^*$ are the same in $l_1$ norm.  We simplify the analysis by imposing $\kappa=\kappa_w, \omega = \omega_x$, and that does not change the results below.
First note that 
\[ \b w_{c}^* = \b w_{cp}^* - (\b a - \b l \frac{B_3}{B_2}).\]
See that by (\ref{A102})
\[ \| \b w_{cp}^* \|_1 = O (p^{1/2}).\]
By  $\b a:=\b \Theta \b 1_p/\b 1_p' \b \Theta \b 1_p$,  and Assumption $8^*$ $\b 1_p' \b \Theta \b 1_p \ge c >0$, use (\ref{*1})
\[ \| \b a \|_1 = O (p^{1/2}).\]
By Lemma B.2(iii)
\[ \|\b  l \|_1 = O (p^{1/2}).\]
Then using triangle inequality and the results above
\[ \| \b w_c^* \|_1 = O ( p^{1/2}) + O ( p^{1/2}) = O (p^{1/2}) = O (\b w_{cp}^*).\]
 
So we will get the same convergence rates in just weight case with TE+weight cases.

Also for the proof of inequality constraints, crucial term is $\epsilon= p \bar{s} l_T K \to 0$ by Assumption $8^*$ by Lemma B2(iv), hence all the proofs of Theorem 3 above will carry over with Theorem B.1, B.2 instead of Theorems 1-2.

\subsection{ Bounded Sharpe Ratio with Scaled Variance of Global Minimum Variance}\label{b1-2}

In this section we restrict the Sharpe Ratio of the maximum Sharpe Ratio portfolio to be bounded as in (\ref{bsrr}), but we relax the condition in variance of GMV compared with previous subsection.   So we change (\ref{bgmv}) to Assumption 7(ii) again which is $|\b 1_p' \b \Theta \b 1_p|/p \ge c >0$. This may happen because the elements of $\Theta$ when added can go to infinity in the precision matrix, but still Sharpe Ratio can be bounded due to differing signs in $\mu_j, \mu_k$ and balancing each other.

Here we first provide an assumption that replaces Assumption 7.  After Assumption  $7^**$ we describe the differences. 

{\bf Assumption $7^{**}$}.{\it 

(i). $p^{1/2} \bar{s} l_T  K^3 \to 0$.

(ii). $\b 1_p' \b \Theta \b 1_p/p \ge c > 0$, $ 0 < c \le \b \mu' \b \Theta \b \mu \le C < \infty$.

(iii). $ \| \b m \|_1 = O ( \| \b w_d^* \|_1)$, and $\b w^{*'} \b \Sigma_y \b w^* \ge c >0$, $| \b w^{*'} \b \mu | \ge c > 0$.}

 There are differences with Assumption 7. First of all, our sparsity assumption is a bit different. We have an extra $p^{1/2}$ rate but we have less $K$ rate compared to Assumption 7(i).
 Then we have a bounded Sharpe Ratio assumption in Assumption $7^{**}$(ii).  Now we provide the main theorem for this subsection.

{\bf Theorem B.3}.{\it 
Under Assumptions 1-6, $7^{**}$ with the following condition $| \b 1_p' \b \Theta \b \mu | /p^{1/2} \ge c > 0$, and $0 < c \le  | \kappa | \le C < \infty$

i). \[ \| \hat{\b w} - \b w^* \|_1 = \| \hat{\b w_d} - \b w_d^* \|_1 = O_p ( p^{1/2} \bar{s} l_T K) =o_p (1).\]

(ii). \[ \left| \frac{\hat{\b w}' \b \Sigma_y \hat{\b w} }{\b w^{*'} \b \Sigma_y \b w^* } -1 
\right| = O_p ( p^{1/2} \bar{s} l_T  K^3) = o_p (1).
\]

(iii). \[ \left| \frac{\hat{SR}^2 }{SR^2} - 1 
\right|  = O_p ( p^{1/2} \bar{s} l_T  K^3) = o_p (1).
\]

}

Remarks.  There are differences between Theorem 1 and Theorem B.3 due to Assumption $7^{**}$ replacing Assumption 7. First we have extra $p^{1/2}$ in estimating weights in the bounded Sharpe Ratio case compared with Theorem 1(i), and for estimating variance and Sharpe Ratio of the portfolio that is analyzed here, we have extra $p^{1/2}$ compared with Theorem 1(ii)(iii).
Hence we can have only consistency with $p<T$ scenario here. One better result compared to Theorem 1(ii)-(iii) is that we have $K$ factor less here, i.e. we have $K^3$ versus $K^4$ rate in Theorem 1(ii)(iii).

{\bf Proof of Theorem B.3}. 

(i). Note that here we use $A, \hat{A}$ definitions in proof of Theorem 1(i), rather than Theorem B.1, since our assumption about GMV variance is the same in the main text Assumption 7(ii), we only assume bounded Sharpe ratio. So $\hat{A}:= \b 1_p' \hat{\b \Theta}' \b 1_p/p$, and $A := \b 1_p' \b \Theta \b 1_p/p$. All the results including 
(\ref{pt1.1})-(\ref{pt1.11}) are valid. Before proceeding further, the term $F$ has to be scaled differently than in proof of Theorem 1(i). The reason is that we assume $\b  \mu' \b \Theta \b \mu$ to be bounded, hence by Cauchy-Schwartz inequality 
\[ | \b 1_p' \b \Theta \b \mu/p^{1/2} | \le [ \b 1_p' \b \Theta \b 1_p/p]^{1/2} [ \b \mu' \b \Theta \b \mu 
 ] ^{1/2} = O(1).\]
 So we define and scale $F:= \b 1_p' \b \Theta \b \mu /p^{1/2}$ and hence it is bounded from above.  We will assume also $|F| \ge c >0$, where $c$ is a positive constant. Note that 
 in proof of Theorem 1(i), this term, $|\b 1_p' \b \Theta \b \mu|$, was scaled by $p$ rather than $p^{1/2}$ here. Also in proof of Theorem B.1(i), this term was not scaled since we imposed boundedness assumptions on both Sharpe Ratio of maximum Sharpe Ratio portfolio, and variance of GMV portfolio. Here we only impose boundedness of Sharpe Ratio of maximum Sharpe Ratio portfolio. Note that we change (\ref{pt1.12}) to following, with $\hat{F}:= \b 1_p' \hat{\b \Theta}' \b \mu/p^{1/2}$,

\begin{equation}
\left\| \left( \frac{\hat{\b \Theta}' \hat{\b \mu}}{\b 1_p' \hat{\b \Theta}' \hat{\b \mu}} - \frac{\b \Theta \b \mu}{\b 1_p' \b \Theta \b \mu}
\right)  \right\|_1 = \frac{ \| (\hat{\b \Theta}' \hat{\b \mu}/p^{1/2}) F - (\b \Theta \b \mu /p^{1/2}) \hat{F}\|_1}{| F \hat{F}|}.\label{B15}
\end{equation}

In (\ref{B15}) consider the denominator first 
\[ | F \hat{F}| \ge F^2 - |F| | \hat{F} - F| \ge c^2 - o_p(1),\]
We now show the details below.
Then under Assumptions $1-7^{**}(i)$, but with new scaling of  F with $p^{1/2}$ here rather than $p$ in the proof of Theorem 1(i)-(\ref{a13a}) 
\begin{equation}
 | \hat{F} - F | = O_p (p^{1/2}K \bar{s} l_T) = o_p (1),\label{B16}
 \end{equation}
with new Assumption $7^{**}$.
Now consider the numerator in (\ref{B15}), by adding and subtracting $(\b \Theta \b \mu/p^{1/2})F$ and triangle inequality
\begin{equation}
\| \frac{ \hat{\b \Theta}' \hat{\b \mu}}{p^{1/2}} F - \frac{\b \Theta \b \mu}{p^{1/2}} \hat{F} \|_1 \le 
\| \frac{ \hat{\b \Theta}' \hat{\b \mu}}{p^{1/2}} F - \frac{\b \Theta \b \mu}{p^{1/2}} F \|_1 + \| \frac{\b \Theta \b \mu}{p^{1/2}}  (\hat{F} - F) \|_1.\label{B17}
\end{equation}
Use the first term on the right side in (A.23)
\begin{eqnarray}
\frac{1}{p^{1/2}} \| (\hat{\b \Theta} - \b \Theta)' \hat{\b \mu} \|_1 & \le & \frac{1}{p^{1/2}} \| (\hat{\b \Theta} - \b \Theta)' \|_{l_1} \| \hat{\b \mu} \|_1 \nonumber \\
& \le & p^{1/2} \| \hat{\b \Theta} - \b \Theta \|_{l_{\infty}} \| \hat{\b \mu} \|_{\infty} = O_p ( p^{1/2} \bar{s} l_T) O_p (K),\label{B18}
\end{eqnarray}
where we use p.345 of \cite{hj2013} for the first inequality and for the rates, we use Theorem 2 of \cite{caner2022} and (B.8) of \cite{caner2022} by Assumptions 1-6,$7^{**}$.
Now consider the second term on the right side of (A.23)
\begin{eqnarray}
\frac{1}{p^{1/2}} \| \b \Theta (\hat{\b \mu} - \b \mu ) \|_1 &\le & p^{1/2} \| \b \Theta \|_{l_1} \| \hat{\b \mu} - \b \mu \|_{\infty} \nonumber \\
& = &p^{1/2}  \| \b \Theta \|_{l_{\infty}} \| \hat{\b \mu} - \b \mu \|_{\infty}  \nonumber \\
&=& O (p^{1/2} \bar{s} r_T K^{3/2}) O_p (max (K \sqrt{\frac{ln T}{T}}, \sqrt{\frac{lnp}{T}}))\label{B19})
\end{eqnarray}
where we use the same analysis as in (\ref{B18}) for the inequality, and the rates are by (A.18) here and Theorem 2(ii) of \cite{caner2022}.
Since $| F| = O(1)$ by our Assumption $7^{**}$ here and using Cauchy-Schwartz inequality, in (\ref{B17}) the first right side term is
\begin{equation}
\| \frac{\hat{\b \Theta}' \hat{\b \mu}}{p^{1/2}} - \frac{\b \Theta \b \mu}{p^{1/2}} \|_1 |F| = O_p ( p^{1/2} \bar{s} l_T K),\label{B20}
\end{equation}
by the slower rate in (\ref{B18}) and by $l_T$ definition in (2).
Next in (\ref{B17}), the second right side term is
\begin{eqnarray}
\| \frac{\b \Theta \b \mu}{p^{1/2}}  (\hat{F} - F) \|_1 & = & \| \frac{\b \Theta \b \mu}{p^{1/2}} \|_1 | \hat{F} - F | \nonumber \\
&=&O_p (p^{1/2} \bar{s} l_T K ),\label{B21}
\end{eqnarray}
where we use 
\begin{eqnarray}
\frac{\| \b \Theta \b \mu \|_1}{p^{1/2}} &\le &\| \b \Theta \b \mu \|_2 \nonumber \\
& \le & [ Eigmax (\b \Theta)]^{1/2} \sqrt{\b \mu' \b \Theta \b \mu } \le C,\label{*2} 
\end{eqnarray}
by $\b \Theta$ being symmetric, and Exercise 7.53a in \cite{abamag2005} and Assumption $7^{**}$.
Use (\ref{*2}) and (\ref{B16}) to get the rate in (\ref{B21}). Clearly the rate in  (\ref{B17})
\begin{equation}
\| \frac{ \hat{\b \Theta}' \hat{\b \mu}}{p} F - \frac{\b \Theta \b  \mu}{p} \hat{F} \|_1 = O_p (p^{1/2} \bar{s} l_T  K).\label{B22}
\end{equation}

Use (\ref{B22}) and the  equation before (\ref{B16}) in (\ref{B15}) to have 
\begin{equation}
\left\| \left( \frac{\hat{\b \Theta}' \hat{\b \mu}}{1_p' \hat{\b \Theta}' \hat{\b \mu}} - \frac{\b \Theta \b \mu}{\b 1_p' \b \Theta \b \mu}
\right)  \right\|_1 =O_p (p^{1/2} \bar{s} l_T  K) = o_p (1),\label{B23}
\end{equation}
by Assumption $7^{**}$(i) to have the last equality. Use (\ref{pt1.11})(\ref{B23}) in the weights definition in (\ref{pt1.1})
\[ \| \hat{\b w}_d - \b w_d^* \|_1 = O_p (p^{1/2} \bar{s} l_T  K) = o_p (1)\].{\bf Q.E.D.}

(ii).  Define $r_{w1c}:= p^{1/2} \bar{s} l_T K$.
Note that (A.30)-(A.32) does not change. Consider each term in (A.31), and see that by Holder's inequality, and $\| \b A \b x \|_{\infty} \le \| \b A \|_{\infty} \| \b x \|_1$ for generic $\b A, \b x$ matrix-vector pair,
\begin{eqnarray}
(\hat{\b w}_d - \b w_d^*)' \b \Sigma_y (\hat{\b w}_d - \b w_d^*) & \le & \| \hat{\b w}_d - \b w_d^* \|_1 \| \b \Sigma_y ( \hat{\b w}_d - \b w_d^* )\|_{\infty} \nonumber \\
& \le & \| \hat{\b w}_d - \b w_d^* \|_1^2 \| \b \Sigma_y \|_{\infty} \nonumber \\
& =& [O_p (r_{w1c})]^2 O (K^2),\label{B24}
\end{eqnarray}
where we use Theorem B.3(i), and (A.32). Then (A.34) is the same. Second right side term in (A.31) will be considered.
In (A.34)
\begin{equation}
\| \frac{\b \Theta \b \mu}{\b 1_p' \b \Theta \b \mu} - \frac{\b \Theta \b 1_p}{\b 1_p' \b \Theta 1_p} \|_1
\le \| \frac{\b \Theta \b \mu/p^{1/2}}{\b 1_p' \b \Theta \b \mu/p^{1/2}} \|_1 + \| \frac{\b \Theta \b 1_p/p}{\b 1_p' \b \Theta \b 1_p/p} \|_1.\label{B25}
\end{equation}

First we consider the denominators in (\ref{B25}).
By Assumption $7^{**}$(ii) and $|\b 1_p' \b \Theta \b \mu |/p^{1/2} \ge c >0$ in the statement of Theorem 1, denominators are bounded away from zero. By (\ref{*2})

\begin{equation}
\| \b \Theta \b \mu/p^{1/2} \|_1 = O(1).\label{B26}
\end{equation}

Clearly using  (\ref{B26})(A.37) in (\ref{B25}), and since $ \kappa$ is a constant,
\begin{equation}
\| \b w_d^* \|_1 = O (1).\label{B27}
\end{equation}
Next (A.39) stays the same.
Combine Theorem B.3(i), (A.32)(\ref{B27}) in (A.39) to have 
\begin{equation}
| \b w_d^{*'} \b \Sigma_y (\hat{\b w}_d - \b w_d^*) |  = O_p (p^{1/2}  \bar{s}  l_T  K^{3})=o_p (1),\label{B28}
\end{equation}
by Assumption $7^{**}$. Clearly rate in (\ref{B28}) is slower than the rate in (\ref{B24})
, so 
\[ |\hat{\b w}_d' \b \Sigma_y \hat{\b w}_d - \b w_d^{*'} \b \Sigma_y \b w_d^* | = O_p (p^{1/2}  \bar{s}  l_T  K^{3})=o_p (1).\]
{\bf Q.E.D.}

(iii). All the equations in (A.43)(A.44) stays the same.Next, 
\begin{eqnarray}
| (\hat{\b w}+ \b w^*)' \b \mu| & \le & \| \hat{\b w} - \b w^* \|_1 \| \b \mu \|_{\infty} + 2 \| \b w^* \|_1 \| \b \mu \|_{\infty} \nonumber \\
& = & O_p (p^{1/2} \bar{s} l_T  K^{2} )   + O ( K) = o_p (1) + O(K) \nonumber \\
&=& O_p ( K),\label{B29}
\end{eqnarray}
where
for the rates we use Theorem B.3(i) here and (B.8) of \cite{caner2022} in which $\| \b \mu \|_{\infty} = O(K)$, and $\b w^*:= \b w_d^* + \b m$, (\ref{B27}), Assumption $7^{**}$(iv).

First consider, by Theorem B.3(i), and Holder's inequality, and by assuming $|\b w^{*'} \b \mu| \ge c >0$, and (B.8) of \cite{caner2022} 
\begin{equation}
\frac{|\hat{\b w}' \b \mu - \b w^{*'} \b \mu|}{ ( \b w^{*'} \b \mu)^2} \le \frac{\| \hat{\b w} - \b w^* \|_1 \| \b \mu \|_{\infty}}{ (\b w^{*'} \b \mu)^2}= O_p (p^{1/2} \bar{s} l_T  K^{2} ).\label{B30}
\end{equation}

So combine (\ref{B29})(\ref{B30}) with (A.44)
\begin{equation}
\frac{ | (\hat{\b w}' \b \mu )^2 - (\b w^{*'} \b \mu)^2|}{(\b w^{*'} \b \mu)^2} = O_p (p^{1/2} \bar{s} l_T   K^{2} ) O_p (K) = 
O_p ( p^{1/2} \bar{s}  l_T K^3) = o_p (1),\label{B31}
\end{equation}
by Assumption $7^{**}(i)$.

Then
\begin{equation}
\left| \frac{\b w^{*'} \b \Sigma_y \b w^* - \hat{\b w}' \b \Sigma_y \hat{\b w}}{\hat{\b w}' \b \Sigma_y \hat{\b w}}\right|
=   O_p (p^{1/2} \bar{s}  l_T K^3),\label{B32}
\end{equation}
by Theorem B.3(ii) proof. Use (\ref{B31})(\ref{B32}) and Sharpe Ratio definition on the left-side of (A.48) to have 
\[ \left| \frac{\widehat{SR}^2}{SR^2} - 1
\right| = O_p ( p^{1/2} \bar{s} l_T  K^3).\]
{\bf Q.E.D.}

Note that Lemma A.1 does not change even with our new Assumptions.  Lemma A.2 will  change but partially.   
We need to change Assumption 8 first.  Two main differences are, sparsity assumption at (i) has extra $p^{1/2}$ here compared with original assumption, but a  factor $K$ less here, and part (ii), we scale by $p^{1/2}$ the very first term in that assumption. That square root p scale is due to bounded Sharpe Ratio of maximum Sharpe Ratio portfolio but GMV variance can be scaled by $p$, and these possibilities are explained at the beginning of this subsection.

{\bf Assumption $8^{**}$}.{\it 

(i). $p^{1/2} \bar{s} l_T  K^{3} =o(1).$ 

(ii). Let $0 < c \le \b \mu' \b \Theta \b\mu \le C < \infty$,  $ \left| \frac{\b 1_p' \b \Theta \b \mu}{p^{1/2}} 
\right| \ge c > 0$, $ \frac{\b 1_p' \b \Theta \b 1_p}{p}
 \ge c >0$,$\left| \frac{\b 1_R' \b \Theta \b 1_p}{p}
\right| \ge c >0$.

 (iii). $0 < r/p <1$, $| w_k - w_a | \ge c > 0$, and $| \omega | \le C < \infty$, $ 0 < c \le | \kappa | \le C < \infty$.
 
 (iv). $| \b w_R^{*'} \b \mu | \ge c > 0$, $ \b w_R^{*'} \b \Sigma_y \b w_R^* \ge c > 0$, $ \| \b m_R \|_1 = O (\b w_{cp}^*)$.
}

Lemma A.2 will be replaced with Lemma B.4 below.

{\bf Lemma B.4}. {\it Under Assumptions 1-6, $8^{**}$ 

(i). \[ \| \hat{\b l }- \b l \|_1 = O_p ( \bar{s} l_T).\]

(ii). \[ | w_u | = O (1).\]

(iii). \[ \| \b l \|_1 = O (1).\]

(iv). \[ | \hat{ w}_u  -  w_u | = O_p ( p^{1/2} \bar{s} l_T  K).\]
}

{\bf Proof of Lemma B.4}.

(i). The proof of this part does not change with new Assumption $8^{**}$ as in Lemma A.2(i).

(ii). We start with a triangle inequality and definition 
\begin{equation}
| w_u | \le \left| \frac{\b 1_R' \b \Theta \b \mu/p^{1/2}}{\b 1_p' \b \Theta \b \mu/p^{1/2}}\right| + \left|\frac{\b 1_R' \b \Theta \b 1_p/p}{\b 1_p' \b \Theta \b 1_p/p}\right|.\label{B33}
\end{equation}
Then the first right side term is, by Cauchy-Schwartz inequality 
\begin{equation}
 | \b 1_R' \b \Theta \b \mu/p^{1/2}| \le \left[\frac{| \b 1_R' \b \Theta  \b 1_R|}{p} \right]^{1/2}  | \b \mu' \b \Theta \b \mu|^{1/2} \le [Eigmax (\Theta)]^{1/2} C = O(1) ,\label{*3}
 \end{equation}
 by Assumption $8^{**}(ii)$, and (B.22) of \cite{caner2022}.
 
 Then by Assumption $8^{**}$,
  $|\b 1_p' \b \Theta \b \mu /p^{1/2}| \ge c >0$
\begin{equation}
\left| \frac{\b 1_R' \b \Theta \b \mu/p^{1/2}}{\b 1_p' \b \Theta \b \mu/p^{1/2}}\right| = O (1).\label{B34}
\end{equation} 
 Use (A.80) and (\ref{B34}) to have 
 \[ | w_u | = O (1).\]

 (iii). The proof is the same as in Lemma A.2(iii).

 (iv). By using $\hat{w}_u, w_u$ definitions
and by triangle inequality
\begin{equation}
| \hat{w}_u - w_u | \le \left|  \frac{\b 1_R' \hat{\b \Theta}' \hat{\b \mu}/p^{1/2}}{\b 1_p' \hat{\b \Theta}' \hat{\b \mu}/p^{1/2}} - \frac{\b 1_R' \b \Theta \b \mu/p^{1/2}}{\b 1_p' \b \Theta \b \mu /p^{1/2}}
\right| + \left| \frac{\b 1_R' \hat{\b \Theta}' \b 1_p/p}{\b 1_p' \hat{\b \Theta}' \b 1_p/p} - \frac{\b 1_R' \b \Theta \b 1_p/p}{\b 1_p' \b \Theta \b 1_p/p}\right|.\label{B35}
\end{equation} 
 Then consider the first right side term in (\ref{B35}), by adding and subtracting and triangle inequality
\begin{eqnarray}
\left|  \frac{\b 1_R' \hat{\b \Theta}' \hat{\b \mu}/p^{1/2}}{\b 1_p' \hat{\b \Theta}' \hat{\b \mu}/p^{1/2}} - \frac{\b 1_R' \b \Theta \b \mu/p^{1/2}}{\b 1_p' \b \Theta \b \mu /p^{1/2}}
\right| 
& \le & \left|  \frac{\b 1_R' \hat{\b \Theta}' \hat{\b \mu}/p^{1/2}}{\b 1_p' \hat{\b \Theta}' \hat{\b \mu}/p^{1/2}} -  \frac{\b 1_R' \b \Theta \b \mu/p^{1/2}}{\b 1_p' \hat{\b \Theta}' \hat{\b \mu}/p^{1/2}} \right|
\nonumber \\
& + & \left| \frac{\b 1_R' \b \Theta \b \mu/p^{1/2}}{\b 1_p' \hat{\b \Theta}' \hat{\b \mu} /p^{1/2}}-\frac{\b 1_R' \b \Theta \b \mu/p^{1/2}}{\b 1_p' \b \Theta \b \mu /p^{1/2}} \right|.\label{B36}
\end{eqnarray} 
Use the same analysis in the proof of Lemma A.2(iv) with scaling by $p^{1/2}$ rather than $p$ there in Lemma A.2, the first right side term in (\ref{B36}) is 
 \begin{equation}
  \left|  \frac{\b 1_R' \hat{\b \Theta} \hat{\b \mu}/p^{1/2}}{\b 1_p' \hat{\b \Theta} \hat{\b \mu}/p^{1/2}} -  \frac{\b 1_R' \b \Theta \b \mu/p^{1/2}}{\b 1_p' \hat{\b \Theta} \hat{\b \mu}/p^{1/2}} \right|
= O_p (p^{1/2} \bar{s} l_T K).\label{B37}
\end{equation} 

In second right side term in (\ref{B36}), use  (A.87) scaled with $p^{1/2}$ with (\ref{*3}), and  with Assumption $8^{**}$, with $r<p$,
\begin{equation}
| \b 1_R' \b \Theta \b \mu /p^{1/2}| \left| \frac{\b 1_p' \hat{\b \Theta}' \hat{\b \mu}/p^{1/2} - \b 1_p' \b \Theta \b \mu/p^{1/2}}{(\b 1_p' \hat{\b \Theta}' \hat{\b \mu}/p^{1/2})(\b 1_p' \b \Theta \b \mu/p^{1/2})}
\right| = O (1) O_p ( p^{1/2} \bar{s} l_T K) = O_p (p^{1/2} \bar{s} l_T K).\label{B38}
\end{equation}

Consider (\ref{B37})(\ref{B38}) in (\ref{B36}) to have 
the first right side term in (\ref{B35}) as 
\begin{equation}
\left|  \frac{\b 1_R' \hat{\b \Theta}' \hat{\b \mu}/p^{1/2}}{\b 1_p' \hat{\b \Theta}' \hat{\b \mu}/p^{1/2}} - \frac{\b 1_R' \b \Theta \b \mu/p^{1/2}}{\b 1_p' \b \Theta \b \mu /p^{1/2}}
\right| =O_p ( p^{1/2} \bar{s} l_T K).\label{B39}
\end{equation}

The second term rate in (\ref{B35}) is the same  as the first  one, via the proof of Lemma A.2(iv),  in (\ref{B39}) so we have 
\[ 
 | \hat{\b w}_u - \b w_u | = O_p ( p^{1/2} \bar{s} l_T  K).\]
 {\bf Q.E.D.}

{\bf Theorem B.4}. {\it Under Assumptions 1-6, $8^{**}$ 

(i). \[ \| \hat{\b w}_R - \b w_R^* \|_1 = O_p ( p^{1/2} \bar{s} l_T K) = o_p (1)
\]

(ii). \[  \| \frac{\hat{\b w}_R' \b \Sigma_y \hat{\b w}_R}{\b w_R^{*'} \b \Sigma_y \b w_R^{*}} - 1 \|_1 =
O_p ( p^{1/2} \bar{s} l_T K^{3}) = o_p (1).
\]

(iii). \[ \left|\left(\frac{\widehat{SR}_R}{SR_R}
\right)^2 -1 
\right| = O_p ( p^{1/2} \bar{s} l_T K^{3}) = o_p (1).\]

}

{\bf Proof of Theorem B.4}

(i). First, all the terms (A.94)-(A.97) will stay the same as in the proof of Theorem 2(i).
 Consider each term in (\ref{pt2.4}) by Lemma B.4
\begin{equation}
 \| \kappa (\hat{w}_u - w_u) (\hat{\b l} -\b  l )\|_1  
\le | \kappa | |\hat{w}_u - w_u | \| \hat{\b l} - \b l \|_1 = O_p (p^{1/2}  \bar{s} l_T K) O_p (\bar{s} l_T ).\label{B40}
\end{equation}
Then consider the second right side term in (\ref{pt2.4}) via Lemma B.4
\begin{equation}
\| \kappa (\hat{w}_u - w_u) \b l \|_1 \le | \kappa | |\hat{w}_u - w_u | \| \b l \|_1 
= O_p ( p^{1/2} \bar{s} l_T K) O (1) = O_p (p^{1/2} \bar{s} l_T K).\label{B41} 
\end{equation}

Then analyze the third term on the right side of (\ref{pt2.4}) by Lemma 	B.4
\begin{equation}
\| (\omega - \kappa w_u ) (\hat{\b l} - \b l )\|_1 \le [ | \omega| + | \kappa | | w_u |] \| \hat{\b l} - \b l \|_1
=O (1) O_p (\bar{s} l_T )
= O_p (\bar{s} l_T ).\label{B42}
\end{equation}

Given our new Assumption $8^{**}(i)$, the rate in (\ref{B41}) is the slowest among (\ref{B40})-(\ref{B42}). Hence
\begin{equation}
 \|(\omega - \kappa \hat{w}_u) \hat{\b l} - (\omega - \kappa w_u )  \b l \|_1
 = O_p (p^{1/2} \bar{s} l_T K).\label{B43}
\end{equation}
Then use (\ref{B43}) and Theorem B.3(i)
\begin{equation}
\| \hat{\b w}_{cp} - \b w_{cp}^* \|_1 = O_p ( p^{1/2} \bar{s} l_T K) + O_p (p^{1/2}  \bar{s} l_T K) = O_p (p^{1/2} \bar{s} l_T K).\label{B44}
\end{equation}
{\bf Q.E.D.}

(ii). Until and including (\ref{pt2.9}) in the proof of Theorem 2(ii) stays the same. 
 Consider the first term on the right side of (\ref{pt2.9}) and using Theorem B.4(i) and (\ref{pt1a-3}) by  Holder's inequality  and $\| \b A \b x \|_{\infty} \le \| \b A \|_{\infty} \| \b x \|_1$ inequality for generic matrix $\b A$, and generic vector $\b x$
 \begin{equation}
 (\hat{\b w}_{cp} - \b w_{cp}^* )' \b \Sigma_y (\hat{\b w}_{cp} - \b w_{cp}^*) \le \| \hat{\b w}_{cp} - \b w_{cp}^* \|_1^2 \| \b \Sigma_y \|_{\infty}
 = O_p (p \bar{s}^2 l_T^2 K^{2}) O (K^2).\label{B45}
 \end{equation}

 We consider $l_1$ norm of $w_{cp}^*$ 
 \begin{eqnarray}
 \| \b w_{cp}^* \|_1 & = & \| (\omega - \kappa w_u ) \b l + \b w_d^* \|_1 \le   \| (\omega - \kappa w_u ) \b l \|_1 + \| \b w_d^* \|_1 \nonumber \\
 & \le & \left[ | \omega | + | \kappa | | w_u| 
 \right]  \| \b l \|_1 + \| \b w_d^* \|_1 \nonumber \\
 & = & O (1) O (1) + O (1) = O ( 1),\label{B46}
  \end{eqnarray}
 where $\omega, \kappa$ are bounded by definition, and we use (\ref{B27}), Lemma B.4(ii)(iii) for the rates. The second term on the right side of (\ref{pt2.9}) becomes
  \begin{eqnarray}
| \b w_{cp}^{*'} \b \Sigma_y (\hat{\b w}_{cp} - \b w_{cp}^*)|  & \le & \| \b w_{cp}^* \|_1 \| \b \Sigma_y \|_{\infty} \| \hat{\b w}_{cp} - \b w_{cp}^* \|_1 \nonumber \\
& = & O ( 1) O (K^2) O_p (p^{1/2} \bar{s}  l_T K) \nonumber \\
&= &O_p ( p^{1/2} \bar{s}  l_T  K^{3}),\label{B47}
\end{eqnarray}
where we use the  Holder inequality and $\| \b A \b x \|_{\infty} \le \| \b A \|_{\infty} \| \b x \|_1$ inequality for generic matrix $\b A$, and generic vector $\b x$ for the inequality and the rates are from (\ref{pt1a-3})(\ref{B46}),  Theorem B.4(i).

\noindent Combine (\ref{B45})(\ref{B47})  in (\ref{pt2.9}) to have,  by Assumption $8^{**}$  
 
 \begin{equation}
 (\hat{\b w}_{cp}'\b  \Sigma_y \hat{\b w}_{cp}) - (\b w_{cp}^{*'} \b \Sigma_y \b w_{cp}^*) = O_p ( p^{1/2}  \bar{s}  l_T  K^{3}).\label{B48}
 \end{equation}
 Next by the analysis in (\ref{B47}) and Assumption $8^{**}$
\begin{equation}
 | \b m_R^{'} \b \Sigma_y (\hat{\b w}_{cp} - \b w_{cp}^*)| = O_p ( p^{1/2}  \bar{s}  l_T  K^{3}).\label{B49}
 \end{equation}
Next combine (\ref{B48})(\ref{B49})

\[ \| \frac{\hat{\b w}_R' \b \Sigma_y \hat{\b w}_R}{\b w_R^{*'} \b \Sigma_y \b w_R^*} -1 \|_1 
= O_p (p^{1/2}  \bar{s}  l_T  K^{3}).\]
{\bf Q.E.D}

(iii). The proof uses (\ref{pt2.15}) in the proof of Theorem 2iii. To evaluate the right side of (\ref{pt2.15}), we need the following results from Theorem B.4(i) above and $\| \b \mu \|_{\infty} = O (K)$ from (B.8) of \cite{caner2022} 
\[ \| \hat{\b w}_R - \b w_R^* \|_1 \| \b \mu \|_{\infty} = O_p ( p^{1/2} \bar{s} l_T  K^{2}).\]
Also by $\b w_R^*:= \b w_{cp}^* + \b m_R$ and (\ref{B46}) and Assumption $8^{**}$
\[ \| \b w_R^* \|_1 \| \b \mu \|_{\infty} = O ( 1) O (K) = O ( K).\]
Use (\ref{pt1a-14}) and the last two results above
\begin{eqnarray}
| ( \hat{\b w}_R + \b w_R^*)' \b \mu | & \le & 
\| \hat{\b w}_R - \b w_R^* \|_1 \| \b \mu \|_{\infty} + 2 \| \b w_R^* \|_1 \| \b \mu \|_{\infty} \nonumber \\
& = & O_p ( p^{1/2} \bar{s}  l_T K^{2}) + O (  K) \nonumber \\
& = & O_p (  K),\label{B50}
\end{eqnarray}
by Assumption $8^{**}$. Then 
\begin{eqnarray}
| \hat{\b w}_R' \b \mu - \b w_R^{*'} \b \mu | & \le & \| \hat{\b w}_R - \b w_R^* \|_1 \| \b \mu \|_{\infty} \nonumber \\
& = & O_p ( p^{1/2} \bar{s}  l_T K^{2}).\label{B51}
\end{eqnarray}
By (\ref{B50})(\ref{B51})
\begin{eqnarray}
\frac{| (\hat{\b w}_R' \b \mu )^2 - (\b w_R^{*'} \b \mu )^2|}{(\b w_R^{*'} \mu )^2} &= &
O_p ( p^{1/2} \bar{s} l_T K^{2}) O_p (  K) \nonumber \\
& = & O_p ( p^{1/2} \bar{s}  l_T K^{3}).\label{B52}
\end{eqnarray}

Then the rest of the proof follows Theorem 1(iii) proof given (\ref{B52}) and Theorem B.4(ii).
\[ \left| \left( \frac{\widehat{SR}_R}{SR_R}
\right)^2 - 1 
\right| = O_p ( p^{1/2} \bar{s}  l_T K^{3}) .\]
{\bf Q.E.D.}

The following lemma replaces Lemma A.3 with new Assumption $8^{**}$.  This  lemma is needed for just weight constraint case.

{\bf Lemma B.5}.{\it 
Under Assumptions 1-6, $8^{**}$, $0<c \le \kappa_w \le C < \infty$, $| w_x | \le C < \infty$, $B_4 B_3 \neq B_2^2$
\[ \| \hat{\b w}_c - \b w_c^* \|_1 = O_p ( p^{1/2} \bar{s} l_T K).\]

}

{\bf Proof of Lemma B.5}. This proof follows the proof of Lemma A.3 under Assumption $8^{**}$. Note that Lemma B.4(i)(iii) are the same as in Lemma A.2(i)(iii), 
(\ref{pla3-1})-(\ref{pla3-14}) are the same. By Theorem B.4(i), we have 
\[ \| \hat{\b w}_{cp} - \b w_{cp}^* \|_1 = O_p ( p^{1/2} \bar{s} l_T  K),\]
and hence this rate dominates the rate in (\ref{pla3-14}) by Assumption $8^{**}(i)$
 so 
\[ \| \hat{\b w}_{c} - \b w_{c}^* \|_1 = O_p ( p^{1/2} \bar{s} l_T  K).\]
{\bf Q.E.D}

Note that without losing any generality we use $\kappa=\kappa_w, w = w_x$, as discussed before Lemma A.3, and same calculations and logic applies here as well. Next, we need to see whether the rate in $l_1$ norm for $\b w_c^*$ is the same as in $\b w_{cp}^*$. From the formula before Lemma A.3, (by having $\kappa=\kappa_w, w=w_x$)

First note that 
\[ \b w_{c}^* = \b w_{cp}^* - (\b a - \b l \frac{B_3}{B_2}).\]
See that by (\ref{B46})
\[ \| \b w_{cp}^* \|_1 = O (1).\]
Then we can use the triangle inequality to have 
\[\| \b a - \b l \frac{B_3}{B_2} \|_1 \le \| \b a \|_1 + \| \b l \|_1 \left| \frac{B_3}{B_2}
\right|.\]
By Lemma A.1
\[ \| \b a \|_1 = O (1).\]
By Lemma B.4(iii)
\[ \|\b  l \|_1 = O (1).\]
Also by definitions of $B_3, B_2$, and using the same analysis in (\ref{3.6}) for $B_3$ with Assumption $8^{**}$,  $\b 1_p' \b \Theta \b 1_p/p \ge c > 0$ so using the triangle inequality
\[ \|\b  a - \b l \frac{B_3/p}{B_2/p}\|_1 = O (1).\]
This last result implies, since $w_{cp}^*$ has the largest rate
\begin{eqnarray*}
  \| \b w_c^* \|_1 &=& \| \b w_{cp}^* - (\b a - \b l \frac{B_3}{B_2}) \|_1 \\
  & \le & \| \b w_{cp}^* \|_1 + \| \b a \|_1 + \| \b l \|_1 \frac{B_3}{B_2} \\
  & = & O_p ( 1)= O (\b w_{cp}^*).
   \end{eqnarray*}
Hence all Theorem B.4(ii)(iii) proofs will follow. We get the same rate of convergence as in Theorem B.4(ii)(iii) for the only weight constraint case.

Inequality constraints will go through with $\epsilon:= p^{1/2} \bar{s}  l_T K$ by Lemma B.4(iv).

\section{Feasible Residual Nodewise Regression}\label{b2}

We cover how to form feasible residual nodewise regression. To describe briefly, let ${\bf \hat{b}}_j$ denote the $K \times 1$ factor loading estimates by OLS
\[ {\bf \hat{b}_j }= ({\b X} {\b X}')^{-1} {\b X} {\b y_j},\]
The residual is
\[ {\bf \hat{u}_j} = {\bf y_j} - {\b X'} {\bf \hat{b}_j},\]
with ${\b y_j}=( y_{j,1}, \cdots, y_{j,t}, \cdots, y_{j, T})': T \times 1$ vector of excess asset return of $j$ th asset across time. Denote ${\b y_{-j}}$ as $\bf y$ matrix without the $j$ th row ($p-1 \times T$), and it can be expressed as
\[ {\b y_{-j}} = {\b B_{-j} } {\b X }+ {\b U_{-j}},\]
with ${\b B_{-j}}$ as factor loadings matrix without $j$ th row ($p-1 \times K$), and ${\b U_{-j}}$ as the error matrix without $j$ th row, $j=1,\cdots, p$. Let ${\bf \hat{B}_{-j}}: (p-1) \times K$ matrix of OLS estimates of factor loadings of ${\b B_{-j}}$. We have the following residual based matrix which will be used in residual based nodewise regression
\[ {\bf \hat{U}_{-j}'} = {\bf y_{-j}'} - {\bf X'} {\bf \hat{B}_{-j}'},\]
with ${\bf \hat{U}_{-j}'}: T \times (p-1)$ matrix, and 
\[ {\bf \hat{B}_{-j}'} = ({\b X } {\b X'})^{-1} {\b X } {\b y_{-j}'},\]
which is $K \times (p-1)$ OLS estimates of factor loadings, except for asset $j$. The feasible residual based nodewise regression depends on the following, for $j=1,\cdots, p$
\begin{equation}
 \hat{\b  \gamma}_j = \argmin_{\b \gamma_j \in R^{p-1}} \left[ \| {\bf \hat{u}_j} - {\bf \hat{U}_{-j}'} \b  \gamma_j \|_T^2 + 2 \lambda \| {\b \gamma_j} \|_1 \right],\label{1.1}
\end{equation}
with $\lambda >0$ as a positive sequence with $\|{\b v} \|_T^2$ as the norm with a generic $T \times 1 $ vector defined as $\|{\b  v} \|_T^2:=\frac{1}{T} \sum_{j=1}^T v_j^2$.

{ \bf Remarks}. 1. The nodewise regression \eqref{1.1} is a graphical lasso method \citep{mein2006}, it can directly compute conditional covariances (precision matrix). 

2. It links sparsity of $\bf \Omega_j$ ($\bf \Omega$) to the sparsity in $\b \gamma_j$, the nodewise regression coefficients. It imposes weaker condition of sparsity of the precision matrix of errors which still allows dense covariance matrix of errors. In the following sections we establish the consistency for constrained portfolios using the residual based nodewise regression estimators.

To build the estimate for the precision matrix of errors, we define each row of the precision matrix of errors ${\bf \hat{\Omega}}_j': 1 \times p$, $j=1,\cdots,p$
\[ {\bf \hat{\Omega}_j'}:= {\bf \hat{C}}_j'/\hat{\tau}_j^2,\]
where ${\bf \hat{C}_j'}: 1 \times p$ row vector with $1$ in the $j$ th cell of that row and the remainder of that row is $- \hat{\b \gamma}_j'$. To give an example
$  \hat{\b C}_1':= (1,  -\hat{\b \gamma}_1'): 1 \times p$ vector. Also 
\[ \hat{\tau}_j^2:= {\bf \hat{u}_j'} ({\bf \hat{u}_j}- {\bf \hat{U}_{-j}'}   \hat{\b \gamma}_j)/T,\]
as in \cite{caner2022}. Stacking all rows of ${\bf \hat{\Omega}_j'}$, we form ${\bf \hat{\Omega}}$, which is the feasible residual based nodewise estimate of the precision matrix of errors.

\section{Consistent Estimation of $\kappa$ risk tolerance parameter}

Note that since Tracking Error is larger than or equal to $ \sqrt{\b w_d^{*'} \b \Sigma_y \b w_d^{*}}$ as noted in Section A.3, we base our estimator for $\kappa$ on that. However, to have a feasible solution, we assume that Tracking Error is  equal to  that term
\begin{equation}
TE = \sqrt{\b w_d^{*'} \b \Sigma_y \b w_d^{*}},\label{b3-1}
\end{equation}
where with (3) in the main text
\[ \b w_d^* = \kappa (\b w_{MSR}^* - \b w_{GMV}^*),\]
and definitions of maximum Sharpe Ratio and Global Minimum Variance Portfolio
\[ \b w_{MSR}^*:= \frac{\b \Theta \b \mu}{\b 1_p' \b \Theta \b \mu}, \quad \b w_{GMV}^*:= \frac{\b \Theta \b 1_p}{\b 1_p' \b \Theta \b 1_p}.\]

So we rewrite (\ref{b3-1}) as, with $\b \Sigma_y$ defined in Section 2, 
\begin{equation}
 \kappa = \frac{TE}{\sqrt{(\b w_{MSR}^* - \b w_{GMV}^*)' \b \Sigma_y (\b w_{MSR}^* - \b w_{GMV}^*)}}.\label{kappa1}
 \end{equation}  
Then setup the estimator by keeping $TE$ the same and fixed
\begin{equation}
\hat{\kappa}= \frac{TE}{\sqrt{(\hat{\b w}_{MSR} - \hat{\b w}_{GMV})'  \hat{\b \Sigma}_y (\hat{\b w}_{MSR} - \hat{\b w}_{GMV})}}.\label{kappa2}
\end{equation}
We define the estimators
\[ \hat{\b w}_{MSR}:= \frac{\hat{\b \Theta}' \hat{\b \mu}}{\b 1_p' \hat{\b \Theta}' \hat{\b \mu}}, \quad \hat{\b w}_{GMV}:= \frac{\hat{\b \Theta}' \b 1_p}{\b 1_p' 
\hat{\b \Theta}' \b 1_p},\]
with
\[ \hat{\b \Sigma}_y: = \hat{\b B} \hat{\b \Sigma}_f \hat{\b B}' + \hat{\b \Sigma}_u,\]
with $\hat{\b \Sigma}_u:= \frac{1}{T} \sum_{t=1}^T \hat{\b u}_t \hat{\b u}_t'$ with $\hat{\b u}_t= \b y_t - \hat{\b B} \b f_t$, hence the OLS residuals, where $\hat{\b B}$ is the OLS estimator for $\b B: p \times K$ matrix. Note that $\hat{\b \Sigma}_f$ is defined in Section 2.

{\bf Lemma B.6}.{\it Under Assumptions 1-7, with the following conditions $| \b 1_p' \b \Theta \b \mu |/p \ge c >0$, with $c$ being a positive constant and $\hat{\b w}_{MSR} \neq \hat{\b w}_{GMV}$, and $\b w_{MSR}^* \neq \b w_{GMV}^*$
\[ | \hat{\kappa} - \kappa | = O_p (\bar{s} l_T K^4) = o_p (1).\]
}

{\bf Proof of Lemma B.6}
Our proof has three steps. First, we have two inequalities that we need, then we show four terms that are critical in our proof, and in the last step, we provide consistency results and rate of convergence for all these four terms and end the proof.

Step 1. 
We provide two inequalities that simplifies our proof. First for a generic vectors $\b x_1, \b x_2$, and a matrix $\b Y$
\begin{equation}
| \b x_1' \b Y \b x_2 | \le \| \b x_1 \|_1 \| \b Y \b x_2 \|_{\infty} \le 
\| \b x_1 \|_1 \| \b x_2 \|_1 \| \b Y \|_{\infty},\label{b3-2}
\end{equation}
where we use Holder's inequality for the first inequality, and use $\| \b Y \b x_2 \|_{\infty} \le \| \b Y \|_{\infty} \| \b x_2 \|_1$. Then have a generic nonrandom vector $\b a$, matrix $\b A$, and their estimators
$\hat{\b a}, \widehat{\b A}$.  We provide the following inequality and then prove that.
\begin{eqnarray}
|\hat{\b a}' \widehat{\b A} \hat{\b a} - \b a' \b A \b a | &\le & 
\| \hat{\b a} - \b a \|_1^2 \| \widehat{\b A} - \b A \|_{\infty} + 2 \| \hat{\b a} - \b a \|_1 \| \b a \|_1 \| \widehat{\b A} - \b A \|_{\infty}\nonumber \\
& + & \| \b a \|_1^2 \| \widehat{\b A} - \b A \|_{\infty} + \| \hat{\b a}- \b a \|_1^2 \| \b A \|_{\infty} \nonumber \\
&+& 2 \| \b a \|_1 \| \hat{\b a} - \b a \|_1 \| \b A \|_{\infty},\label{b3-3} 
\end{eqnarray}
To prove (\ref{b3-3}), see that 
\begin{eqnarray*}
(\hat{\b a}' \widehat{\b A} \hat{\b a} - \b a' \b A \b a )& = & 
[ (\hat{\b a } - \b a ) + \b a ]' [ (\widehat{\b A} - \b A ) + \b A ] [ (\hat{\b a } - \b a ) + \b a ] - (\b a' \b A \b a) \\
& = & (\hat{\b a} - \b a )' (\widehat{\b A} - \b A )(\hat{\b a} - \b a )+ 2 (\hat{\b a} - \b a )' (\widehat{\b A} - \b A) \b a \\
& + & \b a' (\widehat{\b A} - \b A ) \b a + (\hat{\b a} - \b a )' \b A (\hat{\b a} - \b a ) + 2 \b a' \b A (\hat{\b a} - \b a ).
\end{eqnarray*}
Use (\ref{b3-2}) in the equality above to get (\ref{b3-3}).

Step 2.
 Next, there are four terms that we have to learn upper bounds to finish our proof.
These are \[ \| \hat{\b \Sigma}_y - \b \Sigma_y \|_{\infty}, \quad  
 \| \hat{\b w}_{MSR} - \hat{\b w}_{GMV} - (\b w_{MSR}^* - \b w_{GMV}^*) \|_1,\]
and $\| \b \Sigma_y \|_{\infty}$, with $\| \b w_{MSR}^* - \b w_{GMV}^* \|_1$. Start with 
\begin{equation}
\| \hat{\b \Sigma}_y - \b \Sigma_y \|_{\infty} \le \| \hat{\b B} \hat{\b \Sigma}_f \hat{\b B}' - \b B \b \Sigma_f \b B' \|_{\infty} 
+ \| \hat{\b \Sigma}_u - \b \Sigma_u \|_{\infty}.\label{b3-4}
\end{equation}

Step 3.
First on (\ref{b3-4}) start with the second term on the right side, and also define the (infeasible) sample covariance matrix of errors
as $\bar{\b \Sigma}_u:= \frac{1}{T} \sum_{t=1}^T \b u_t \b u_t'$. Then 
\begin{equation}
\| \hat{\b \Sigma}_u - \b \Sigma_u \|_{\infty} \le \| \hat{\b \Sigma}_u - \bar{\b \Sigma}_u \|_{\infty} + 
\| \bar{\b \Sigma}_u - \b \Sigma \|_{\infty}.\label{b3-5}
\end{equation}
Then clearly with using the residuals $\hat{\b u}_t= \b u_t - (\hat{\b B} - \b B) \b f_t$
\begin{eqnarray}
\| \hat{\b \Sigma}_u - \bar{\b \Sigma}_u \|_{\infty} & \le & \| (\hat{\b B} - \b B ) \frac{\sum_{t=1}^T \b f_t \b u_t' }{T} \|_{\infty} +
+  \| \frac{\sum_{t=1}^T \b u_t \b f_t' }{T} (\hat{\b B} - \b B )' \|_{\infty} \nonumber \\
& + & \| (\hat{\b B} - \b B) [ \frac{\sum_{t=1}^T \b f_t \b f_t' }{T}] (\hat{\b B} - \b B)'\|_{\infty}.\label{b3-6}
\end{eqnarray}

Then by dual norm inequality in Section 4.3 of \cite{vdg2016}, the second term on the right side  of (\ref{b3-6}) is bounded by 
\begin{eqnarray}
\| (\frac{1}{T} \sum_{t=1}^T \b u_t \b f_t' )(\hat{\b B} - \b B)' \|_{\infty} & \le & 
\| \frac{1}{T} \sum_{t=1}^T \b u_t \b f_t' \|_{\infty} \| (\hat{\b B} - \b B)' \|_{l_1} \nonumber \\
& = & 
\| \frac{1}{T} \sum_{t=1}^T \b u_t \b f_t' \|_{\infty} \| \hat{\b B} - \b B \|_{l_{\infty}} \nonumber \\
& = & O_p ( \sqrt{ln p/T}) O_p ( K^{3/2} \sqrt{ln p/T}) \nonumber \\
&=& O_p ( K^{3/2} lnp/T),\label{b3-7}
\end{eqnarray}
where the first equality is by $l_1$ and $l_{\infty}$ matrix norm definitions, and the rates are by Lemma A.3(iii) and Lemma A.10(i) of \cite{caner2022}. The first term on the right side of (\ref{b3-6}) has the same rate as in (\ref{b3-7}). Consider the last term on the right side of (\ref{b3-6})
, first use Lemma A.1(ii) inequality in \cite{caner2022} for the first inequality below
\begin{eqnarray}
\| (\hat{\b B} - \b B) [ \frac{\sum_{t=1}^T \b f_t \b f_t' }{T}] (\hat{\b B} - \b B)'\|_{\infty} & \le & 
K^2 \| (\hat{\b B} - \b B)\|_{\infty}^2 \| \frac{1}{T} \sum_{t=1}^T \b f_t \b f_t' \|_{\infty} \nonumber \\
& = & K^2 O_p ( K \frac{ln p}{T}) O_p(1) = O_p ( K^3 \frac{ln p}{T}),\label{b3-8}
\end{eqnarray}
and the rates are via the proof of (A.70) of \cite{caner2022}, Lemma A.3 of \cite{caner2022}, and by Assumption 4 here.So by (\ref{b3-7})(\ref{b3-8})
\[ \| \hat{\b \Sigma}_u - \bar{\b \Sigma}_u \|_{\infty} = O_p ( K^3 \frac{ln p}{T}).\]
But by Lemma A.3 (i) of \cite{caner2022} 
\[ \| \bar{\b \Sigma}_u - \b \Sigma_u \|_{\infty} = O_p ( \sqrt{ln p/T}).\]
Hence since $K^3 \sqrt{lnp/T} \to 0$ by Assumption 7 here, we have the rate
\begin{equation}
\| \hat{\b \Sigma}_u - \b \Sigma_u \|_{\infty} = O_p ( \sqrt{lnp/T}).\label{b3-9}
\end{equation}
Now in (\ref{b3-4}), take the first right side term and by adding and subtracting and triangle inequality
\begin{eqnarray}
\| \hat{\b B} \hat{\b \Sigma}_f \hat{\b B}' - \b B \b \Sigma_f \b B' \|_{\infty} 
& = & \| (\hat{\b B} - \b B + \b B) [\hat{\b \Sigma}_f - \b \Sigma_f + \b \Sigma_f] (\hat{\b B}- \b B + \b B)' - \b B \b \Sigma_f \b B' \|_{\infty} \nonumber \\
& \le & 
2 \| (\hat{\b B} - \b B ) \b \Sigma_f \b B' \|_{\infty} + 
\| \b B (\hat{\b \Sigma}_f  - \b \Sigma_f ) \b B' \|_{\infty} \nonumber \\
& + & \| (\hat{\b B} - \b B) \b \Sigma_f (\hat{\b B} - \b B)' \|_{\infty} + 
2 \| \b B (\hat{\b \Sigma}_f - \b \Sigma_f ) (\hat{\b B} - \b B)' \|_{\infty} \nonumber \\
& + & \| (\hat{\b B} - \b B) (\hat{\b \Sigma}_f - \b \Sigma_f) (\hat{\b B} - \b B)' \|_{\infty} \nonumber \\
& = & O_p ( K \sqrt{lnp/T}) + O_p ( K^2 \sqrt{ln T/T}) \nonumber \\
& + & O_p ( K^2 ln p/T) + o_p ( K^2 \sqrt{ln T/T}) + o_p (K^2 \sqrt{ln T/T}) \nonumber \\
& = & O_p (max (K \sqrt{lnp/T}, K^2 \sqrt{ln T/T})) = o_p (1),\label{b3-10} 
\end{eqnarray}
where the rates are from (B.13)-(B.18) of \cite{fan2011}. Clearly the rate in (\ref{b3-4})  is
by (\ref{b3-9})(\ref{b3-10})
\begin{equation}
\| \hat{\b \Sigma}_y - \b \Sigma_y \|_{\infty} = O_p (max (K \sqrt{lnp/T}, K^2 \sqrt{ln T/T})) = o_p (1), \label{b3-11}
\end{equation}
where the last equality is by Assumption 7. Next the second term is 
\begin{equation}
\| \hat{\b w}_{MSR} - \hat{\b w}_{GMV} - (\b w_{MSR}^* - \b w_{GMV}^*) \|_1 = O_p (\bar{s} l_T K^{3/2}),\label{b3-12}
\end{equation}
where we use the proof of Theorem 1(i) in (A.2) (the square bracketed term). Next the third term that we are interested is 
$ \| \b \Sigma_y \|_{\infty} = O (K^2)$ by (A.32) in the proof. Next we consider 
\begin{equation}
\| \b w_{MSR}^* - \b w_{GMV}^* \|_1 = O (K^{1/2}),\label{b3-13}
\end{equation}
by (A.34)-(A.38), since the term on the right side of (A.34) (without $\kappa$)  is the left side term in (\ref{b3-13}). Now  we use all (\ref{b3-9})-(\ref{b3-13}) in the inequality  in (\ref{b3-3})
for the following term 
\begin{eqnarray*}
& & | (\hat{\b w}_{MSR} - \hat{\b w}_{GMV})' \hat{\b \Sigma}_y (\hat{\b w}_{MSR} - \hat{\b w}_{GMV})  - 
(\b w_{MSR}^* - \b w_{GMV}^* )' \b \Sigma_y (\b w_{MSR}^* - \b w_{GMV}^*)|\nonumber \\
& \le& 
\| (\hat{\b w}_{MSR} - \hat{\b w}_{GMV}) - (\b w_{MSR}^* - \b w_{GMV}^*) \|_1^2
\| \hat{\b \Sigma}_y - \b \Sigma_y \|_{\infty} \nonumber \\
& + & 2 \| (\hat{\b w}_{MSR} - \hat{\b w}_{GMV}) - (\b w_{MSR}^* - \b w_{GMV}^*) \|_1 
\| \b w_{MSR}^* - \b w_{GMV}^* \|_1
\| \hat{\b \Sigma}_y - \b \Sigma_y \|_{\infty} \nonumber \\
& + & \| \b w_{MSR}^* - \b w_{GMV}^* \|_1^2 \| \hat{\b \Sigma}_y - \b \Sigma_y \|_{\infty} \nonumber \\
& + & \| (\hat{\b w}_{MSR} - \b w_{MSR}^* ) - (\hat{\b w}_{GMV}- \b w_{GMV}^*) \|_1^2 \| \b \Sigma_y \|_{\infty} \nonumber \\
& + & 2 \| \b w_{MSR}^* - \b w_{GMV}^* \|_1
\| (\hat{\b w}_{MSR} - \hat{\b w}_{GMV}) - (\b w_{MSR}^* - \b w_{GMV}^*) \|_1 \| \b \Sigma_y \|_{\infty} \nonumber \\
& = & O_p (\bar{s}^2 l_T^2 K^3) O_p (\max (K \sqrt{lnp/T}, K^2 \sqrt{ln T/T})) \nonumber \\ 
&+& O_p ( \bar{s} l_T K^{3/2}) O (K^{1/2}) O_p (\max (K \sqrt{ln p/T}, K^2 \sqrt{ln T/T})) \nonumber \\
& + & O (K) O_p (\max (K \sqrt{ln p/T}, K^2 \sqrt{ln T/T})) \nonumber \\
& + & O_p ( \bar{s}^2 l_T^2 K^3) O (K^2) \nonumber \\
& + &  O (K^{1/2}) O_p (\bar{s} l_T K^{3/2}) O (K^2),\label{b3-14}
\end{eqnarray*}
and since there are many terms, we consider how each rate on the right side of (\ref{b3-14}). The first rate corresponds to the first term on the right side of the inequality and obtained by 
(\ref{b3-11})(\ref{b3-12}). The second right side term in the inequality has the rate by (\ref{b3-11})-(\ref{b3-13}) as the second rate in the equality. The third right side term in the inequality above, has the rate obtained by (\ref{b3-12})(\ref{b3-13}) which is the third rate in the equality. The fourth right side term in (\ref{b3-14}) has the rate obtained by (A.32) and (\ref{b3-12}) which is the fourth rate in the equality in (\ref{b3-14}). The fifth term on the right side of inequality in (\ref{b3-14}) has the rate obtained by (\ref{b3-12})(\ref{b3-13}) and (A.32) which is the fifth rate.
The fifth rate is the rate that converges to the zero slowest, by $l_T$ definition in (2). To see this, by Assumption A.7  $\bar{s} l_T K^4 = o(1)$, and hence we have (\ref{b3-11}) which is the asymptotically negligible rate. So we have 
\begin{eqnarray}
| (\hat{\b w}_{MSR} - \hat{\b w}_{GMV})' \hat{\b \Sigma}_y (\hat{\b w}_{MSR} - \hat{\b w}_{GMV}) & -& 
(\b w_{MSR}^* - \b w_{GMV}^* )' \b \Sigma_y (\b w_{MSR}^* - \b w_{GMV}^*)| \nonumber \\
&=& O_p (\bar{s} l_T K^4) = o_p (1).\label{b3-15}
\end{eqnarray}

Now write $\hat{\kappa}:= \frac{TE}{\hat{z}}$, and $\kappa:= \frac{TE}{z}$ where we use Section A.3, and (\ref{kappa1})(\ref{kappa2}) by defining 
\[ \hat{z}:= (\hat{\b w}_{MSR} - \hat{\b w}_{GMV})' \hat{\b \Sigma}_y (\hat{\b w}_{MSR} - \hat{\b w}_{GMV}),\]
and 
\[ z:= (\b w_{MSR}^* - \b w_{GMV}^* )' \b \Sigma_y (\b w_{MSR}^* - \b w_{GMV}^*).\]

Clearly
\begin{eqnarray*}
|\hat{\kappa} - \kappa| = \left| \frac{TE}{\hat{z}} - \frac{TE}{z} 
\right|   & \le & \frac{ (TE) z - (TE) \hat{z}}{| \hat{z} | | z | } \nonumber \\
& \le & \frac{ TE | \hat{z} - z|}{z^2 - | z| | \hat{z} - z |} = O_p ( \bar{s} l_T K^4) = o_p (1), 
\end{eqnarray*}
with (\ref{b3-15}) with $z, \hat{z}$ definitions and also 
\[ z \ge Eigmin ( \b \Sigma_y) \| \b w_{MSR}^* - \b w_{GMV}^* \|_2^2 \ge c > 0,\]
by $Eigmin (\b \Sigma_y) \ge c > 0$ by Assumptions 1 and 4, and $\b w_{MSR}^* \neq \b w_{GMV}^*$.
{\bf Q.E.D.}

\section{Discussion on Rates for Different Restrictions}

In this part of the Appendix B, we discuss in detail, why the rate of convergence to the targets are the same under different restrictions. The issue is that why the rates in Theorems 1-3, and Lemma A.3 and Lemma A.4 are the same. Note that Theorem 1 results depend on the rate of convergence of (unconstrained) maximum Sharpe Ratio portfolio weights, 
\begin{equation}
w_{MSR}^*:=
\frac{\b \Theta \b \mu }{\b 1_p' \b \Theta \b \mu},\label{wmsr}
\end{equation}

Going through the proof of Theorem 1(i), the rate of convergence corresponds to the rate in (A.29), which is the rate corresponding to $w_{MSR}^*$ estimation. It is much slower than the Global Minimum Variance (GMV) portfolio weights in (A.14). Hence, this is the main factor contributing to the rate in Tracking Error constraint. For Theorem 1(ii)(iii), the same logic applies.
To see why Theorem 2(i) (joint Tracking Error and Equality Weight case) rate is the same as in Theorem 1(i) (Tracking Error restriction only), we observe that optimal weight for joint Tracking Error and Equality Weight restriction has the optimal Tracking Error restricted weight $w_d^*$ in (7) and an extra part $(\omega - \kappa w_u) \b l$, which is a weighted difference between unrestricted and restricted GMV portfolio. The proofs of Theorem 2(i) clearly shows that in (A.97)-(A.100), the rate is from (A.99) and this is the same rate as in (A.29) as in Theorem 1(i). So the added equality weight restriction did not change the rate of convergence. Theorems 2(ii)(iii) and 1(ii)(iii) rates are the same because of the same reason discussed above.
Theorem 3  explanation is the same.

In Lemma A.4, for the unconstrained portfolio weight scenario, the target weight is in Proposition A.2 which is 
\[ \b w_n^* = \kappa_w ( \b w_{MSR}^* - \b a ) + \b a ,\]
where we use (\ref{wmsr}) and $\b a $ is the GMV portfolio weights (unconstrained). Note that the rate for estimation error  of $\b w_{MSR}^*$  is the first right side rate of (A.144) and it is 
$O_p ( \bar{s} l_T K^{3/2})$  and the second rate in (A.144) is for estimation error of $\b a$ and that rate is $O_p ( \bar{s} l_T)$. Hence the rate for $\b w_{MSR}^*$ converges to zero slower than the GMV portfolio estimation error rate.  So all rates are driven by MSR portfolio weights. For the only weight restriction going through the proofs  of Lemma A.3, and Proposition A.1, we have the same explanation as above.

\section{Simulations for Non-sparse Precision Matrix of Errors}
In this section we report the simulation results for non-sparse precision matrix of errors. Specifically, we consider the `dense' design for $\b \Sigma_u$ as follows (holding other parameters' DGP the same as in the Section 6 of the main text, Tables 1-2, specifically): for the $(i,j)$-th element of $\b \Sigma_u$, it has the form of $\frac{1}{(1+|i-j|)^\tau}$, where $\tau$ is a parameter that controls the rate of decay.  In the simulations, we set $\tau=0.001$, 0.0005, and 0.0001 for $p=80$, 160, and 320, respectively. Accordingly, the minimum elements of the precision matrix of errors  $\b \Omega$ are 0.0891, 0.0448, and 0.0564 for $p=80$, 160, and 320, respectively. In this way, we create dense precision matrices for errors that help us evaluate the robustness of CROWN under  conditions deviate from the theoretical assumptions.

Each Monte Carlo experiment is repeated 200 times. We report the same statistics as in the main text Section 6 and appendix Section A.1.

Our simulation results show that CROWN generally performs better than other methods, even in the dense precision matrix design. The sparsity assumption is to expedite the discussions in theoretical properties, but our proposed method also performs well in the non-sparse cases. We would leave the theoretical discussions of the non-sparse precision matrix of errors for future studies.

\clearpage

\begin{table}
    	\scriptsize
        \centering
        \caption{Simulations with TE constraint, p=80}
\begin{tabular}{lccccccccccc}		\noalign{\global\arrayrulewidth1pt}\hline \noalign{\global\arrayrulewidth0.4pt}
            & \multicolumn{5}{c}{p=80, T=40}            &  & \multicolumn{5}{c}{p=80, T=100}        \\
           \cline{2-6} \cline{8-12}
& \multicolumn{11}{c}{TE = 0.1} \\ \hline
               & TE	& Weight-ER	& Risk-ER & SR-ER &	SR & & TE	& Weight-ER	& Risk-ER & SR-ER &	SR\\
Oracle & 0.1000 &0.0000 &0.0000 &0.0000 &0.0292&   &0.1000 &0.0000 &0.0000 &0.0000 &0.0292\\
NCON & 0.6251&	3.3985&	0.3624&	0.2101&	0.0321 &   &0.6411 &3.4545 &0.3645 &0.2106 &0.0321\\
Index & -&0.7208&	0.0863&	0.0541&	0.0284&   &- &0.7208 &0.0863 &0.0541 &0.0284\\ \hline
NW &0.0967 &0.5810 &0.0119 &0.0074 &0.0291 &   &0.1080 &0.5912 &0.0133 &0.0056 & \textbf{0.0292}\\
CROWN &0.0961 & \textbf{0.3066} & \textbf{0.0009} & \textbf{0.0006} & \textbf{0.0292} &   &0.0991 & \textbf{0.2730} & \textbf{0.0007} & \textbf{0.0005} & \textbf{0.0292}\\
\end{tabular}
        \end{table}

\clearpage

\begin{table}
    	\scriptsize
        \centering
        \caption{Simulations with TE constraint, p=160}
        \begin{tabular}{lccccccccccc}		\noalign{\global\arrayrulewidth1pt}\hline \noalign{\global\arrayrulewidth0.4pt}
            & \multicolumn{5}{c}{p=160, T=80}            &  & \multicolumn{5}{c}{p=160, T=200}        \\
           \cline{2-6} \cline{8-12}
& \multicolumn{11}{c}{TE = 0.1} \\ \hline
               & TE	& Weight-ER	& Risk-ER & SR-ER &	SR & & TE	& Weight-ER	& Risk-ER & SR-ER &	SR\\
Oracle & 0.1000 &0.0000 &0.0000 &0.0000 &0.0293&   &0.1000 &0.0000 &0.0000 &0.0000 &0.0293\\
NCON & 0.6192 &3.5063 &0.3551 &0.1991 &0.0321&   & 0.6191 &3.5335 &0.3553 &0.1992 &0.0321\\
Index & - &0.7583 &0.0865 &0.0528 &0.0286&   &- &0.7583 &0.0865 &0.0528 &0.0286\\ \hline
NW & 0.0994 &0.6297 &0.0088 &0.0070 &0.0292&   &0.1128 &0.6656 &0.0150 &0.0038 & \textbf{0.0293}\\
CROWN & 0.0984 & \textbf{0.3440} & \textbf{0.0007} & \textbf{0.0005} & \textbf{0.0293}&   &0.0996 & \textbf{0.3117} & \textbf{0.0007} & \textbf{0.0005} & \textbf{0.0293}\\
\end{tabular}
        \end{table}

\clearpage

\begin{table}
    	\scriptsize
        \centering
        \caption{Simulations with TE constraint, p=320}
        \begin{tabular}{lccccccccccc}		\noalign{\global\arrayrulewidth1pt}\hline \noalign{\global\arrayrulewidth0.4pt}
            & \multicolumn{5}{c}{p=320, T=160}            &  & \multicolumn{5}{c}{p=320, T=400}        \\
           \cline{2-6} \cline{8-12}
& \multicolumn{11}{c}{TE = 0.1} \\ \hline
               & TE	& Weight-ER	& Risk-ER & SR-ER &	SR & & TE	& Weight-ER	& Risk-ER & SR-ER &	SR\\
Oracle &0.1000 &0.0000 &0.0000 &0.0000 &0.0294&   &0.1000 &0.0000 &0.0000 &0.0000 &0.0294\\
NCON &0.6231 &3.6331 &0.3498 &0.1984 &0.0321&   &0.6279 &3.6456 &0.3502 &0.1985 &0.0321\\
Index & - &0.8129 &0.0863 &0.0526 &0.0286&   & - &0.8129 &0.0863 &0.0526 &0.0286\\ \hline
NW &0.1047 &0.7095 &0.0087 &0.0035 &\textbf{0.0293} &   &0.1132 &0.7478 &0.0143 &0.0029 &\textbf{0.0294}\\
CROWN &0.0990 &\textbf{0.4435} &\textbf{0.0005} &\textbf{0.0003} &\textbf{0.0293} &   &0.0999 &\textbf{0.4350} &\textbf{0.0005} &\textbf{0.0003} &0.0293\\
\end{tabular}
        \end{table}

\clearpage

\begin{table}
    	\scriptsize
        \centering
        \caption{Simulation results for weight constraint, p=80}
        \begin{tabular}{lccccccccccc}		\noalign{\global\arrayrulewidth1pt}\hline \noalign{\global\arrayrulewidth0.4pt}
            & \multicolumn{5}{c}{p=80, T=40}            &  & \multicolumn{5}{c}{p=80, T=100}        \\
           \hline
& TE	& Weight-ER	& Risk-ER & SR-ER &	SR & & TE	& Weight-ER	& Risk-ER & SR-ER &	SR \\
Oracle & 0.4164 &0.0000 &0.0000 &0.0000 &0.0303&   &0.4214 &0.0000 &0.0000 &0.0000 &0.0303\\
NCON &0.6251 &2.5902 &0.3763 &0.1268 &0.0321&   &0.6411 &2.5273 &0.3784 &0.1273 &0.0321\\
Index & - &2.4347 &0.0770 &0.1192 &0.0284&   & - &2.4347 &0.0770 &0.1192 &0.0284\\ \hline
NW &\textbf{0.0592} &2.5603 &0.1042 &0.1252 &0.0283&   &\textbf{0.1582} &2.1374 &0.0646 &0.0413 &0.0296\\
CROWN &0.4139 &\textbf{1.2575} &\textbf{0.0024} &\textbf{0.0012} &\textbf{0.0302}&   &0.4196 &\textbf{1.1920} &\textbf{0.0017} &\textbf{0.0009} &\textbf{0.0303}\\
POET &1.9103 &34.7183 &0.3529 &0.2367 &0.0264&   &1.1985 &33.7661 &0.8045 &0.7369 &0.0150\\
NLS &1.0607 &52.9596 &0.3632 &0.5282 &0.0204&   &1.1244 &52.3220 &0.5452 &0.7622 &0.0144\\
SFNL &1.0958 &50.6301 &0.3990 &0.5912 &0.0188&   &1.1089 &52.8416 &0.5357 &0.7422 &0.0150   \\ \noalign{\global\arrayrulewidth1pt}\hline \noalign{\global\arrayrulewidth0.4pt}    
        \end{tabular}
        \end{table}

\begin{table}
    	\scriptsize
        \centering
        \caption{Simulation results for weight constraint, p=160}
        %
        \end{table}

\begin{table}
    	\scriptsize
        \centering
        \caption{Simulation results for weight constraint, p=320}
        %
        \end{table}

\clearpage

\begin{table}
    	\scriptsize
        \centering
        \caption{Simulations with TE and weight constraint, p=80}
        %
        \end{table}

\clearpage
        
\begin{table}
    	\scriptsize
        \centering
        \caption{Simulations with TE and weight constraint, p=160}
        %
        \end{table}

\clearpage

\begin{table}
    	\scriptsize
        \centering
        \caption{Simulations with TE and weight constraint, p=320}
        %
        \end{table}

\clearpage
       
\begin{table}
    	\scriptsize
        \centering
        \caption{Simulations with TE constraint and inequality constraint, p=80}
        %
        \end{table}

\clearpage
   
\begin{table}
    	\scriptsize
        \centering
        \caption{Simulations with TE constraint and inequality constraint, p=160}
        %
        \end{table}

\clearpage

\begin{table}
    	\scriptsize
        \centering
        \caption{Simulations with TE constraint and inequality constraint, p=320}
        %
        \end{table}

\clearpage

\begin{table}
	\scriptsize
    \centering
    \caption{Simulation Average: TE constraint, p=80}
    %
\end{table}

\clearpage

\begin{table}
\scriptsize
    \centering
    \caption{Simulation Average: TE constraint, p=160}
    %
        \end{table}

\clearpage

\begin{table}
	\scriptsize
    \centering
    \caption{Simulation Average: TE constraint, p=320}
    %
        \end{table}

\clearpage

\begin{table}
    	\scriptsize
        \centering
        \caption{Simulation results for weight constraint, p=80}
        %
        \end{table}

\begin{table}
    	\scriptsize
        \centering
        \caption{Simulation results for weight constraint, p=160}
        %
        \end{table}

\begin{table}
    	\scriptsize
        \centering
        \caption{Simulation results for weight constraint, p=320}
        %
        \end{table}

\clearpage
        
\begin{table}
    	\scriptsize
        \centering
        \caption{Simulation Average: TE and weight constraints, p=80}
        %
        \end{table}

\clearpage

\begin{table}
    	\scriptsize
        \centering
        \caption{Simulation Average: TE and weight constraints, p=160}
        %
        \end{table}

\clearpage

\begin{table}
    	\scriptsize
        \centering
        \caption{Simulation Average: TE and weight constraints, p=320}
        %
        \end{table}

\clearpage

\begin{table}
    	\scriptsize
        \centering
        \caption{Simulation Average: TE and inequality weight constraint, p=80}
        %
        \end{table}

\clearpage

\begin{table}
    	\scriptsize
        \centering
        \caption{Simulation Average: TE and inequality weight constraint, p=160}
        %
        \end{table}

\clearpage

\begin{table}
    	\scriptsize
        \centering
        \caption{Simulation Average: TE and inequality weight constraint, p=320}
        %
        \end{table}

 \clearpage

\bibliographystyle{chicagoa}
\bibliography{cp}
\end{document}